\documentclass{aa}  

\usepackage{graphicx}
\usepackage{txfonts}
\usepackage{lscape}
\usepackage{color}
\usepackage{ulem}

\newcommand{\rasecp}{\mbox{\rlap{.}$^{\rm s}$}} 
\newcommand{\tablenotea}[1]{\parbox{8.8cm}{\indent \footnotesize{#1}}}
\newcommand{\tablenoteb}[1]{\parbox{18.4cm}{\indent \footnotesize{#1}}}

\newcommand{\fdis}{Faraday Discuss.}
\newcommand{\ieee}{IEEE Trans. Antennas Propag.}
\newcommand{\jms}{J. Mol. Spectr.}
\newcommand{\jmst}{J. Mol. Struct.}

\newcommand{\chemrev}{Chem. Rev.}

\newcommand{\rpp}{Rep. Prog. Phys.}
\newcommand{\nature}{Nature}
\newcommand{\znata}{Z. Naturforsch. A}

\begin{document}

\title{A sensitive $\lambda$\,3 mm line survey of L483\thanks{Based on observations carried out with the IRAM 30m Telescope. IRAM is supported by INSU/CNRS (France), MPG (Germany) and IGN (Spain).}}

\subtitle{A broad view of the chemical composition of a core around a Class\,0 object}

\titlerunning{A $\lambda$\,3 mm line survey of L483}
\authorrunning{Ag\'undez et al.}

\author{
M.~Ag\'undez\inst{1},
N.~Marcelino\inst{1},
J.~Cernicharo\inst{1},
E.~Roueff\inst{2}, \and
M.~Tafalla\inst{3}
}

\institute{
Instituto de F\'isica Fundamental, CSIC, C/ Serrano 123, E-28006 Madrid, Spain \and
Sorbonne Universit\'e, Observatoire de Paris, Universit\'e PSL, CNRS, LERMA, F-92190 Meudon, France \and
Observatorio Astron\'omico Nacional, C/ Alfonso XII 3, E-28014 Madrid, Spain
}

\date{Received; accepted}

 
\abstract
{An exhaustive chemical characterization of dense cores is mandatory to our understanding of chemical composition changes from a starless to a protostellar stage. However, only a few sources have had their molecular composition characterized in detail. Here we present a $\lambda$ 3 mm line survey of L483, a dense core around a Class\,0 protostar, which was observed with the IRAM 30m telescope in the 80-116 GHz frequency range. We detected 71 molecules (140 including different isotopologs), most of which are present in the cold and quiescent ambient cloud according to their narrow lines (FWHM $\sim$0.5 km s$^{-1}$) and low rotational temperatures ($\lesssim$10 K). Of particular interest among the detected molecules are the $cis$ isomer of HCOOH, the complex organic molecules HCOOCH$_3$, CH$_3$OCH$_3$, and C$_2$H$_5$OH, a wide variety of carbon chains, nitrogen oxides like N$_2$O, and saturated molecules like CH$_3$SH, in addition to eight new interstellar molecules (HCCO, HCS, HSC, NCCNH$^+$, CNCN, NCO, H$_2$NCO$^+$, and NS$^+$) whose detection has
already been reported. In general, fractional molecular abundances in L483 are systematically lower than in TMC-1 (especially for carbon chains), tend to be higher than in L1544 and B1-b, and are similar to those in L1527. Apart from the overabundance of carbon chains in TMC-1, we find that L483 does not have a marked chemical differentiation with respect to starless/prestellar cores like TMC-1 and L1544, although it does chemically differentiate from Class\,0 hot corino sources like IRAS\,16293$-$2422. This fact suggests that the chemical composition of the ambient cloud of some Class\,0 sources could be largely inherited from the dark cloud starless/prestellar phase. We explore the use of potential chemical evolutionary indicators, such as the HNCO/C$_3$S, SO$_2$/C$_2$S, and CH$_3$SH/C$_2$S ratios, to trace the prestellar/protostellar transition. We also derived isotopic ratios for a variety of molecules, many of which show isotopic ratios close to the  values for the local interstellar medium (remarkably all those involving $^{34}$S and $^{33}$S), while there are also several isotopic anomalies like an extreme depletion in $^{13}$C for one of the two isotopologs of $c$-C$_3$H$_2$, a drastic enrichment in $^{18}$O for SO and HNCO (SO being also largely enriched in $^{17}$O), and different abundances for the two $^{13}$C substituted species of C$_2$H and the two $^{15}$N substituted species of N$_2$H$^+$. We report the first detection in space of some minor isotopologs like $c$-C$_3$D. The exhaustive chemical characterization of L483 presented here, together with similar studies of other prestellar and protostellar sources, should allow us to identify the main factors that regulate the chemical composition of cores along the process of formation of low-mass protostars.}


\maketitle

\section{Introduction}

The process by which the chemical composition of a dense core changes as it evolves from a starless to a protostellar stage is not yet well understood (e.g., \citealt{Aikawa2013,Ceccarelli2017,Lefloch2018}). In particular, we do not know whether the prestellar/protostellar transition drives a clear change in the chemical composition of the host ambient cloud or whether there is no such chemical differentiation between prestellar and protostellar cores. Furthermore, it is unsure whether chemical variations from source to source arise from the different evolutionary status or are rather dominated by environmental characteristics and/or the particular history of each source.

It has been long thought that unsaturated carbon chains are associated to early starless evolutionary stages, while saturated complex organic molecules become dominant in more evolved protostellar phases \citep{Suzuki1992,Herbst2009}. In recent years however, radioastronomical observations have revealed unexpected results that challenge our understanding of the chemistry at the earliest stages of low-mass star formation. On the one hand, abundant carbon chains have been observed in various dense cores around protostars \citep{Sakai2008a,Sakai2009a,Agundez2008,Gupta2009,Hirota2009,Cordiner2013,Graninger2016a,Lindberg2016,Law2018}, and on the other, complex organic molecules such as methyl formate and dimethyl ether have been detected toward prestellar cores \citep{Bacmann2012,Jimenez-Serra2016}. These observational results are at the heart of a profound revision of the chemistry at the earliest stages of low-mass star formation that is still underway (e.g., \citealt{Aikawa2008,Aikawa2013,Hassel2008,Ruaud2015,Balucani2015,Kalvans2015,Hincelin2016,Vasyunin2017,Shingledecker2018}).

To shed light on the scenario that governs the chemical composition of dense cores during low-mass star formation it is mandatory to increase the number of prestellar and protostellar sources for which the chemical composition (i.e., the inventory of molecules and their abundances) is known in detail. Line surveys at millimeter wavelengths are an ideal tool for this purpose, since they allow for all the molecular lines lying within a relatively broad frequency range to be observed in a homogeneous and unbiased way (e.g., \citealt{Lefloch2018}).

Here we present a sensitive $\lambda$ 3 mm line survey carried out with the IRAM 30m telescope toward L483, a dark cloud core which contains a Class\,0 low-mass protostar that powers a bipolar outflow. The lack of important abundance enhancements of CH$_3$OH and SiO in the outflow, together with the presence of an infrared (IR) reflection nebula coincident with the outflow suggests that L483 may be in transition to Class\,I \citep{Tafalla2000}. However, the relatively high degree of collimation of the outflow and the fact that the protostar is still deeply embedded in the parental cloud may indicate that it is not a particularly evolved object within the Class\,0 phase. Apart from these peculiarities, there are various interesting chemical features. First, L483 is, together with a few other dense cores like L1527, one of the low-mass protostellar sources where brighter emission from carbon chains such as C$_4$H has been observed \citep{Agundez2008,Sakai2009a}. Second, observations with ALMA have recently unveiled the existence of a hot corino (i.e., warm complex organic molecules concentrated around the protostar) in this source \citep{Oya2017,Jacobsen2018}. And third, single-dish observations indicate that L483 has an exceptionally rich chemical composition, as indicated by the recent discoveries of various new interstellar molecules: the radical HCCO \citep{Agundez2015a}, the protonated form of cyanogen (NCCN) and its metastable isomer CNCN \citep{Agundez2015b,Agundez2018b}, the ion NS$^+$ \citep{Cernicharo2018}, the radical NCO and the ion H$_2$NCO$^+$, which are precursors of isocyanid acid and its isomers \citep{Marcelino2018a}, and the S-bearing radicals HCS and HSC \citep{Agundez2018a}. These chemical features make L483 an ideal target to carry out a line survey to study its chemical composition and evaluate a possible connection with evolutionary status.

\section{The source}

The source L483 is an optical dark cloud core located in the Aquila Rift star-forming region \citep{Lee1999}, which is estimated to lie at a distance of 200 pc \citep{Dame1985}. The core hosts an embedded infrared source, IRAS\,18148$-$0440, which is classified as a Class\,0 object based on its low bolometric temperature of 46-56 K \citep{Fuller1995}. The core is in a state of gravitational collapse, as evidenced by the line profiles of molecules like H$_2$CO and CS, which show the typical signature of infall motions, that is, enhanced self-absorption at redshifted velocities \citep{Myers1995,Mardones1997,Tafalla2000}. Interferometric observations of molecules like CS show hints of rotation of the envelope at different spatial scales \citep{Jorgensen2004,Leung2016,Oya2017,Jacobsen2018}. One of the most remarkable components present in L483 is a collimated bipolar outflow, well traced by CO and HCO$^+$, which extends out to 50-100$''$ from the IRAS source \citep{Fuller1995,Hatchell1999,Tafalla2000,Park2000,Oya2018}. Neither CH$_3$OH nor SiO show important abundance enhancements in the outflow \citep{Tafalla2000}, unlike in other Class\,0 sources like L1448 \citep{Martin-Pintado1992,Jimenez-Serra2005} or L1157 \citep{Bachiller1997}. The source also shows H$_2$ emission and a near-IR reflection nebula which coincide with the bipolar outflow and probe shocked gas present in the walls of the cavity opened by the outflow \citep{Fuller1995,Velusamy2014}. In summary, the evolutionary status of L483 is that of a Class\,0 source in which the protostar has disrupted the ambient infalling envelope through a collimated bipolar outflow.

\begin{figure}
\centering
\includegraphics[angle=0,width=\columnwidth]{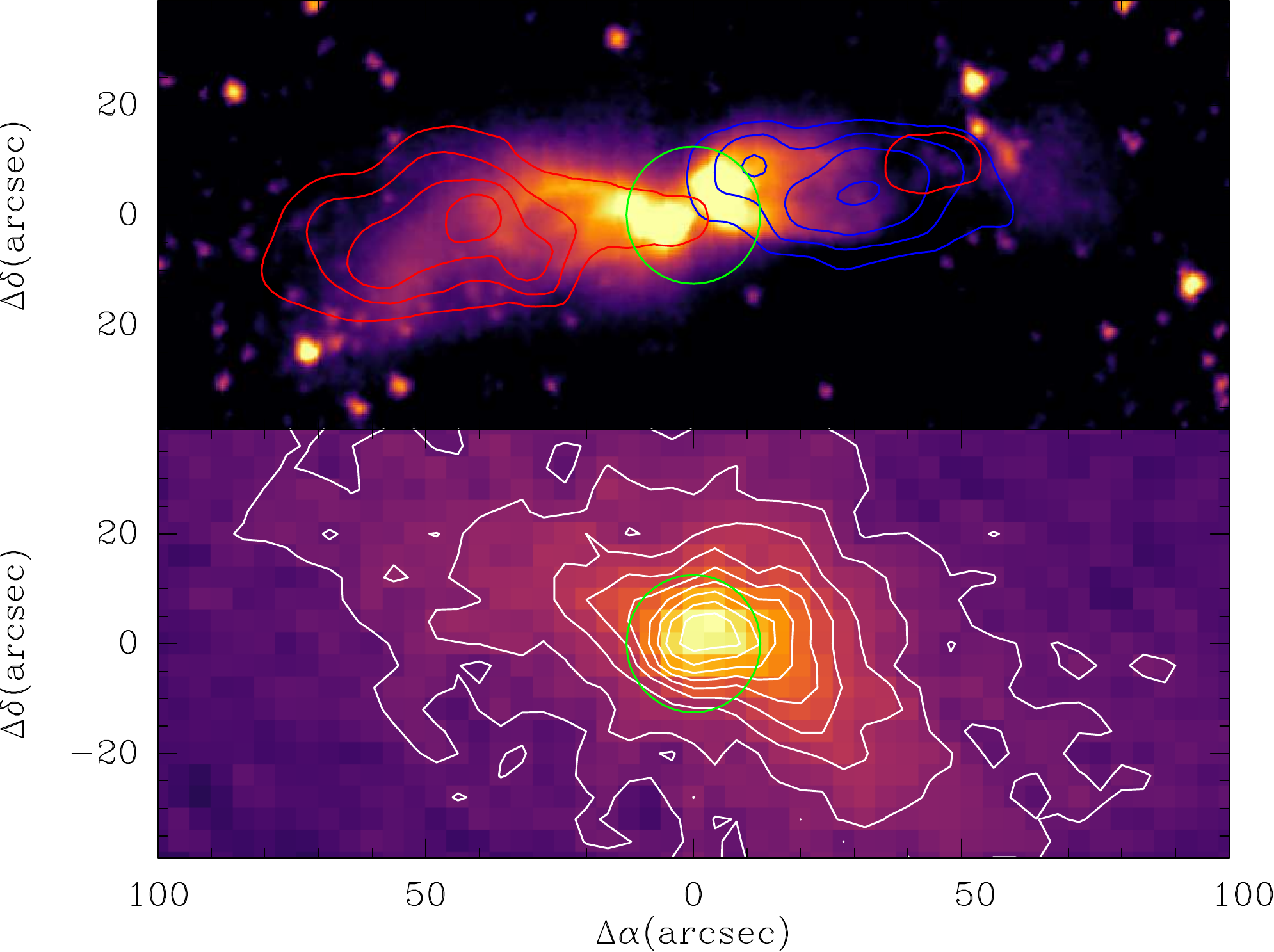}
\caption{\textit{Upper panel}: High-velocity CO $J$=2-1 emission (red/blue contours correspond to red/blue-shifted velocities; from \citealt{Tafalla2000}) is shown superimposed on the 3.6 $\mu$m \textit{Spitzer}/IRAC image. \textit{Lower panel}: Map of the $\lambda$ 1.3 mm dust continuum emission observed with MAMBO (Tafalla et al., unpublished data) is shown in both contours and color. First contour and contour interval are 20 mJy (11$''$-beam)$^{-1}$. The green circle in both panels corresponds to a size of 25$''$, representative of the region probed by the IRAM 30m telescope in the $\lambda$ 3 mm band.} \label{fig:image}
\end{figure}

Concerning the chemical properties of L483, we may summarize them as follows. The source shows abundant carbon chains distributed over the ambient cloud, as seen at low angular resolution ($\sim30''$) with single-dish telescopes \citep{Agundez2008,Sakai2009a}, and complex organic molecules concentrated around the protostar, as seen at sub-arcsecond angular resolution with ALMA \citep{Oya2017,Jacobsen2018}. The observations presented here, in which the beam size is 21-30$''$, cover three main components: the ambient cloud, the bipolar outflow, and the warm surroundings of the protostar. This latter component is severely diluted in the IRAM 30m beam at $\lambda$ 3 mm, while the outflow is only traced by the high-velocity emission of a few molecules like CO and HCO$^+$. Our observations therefore mainly probe the ambient quiescent cloud. We note that most of the mass in L483 is present in the ambient cloud rather than in the outflow. This is strongly suggested by the distribution of the continuum emission from dust (see lower panel in Fig.~\ref{fig:image}) and by the fact that the emission at high velocities only makes a small fraction of the total emission in the minor isotopologs of CO such as C$^{18}$O, which are less affected by optical depth effects (see Sect.~\ref{sec:line_profiles} for the $J$=1-0 line and \citealt{Tafalla2000} for the $J$=2-1 line). Therefore, for the great majority of molecules detected in this $\lambda$ 3 mm line survey, the observed emission arises exclusively from the ambient cloud.

\section{Observations}

\begin{figure*}
\centering
\includegraphics[angle=0,width=\textwidth]{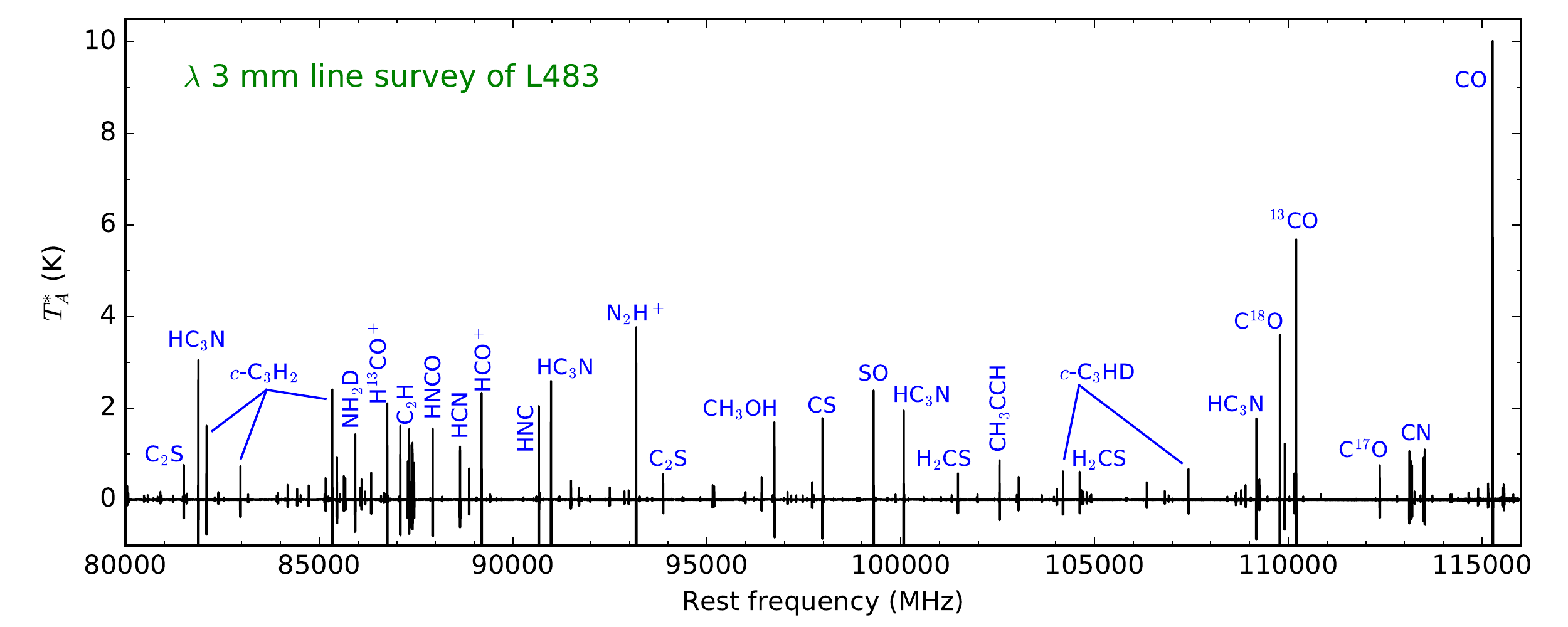}
\caption{Overview of the $\lambda$ 3 mm line survey of L483 covering the frequency range 80-116 GHz. The carriers of the most intense lines are indicated. Negative signals correspond to the artifacts caused by the frequency-switching technique.} \label{fig:overview}
\end{figure*}

The observations were carried out with the IRAM 30m telescope, located in Pico Veleta (Granada, Spain), in several sessions from 2016 August to 2018 April and consisted of a spectral scan of L483 in the $\lambda$ 3 mm band, from 80 to 116 GHz. The telescope was pointed to the position of the infrared source IRAS\,18148$-$0440 \citep{Fuller1993}, with coordinates $\alpha_{2000.0}$ = 18$^{\rm h}$17$^{\rm m}$29\rasecp8, $\delta_{2000.0}$ = $-04^{\circ}$39$'$38$''$. This position coincides with the maximum of intensity of CH$_3$OH emission \citep{Tafalla2000}.

We used the EMIR receiver E090 in dual sideband mode, with image rejections of 10-20 dB (depending on the local oscillator frequency and polarization). These values were measured using intense lines. The E090 receiver was connected to a Fast Fourier Transform Spectrometer (FTS), which was operated in its narrow mode, providing a spectral resolution of 50 kHz, which translates to velocity resolutions of 0.19 km\,s$^{-1}$ at 80 GHz and 0.13 km\,s$^{-1}$ at 116 GHz. This spectral resolution is good enough to partially resolve the lines in the spectrum of L483, most of which have line widths of the order of 0.5 km s$^{-1}$. Thanks to the versatility of EMIR and FTS it is possible to cover a bandwidth of $4\times1.8$ GHz in a single tuning setup, the four parts corresponding to the Lower Outer, Lower Inner, Upper Inner, and Upper Outer bands. This allowed us to completely cover the 80-116 GHz frequency range using six tuning setups.

We employed the frequency-switching technique, in which the telescope is most efficient since no off position is needed to subtract the signal from the sky. This results in a sensitivity improvement of a factor of $\sqrt{2}$ with respect to the wobbler- and position-switching observing modes. A frequency throw of 7.2 MHz was adopted to minimize the effect of standing waves on the spectral baselines. Still, the obtained spectra showed baseline ripples with typical periods of 15 MHz and amplitudes of 0.1 K. This fact does not create a problem to distinguish astronomical lines, which in L483's spectrum are narrow, although it can make it more difficult to distinguish weak lines present at a level of a few $\sigma$ compared to spectra obtained with the wobbler-switching technique, which show rather flat baselines.

The intensity scale is calibrated using two absorbers at different temperatures and the atmospheric transmission model ATM \citep{Cernicharo1985,Pardo2001}. We express intensities in terms of $T_A^*$, the antenna temperature corrected for atmospheric absorption and for antenna ohmic and spillover losses. The uncertainty in $T_A^*$ due to calibration is estimated to be around 10 \%. To convert to main beam brightness temperature ($T_{mb}$) one has to divide $T_A^*$ by $B_{\rm eff}/F_{\rm eff}$, where $B_{\rm eff}$ = 0.863 $\exp{[-(\nu{\rm (GHz)}/361)^2]}$ and $F_{\rm eff}$ = 0.95\footnote{\texttt{http://www.iram.es/IRAMES/mainWiki/Iram30mEfficiencies}}. The telescope focus was checked on planets at the beginning of each session, which typically run for 3-6 h. The telescope pointing was regularly checked (every one and a half hours) by observing the nearby radio source 1741$-$038. Pointing errors were typically 2-3$''$, while the half power beam width (HPBW) of the IRAM 30m telescope ranges between 30$''$ at 80 GHz and 21$''$ at 116 GHz.

Weather conditions between the different observing sessions ranged from fairly good, with amounts of precipitable water vapor (PWV) of only 1-2 mm, to bad, with clouds and high amounts of water vapor. We nevertheless only included data obtained under reasonably good weather conditions, typically with PWV $<$ 10 mm. Average system temperatures were in the range 80-120 K for frequencies below 110 GHz and between 130 and 270 K at higher frequencies.

The data reduction was carried out with the program CLASS of the GILDAS software package\footnote{\texttt{http://www.iram.fr/IRAMFR/GILDAS}}. The raw data obtained at the telescope consist of frequency-switching spectra which are already folded. We found it necessary to go back to the unfolded spectra to verify a few suspicious signals which turned out to arise from spikes. After visual inspection of individual spectra, bad channels were removed and all spectra with the same frequency range, polarization, and observing date were averaged. Lines arising from the image side band were easily identified since data from the lower and upper side bands are available and also because in frequency-switching spectra image lines appear as negative signals once spectra have been folded. Telluric lines from stratospheric ozone were also easily identified due to their broad nature and thanks to the line list in the JPL Molecular Spectroscopy Catalogue\footnote{\texttt{https://spec.jpl.nasa.gov/}} \citep{Pickett1998}. Their presence however does not prevent from detecting overlapping astronomical lines, which are much narrower, although this was found to occur rarely. After spectra had been cleaned from image and telluric lines, all spectra with the same frequency range, independently of the polarization and observing date, were averaged to reduce the noise using the command {\small STITCH} of CLASS. Since spectra observed in different epochs are not perfectly aligned in frequency in the LSR frame, spectra need to be resampled to place all of them in the same rest frequency scale before averaging them. The action of resampling introduces a correlation in adjacent channels causing the noise to be slightly reduced. Lines also tend to be artificially broadened, especially the narrowest ones, typically by no more than 10 \%. The center frequency and area however remain unchanged. At this point of the data-reduction process we have $6\times4$ spectra which are 1.8 GHz wide, each covering a different frequency range.

We carried out the search for astronomical lines before subtracting a baseline from the data, with all spectra averaged to reduce the
noise as much as possible. Given the rather irregular baselines of the frequency-switching spectra, baseline subtraction is a delicate issue, and weak lines are better distinguished by visual inspection before a baseline has been subtracted. The assignment of the detected lines to known rotational transitions of molecules was done by checking the Cologne Database for Molecular Spectroscopy $\footnote{\texttt{https://cdms.astro.uni-koeln.de/}}$ \citep{Muller2005}, the JPL Molecular Spectroscopy Catalogue \citep{Pickett1998}, and the private catalogue of J. Cernicharo generated from the MADEX code\footnote{\texttt{https://nanocosmos.iff.csic.es/?page\_id=1619}} \citep{Cernicharo2012:madex}. 

Once the position of all astronomical lines was known, we subtracted a baseline from each of the $6\times4$ 1.8 GHz-wide spectra. Since polynomials are not adequate for such a wide frequency range, the baseline is generated through a procedure involving the smoothing of the channels that are free of lines. After baseline subtraction, the $6\times4$ spectra, some of which overlap in certain frequency ranges, were averaged again using the command {\small STITCH} to increase the sensitivity in the overlapping regions.

The observed line profiles were fitted to Gaussian functions. Some lines showed nonGaussian line shapes, mainly consisting of wings in lines that arise to some extent from the outflow or self-absorption in some optically thick lines. In these cases no attempt was made to fit a Gaussian and only the observed area and peak intensity were retrieved. An overview of the line survey with the most intense lines is shown in Figure~\ref{fig:overview}, while the whole data set is shown in Fig.~\ref{fig:spectrum} with the frequency and intensity scales chosen to permit the visualization of the weakest lines. The frequency range 80-116 GHz has been completely covered. The blank spaces that appear from time to time in Fig.~\ref{fig:spectrum} correspond to spectral regions that have been removed due to contamination with image side-band lines or telluric ozone lines. The negative artifacts produced by the frequency-switching technique, which consist of two negative signals located at $\pm$ 7.2 MHz of each line with half its intensity, have not been removed. As can be seen in Fig.~\ref{fig:spectrum}, the sensitivity achieved is very good, with $T_A^*$ rms noise levels of 1-3 mK per 50 kHz channel below 110 GHz, increasing up to noise levels of 10 mK at the high-frequency edge of 115-116 GHz.

\begin{figure*}
\centering
\includegraphics[angle=0,width=\textwidth]{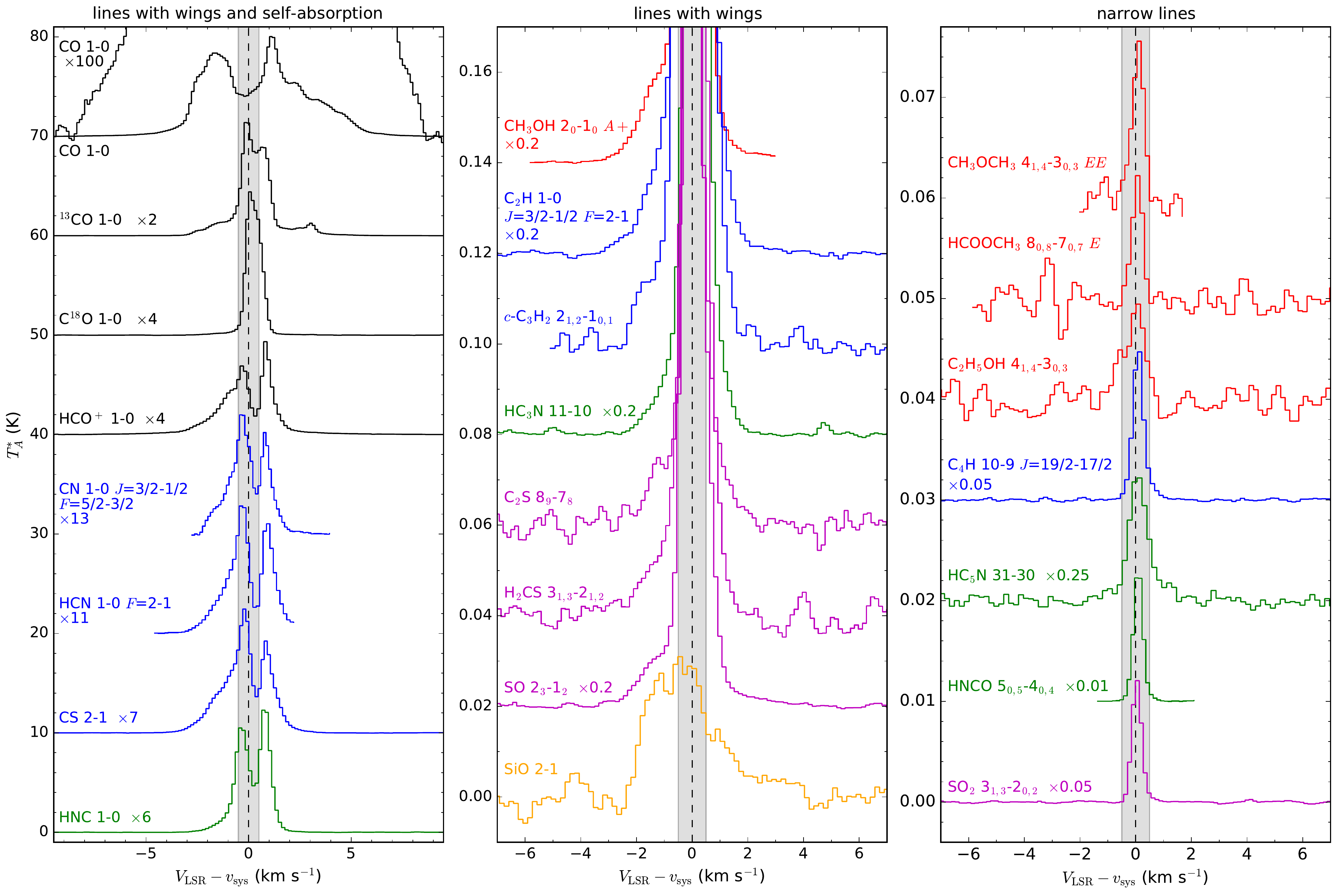}
\caption{Characteristic line profiles observed with the IRAM 30m telescope in the $\lambda$ 3 mm band toward L483. The left panel shows lines with wings and self-absorption (except C$^{18}$O $J$=1-0), the middle panel lines that show wings, and the right panel a sample of narrow lines, which form the majority in our line survey. The gray shaded area indicates the velocity range $\pm$ 0.5 km s$^{-1}$ around the systemic velocity of the source.} \label{fig:line_profiles}
\end{figure*}

\section{Results and discussion} \label{sec:results}

The $\lambda$ 3 mm line survey between 80 and 116 GHz resulted in the detection of 631 lines. Among them, 613 lines were assigned to known rotational transitions of molecules and are listed in Table~\ref{table:lines} together with their associated parameters\footnote{We use the $e$/$f$ labeling of parity levels for linear molecules with a $^2\Pi$ ground electronic state \citep{Brown1975}}. The total number of molecules detected is 71 (140 if different isotopologs are considered). These molecules can be grouped into the chemical families of O-bearing molecules, hydrocarbons, N-bearing molecules, S-containing species, and one Si-bearing molecule (SiO). There are also 18 lines which could not be assigned to known molecular transitions and therefore remain unidentified. Some of them have previously been observed in other sources; these are presented in Table~\ref{table:ulines}.

\subsection{Line profiles} \label{sec:line_profiles}

All the lines are observed in emission, at the exception of the CH$_3$OH transition 3$_1$-4$_0$ $A+$ lying at 107013.831 MHz, which is observed in absorption. There are several types of profiles among the emission lines detected. Some lines show more or less bright wings, which indicate that part of the emission arises from the outflow. Among them, a few optically thick lines show self-absorption at the line center. From the 71 molecules detected, only 14 (those shown in the left and middle panels of Fig.~\ref{fig:line_profiles}) show wings in their line profiles. The rest, that is the vast majority of the molecules detected (see a sample in the right panel of Fig.~\ref{fig:line_profiles}), show narrow line profiles, with the emission restricted to the velocity range $V_{\rm LSR}-v_{\rm sys}$ between $-0.5$ and $+0.5$ km s$^{-1}$ (gray shadowed area in Fig.~\ref{fig:line_profiles}). The emission of these narrow lines arises exclusively from the ambient quiescent cloud. From the lines of all these species for which we could fit a Gaussian function, we derive an average systemic velocity $v_{\rm sys} = 5.30 \pm 0.11$ km s$^{-1}$ and a mean line width FWHM = $0.52 \pm 0.15$ km s$^{-1}$. The systemic velocity is in good agreement with previously reported values. For example, \cite{Fuller1993} derive values of $v_{\rm sys}$ between 5.237 and 5.460 km s$^{-1}$ from high-spectral-resolution observations of HC$_3$N and NH$_3$.

The line that shows emission at the highest velocities is CO $J$=1-0, the emission of which extends up to $\sim$8 km s $^{-1}$ from the systemic velocity of the source on both the blue and red sides (see the top line in the left panel of Fig.~\ref{fig:line_profiles}). The CO $J$=1-0 line has a complex shape, very similar to that of the CO $J$=2-1 line observed with the IRAM 30m by \cite{Tafalla2000}, showing asymmetric wings, with more high-velocity emission in the red side, and self-absorption around the systemic velocity ($V_{\rm LSR}-v_{\rm sys}$ = $\pm$ 1 km s$^{-1}$) that results from cold CO gas in the outer parts of the envelope. The second line with the highest-velocity wings is HCO$^+$ $J$=1-0, for which emission extends up to $\sim$7 km s$^{-1}$ from the velocity of the ambient cloud. This is consistent with the fact that both CO and HCO$^+$ are good tracers of the outflow (e.g., \citealt{Fuller1995,Park2000}). We note however that as we move to less optically thick lines (e.g., CO $\rightarrow$ $^{13}$CO $\rightarrow$ C$^{18}$O; see left panel in Fig.~\ref{fig:line_profiles}) the contribution of the wings becomes vanishingly small, which indicates that the fraction of mass in the outflow is much lower than in the quiescent ambient cloud.

Lines from other molecules also show more or less bright wings extending out to velocities of 2-4 km s$^{-1}$. For example, the molecules HCN, HNC, CN, and CS show relatively bright wings (see left panel in Fig.~\ref{fig:line_profiles}), which indicates that part of the detected emission comes from the bipolar outflow. Other molecules like CH$_3$OH, C$_2$H, $c$-C$_3$H$_2$, HC$_3$N, C$_2$S, H$_2$CS, and SO display in some of their lines weak wings (see middle panel in Fig.~\ref{fig:line_profiles}), which indicates that most of the detected emission comes from the ambient cloud with a minor contribution from the outflow. The low abundance of methanol in the outflow is consistent with the map by \cite{Tafalla2000}, which indicates that CH$_3$OH does not trace the outflow but presents an extended distribution over the core. For SiO we detect weak emission from the $J$=2-1 line ($T_A^*$ $\sim$0.03 K, with a velocity-integrated intensity of 0.086 $\pm$ 0.008 K km s$^{-1}$, consistent with the upper limit of 0.1 K km s$^{-1}$ by \citealt{Tafalla2000}). The line however is wide (see bottom line in middle panel of Fig.~\ref{fig:line_profiles}), which indicates that most of the emission does not arise from the ambient cloud, but probably from the outflow. Recently, \cite{Oya2017} detected and imaged with ALMA at subarcsecond resolution the $J$=6-5 line of SiO. Emission in this line does not extend along the outflow but shows a compact distribution ($<$1$''$) around the protostar, which in part can be explained by the high excitation requirements of the line ($E_{\rm up}$ = 43.8 K) and the lack of short spacings in the ALMA data, which causes any extended emission to be filtered out. It is therefore unclear whether SiO is mostly present in the outflow or around the protostar. In any case, it is clear that neither SiO nor CH$_3$OH reaches high abundances in the L483 outflow, unlike in other Class\,0-powered outflows like L1157 or L1448.

Apart from the wings, some of the brightest lines show evidence of self-absorption near the cloud systemic velocity (see left panel in Fig.~\ref{fig:line_profiles}). In the lines of the dense gas tracers, the self-absorption feature lies at higher velocity than the systemic component, suggesting that the self-absorption is caused by a foreground layer of gas that is moving toward the cloud. This interpretation is in agreement with evidence for infall motions previously reported from H$_2$CO and CS observations \citep{Myers1995,Mardones1997,Tafalla2000}. Our data however show that the lines of CN, HCN, and CS present a brighter blue peak, while the lines of HCO$^+$ and HNC have a brighter red peak. This variety of profile shapes may be caused by the varying contribution of the outflow wings, or may indicate that the line-of-sight motions in the cloud are more complex than those expected for a simple collapse pattern. Similar line-shape differences between dense gas tracers have previously been seen toward other star-forming regions \citep{Lee2004}, meaning that they are not unique to L483.

\subsection{Column densities and temperatures} \label{sec:column_densities}

From the observed lines we derived beam-averaged column densities assuming local thermodynamic equilibrium (LTE). For some species, only lower limits to the column density could be obtained because the observed lines are optically thick. In general, we adopted the same rotational temperature for all the isotopologs of a given molecule to ensure that isotopic ratios are not affected by different rotational temperatures. When a precise determination of the rotational temperature is possible for various isotopologs (because a high-enough number of lines spanning a wide-enough range of upper level energies is available), we adopt the value that is most accurately derived. If the rotational temperature cannot be precisely determined for any isotopolog, we adopt the value derived for a chemically related species; for example, for C$_5$H we adopt the rotational temperature derived for C$_4$H. When there was no obvious choice for the rotational temperature, we adopted a value of 10 K, which is close to the gas kinetic temperature in L483 (see below).

\begin{table}
\caption{Beam-averaged column densities and rotational temperatures} \label{table:column_densities}
\small
\centering
\begin{tabular}{lrrcr}
\hline \hline
\multicolumn{1}{l}{Species} & \multicolumn{1}{c}{$N$ (cm$^{-2}$)$^a$} & \multicolumn{1}{c}{$T_{\rm rot}$ (K)} & \multicolumn{1}{c}{$E_{\rm up}$ (K)} & \multicolumn{1}{c}{$N_{\rm lines}$$^b$} \\
\hline
\multicolumn{5}{c}{O-bearing molecules} \\
\hline
CO                                  & $>$$6.2\times10^{16}$    & 10$^c$          & 5.5 - 5.5        & 1 \\
$^{13}$CO                      & $>$$1.4\times10^{16}$    & 10$^c$          & 5.3 - 5.3        & 1 \\
C$^{18}$O                      & $>$$4.8\times10^{15}$    & 10$^c$          & 5.3 - 5.3        & 1 \\ 
C$^{17}$O                      & $1.7\times10^{15}$          & 10$^c$          & 5.4 - 5.4        & 2 \\ 
$^{13}$C$^{18}$O          & $1.3\times10^{14}$          & 10$^c$          & 5.0 - 5.0        & 1 \\ 
$^{13}$C$^{17}$O          & $4.1\times10^{13}$          & 10$^c$          & 5.1 - 5.1        & 2 \\ 
C$_2$O                                  & $5.4\times10^{11}$           & 10$^c$           & 11.0 - 11.5    & 2 \\
C$_3$O                                  & $7.6(49)\times10^{11}$     & 10.2(21)         & 20.8 - 36.0   & 4 \\
HCO                                       & $2.8\times10^{12}$           & 10$^c$          & 4.2 - 4.2       & 3 / 4 \\ 
HCCO                              & $3.1\times10^{11}$           & 10$^c$           & 10.4 - 10.4  & 4 \\
CH$_3$O                        & $8.0\times10^{11}$           & 10$^c$           & 4.0 - 4.0      & 3 / 4 \\ 
D$_2$CO                        & $1.4\times10^{12}$           & 10$^c$           & 5.3 - 5.3      & 1 \\ 
%
CH$_3$OH$^d$           &                                         &                      & 4.6 - 32.5       & 9 \\ 
CH$_2$DOH               & $5.5(18)\times10^{12}$   & 4.3(3)            & 6.4 - 25.8       & 7 / 8 \\
CH$_3$OD                 & $4.0\times10^{12}$          & 4.3$^c$         & 6.5 - 6.5         & 1 \\ 
CHD$_2$OH               & $8.2\times10^{11}$          & 4.3$^c$         & 6.0 - 6.0         & 1 \\ 
$^{13}$CH$_3$OH     & $4.3\times10^{12}$          & 4.3$^c$         & 4.5 - 6.8         & 2 \\ 
H$_2$CCO                      & $4.1\times10^{12}$           & 10$^c$           & 8.6 - 14.5   & 6 \\ 
HDCCO                           & $3.7\times10^{11}$           & 10$^c$           & 13.5 - 19.0  & 2 \\ 
CH$_3$CHO                    & $3.8(10)\times10^{12}$    & 5.8(5)             & 4.9 - 23.0    & 18 \\ 
HCCCHO                         & $8.3(17)\times10^{11}$     & 7.6(5)             & 4.4 - 27.2    & 9 / 10 \\
$c$-C$_3$H$_2$O          & $2.3(13)\times10^{11}$     & 7.9(13)           & 14.0 - 28.0  & 11 \\ 
$t$-HCOOH             & $1.2(8)\times10^{12}$       & 7.0(17)          & 10.8 - 18.8   & 5 \\
$c$-HCOOH                 & $7.2\times10^{10}$           & 7.0$^c$         & 4.6 - 14.5     & 3 \\
C$_2$H$_5$OH         & $2.0\times10^{12}$          & 4.3$^c$         & 9.3 - 9.3         & 1 \\
HCOOCH$_3$            & $2.3\times10^{12}$           & 10$^c$          & 17.4 - 24.9     & 16 / 17 \\ 
CH$_3$OCH$_3$       & $5.3\times10^{12}$           & 10$^c$          & 6.7 - 19.0       & 6 / 9 \\ 
HCO$^+$                       & $>$$6.7\times10^{12}$     & 10$^c$          & 4.3 - 4.3        & 1 \\
H$^{13}$CO$^+$           & $>$$2.5\times10^{12}$     & 10$^c$          & 4.2 - 4.2        & 1 \\
HC$^{18}$O$^+$           & $4.0\times10^{11}$           & 10$^c$          & 4.1 - 4.1        & 1 \\
HC$^{17}$O$^+$           & $1.3\times10^{11}$           & 10$^c$          & 4.2 - 4.2        & 3 \\
HCO$_2$$^+$                   & $3.1\times10^{11}$            & 10$^c$         & 10.3 - 15.4   & 2 \\ 
DCO$_2$$^+$                   & $4.1\times10^{10}$            & 10$^c$         & 9.6 - 9.6      & 1 / 2 \\ 
\hline
\multicolumn{5}{c}{Hydrocarbons} \\
\hline
C$_2$H                         & $>$$5.1\times10^{14}$     & 10$^c$          & 4.2 - 4.2        & 2 / 6 \\ 
$^{13}$CCH                  & $3.1\times10^{12}$           & 10$^c$          & 4.0 - 4.0        & 7 / 8 \\ 
C$^{13}$CH                  & $7.2\times10^{12}$           & 10$^c$          & 4.1 - 4.1        & 6 / 7 \\ 
$c$-C$_3$H                      & $7.9\times10^{12}$            & 10$^c$         & 4.4 - 4.4       & 9 \\ 
$c$-C$_3$D                      & $3.5\times10^{11}$            & 10$^c$         & 5.4 - 5.4       & 1 \\ 
$l$-C$_3$H                       & $6.4(18)\times10^{11}$      & 10.5(15)       & 12.5 - 28.0   & 6 \\
C$_4$H                                   & $1.2(5)\times10^{14}$       & 7.9(8)            & 20.5 - 35.6   & 8 \\ 
C$_4$D                                   & $2.3\times10^{12}$           & 7.9$^c$         & 23.3 - 28.0   & 4 \\ 
C$_5$H                                   & $7.2\times10^{11}$           & 7.9$^c$         & 37.0 - 37.0   & 2 \\
$c$-C$_3$H$_2$              & $2.1\times10^{14}$           & 4.1$^c$         & 26.7 - 28.8  & 2 / 6 \\ 
$c$-HCC$^{13}$CH          & $4.1(15)\times10^{12}$     & 4.1(3)            & 6.3 - 15.9    & 9 \\ 
$c$-HC$^{13}$CCH          & $4.7\times10^{11}$           & 4.1$^c$         & 3.9 - 9.4      & 3 \\ 
$c$-C$_3$HD                   & $2.2(7)\times10^{13}$       & 4.1(2)            & 7.6 - 26.6    & 7 / 10 \\ 
$c$-C$_3$D$_2$             & $2.1\times10^{12}$            & 4.1$^c$         & 6.1 - 20.2    & 5 \\ 
$c$-H$^{13}$CCCD         & $2.0\times10^{11}$            & 4.1$^c$         & 10.7 - 10.7  & 2 \\ 
$c$-HCC$^{13}$CD         & $2.5\times10^{11}$            & 4.1$^c$         & 10.6 - 10.6  & 1 / 2 \\ 
$c$-HC$^{13}$CCD $^e$ & $2.5\times10^{11}$            & 4.1$^c$         & 10.6 - 10.6  & 1 \\ 
$l$-C$_3$H$_2$               & $1.1(7)\times10^{12}$        & 5.4(12)         & 8.9 - 15.0     & 6 \\ 
$l$-C$_3$HD                    & $8.1\times10^{10}$            & 5.4$^c$        & 14.0 - 22.9   & 2 \\ 
H$_2$C$_4$                          & $3.5(14)\times10^{11}$     & 9.6(13)           & 18.8 - 33.4  & 12 \\ 
%
CH$_3$CCH                          & $9.3\times10^{13}$            & 10.2$^c$      & 11.5 - 82.3   & 8 \\ 
CH$_2$DCCH                       & $1.8(3)\times10^{13}$        & 10.2(7)         & 11.6 - 43.6   & 15 \\ 
$^{13}$CH$_2$DCCH           & $7.9\times10^{11}$            & 10.2$^c$      & 15.9 - 15.9   & 1 \\ 
CH$_3$CCD                          & $5.3\times10^{12}$            & 10.2$^c$      & 15.0 - 20.9   & 4 \\ 
$^{13}$CH$_3$CCH              & $1.5\times10^{12}$            & 10.2$^c$      & 11.2 - 16.8   & 4 \\ 
CH$_3$$^{13}$CCH              & $1.7\times10^{12}$            & 10.2$^c$      & 11.5 - 17.2   & 4 \\ 
CH$_3$C$^{13}$CH              & $1.6\times10^{12}$            & 10.2$^c$         & 11.1 - 16.7   & 4 \\ 
CH$_3$C$_4$H                    & $7.6\times10^{12}$            & 10$^c$         & 40.8 - 45.1    & 4 \\ 
\hline
\end{tabular}
\end{table}

\setcounter{table}{0}
\begin{table}
\caption{Continued}
\small
\centering
\begin{tabular}{lrrcr}
\hline \hline
\multicolumn{1}{l}{Species} & \multicolumn{1}{c}{$N$ (cm$^{-2}$)$^a$} & \multicolumn{1}{c}{$T_{\rm rot}$ (K)} & \multicolumn{1}{c}{$E_{\rm up}$ (K)} & \multicolumn{1}{c}{$N_{\rm lines}$$^b$} \\
\hline
\multicolumn{5}{c}{N-bearing molecules} \\
\hline
NH$_2$D                     & $7.1\times10^{13}$            & 10$^c$          & 20.1 - 21.3    & 8 / 10 \\
CN                                 & $>$$6.6\times10^{13}$     & 10$^c$          & 5.4 - 5.4        & 2 / 9 \\ 
$^{13}$CN                     & $6.4\times10^{12}$           & 10$^c$          & 5.2 - 5.2        & 10 / 21 \\
C$^{15}$N                     & $7.6\times10^{11}$           & 10$^c$          & 5.3 - 5.3        & 6 \\
HCN                               & $>$$2.4\times10^{13}$     & 10$^c$          & 4.3 - 4.3        & 1 / 3 \\ 
H$^{13}$CN                   & $3.9\times10^{12}$           & 10$^c$          & 4.1 - 4.1        & 1 / 3 \\ 
HC$^{15}$N                   & $4.1\times10^{11}$           & 10$^c$          & 4.1 - 4.1        & 1 \\ 
H$^{13}$C$^{15}$N       & $1.2\times10^{10}$           & 10$^c$          & 4.0 - 4.0        & 1 \\ 
HNC                               & $>$$7.8\times10^{12}$     & 10$^c$          & 4.4 - 4.4        & 1 \\
HN$^{13}$C                   & $>$$3.7\times10^{12}$     & 10$^c$          & 4.2 - 4.2        & 1 \\ 
H$^{15}$NC                   & $8.2\times10^{11}$           & 10$^c$          & 4.3 - 4.3        & 1 \\ 
H$^{15}$N$^{13}$C       & $2.8\times10^{10}$           & 10$^c$          & 4.1 - 4.1        & 1 \\ 
H$_2$CN $^f$                       & $2.4\times10^{12}$           & 10$^c$          & 3.5 - 3.5      & 9 / 13 \\ 
C$_3$N                                   & $2.9(18)\times10^{12}$     & 8.5(15)          & 21.4 - 31.3   & 6 \\
HC$_3$N                                & $4.2\times10^{13}$            & 9.1$^c$         & 19.6 - 28.8   & 6 / 10 \\ 
H$^{13}$CCCN                       & $4.6\times10^{11}$            & 9.1$^c$         & 23.3 - 33.0   & 3 \\ 
HC$^{13}$CCN                       & $4.5\times10^{11}$            & 9.1$^c$         & 19.6 - 33.9   & 4 \\ 
HCC$^{13}$CN                       & $5.3(3.1)\times10^{11}$     & 9.1(16)          & 19.6 - 33.9   & 4 \\ 
HC$_3$$^{15}$N                    & $8.6\times10^{10}$             & 9.1$^c$        & 23.3 - 28.0   & 2 \\ 
DC$_3$N                                & $1.2\times10^{12}$            & 9.1$^c$         & 22.3 - 36.9   & 4 \\ 
DC$^{13}$CCN                       & $3.9\times10^{10}$            & 9.1$^c$         & 22.2 - 22.2   & 1 \\ 
HCCNC                                   & $5.7\times10^{11}$            & 9.1$^c$         & 21.5 - 31.5   & 3 \\ 
HNC$_3$                                & $5.0\times10^{10}$            & 9.1$^c$         & 20.2 - 24.6   & 2 \\ 
HC$_5$N                                & $7.6(28)\times10^{11}$      & 28(3)             & 63.4 - 110.0  & 11 \\
CH$_2$CN                     & $1.2\times10^{12}$           & 10$^c$          & 8.8 - 13.7     & 8 / 37 \\ 
CH$_3$CN                     & $4.1\times10^{11}$           & 10$^c$          & 12.4 - 18.5    & 4 / 5 \\ 
CH$_2$DCN                  & $5.4\times10^{10}$           & 10$^c$          & 12.5 - 18.0    & 4 \\ 
CH$_3$NC                     & $3.7\times10^{10}$           & 10$^c$          & 13.5 - 14.5    & 2 / 3 \\ 
CH$_3$C$_3$N             & $5.0\times10^{11}$           & 10$^c$          & 41.6 - 45.8    & 2 / 4 \\ 
C$_2$H$_3$CN             & $1.9\times10^{11}$           & 10$^c$          & 20.4 - 29.2    & 7 \\
CNCN                             & $1.9\times10^{12}$           & 10$^c$          & 17.9 - 27.3   & 3 \\
HNO                                & $1.2\times10^{12}$           & 10$^c$          & 3.9 - 3.9       & 1 \\
N$_2$O                           & $5.8\times10^{12}$           & 10$^c$           & 12.1 - 12.1   & 1 \\
NCO                                    & $2.2\times10^{12}$            & 10$^c$         & 6.6 - 11.7     & 10 \\
HNCO                                  & $1.7\times10^{13}$            & 10$^c$         & 10.5 - 15.8   & 4 / 6 \\ 
HN$^{13}$CO                      & $2.7\times10^{11}$            & 10$^c$         & 10.5 - 15.8   & 2 \\ 
HNC$^{18}$O                      & $7.2\times10^{10}$            & 10$^c$         & 10.0 - 10.0   & 1 \\ 
DNCO                                  & $6.4\times10^{11}$            & 10$^c$         & 9.8 - 14.7     & 2 \\ 
HOCN                                  & $1.5\times10^{11}$            & 10$^c$         & 10.1 - 15.1   & 2 \\
HCNO                                  & $7.0\times10^{10}$            & 10$^c$         & 11.0 - 16.5   & 2 \\
N$_2$H$^+$                 & $>$$6.0\times10^{13}$     & 10$^c$          & 4.5 - 4.5        & 1 / 7 \\ 
$^{15}$NNH$^+$          & $8.0\times10^{10}$           & 10$^c$          & 4.3 - 4.3        & 3 \\ 
N$^{15}$NH$^+$          & $1.3\times10^{11}$           & 10$^c$          & 4.4 - 4.4        & 3 \\ 
HCNH$^+$ $^f$                   & $2.7\times10^{13}$           & 10$^c$          & 3.6 - 3.6      & 1 \\
HC$_3$NH$^+$                      & $2.3\times10^{11}$            & 9.1$^c$         & 22.9 - 27.4   & 2 \\ 
NCCNH$^+$                   & $1.5\times10^{10}$           & 10$^c$          & 23.4 - 28.1   & 2 \\
H$_2$NCO$^+$                  & $2.9\times10^{10}$            & 10$^c$         & 8.7 - 14.6     & 6 \\
\hline
\multicolumn{5}{c}{S-bearing molecules} \\
\hline
CS                                      & $>$$2.3\times10^{13}$       & 10$^c$        & 7.1 - 7.1       & 1 \\ 
$^{13}$CS                          & $1.7\times10^{12}$             & 10$^c$        & 6.7 - 6.7       & 1 \\ 
C$^{34}$S                          & $3.2\times10^{12}$             & 10$^c$        & 6.9 - 6.9       & 1 \\ 
C$^{33}$S                          & $8.1\times10^{11}$             & 10$^c$        & 7.0 - 7.0       & 4 \\ 
$^{13}$C$^{34}$S              & $5.5\times10^{10}$             & 10$^c$        & 6.5 - 6.5       & 1 \\ 
C$_2$S                               & $4.9(39)\times10^{12}$      & 8.8(19)         & 23.3 - 33.6   & 5 / 8 \\ 
C$_2$$^{34}$S                   & $2.9\times10^{11}$            & 8.8$^c$        & 19.5 - 24.5   & 2 / 3 \\ 
C$^{13}$CS                        & $1.7\times10^{11}$             & 8.8$^c$       & 15.3 - 15.3   & 2 / 4 \\ 
C$_3$S                               & $1.2(6)\times10^{12}$        & 9.6(11)         & 29.1 - 52.7   & 5 / 6 \\ 
HCS                                      & $7.3\times10^{12}$            & 10$^c$         & 5.8 - 5.8       & 5 \\
HSC                                      & $2.0\times10^{11}$            & 10$^c$         & 5.9 - 5.9       & 1 / 2 \\ 
H$_2$CS                              & $1.4\times10^{13}$            & 8.0$^c$           & 8.1 - 9.9       & 3 \\ 
H$_2$$^{13}$CS                  & $1.2\times10^{11}$            & 8.0$^c$           & 7.8 - 9.5       & 3 \\ 
H$_2$C$^{34}$S                  & 4.6$\times10^{11}$            & 8.0$^c$           & 8.0 - 8.3       & 2 / 3 \\ 
HDCS                                   & $1.9(10)\times10^{12}$      & 8.0(18)         & 8.9 - 18.1     & 3 \\ 
D$_2$CS                              & $6.2\times10^{11}$            & 8.0$^c$           & 6.6 - 13.6     & 5 \\ 
\hline
\end{tabular}
\end{table}

\setcounter{table}{0}
\begin{table}
\caption{Continued}
\small
\centering
\begin{tabular}{lrrcr}
\hline \hline
\multicolumn{1}{l}{Species} & \multicolumn{1}{c}{$N$ (cm$^{-2}$)$^a$} & \multicolumn{1}{c}{$T_{\rm rot}$ (K)} & \multicolumn{1}{c}{$E_{\rm up}$ (K)} & \multicolumn{1}{c}{$N_{\rm lines}$$^b$} \\
\hline
CH$_3$SH                         & $2.4\times10^{12}$            & 10$^c$        & 8.4 - 8.5       & 2 / 4 \\ 
SO                                           & $2.0\times10^{14}$        & 4.5$^c$        & 19.3 - 21.1   & 2 / 4 \\ 
$^{34}$SO                               & $6.5(36)\times10^{12}$      & 4.5(5)           & 9.1 - 20.9     & 3 \\ 
$^{33}$SO                               & $1.4\times10^{12}$            & 4.5$^c$        & 9.2 - 9.2       & 4 / 5 \\ 
S$^{18}$O~~~~~~~~~~~ & $1.3\times10^{12}$            & 4.5$^c$        & 8.7 - 8.7       & 1 \\ 
S$^{17}$O                               & $1.6\times10^{12}$            & 4.5$^c$        & 9.0 - 9.0       & 1 \\ 
SO$_2$                              & $4.0(12)\times10^{12}$      & 7.9(7)          & 7.7 - 36.7     & 2 \\ 
$^{34}$SO$_2$                  & $1.3\times10^{11}$            & 7.9$^c$       & 7.6 - 7.6       & 1 \\ 
OCS                                    & $1.6(12)\times10^{13}$      & 7.8(17)         & 16.3 - 26.3   & 3 \\
NS                                           & $5.4\times10^{12}$            & 10$^c$         & 8.8 - 8.9       & 10 \\ 
N$^{34}$S                               & $2.4\times10^{11}$           & 10$^c$           & 8.7 - 8.7       & 2 \\ 
HNCS                                   & $2.1\times10^{11}$            & 10$^c$         & 15.8 - 25.3   & 3 \\
HSCN                                   & $1.1\times10^{11}$            & 10$^c$         & 15.4 - 19.8   & 2 \\
HCS$^+$                            & $1.3\times10^{12}$             & 10$^c$        & 6.1 - 6.1       & 1 \\ 
HC$^{34}$S$^+$                & $6.3\times10^{10}$             & 10$^c$        & 6.0 - 6.0       & 1 \\ 
SO$^+$                                   & $1.6\times10^{12}$            & 4.5$^c$        & 8.9 - 8.9       & 1 \\ 
NS$^+$                                   & $2.1\times10^{11}$            & 10$^c$         & 7.2 - 7.2       & 4 \\ 
\hline
\multicolumn{5}{c}{Si-bearing molecules} \\
\hline
SiO~~~~~~~~~~~~~~~~~ & $2.6\times10^{11}$           & 10$^c$          & 6.3 - 6.3        & 1 \\
\hline
\end{tabular}
\tablenotea{Numbers in parentheses are 1$\sigma$ uncertainties in units of the last digits. Fractional abundances relative to H$_2$ can be directly computed from the column densities listed using a column density of H$_2$ of $4\times10^{22}$ cm$^{-2}$ (see Sect.~\ref{sec:column_densities}). \\
$^a$ The error in the column density for those species for which the rotational temperature has been fixed is estimated to be 50 \% (see text).\\
$^b$ Number of lines observed. The notation $x/y$ means that $x$ lines, out of $y$ observed lines, were included in the determination of the column density.\\
$^c$ Rotational temperature has been fixed.\\
$^d$ Impossible to fit lines to a rotational diagram. The column density of CH$_3$OH is estimated to be $2.9\times10^{14}$ cm$^{-2}$ based on $^{13}$CH$_3$OH and adopting a $^{12}$C/$^{13}$C isotopic ratio of 68 \citep{Milam2005}.\\
$^e$ Tentative detection. \\
$^f$ Molecule observed at frequencies below 80 GHz (Ag\'undez et al., unpublished data).}
\end{table}

The column densities of the 140 species detected, together with the rotational temperatures assumed or derived, are given in Table~\ref{table:column_densities}. We also give column densities for H$_2$CN and HCNH$^+$, molecules that are detected at frequencies below 80 GHz (Ag\'undez et al., unpublished data). For the species for which the rotational temperature was derived, an uncertainty in both the column density and the rotational temperature was calculated from the least-squares fit. These uncertainties include the 10~\% error in the observed line intensities due to calibration. For the species for which uncertainties in the column densities were calculated, these range from 20 to 80~\%. Based on these numbers, we estimate that the column densities given in Table~\ref{table:column_densities} have errors of the order of 50~\%.

The gas kinetic temperature in L483 has been determined to be 10 K in a NH$_3$ condensation which appears centered onto the protostar and extends over a region of 40-60$''$ in diameter, about twice the size of the IRAM 30m main beam at $\lambda$ 3 mm \citep{Anglada1997}. From our CO data we can obtain an estimation of the gas kinetic temperature. The CO $J$=1-0 line is optically thick and shows self-absorption around the velocity of the ambient cloud (see top line in left panel of Fig.~\ref{fig:line_profiles}). In the region of self-absorption the line has $T_A^*$ = 4 K ($T_{\rm mb}$ = 4.9 K). Given that in this velocity range the line is optically thick, the excitation temperature of the absorbing CO gas located in the outer parts of the cloud has to be 8.2 K, a value which is probably close to the kinetic temperature of the gas in that region. Similar numbers (8.5-9.5 K) were inferred from observations of the CO $J$=2-1 line by \cite{Tafalla2000}. The $^{13}$CO $J$=1-0 line is also optically thick and has an intensity of $T_A^*$ = 5.6 K ($T_{\rm mb}$ = 6.7 K) around the systemic velocity of the ambient cloud (see left panel in Fig.~\ref{fig:line_profiles}), which implies an excitation temperature of 10 K. This value is a good estimate of the gas kinetic temperature in the region of the cloud where $^{13}$CO $J$=1-0 becomes optically thick. An independent estimate of the gas kinetic temperature is provided by the relative intensities of the $K$=0-3 components of CH$_3$CCH observed in the $J$=5-4 and $J$=6-5 transitions, which yield a gas kinetic temperature of 15 $\pm$ 2 K. This value is somewhat higher than the above estimates. It is likely that the different gas kinetic temperature tracers probe different regions, with $^{13}$CO and NH$_3$ probing a similar region, while the optically thick lines of CO probe outer and colder parts of the cloud and CH$_3$CCH traces inner and warmer regions.

The rotational temperatures derived for the different molecules range from 4.1 to 10.5 K. A notable outsider is HC$_5$N, for which we derive $T_{\rm rot}$ = 28 $\pm$ 3 K. Line widths for this molecule are also slightly higher
(0.87 km s$^{-1}$ on average) than for most other species showing narrow lines. These two facts suggest that HC$_5$N is probably distributed in a warmer region than the other molecules, probably closer to the protostar. Apart from HC$_5$N, the low excitation temperatures derived together with the small line widths are consistent with emission arising from a cold and extended part of the dense core. The gas densities at these scales are probably not too high, as indicated by the low rotational temperatures (down to 4 K) of some molecules, which strongly suggest subthermal excitation in a low-density gas at a kinetic temperature around 10 K. From a model of the continuum emission in L483, \cite{Jorgensen2002} derive a volume density of H$_2$ of $3.4\times10^4$ cm$^{-3}$ at the radius at which the gas and dust temperature become 10 K, which in their model occurs at $7.8\times10^3$ au (or 40$''$ at 200 pc) from the protostar. From ammonia observations, \cite{Anglada1997} derive a similar H$_2$ volume density of $3\times10^4$ cm$^{-3}$. The analysis of CH$_3$OH lines observed also provides some constraints on the H$_2$ volume density. For this molecule we detect nine lines with a wide range of optical depths and excitation temperatures, which makes it impossible to fit them in a rotational diagram. The lines at 96739.358 MHz and 96741.371 MHz are probably optically thick (inferred optical depths are in slight excess of one). The remaining lines are optically thin and one of them, lying at 107013.831 MHz, appears in absorption against the cosmic microwave background, and therefore has an excitation temperature below 2.7 K. Statistical equilibrium calculations using the large velocity gradient (LVG) formalism do not allow for the intensities of all observed lines to be globally
reproduced, but the main features can be reproduced adopting a column density of $2.9\times10^{14}$ cm$^{-2}$ (estimated from $^{13}$CH$_3$OH), a gas kinetic temperature of 10 K (in agreement with the above quoted values), and a volume density of H$_2$ of $\sim$$3\times10^4$ cm$^{-3}$, which is also in line with the values derived by \cite{Anglada1997} and \cite{Jorgensen2002}.

From the observed C$^{17}$O $J$=1-0 line, which is optically thin, we derive a column density for C$^{17}$O of $1.7\times10^{15}$ cm$^{-2}$ assuming a rotational temperature of 10 K (see Table~\ref{table:column_densities}). If we adopt a standard $N$(C$^{18}$O)/$N$(H$_2$) for dense cores, $1.7\times10^{-7}$ \citep{Frerking1982}, and a $^{18}$O/$^{17}$O isotopic ratio of 4.16 \citep{Wouterloot2008}, the resulting column density of H$_2$ in L483 is $4\times10^{22}$ cm$^{-2}$, which is close to the value of $3\times10^{22}$ cm$^{-2}$ reported by \cite{Tafalla2000} based on the same C$^{17}$O line observed also with the IRAM 30m more than 20 years ago. The main source of uncertainty in the value derived from C$^{17}$O is the degree of depletion of CO. In fact, by modeling the dust continuum emission, \cite{Jorgensen2002} derive $N$(H$_2$) = $9.3\times10^{23}$ cm$^{-2}$ for a size of 40$''$ from the protostar, which would imply that CO is severely depleted from the gas phase in L483. The main source of uncertainty in this latter value is related to dust parameters such as the size and optical properties of dust grains and the gas-to-dust-mass ratio. A detailed model of multi-wavelength continuum and CO spatially resolved data will help to better constrain the physical structure of L483. For the purpose of computing fractional abundances relative to H$_2$, here we adopt the H$_2$ column density obtained from C$^{17}$O, bearing in mind that CO depletion may be an issue in L483 and that $N$(H$_2$) could be significantly higher.

\subsection{The chemical composition of L483} \label{sec:abundances}

\begin{figure*}[!ht]
\centering
\includegraphics[angle=0,width=\textwidth]{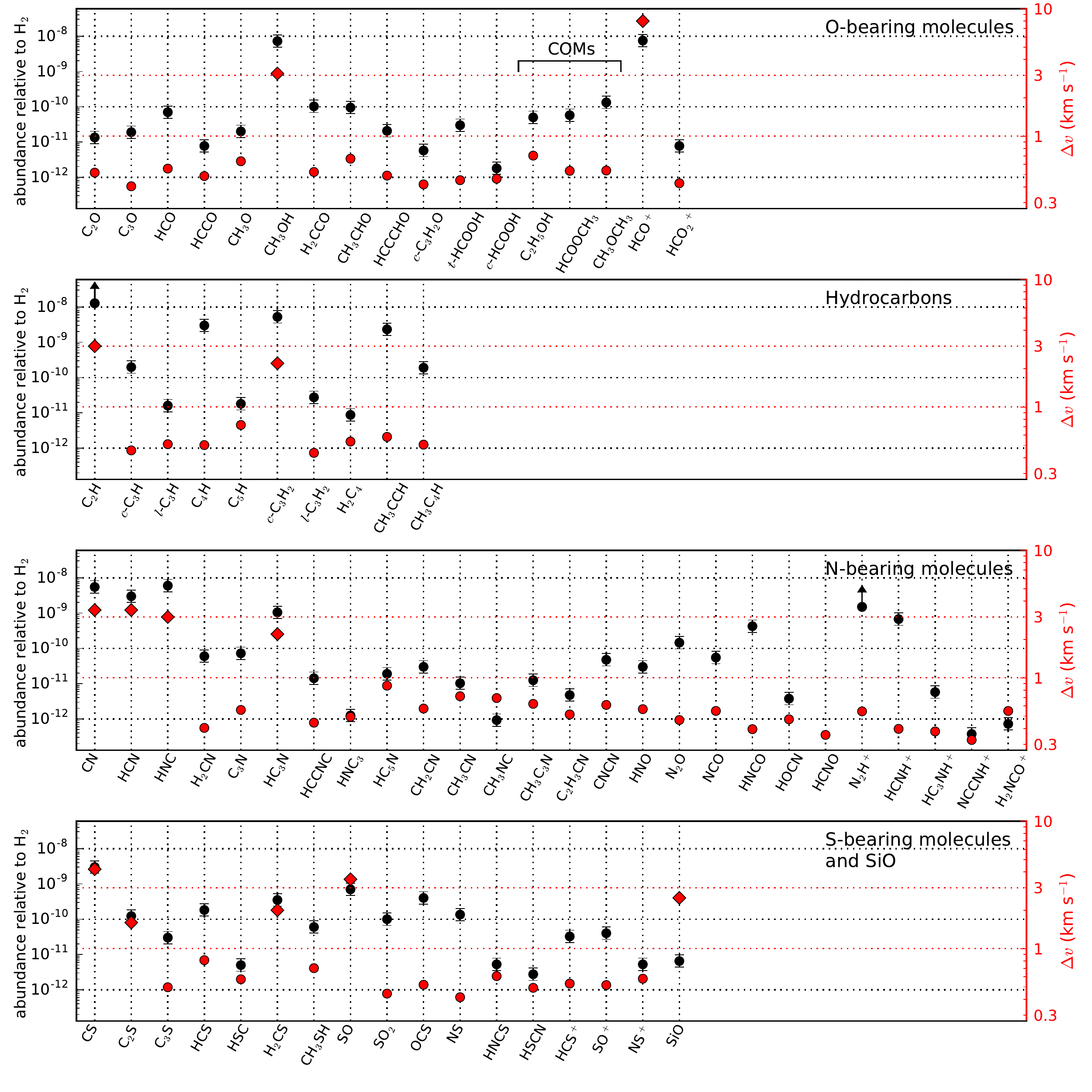}
\caption{Visualization of molecular fractional abundances (in black and referred to the left axis) and line widths (in red and referred to the right axis) in L483 for different chemical families. Abundances are obtained from the beam-averaged column densities listed in Table~\ref{table:column_densities}, adopting a column density of H$_2$ of $4\times10^{22}$ cm$^{-2}$ (see Sect.~\ref{sec:column_densities}). CO is not plotted. For molecules with optically thick lines we used the column densities of optically thin isotopologs and scaled up adopting local ISM isotopic ratios as follows: for HCO$^+$ the $^{17}$O isotopolog was used adopting $^{16}$O/$^{17}$O = 2317 \citep{Wouterloot2008}; for CN, HCN, and HNC we used the $^{15}$N isotopolog assuming $^{14}$N/$^{15}$N = 290 \citep{Adande2012}; and for CH$_3$OH and CS we used the $^{13}$C isotopolog and assumed $^{12}$C/$^{13}$C = 68 \citep{Milam2005}. In the cases of C$_2$H and N$_2$H$^+$, scaling from $^{13}$C and $^{15}$N isotopologs, respectively, is probably not valid due to important fractionation effects (see Sect.~\ref{sec:isotopic_ratios_discussion_13c} and \ref{sec:isotopic_ratios_discussion_rest}), and therefore only lower limits are given. H$_2$CO and NH$_3$ are not shown because we only detect deuterated forms and  the deuterium fractionation is uncertain for these species. Line widths represented as red circles are averages over the FWHM of all those lines fitted to a Gaussian function ($\Delta v$ in Table~\ref{table:lines}). For those molecules showing complex profiles (e.g., wings) the values plotted as red diamonds correspond to the highest velocity $|{V_{\rm LSR}-v_{\rm sys}}|$ at which emission is detected.} \label{fig:abundances}
\end{figure*}

In Fig.~\ref{fig:abundances} we show the fractional abundances relative to H$_2$\footnote{Hereafter fractional abundances are always expressed relative to H$_2$.} of the parent molecules (i.e., excluding minor isotopologs) detected in this $\lambda$ 3 mm line survey. We also show information on the mean line width derived for each molecule. Apart from the species that show line wings (CO, HCO$^+$, CN, HCN, HNC, CS, and SiO show bright wings relative to the emission at the systemic velocity, while CH$_3$OH, C$_2$H, $c$-C$_3$H$_2$, HC$_3$N, C$_2$S, H$_2$CS, and SO show weak wings; see discussion in Sect.~\ref{sec:line_profiles}), the vast majority of molecules show narrow lines, with widths around 0.5 km s$^{-1}$. Therefore, while for a few molecules part of the detected emission arises from the bipolar outflow, for most molecules the emission comes either exclusively or mostly from the ambient quiescent cloud. We are therefore probing the chemical composition of the dense core at a relatively large scale and not in the surroundings of the protostar. We note that the emission associated with the hot corino unveiled by ALMA is very compact, with sizes well below 1$''$ and thus severely diluted in the 21-30$''$ beam of the IRAM 30m telescope, and wide in velocity as it extends over several kilometres per second \citep{Oya2017,Jacobsen2018}.

One of the most remarkable characteristics of L483 is that its large-scale chemical composition displays a very rich variety of molecules. Many of them are widely known interstellar molecules. However, some of the detected molecules have rarely been found toward similar sources and, moreover, a few of them had not been observed in space before their discovery in L483. Among the detected molecules, there is a good variety of O-bearing molecules, various unsaturated hydrocarbons, a large number of N-bearing molecules, a diverse sample of S-bearing molecules, and only one molecule containing silicon: SiO. It is also worth noting that L483 contains a rich variety of deuterated molecules, which is evidence that deuterium fractionation has proceeded efficiently in this source (see more details in Sect.~\ref{sec:deuterium}).

\subsubsection{O-bearing molecules} \label{sec:o-bearing}

The fractional abundances of the oxygen-bearing molecules detected in the line survey are shown in the upper panel of Fig.~\ref{fig:abundances}. We see that, apart from CO, the most abundant molecules are CH$_3$OH and HCO$^+$ ($\sim$10$^{-8}$), while the rest can be considered as minor as they have abundances around or below 10$^{-10}$. Other O-bearing molecules not covered in this line survey that could be abundant are water, carbon dioxide, molecular oxygen, and formaldehyde. In fact, \textit{Herschel} observed abundant and warm gaseous water toward L483, the origin of which is the close surroundings of the protostar \citep{Mottram2014}. This strongly suggests that at the colder and more extended scales of the ambient cloud, water ice probably makes an important part of the oxygen budget. In the case of CO$_2$, a rough estimate of its abundance can be obtained from that derived for its protonated form. Using a simplified chemical scheme, the abundance of CO$_2$ is inferred to be $\sim$10$^4$ times that of HCO$_2$$^+$ for dense clouds with parameters similar to those of L483 \citep{Sakai2008b,Vastel2016}, which would imply that in L483, CO$_2$ has an abundance as large as $8\times10^{-8}$. The abundance of O$_2$ in L483 is very uncertain. The recently discovered isocyanate radical could bring constraints on the abundance of O$_2$ as NCO is thought to be essentially formed in the reaction between CN and O$_2$. In the chemical model of L483 presented in \cite{Marcelino2018a}, which successfully reproduced the observed abundances of NCO, HNCO, and various other related species, the abundance of O$_2$ is $7\times10^{-7}$. Although there is no observational constraint on the abundance of O$_2$ in L483, this value is probably too high given the upper limits derived from \textit{Herschel} data to the abundance of O$_2$ in other similar sources, $<$(0.6-1.6)$\times10^{-7}$ in cold prestellar cores \citep{Wirstrom2016}, and $<$$6\times10^{-9}$ toward the low-mass Class\,0 protostar NGC\,1333-IRAS\,4A \citep{Yildiz2013}. Formaldehyde is certainly an important species in the ambient cloud of L483, where it is present with an abundance of $1.5\times10^{-9}$ \citep{Tafalla2000}, around ten times less abundant than methanol.

All the hydrogenation derivatives of CO (i.e., H$_x$CO, $x$ = 1-4) are observed in L483. The observed H$_2$CO/HCO and CH$_3$OH/CH$_3$O abundance ratios are $\sim$20 and $\sim$360, respectively, consistent with the general trend found in other cold dense cores, regardless of whether or not they host a protostar, where H$_2$CO is found to be around ten times more abundant than HCO while CH$_3$OH is about 100 times more abundant than CH$_3$O \citep{Antinolo2016,Bacmann2016,Ocana2017}. The origin of these species in cold dense clouds is still under discussion. Although they can be naturally formed upon CO adsorption on dust grains and consecutive additions of H atoms, the efficiency of the different hydrogenation and chemical desorption channels is not yet well constrained (e.g., \citealt{Minissale2016}). Moreover, with the exception of CH$_3$OH, all these species have efficient formation routes in the gas phase (e.g., \citealt{Antinolo2016,Ocana2017}).

The carbon chain oxides C$_2$O and C$_3$O are present with similar abundances ($\sim$10$^{-11}$) in L483. In TMC-1, both species are also found with similar abundances (see \citealt{Agundez2013}). While C$_2$O has not been very widely observed in interstellar space (the only reported detection is toward TMC-1; \citealt{Ohishi1991}), C$_3$O has been observed in a few other cold dense clouds like TMC-1, Elias\,18, and L1544 \citep{Matthews1984,Palumbo2008,Vastel2014}. The formation of these two molecules in cold dense clouds is well explained by gas-phase chemistry (e.g., \citealt{Agundez2013}), although laboratory experiments have shown that irradiation of CO ice with energetic protons can also result in the formation of carbon oxides like C$_2$O, C$_3$O, and C$_3$O$_2$ \citep{Palumbo2008}; the latter species being nonpolar it cannot be observed through radioastronomical techniques.

The ketenyl radical (HCCO), the hydrogenated descendant of C$_2$O, was recently discovered in space toward the dense cores Lupus-1A and L483 \citep{Agundez2015a}. In L483 we find that the different hydrogenation forms of C$_2$O have relative abundances HCCO:H$_2$CCO:CH$_3$CO:CH$_3$CHO $\sim$ 1:10:$?$:10. These values are interestingly similar to those typically found in cold dense clouds for the hydrogenated forms of CO, in which case HCO:H$_2$CO:CH$_3$O:CH$_3$OH are about 1:10:0.1:10 \citep{Bacmann2016}. This fact may be accidental, although it is also possible that it reflects similar efficiencies along the different hydrogenation steps for the two types of molecules. If the relative abundances of H$_x$CO hold for H$_x$CCO, then the acetyl radical (CH$_3$CO), which has not  yet been observed in space, would be ten times less abundant than HCCO.

Propynal (HCCCHO) and cyclopropenone ($c$-C$_3$H$_2$O) are two isomers which are detected in L483 with abundances of the order of 10$^{-11}$, propynal being somewhat more abundant. This is in line with the abundances observed for these two species in other cold dense clouds, where their synthesis is accounted for by gas-phase chemistry \citep{Loison2016}.

An interesting result is also the detection of the $trans$ and $cis$ conformers of the organic molecule HCOOH. The $trans$ species, which is the most stable, is widely observed in cold and hot cores. Its formation in cold cores can be explained by gas-phase chemistry \citep{Vigren2010}. The $cis$ conformer however has only been observed in a few interstellar sources. It was discovered toward the Orion Bar photodissociation region, where it is present with a high abundance (only 2.8 times less than the $trans$ conformer) in the region illuminated by ultraviolet (UV) photons \citep{Cuadrado2016}. This fact led these authors to suggest that the $cis$ conformer is formed from the $trans$ form in a photoswitching process induced by UV photons. More recently, $cis$-HCOOH has been detected in a different environment: the cold dark cloud B5, where the $cis$/$trans$ abundance ratio is found to be 6 \% \citep{Taquet2017}. In L483 we derive a similar $cis$/$trans$ ratio of 6 \% for HCOOH. The lack of a strong UV field in B5 and L483, together with the much lower $cis$/$trans$ ratios derived compared to that found in the Orion Bar, suggests that in these sources $cis$-HCOOH does not originate from a photoswitching mechanism. Moreover, the identical $cis$/$trans$ ratios derived in B5 and L483 strongly suggest a common formation mechanism. Whether such a mechanism occurs in the gas phase (through, e.g., the dissociative recombination of HCOOH$_2^+$ or a chemical switching process) or in the surface of dust grains (via hydrogenation of HO-CO radicals; \citealt{Ioppolo2011}) is still to be investigated.

One of the most remarkable results obtained in this line survey is the detection of the complex organic molecules (COMs) methyl formate (HCOOCH$_3$), dimethyl ether (CH$_3$OCH$_3$), and ethanol (C$_2$H$_5$OH). These molecules are often observed toward hot cores and hot corinos with single-dish telescopes (e.g., \citealt{Cazaux2003,Bottinelli2007,Bianchi2019}) or interferometers (e.g., \citealt{Tercero2015,Imai2016}); they have been observed toward L483 with ALMA \citep{Oya2017,Jacobsen2018}, although in those observations the detected emission arises from a compact and warm region around the protostar. In this line survey however the lines of COMs observed arise from the large-scale cold ambient cloud, similarly to previous single-dish detections of COMs in prestellar cores like L1689B and L1544 \citep{Bacmann2012,Jimenez-Serra2016} and cold dense clouds around low-mass protostars such as B1-b and B5 \citep{Oberg2010,Cernicharo2012,Taquet2017}. Extended emission from HCOOCH$_3$ and CH$_3$OCH$_3$ has been detected in all or some of the above quoted sources. Our detection of C$_2$H$_5$OH is thus the first detection of this molecule in the cold, quiescent, and extended component of a prestellar or protostellar dense core, and adds one further piece to the puzzle of the presence of COMs in cold dense clouds \citep{Ruaud2015,Balucani2015,Kalvans2015,Vasyunin2017,Shingledecker2018}.

\subsubsection{Hydrocarbons} \label{sec:hydrocarbons}

There are various hydrocarbons detected in L483 in the $\lambda$ 3 mm band (see second panel in Fig.~\ref{fig:abundances}). Most of them are highly unsaturated and some of them, like the radicals C$_2$H and C$_4$H, the ring molecule $c$-C$_3$H$_2$, and methyl acetylene (CH$_3$CCH), are quite abundant (between 10$^{-9}$ and $>$10$^{-8}$).

We detect all the members of the series of radicals C$_n$H with $n$=2-5. Larger members are probably not detected because for rotational temperatures around 10 K the most favorable transitions lie below 80 GHz. For example, for C$_6$H we derive a 3$\sigma$ upper limit to its column density of $4\times10^{12}$ cm$^{-2}$ (adopting the rotational temperature of 7.9 K derived for C$_4$H; see Table~\ref{table:column_densities}). This implies a C$_6$H/C$_4$H ratio of $<$3 \%, which is still above the values observed in TMC-1 and L1527, 0.3 \% and 1 \%, respectively (e.g., \citealt{Sakai2008a}). Therefore, it would not be strange to detect carbon chains
in L483 at cm wavelengths that are longer than those observed here. In this line it is interesting to note that already relatively heavy hydrocarbons like C$_5$H and CH$_3$C$_4$H are detected in the $\lambda$ 3 mm band through lines with upper-level energies around 40 K, well in excess of the gas kinetic temperature of $\sim$10 K. This implies that in L483, long carbon chains are either relatively abundant or are significantly warmer than 10 K. Unfortunately the low number of lines of C$_5$H and CH$_3$C$_4$H detected and their low signal-to-noise ratios prevent us from precisely constraining their excitation temperatures. This is not the case for HC$_5$N, for which a relatively high rotational temperature is inferred (see Sect.~\ref{sec:n-bearing}), supporting the scenario that long carbon chains are warmer than shorter carbon chains in L483.

Among the hydrocarbons observed, we detect the cyclic and linear isomers of C$_3$H and C$_3$H$_2$. The chemistry of these species in cold dense clouds has been discussed by \cite{Loison2017}, highlighting the importance of the dissociative recombination of C$_3$H$_2^+$ and C$_3$H$_3^+$ isomers in establishing the cyclic-to-linear abundance ratios. From the column densities derived in this line survey, we find an abundance ratio $c$-C$_3$H/$l$-C$_3$H of 12, which is in the range of values derived in TMC-1 and B1-b \citep{Fosse2001,Loison2017}. Cyclic C$_3$H$_2$ is found to be substantially more abundant than its linear isomer, in line with observations of other cold dense clouds \citep{Fosse2001,Loison2017}.

The presence of carbon chain hydrocarbons like C$_4$H in cold dark clouds has long been known. These species are particularly abundant toward the cyanopolyyne peak of the starless source TMC-1 (e.g., \citealt{Ohishi1998}), but they are also relatively abundant in other cold dense clouds, some of which are starless while others contain protostars (e.g., \citealt{Agundez2008}). The detection of bright C$_4$H lines in some dense cores around low-mass protostars like L1527 and Lupus-1A \citep{Sakai2008a,Sakai2009a} motivated the proposal of a scenario called warm carbon chain chemistry (WCCC) in which carbon chains would form by gas phase chemistry from the evaporation of CH$_4$ ice \citep{Sakai2013}. If this scenario is accurate, methane should be an important hydrocarbon in these protostellar sources rich in carbon chains. Still, it is not clear whether the presence of abundant carbon chains in these sources is an inheritance from the dark cloud prestellar phase, a consequence of a specific type of chemistry triggered by CH$_4$ ice evaporation upon the switch on of the protostar, or the result of environmental effects or the particular history of the cloud. In any case, L483 shares a commonality with L1527 and Lupus-1A in that it contains abundant carbon chain hydrocarbons. It is also interesting to note that while in the case of L1527, the presence of carbon chains and COMs seems to be mutually exclusive \citep{Sakai2008a,Yoshida2019}, this is not the case for L483, where both types of molecules coexist in the ambient cloud.

\subsubsection{N-bearing molecules} \label{sec:n-bearing}

There is a great diversity of N-bearing molecules detected in this line survey (see third panel in Fig.~\ref{fig:abundances}). A great fraction of them are unsaturated carbon chains that contain only N, C, and H, although there are also several N-bearing molecules containing oxygen.

The most abundant (above 10$^{-9}$) N-bearing molecules detected are CN, HCN, HNC, HC$_3$N, HCNH$^+$, and N$_2$H$^+$. The latter is a proxy of molecular nitrogen, which is probably one of the most important reservoirs of nitrogen. The detection of NH$_2$D also points to ammonia as an important N-containing molecule. In fact, \cite{Anglada1997} derive a column density of NH$_3$ of $1.4\times10^{15}$ cm$^{-2}$, which translates to an abundance of several times 10$^{-8}$. Another potentially abundant N-bearing molecule is cyanogen (NCCN), the presence of which is supported by the detection of NCCNH$^+$ and CNCN, previously reported by \cite{Agundez2015b} and \cite{Agundez2018b}, respectively. It is estimated that NCCN could have an abundance between 10$^{-9}$ and 10$^{-7}$ (see \citealt{Agundez2018b}). The abundance ratio HNC/HCN derived from the $^{15}$N isotopologs is found to be 2.0 in L483, in the range of values found in other cold dense cores and consistent with formation of both species from the dissociative recombination of HCNH$^+$ \citep{Hirota1998}.

The existence of abundant unsaturated carbon chains of the family of cyanopolyynes like HC$_3$N is in line with the presence of abundant carbon chains of the family of polyynes like C$_4$H and S-bearing carbon chains like C$_2$S (see Sect.~\ref{sec:s-bearing}). In this sense, L483 is a source rich in carbon chains, as in other starless or protostellar dense cores \citep{Hirota2009}. Apart from HC$_3$N, we also detected the related radical C$_3$N and the longer cyanopolyyne HC$_5$N. The latter is the only molecule for which we derive a rotational temperature that is without doubt well in excess of 10 K, that is, 28 $\pm$ 3 K. This fact indicates that HC$_5$N is probably distributed over a more compact region around the protostar than HC$_3$N, which should have a more extended distribution according to the lower  derived rotational temperature of 9.1 $\pm$ 1.6 K. We also detected the two metastable isomers of cyanoacetylene, HCCNC and HNC$_3$, as well as its protonated form. We find that HCCNC is approximately 70 times less abundant than HC$_3$N and approximately 10 times more abundant than HNC$_3$, while HC$_3$NH$^+$ is about 200 times less abundant than HC$_3$N. These relative abundances are similar to those found in TMC-1 and L1544 \citep{Vastel2018a}.
Just as CH$_3$CCH and CH$_3$C$_4$H are detected in L483, the cyanide analogs CH$_3$CN and CH$_3$C$_3$N are also observed. We find that the two molecules have nearly the same abundance, very similarly to what is found in TMC-1 \citep{Broten1984}. The cyanomethyl radical (CH$_2$CN) is observed to be four times more abundant than CH$_3$CN in L483, while in TMC-1, CH$_2$CN is also found to be more abundant than CH$_3$CN by a factor of approximately ten \citep{Irvine1988}. We also detected methyl isocyanide (CH$_3$NC), a metastable isomer of methyl cyanide, which has only been observed in a few astronomical sources. This isomer has been observed to be 7-50 times less abundant than CH$_3$CN in interstellar space, concretely toward Sgr\,B2 and in the Horsehead photodissociation region \citep{Cernicharo1988,Remijan2005,Gratier2013}, while it is much less abundant than CH$_3$CN, by factors of 200-500, in hot cores and hot corinos \citep{Lopez2014,Calcutt2018a}. In L483 we find a CH$_3$CN/CH$_3$NC abundance ratio of 11, which is more in line with the interstellar values than with the protostellar ones. The partially saturated cyanide C$_2$H$_3$CN, which has been observed in both cold and hot cores, is also detected in L483, where the emission probably arises from the cold and quiescent cloud.

We detected several N-bearing molecules that contain oxygen, many of which are related to HNCO, which is a relatively abundant molecule in L483. We detected two metastable isomers of isocyanic acid, cyanic acid (HOCN), and fulminic acid (HCNO), which are about 100 and 200 times less abundant than HNCO, respectively. Therefore, HOCN and HCNO have similar abundances in L483  (within a factor of two), similarly to what is found in the prestellar cores L183 and L1544 and in the young low-mass protostellar cores B1-b and L1527 \citep{Marcelino2010}. In L483 we also detected the radical isocyanate (NCO) and protonated isocyanic acid (H$_2$NCO$^+$), which are thought to be precursors of HNCO and its isomers. In fact, L483 is the only source where all these chemically related species have been detected to date, which makes it a perfect test-bed to study the chemistry of this chemical family (see \citealt{Marcelino2018a}).

We also detected nitroxyl (HNO) and nitrous oxide (N$_2$O), two species that belong to the chemical family of nitrogen oxides and their derivatives. Nitric oxide (NO), the most widespread and abundant member of this family, has no lines in the $\lambda$ 3 mm band but it is very likely abundant in L483. Since the discovery of HNO and N$_2$O toward Sgr\,B2 \citep{Ulich1977,Ziurys1994}, none of these molecules have been widely observed in space. In the case of HNO, it has been detected in various massive star-forming regions and in the cold dense cloud L134N, where the derived NO/HNO abundance ratio is approximately 800 \citep{Snyder1993}. If the same ratio holds in L483, then NO would have an abundance of a few times 10$^{-8}$, a value similar to those found in TMC-1 and L134N (see, e.g., \citealt{Agundez2013}). Concerning N$_2$O, it has only been observed toward Sgr\,B2 \citep{Ziurys1994,Halfen2001} and more recently in the hot corino IRAS\,16293$-$2422 \citep{Ligterink2018}. In these two sources, N$_2$O is present in warm gas. In L483, both HNO and N$_2$O are detected through a single line, and therefore we cannot constrain their excitation temperatures. However, the small line width observed for both species ($\sim$0.5 km s$^{-1}$) indicates that their emission must arise from the cold and quiescent cloud. Our detection of N$_2$O in L483 is thus the first one in a cold dense cloud and provides an additional constraint to understand the chemistry of nitrogen oxides in these environments. Interestingly, we find that the N$_2$O/HNO abundance ratio in L483 is approximately five, which is close to that found in Sgr\,B2 ($\sim$3; \citealt{Ziurys1994}), despite the different nature of these two environments.

\subsubsection{S-bearing molecules} \label{sec:s-bearing}

The $\lambda$ 3 mm scan of L483 revealed a great variety of S-bearing molecules. Many of them are not simply minor species but are present at an important abundance level. The most abundant S-bearing molecule is CS ($>$10$^{-9}$), but there are also several relatively abundant ($>$10$^{-10}$) species like C$_2$S, HCS, H$_2$CS, SO, SO$_2$, OCS, and NS.

The carbon chains C$_2$S and C$_3$S are detected through rather intense lines, which together with the detection of bright emission from C$_4$H and cyanopolyynes like HC$_3$N, stresses the nature of L483 as a source rich in carbon chains. This distinctive characteristic is also found in other cold dense clouds like TMC-1 and L1527, although it is not universal to dense cores, which show different degrees of richness in carbon chains \citep{Hirota2009}.

An interesting aspect of the chemistry of sulfur unveiled by the $\lambda$ 3 mm line survey is that the detection of HCS and HSC (see \citealt{Agundez2018a}), together with that of CH$_3$SH, permits a relatively complete description of the different intermediates along the successive hydrogenation steps of CS. It is interesting to compare the relative abundances between these species with the oxygen analog case, for which we are informed that HCO:H$_2$CO:CH$_3$O:CH$_3$OH are about 1:10:0.1:10 in cold dense clouds \citep{Bacmann2016}. For sulfur we find a quite different behavior in L483, where HCS:H$_2$CS:CH$_3$S:CH$_3$SH are 1:2:$<$0.1:0.3, where the upper limit for CH$_3$S is computed using the entry in MADEX, which is based on the laboratory spectrum by \cite{Endo1986}. That is, while the electronic closed shell molecules H$_2$CO and CH$_3$OH are substantially more abundant than the radical HCO, the radical HCS is almost as abundant as H$_2$CS and significantly more abundant than CH$_3$SH. These relative abundances hold key information about the underlying chemical processes at work along the hydrogenation sequences of CO and CS, which must be quite different.

The sulfur oxides SO and SO$_2$ and carbonyl sulfide (OCS) are commonly observed in cold dense clouds with abundances between 10$^{-10}$ and 10$^{-8}$ (e.g., \citealt{Agundez2013}). These molecules are substantially enhanced in hot cores and hot corinos (e.g., \citealt{Schoier2002,Tercero2010,Esplugues2013}). The abundances derived in L483 for SO, SO$_2$, and OCS are $5\times10^{-9}$, $10^{-10}$, $4\times10^{-10}$, respectively, which are in line with the values typically found in cold dense clouds like TMC-1 \citep{Matthews1987,Lique2006,Cernicharo2011}, L1544 \citep{Vastel2018b}, and B1-b \citep{Fuente2016}.

\begin{figure*} [!ht]
\centering
\includegraphics[angle=0,width=\textwidth]{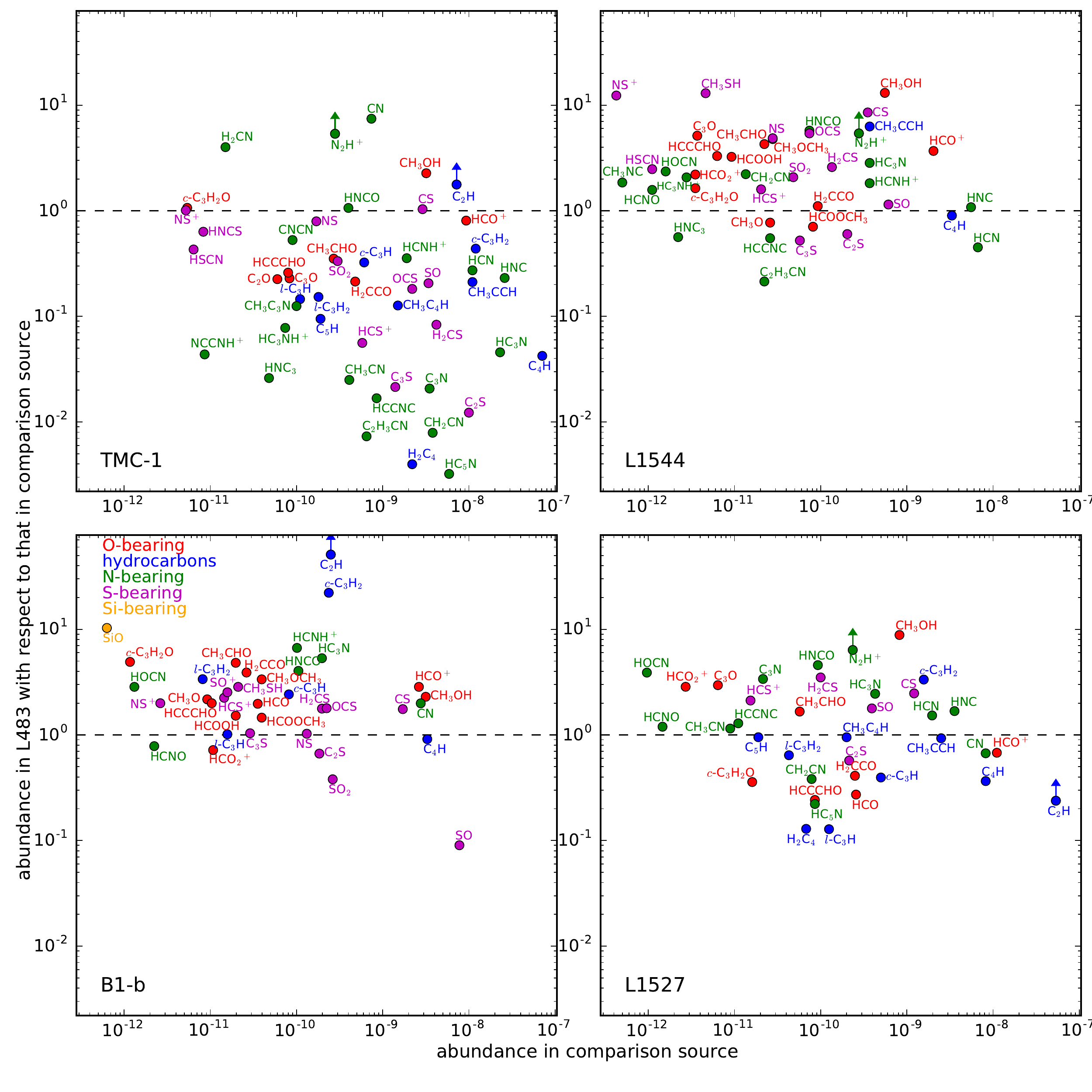}\caption{Fractional abundances relative to H$_2$ in L483 are compared with those in another four reference sources: TMC-1, L1544, B1-b, and L1527. The abundance ratios between L483 and the reference source are plotted against the fractional abundance in the corresponding reference source. Upper and lower limits are indicated by arrows. Different chemical families are plotted in different colors (see legend in bottom-left panel). Beam-averaged column densities derived from single-dish observations in the reference sources are taken from the literature. TMC-1: \cite{Adande2010,Agundez2013,Agundez2015b,Agundez2018b,Gratier2016,Loison2016,Loison2017,Cernicharo2018}; Marcelino et al. (in preparation). L1544: \cite{Caselli2002,Gupta2009,Marcelino2009,Marcelino2010,Vastel2014,Vastel2015,Vastel2016,Vastel2018a,Vastel2018b,Jimenez-Serra2016,Quenard2017,Cernicharo2018}. B1-b: \cite{Marcelino2005,Marcelino2009,Marcelino2010,Agundez2008,Oberg2010,Cernicharo2012,Loison2016,Loison2017,Fuente2016,Widicus-Weaver2017}; Marcelino et al. (in preparation). L1527: \cite{Sakai2008b,Marcelino2009,Marcelino2010,Araki2017,Yoshida2019}. Adopted H$_2$ column densities (needed to convert molecular column densities to fractional abundances) are $1\times10^{22}$ cm$^{-2}$ for TMC-1 \citep{Cernicharo1987}, $5.4\times10^{22}$ cm$^{-2}$ for L1544 \citep{Jimenez-Serra2016}, $7.6\times10^{22}$ cm$^{-2}$ for B1-b \citep{Daniel2013}, and $2.8\times10^{22}$ cm$^{-2}$ for L1527 \citep{Jorgensen2002}.} \label{fig:abundances_comparison}
\end{figure*}

We also detected HNCS and its metastable isomer HSCN, with abundances of several times 10$^{-12}$ and a HNCS/HSCN abundance ratio of approximately two. These values are very similar to those found in TMC-1, where the HNCS/HSCN ratio is approximately one \citep{Adande2010}. Clearly these species have a different behavior from their oxygen analogs, in which case the most stable isomer HNCO is about 100 times more abundant than the metastable isomer HOCN. This marked differentiation in the abundance ratios of the sulfur and oxygen cases is also seen in TMC-1 \citep{Marcelino2010,Adande2010}.

A last aspect concerning S-bearing molecules that is worth commenting on  is the detection of the cations SO$^+$ and NS$^+$, for which we find abundance ratios of SO/SO$^+$ = 125 and NS/NS$^+$ = 26, which are in line with the values typically found in cold dense clouds. For example, the SO/SO$^+$ ratio is around 100 in cold dense cores \citep{Turner1995,Turner1996}. In B1-b, the SO/SO$^+$ ratio is  approximately 250 when the abundance of SO is evaluated from $^{33}$SO \citep{Fuente2016}. The value of about 1000 given by these authors is probably too high as it is obtained from S$^{18}$O, for which oxygen fractionation is important \citep{Loison2019b}. In the case of NS$^+$, the NS/NS$^+$ ratio has been found to be remarkably uniform, in the range 29-52, across different starless and protostellar cores \citep{Cernicharo2018}.
 
Overall, the inventory of S-bearing molecules observed in L483 in the $\lambda$ 3 mm band and the abundances derived do not differ much from what is found in prestellar cores like TMC-1 and L1544 and low-mass protostellar cores like B1-b (see Sect.~\ref{sec:comparison}).

\subsection{Comparative chemical composition} \label{sec:comparison}

A pertinent question concerning the chemical composition of dense cores is whether the evolution of a source along the star-formation process drives a change in the chemical composition of the cloud and whether this change is somehow universal. In other words, to what extent is the chemical composition of a dense core determined by the particular evolutionary stage of the object$?$ The exhaustive chemical characterization of L483 provided by this line survey makes it possible to perform a comparison with other sources in nearby evolutionary stages. There are not many such sources for which the chemical composition has been exhaustively characterized however; that is, there are few unbiased line surveys. Here we have selected four sources, two of them with no sign of star formation (TMC-1 and L1544) and two that contain protostellar activity (B1-b and L1527). The object TMC-1 is a starless dark cloud in an earlier evolutionary stage than the prestellar phase, with no evidence of ongoing gravitational collapse (e.g., \citealt{Pratap1997}). The source L1544 however is a prestellar core with evidence of infall and a marked density gradient, with a high density ($>$10$^6$ cm$^{-3}$) and low temperature ($<$10 K) at the core center (e.g., \citealt{Caselli2012}). The dark cloud Barnard 1b (B1-b) harbors two extremely young protostellar objects separated by 20$''$ \citep{Hirano1999}, one of which is a candidate for the first hydrostatic core while the other is consistent with a very early Class\,0 source with an incipient hot corino \citep{Marcelino2018b}. Finally, L1527 is a dense core containing an embedded IRAS source, consistent with a protostar in the Class\,0 stage (e.g., \citealt{Sakai2008a}). Therefore, the four sources can be ordered in a evolutionary sequences as TMC-1 $\rightarrow$ L1544 $\rightarrow$ B1-b $\rightarrow$ L1527, with L483 being in a similar evolutionary stage to L1527.

To compare the chemical composition of L483 with that of TMC-1, L1544, B1-b, and L1527 we have collected molecular column densities from the literature. The four sources are included in the IRAM 30m Large Program ASAI \citep{Lefloch2018}, although at the time of writing, column densities have not been systematically reported. We focus on column densities derived from single-dish data (mostly IRAM 30m, Nobeyama 45m, and GBT), meaning that the comparison is made between abundances averaged over a similar scale, typically 15-45$''$ in diameter. In the case of TMC-1, column densities are mostly based on the line survey carried out with the Nobeyama 45m telescope \citep{Ohishi1998}, recently revised by \cite{Gratier2016}. For L1544 and B1-b, most data are based on IRAM 30m observations, some of which were taken in the context of ASAI (e.g., \citealt{Fuente2016,Vastel2018b}). In the case of L1527, column densities are mostly taken from the $\lambda$ 3 mm line survey carried out with the Nobeyama 45m telescope recently published by \cite{Yoshida2019}. All the literature sources used are given in the caption of Fig.~\ref{fig:abundances_comparison}.

In Fig.~\ref{fig:abundances_comparison} we compare the abundances of the molecules detected in the $\lambda$ 3 mm line survey of L483 with those in each of the four comparison sources: TMC-1, L1544, B1-b, and L1527. We compare fractional abundances relative to H$_2$ rather than column densities because this is a more meaningful quantity when comparing between sources with different column densities of material. We however note that the adopted column density of H$_2$ in each source becomes crucial as changes on it cause all relative abundances to shift up or down by the same amount, thus making all molecules appear more or less overabundant (or underabundant) in L483 than in the comparison source. A quick look at Fig.~\ref{fig:abundances_comparison} shows that, in general, molecules in L483 are underabundant with respect to TMC-1, overabundant with respect to L1544 and B1-b, and have abundances of the same order as in L1527. Assuming that the adopted H$_2$ column densities are not drastically different from the true ones, this behavior could be explained in terms of the specific physical and chemical properties of each comparison source. The source TMC-1 is known to be particularly rich in molecules, especially in unsaturated organic molecules like carbon chains. Still, it is not clear whether such chemical richness is the result of a very peculiar history of the source or is a short-lived feature that is observed in TMC-1 but not in other starless cores that may have different ages (see, e.g., \citealt{Agundez2013}). In any case, the lower molecular abundances observed in L483 with respect to TMC-1 are very likely a consequence of the exceptional chemical richness of the latter. The higher abundances derived in L483 compared to L1544 and B1-b may result from the fact that in these two latter sources the high volume densities, together with the very low temperatures, have accelerated the condensation of species onto dust grains resulting in an overall depletion of all gas-phase molecules except H$_2$. In the case of L1544, this phenomenon is well documented (e.g., \citealt{Caselli1999}). One may be tempted to interpret this behavior in evolutionary terms by saying that the chemical richness present at some moment of the starless phase is suppressed to an important extent during the prestellar and very early protostellar phases to then reappear to some degree at the Class\,0 stage. We however warn that this interpretation is not sufficiently robust because of the low statistics of sources analyzed here.

\begin{figure*}
\centering
\includegraphics[angle=0,width=\textwidth]{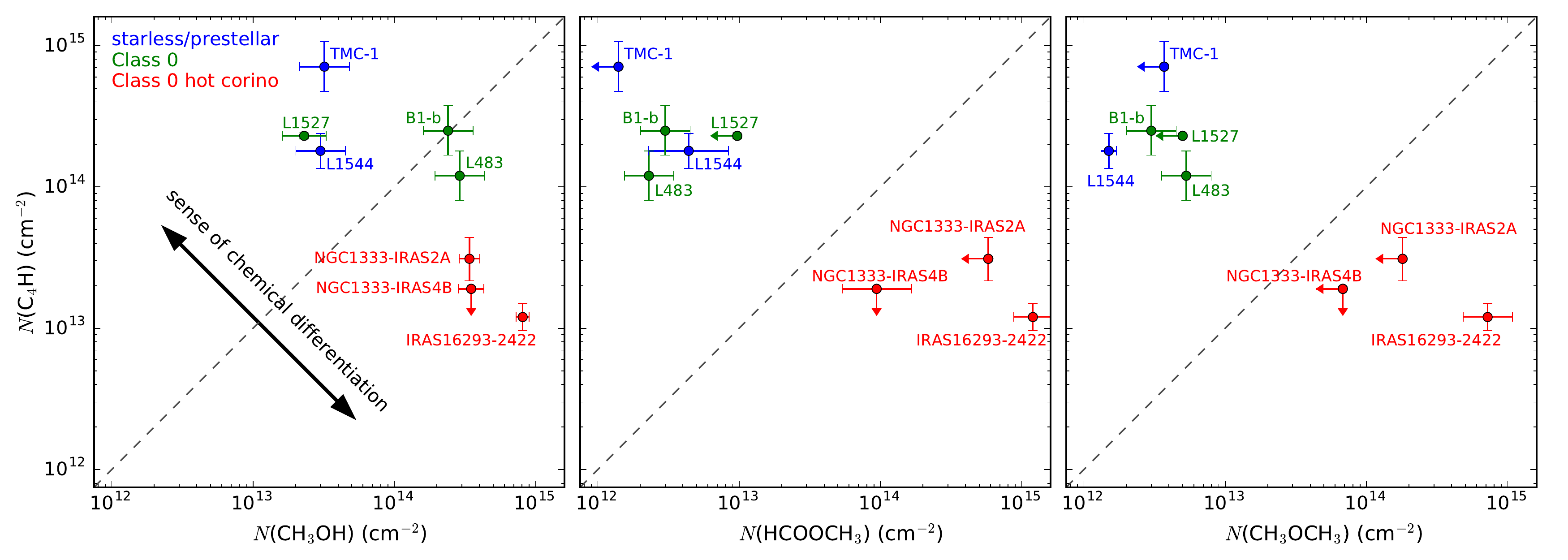}
\caption{Column density of C$_4$H represented as a function of the column density of CH$_3$OH (left panel), HCOOCH$_3$ (middle panel), and CH$_3$OCH$_3$ (right panel) in various starless/prestellar cores, Class\,0 sources, and hot corinos. All column densities are beam-averaged values obtained with single-dish telescopes. See references in the caption of Fig.~\ref{fig:abundances_comparison} for the column densities of C$_4$H and CH$_3$OH in TMC-1, L1544, B1-b, and L1527. Column densities of HCOOCH$_3$ and CH$_3$OCH$_3$ are taken from Marcelino et al. (in preparation) for TMC-1 (upper limits), \cite{Jimenez-Serra2016} for L1544, \cite{Cernicharo2012} for B1-b, and \cite{Yoshida2019} for L1527 (upper limits). For the three hot corino sources, column densities for C$_4$H are from \cite{Sakai2009a}, for CH$_3$OH from \cite{Maret2005}, and for HCOOCH$_3$ and CH$_3$OCH$_3$ from \cite{Cazaux2003} (values have been averaged for a 10$''$ source size) and from \cite{Bottinelli2007}.} \label{fig:c4h_vs_ch3oh}
\end{figure*}

Apart from the systematic abundance enhancement (or decline) of all molecules seen between L483 and some of the comparison sources, it is interesting to discuss whether or not there is some chemical differentiation between L483 and the other objects; that is, whether different types of molecules behave differently. The most clear chemical differentiation is seen between L483 and TMC-1 for two types of molecules. On the one hand, we have carbon chains comprising unsaturated hydrocarbons of the family of polyynes like C$_4$H, C$_5$H, and H$_2$C$_4$, N-bearing molecules of the family of cyanopolyynes like C$_3$N, HC$_3$N, and HC$_5$N, and the S-bearing molecules C$_2$S and C$_3$S. On the other hand, there are several molecules which are present in cold dense clouds but tend to experience important abundance enhancements in hot cores and hot corinos. In this group we can include species like CH$_3$OH, HCOOCH$_3$, CH$_3$OCH$_3$, HCOOH, CH$_3$CHO, HNCO, SO, SO$_2$, OCS, HNCS, and CH$_3$SH. From Fig.~\ref{fig:abundances_comparison} it is clear that when moving from TMC-1 to L483, carbon chains experience a more drastic abundance depletion than hot core-like molecules. In other words, the relative importance of carbon chains with respect to hot core-like molecules is higher in TMC-1 than in L483. Along the same lines, \cite{Law2018} recently found that carbon chains are significantly less abundant in dense cores around Class\,0/I low-mass protostars than in TMC-1. A similar behavior can be tentatively found when comparing L483 with L1544. From Fig.~\ref{fig:abundances_comparison} it can be seen that, in general, hot core-like molecules tend to increase their abundance more than carbon chains when moving from L1544 to L483; although this does not happen systematically for all species of each type and in any case the chemical differentiation is less clear than in the case of TMC-1.

Therefore, the relative importance of hot core-like molecules with respect to carbon chains is higher in L483 than in TMC-1 and L1544 (tentative for the latter). When comparing L483 with B1-b and L1527 we do not see any clear chemical differentiation between carbon chains and hot core-like molecules or any other type of chemical family; that is, when moving from any of these two sources to L483, there is no obvious differentiation in the abundance variation of different types of molecules. In this sense, L483 is significantly different from TMC-1, barely different from L1544, and quite similar to B1-b and L1527 regarding the chemical composition of the ambient cloud. This conclusion is based on an overall look at the chemical composition of each source. In any case, the fact that we see a chemical differentiation between L483 and the starless/prestellar sources TMC-1 and L1544 and a lack of it between L483 and the protostellar sources B1-b and L1527 raises the question of whether starless and prestellar sources on the one side, and protostellar sources on the other, have differences in the chemical composition of their host cloud which could be associated to the switch on of a protostar at the center of the core.
To get insight into the potential chemical differentiation between sources with different evolutionary stages, we now move from looking at the overall chemical composition to the use of chemical indicators. We first focus on the column density of C$_4$H versus that of CH$_3$OH, the former being a typical carbon chain, the latter being a classical hot core-like molecule. In the left panel of Fig.~\ref{fig:c4h_vs_ch3oh} we plot these two quantities for a few well-studied dense cores, some of which have no protostar while others have a Class\,0 object. We see that the Class\,0 sources B1-b, L1527, and L483 cluster around the same region as the starless/prestellar sources, while the three other Class\,0 sources classified as hot corinos based on single-dish observations of abundant and warm COMs around the protostar \citep{Cazaux2003,Bottinelli2004,Bottinelli2007} appear in a different region of the diagram.

Recently, the C$_4$H/CH$_3$OH ratio was surveyed in a sample of 40 embedded protostars \citep{Graninger2016a,Graninger2016b,Lindberg2016}. These studies mostly populate the region of low column densities along the band of C$_4$H/CH$_3$OH close to one, where B1-b, L1527, and L483 reside, bridging part of the gap between these three latter sources and the hot corinos, which still appear chemically differentiated from the rest of sources. The underlying cause of such chemical differentiation between sources that are supposed to be in a similar, relatively short-lived (a few 10$^4$ yr; \citealt{Andre2000}) evolutionary stage like the Class\,0 one remains unclear. This may be related to the way the switch on of the protostar affects the ambient cloud. In some sources the cloud would preserve much of the characteristics of the starless/prestellar phase, displaying high C$_4$H/CH$_3$OH ratios and being rich in carbon chains, while in others the chemical content of the ambient cloud would be strongly processed by the injection of energy from the protostar, thus showing low C$_4$H/CH$_3$OH ratios and abundant COMs (hot corinos). However, a deeper understanding of why this would happen is still lacking. A possible explanation of the chemical differentiation between Class\,0 sources rich in carbon chains and rich in COMs (hot corinos) has been suggested by \cite{Sakai2008a}. These authors proposed that sources rich in carbon chains would have had a short lifetime of the prestellar collapse, in which case ices would be rich in CH$_4$ and the evaporation of these during the switch on of the protostar would lead to the formation of carbon chains by gas phase chemistry. On the other hand, hot corinos would result from a longer prestellar collapse phase, meaning that ices would be rich in CO, which upon hydrogenation and further processing driven by protostar heating would lead to abundant COMs in the gas phase. Chemical models following this idea have been carried out \citep{Aikawa2008,Aikawa2012,Hassel2008,Hassel2011} but still a definitive theoretical validation of the chemical differentiation between Class\,0 sources rich in carbon chains and hot corinos is lacking. More recently, a survey of the C$_2$H/CH$_3$OH in 36 Class\,0/I protostars by \cite{Higuchi2018} pointed to the environmental effect of the dense core being located in a more inner or outer region of the molecular cloud complex as a possible driver of the diversity of C$_2$H/CH$_3$OH ratios observed.

\begin{figure}
\centering
\includegraphics[angle=0,width=\columnwidth]{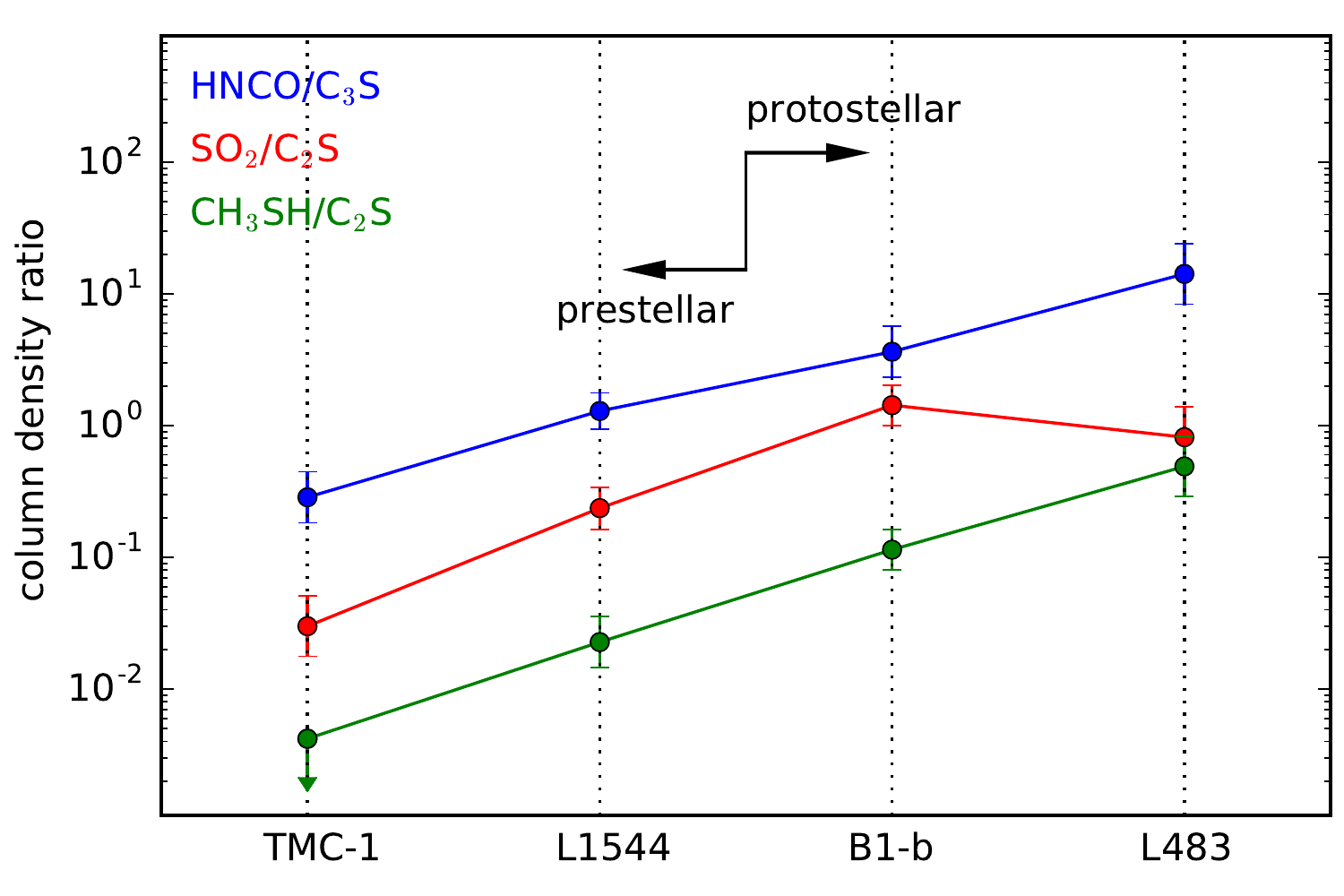}
\caption{Some column density ratios which are potentially good chemical evolutionary indicators of the prestellar/protostellar transition. The upper limit to the column density of CH$_3$SH in TMC-1 is taken from Marcelino et al. (in preparation).} \label{fig:column_density_ratios}
\end{figure}

We also investigated whether the use of more specific proxies of COMs like HCOOCH$_3$ and CH$_3$OCH$_3$, rather than the more generic one of CH$_3$OH, could provide some insight into the chemical differentiation between starless/prestellar and protostellar sources. Therefore, in the middle and right panels of Fig.~\ref{fig:c4h_vs_ch3oh} we show the column density of C$_4$H versus that of HCOOCH$_3$ and CH$_3$OCH$_3$, respectively, for the same sources considered before. We see that the Class\,0 sources B1-b, L1527, and L483 cluster around the same region in the diagram as the starless/prestellar sources, while they can be clearly differentiated from the Class\,0 hot corino sources. Therefore, the use of the CH$_4$/HCOOCH$_3$ and CH$_4$/CH$_3$OCH$_3$ chemical indicators makes it impossible to distinguish L483, B1-b, and L1527 from starless and prestellar sources. In other words, COMs are present at similar level in all these sources, despite the fact that a protostar has switched on in some of them while not in others.

Pursuing the idea of using column density ratios between different pairs of molecules as indicators of chemical evolution along the prestellar-protostellar transition, we have examined various potential indicators using molecules for which we have column densities at hand for TMC-1, L1544, B1-b, and L483. In Fig.~\ref{fig:column_density_ratios} we plot some of these ratios for the above four sources, which have been ordered according to their evolutionary status. The ratios have been chosen between different pairs of molecules so that one is a hot core-like molecule and the other a carbon chain. There is a clear trend in which the HNCO/C$_3$S, SO$_2$/C$_2$S, and CH$_3$SH/C$_2$S ratios increase with the age of the object, with a marked difference at the prestellar-protostellar transition. Therefore, these are promising tracers of chemical evolution of the cloud along the protostar formation process. The observed trends however have to be considered as tentative for the moment due to the low number of sources included.

\subsection{Isotopic ratios} \label{sec:isotopic_ratios}

The high sensitivity of the line survey permitted us to detect a good number of isotopologs substituted with rare isotopes (D, $^{13}$C, $^{15}$N, $^{18}$O, $^{17}$O, $^{34}$S, and $^{33}$S), and even some doubly substituted species. Isotopolog abundance ratios have been obtained from the column densities derived and are presented in Tables~\ref{table:isotopic_ratios} and \ref{table:deuterated_ratios}. The uncertainties in these ratios are estimated to be 30~\%. We note that although the estimated uncertainties in the column densities are of the order of 50~\% (see above in Sect.~\ref{sec:column_densities}), we can reasonably expect lower uncertainties for isotopic ratios. An important part of the uncertainty that affects the determination of a column density is dominated by the error in the rotational temperature, which becomes less important when computing column density ratios between different isotopologs of a given molecule because these are expected to have similar excitation conditions. This latter statement has been verified in the few cases in which we have enough lines to estimate the rotational temperature for different isotopologs of the same molecule. For example, for $c$-HCC$^{13}$CH and $c$-C$_3$HD we derive rotational temperatures of 4.1 $\pm$ 0.3 and 4.1 $\pm$ 0.2, respectively (see Table~\ref{table:column_densities}). In the case of the isotopologs of HC$_3$N, rotational temperatures could be derived with good accuracy for HC$^{13}$CCN, HCC$^{13}$CN, and DC$_3$N and all them fall in the narrow range 7.5-9.1 K. 

The abundance ratios between isotopologs can be merely statistical, in which case they would reflect the corresponding isotopic ratio of the parental cloud, or they can be affected by fractionation processes, in which case they would deviate from the isotopic ratios of the ambient cloud. These deviations or anomalies can consist in either an enrichment or a dilution in the heavy isotope, or even in different abundances for isotopologs resulting from different isotopic substitutions of the same type of atom (e.g., $^{13}$CCH and C$^{13}$CH). In L483, some molecules show no evidence of fractionation while other species show the aforementioned anomalies. In general, at low temperatures isotopic exchange gas phase reactions tend to enhance the abundance of the heavy isotopolog due to the lowering of the zero-point energy, as first suggested by \cite{Solomon1973} to explain the surprising large abundance of DCN in the Orion molecular cloud. Such considerations allow for us to account for enrichments of deuterated molecules of several orders of magnitude, compared to the elemental D/H ratio of 1.5$\times$10$^{-5}$, due to differences in the zero-point energy of several hundreds of degrees Kelvin \citep{Gerin1987,Roueff2003,Roueff2005,Albertsson2013}. The isotopic enhancement is expected to decrease for heavier atoms as the zero-point energy variation is much smaller, typically of the order of 30 K for $^{12}$C/$^{13}$C-containing molecules, 20 K for $^{14}$N/$^{15}$N species, and 7 K in the case of $^{32}$S/$^{34}$S. In the absence of any other specific mechanism, fractionation is expected to decrease in this order $^{13}$C, $^{15}$N, and $^{34}$S, which in general is in agreement with observational results.

In the rest of the section we discuss the available information in the literature on the isotopic ratios in the local interstellar medium (ISM), the molecular isotopic ratios derived in L483 involving the rare isotopes $^{13}$C, $^{15}$N, $^{18}$O, $^{17}$O, $^{34}$S, and $^{33}$S, and the results obtained for deuterated molecules.

\subsubsection{Local ISM isotopic ratios} \label{sec:local_ism_isotopic_ratios}

Measurements of isotopic ratios in the local ISM have been carried out over the years, mostly by observing molecules at millimeter wavelengths in diffuse and dense clouds, but also from optical observations of molecules in diffuse media.

The $^{12}$C/$^{13}$C ratio has been one of the most studied. Reported values are 59 $\pm$ 2 (from HCO$^+$, HCN, and HNC mm data of diffuse clouds; \citealt{Lucas1998}), 69 $\pm$ 6 (from CO and H$_2$CO mm data of dense clouds; \citealt{Wilson1999}), 68 $\pm$ 15 (from CN mm observations of dense clouds; \citealt{Milam2005}), 70 $\pm$ 2 (from optical data of CO in diffuse clouds; \citealt{Sheffer2007}), 76 $\pm$ 2 (from CH$^+$ optical measurements of diffuse clouds; \citealt{Stahl2008}), and 74.4 $\pm$ 7.6 (from CH$^+$ data at visible wavelengths; \citealt{Ritchey2011}). Although fractionation may be an issue for some of these molecules, it is worth noting that all these studies show good agreement, with $^{12}$C/$^{13}$C ratios in the range 59-76.

The study of the $^{14}$N/$^{15}$N ratio has also been the subject of intense research activity. From HCN millimeter data of dense clouds, \citet{Dahmen1995} established a Galactic gradient with a value of 450 for the local ISM, while \citet{Wilson1999} reported an average value of 388 $\pm$ 32. \citet{Adande2012} used CN and HNC millimeter data and found 290 $\pm$ 40 for the local ISM. \citet{Lucas1998} found a value of 237$^{+27}_{-21}$ in one local diffuse cloud, while more recently, \citet{Ritchey2015} found 274 $\pm$ 18 from optical measurements of CN in diffuse clouds. Literature $^{14}$N/$^{15}$N ratios differ by up to a factor of two, with values in the range 237-450.

Oxygen and sulfur isotopic ratios have not been so widely studied as those of carbon and nitrogen. Reported values of the $^{16}$O/$^{18}$O ratio in the local ISM are 672 $\pm$ 110 (from HCO$^+$ mm data of one local diffuse cloud; \citealt{Lucas1998}) and 557 $\pm$ 30 (from H$_2$CO mm data of dense clouds; \citealt{Wilson1999}). The $^{18}$O/$^{17}$O ratio is found to be 3.6 $\pm$ 0.2 by \cite{Wilson1999} and 4.16 $\pm$ 0.09 by \cite{Wouterloot2008} from CO mm data, values that translate to $^{16}$O/$^{17}$O ratios of 2005 $\pm$ 155 and 2317 $\pm$ 134, respectively, adopting a $^{16}$O/$^{18}$O ratio of 557 $\pm$ 30 \citep{Wilson1999}. Concerning sulfur, \citet{Chin1996} found $^{32}$S/$^{34}$S = 24.4 $\pm$ 5.0 and $^{32}$S/$^{33}$S = 153 $\pm$ 40 from CS mm observations in local star-forming regions. Further reported values of the $^{32}$S/$^{34}$S ratio in the local ISM are 19 $\pm$ 8 (from CS in diffuse clouds; \citealt{Lucas1998}) and $\sim$22 \citep{Wilson1999}, values which are in agreement with that of \citet{Chin1996}.

\begin{table*}
\caption{Direct isotopic ratios} \label{table:isotopic_ratios}
\small
\centering
\begin{tabular}{lcccccccc}
\hline \hline
\multicolumn{1}{l}{Ratio} & \multicolumn{1}{c}{L483} & \multicolumn{1}{c}{TMC-1} & \multicolumn{1}{c}{L1521E} & \multicolumn{1}{c}{L1521B} & \multicolumn{1}{c}{L134N} & \multicolumn{1}{c}{L1544} & \multicolumn{1}{c}{B1-b} & \multicolumn{1}{c}{L1527} \\
\hline
\multicolumn{9}{c}{$^{12}$C/$^{13}$C} \\
\hline
C$^{17}$O/$^{13}$C$^{17}$O                      & 42 $\pm$ 13 & ... & ... & ... & ... & ... & ... & ... \\
C$_2$H/$^{13}$CCH                                   & $>$162 & $>$250 $^a$ & ... & ... & ... & ... & ... & 210 $\pm$ 60 $^b$\\
C$_2$H/C$^{13}$CH                                   & $>$70 & $>$170 $^a$ & ... & ... & ... & ... & ... & 140 $\pm$ 40 $^b$ \\
$c$-C$_3$H$_2$/$c$-HCC$^{13}$CH~~~$\times$2   & 106 $\pm$ 32 & ... & ... & ... & ... & ... & ... & 82 $\pm$ 16 $^b$ \\
$c$-C$_3$H$_2$/$c$-HC$^{13}$CCH         & 458 $\pm$ 138 & ... & ... & ... & ... & ... & ... & 200 $\pm$ 30 $^b$ \\ 
$c$-C$_3$HD/$c$-H$^{13}$CCCD              & 112 $\pm$ 34 & ... & ... & ... & ... & ... & ... & ... \\
$c$-C$_3$HD/$c$-HCC$^{13}$CD              & 87 $\pm$ 26 & ... & ... & ... & ... & ... & ... & ... \\
CH$_3$CCH/$^{13}$CH$_3$CCH               & 60 $\pm$ 18 & ... & ... & ... & ... & ... & ... & ... \\
CH$_3$CCH/CH$_3$$^{13}$CCH               & 53 $\pm$ 16 & ... & ... & ... & ... & ... & ... & ... \\
CH$_3$CCH/CH$_3$C$^{13}$CH               & 58 $\pm$ 17 & ... & ... & ... & ... & ... & ... & ... \\
CH$_2$DCCH/$^{13}$CH$_2$DCCH          & 22 $\pm$ 7 & ... & ... & ... & ... & ... & ... & ... \\
HC$^{15}$N/H$^{13}$C$^{15}$N                & 34 $\pm$ 10 & ... & ... & ... & ... & ... & ... & ... \\
H$^{15}$NC/H$^{15}$N$^{13}$C                & 29 $\pm$ 9 & ... & ... & ... & ... & ... & ... & ... \\
HC$_3$N/H$^{13}$CCCN                           & 91 $\pm$ 27 & 79 $\pm$ 11 $^c$ & ... & 117 $\pm$ 16 $^d$ & 61 $\pm$ 9 $^d$ & ... & ... & 86.4 $\pm$ 1.6 $^e$ \\
HC$_3$N/HC$^{13}$CCN                           & 93 $\pm$ 28 & 75 $\pm$ 10 $^c$ & ... & 117 $\pm$ 16 $^d$ & 94 $\pm$ 26 $^d$ & ... & ... & 85.4 $\pm$ 1.7 $^e$ \\
HC$_3$N/HCC$^{13}$CN                           & 79 $\pm$ 24 & 55 $\pm$ 7 $^c$ & ... & 76 $\pm$ 6 $^d$ & 46 $\pm$ 9 $^d$ & ... & ... & 64.2 $\pm$ 1.1 $^e$ \\
DC$_3$N/DC$^{13}$CCN                           & 30 $\pm$ 9 & ... & ... & ... & ... & ... & ... & ... \\
HNCO/HN$^{13}$CO                                  & 62 $\pm$ 19 & ... & ... & ... & ... & ... & ... & ... \\
C$^{34}$S/$^{13}$C$^{34}$S                     & 58 $\pm$ 18 & ... & ... & ... & ... & ... & ... & ... \\
H$_2$CS/H$_2$$^{13}$CS                        & 113 $\pm$ 34 & 79 $\pm$ 26 $^{f \star}$ & ... & ... & ... & ... & ... & ... \\
C$_2$S/C$^{13}$CS                                  & 28 $\pm$ 8 & 54 $\pm$ 5 $^{g \star}$ & 51 $\pm$ 13 $^{g \star}$ & ... & ... & ... & ... & ... \\ 
C$_2$S/$^{13}$CCS                                  & $>$25 & 230 $\pm$ 130 $^{g \star}$ & $>$130 $^{g \star}$ & ... & ... & ... & ... & ... \\ 
\hline
\multicolumn{9}{c}{$^{14}$N/$^{15}$N} \\
\hline
H$^{13}$CN/H$^{13}$C$^{15}$N              & 321 $\pm$ 96 & ... & ... & ... & ... & ... & ... & ... \\
HC$_3$N/HC$_3$$^{15}$N                      & 490 $\pm$ 147 & 257 $\pm$ 54 $^{h \star}$ & ... & ... & ... & 400 $\pm$ 20 $^i$ & ... & ... \\
N$_2$H$^+$/$^{15}$NNH$^+$                 & $>$747 & ... & ... & ... & ... & 1110 $\pm$ 240 $^j$ & $>$600 $^k$ & ... \\
N$_2$H$^+$/N$^{15}$NH$^+$                 & $>$450 & ... & ... & ... & 670$^{+150}_{-230}$ $^l$ & 1050 $\pm$ 220 $^j$ & 400$^{+100}_{-65}$ $^j$ & ... \\
\hline
\multicolumn{9}{c}{$^{16}$O/$^{18}$O} \\
\hline
HNCO/HNC$^{18}$O                                 & 232 $\pm$ 70 & ... & ... & ... & ... & ... & ... & ... \\
SO/S$^{18}$O                                            & 158 $\pm$ 47 & ... & ... & ... & ... & ... & ... & ... \\
\hline
\multicolumn{9}{c}{$^{16}$O/$^{17}$O} \\
\hline
SO/S$^{17}$O                                            & 128 $\pm$ 38 & ... & ... & ... & ... & ... & ... & ... \\
\hline
\multicolumn{9}{c}{$^{32}$S/$^{34}$S} \\
\hline
$^{13}$CS/$^{13}$C$^{34}$S                   & 31 $\pm$ 9 & ... & ... & ... & ... & ... & ... & ... \\
H$_2$CS/H$_2$C$^{34}$S                      & 29 $\pm$ 9 & ... & ... & ... & ... & ... & ... & ... \\
C$_2$S/C$_2$$^{34}$S                           & 17 $\pm$ 5 & ... & ... & ... & ... & ... & ... & ... \\
SO/$^{34}$SO                                          & 31 $\pm$ 9 & ... & ... & ... & ... & ... & ... & ... \\
SO$_2$/$^{34}$SO$_2$                          & 31 $\pm$ 9 & ... & ... & ... & ... & ... & ... & ... \\
NS/N$^{34}$S                                          & 22 $\pm$ 7 & ... & ... & ... & ... & ... & ... & ... \\
HCS$^+$/HC$^{34}$S$^+$                      & 21 $\pm$ 6 & ... & ... & ... & ... & ... & ... & ... \\
\hline
\multicolumn{9}{c}{$^{32}$S/$^{33}$S} \\
\hline
SO/$^{33}$SO                                          & 151 $\pm$ 45 & ... & ... & ... & ... & ... & ... & ... \\
\hline
\end{tabular}
\tablenoteb{\\
$^a$~\cite{Sakai2010}. $^b$~\cite{Yoshida2019}. $^c$~\cite{Takano1998}. $^d$~\cite{Taniguchi2017}. $^e$~\cite{Araki2016}. $^f$~\cite{Liszt2012}. $^g$~\cite{Sakai2007}. $^h$~\cite{Taniguchi-Saito2017}. $^i$~\cite{Hily-Blant2018}. $^j$~\cite{Bizzocchi2013}. $^k$~\cite{Daniel2013}. $^l$~\cite{Redaelli2018}. \\
$^{\star}$ Isotopic ratio is not direct but derived through the double isotope method.
}
\end{table*}

\begin{figure*}
\centering
\includegraphics[width=0.95\textwidth]{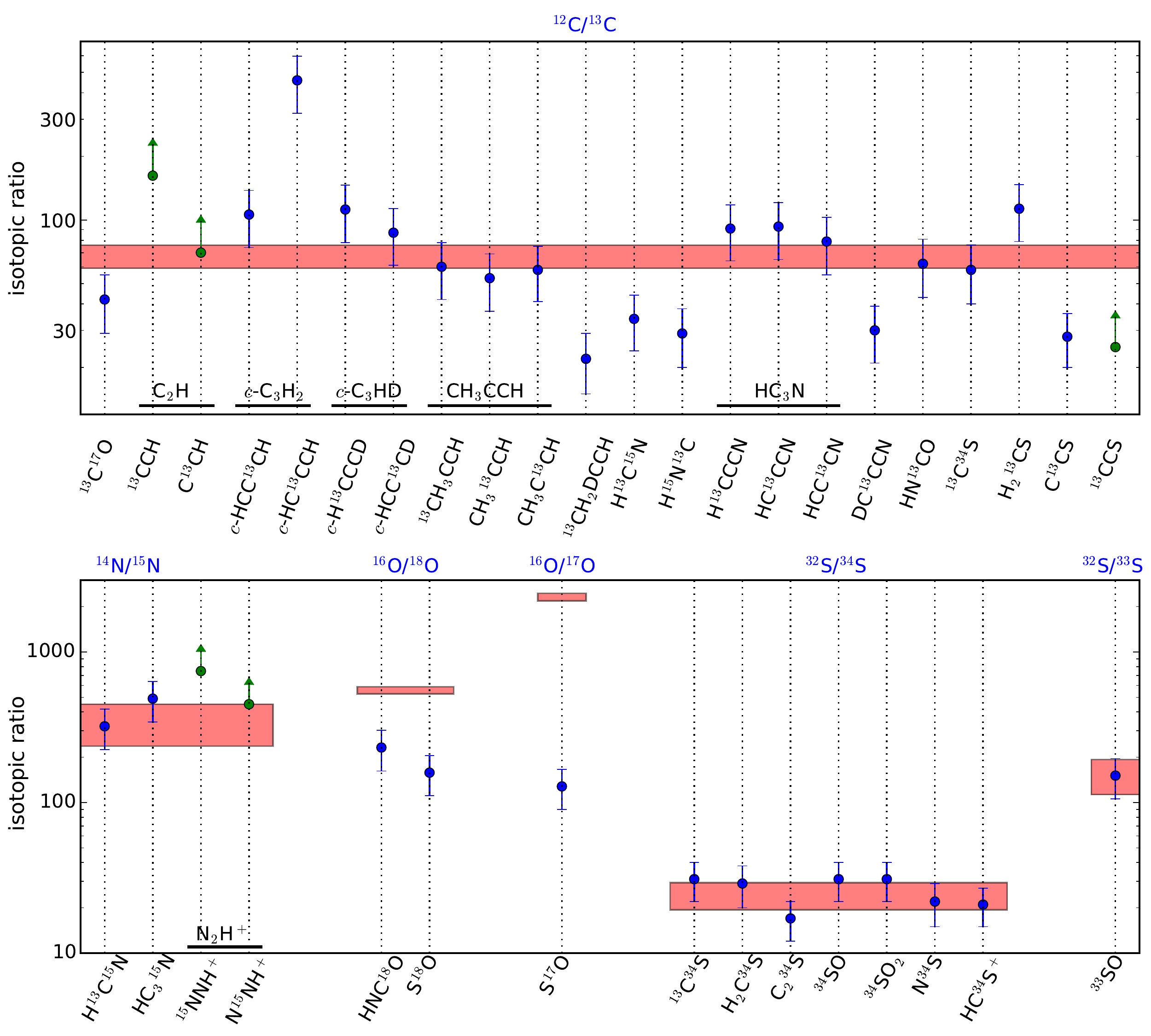}
\caption{Direct isotopic ratios derived in L483 (see Table~\ref{table:isotopic_ratios}) are compared with local ISM values. Lower limits are indicated in green. In the case of $c$-HCC$^{13}$CH, the $c$-C$_3$H$_2$/$c$-HCC$^{13}$CH ratio is multiplied by two to account for the enhanced probability of substitution of either of the two equivalent carbon atoms of $c$-C$_3$H$_2$ (see Table~\ref{table:isotopic_ratios}). This way, the plotted value can be directly compared with the $^{12}$C/$^{13}$C ratio in the local ISM. Isotopic ratios representative of the local ISM are indicated by red horizontal rectangles and are taken as $^{12}$C/$^{13}$C = 59-76, $^{14}$N/$^{15}$N = 237-450, $^{16}$O/$^{18}$O = 557 $\pm$ 30, $^{16}$O/$^{17}$O = 2317 $\pm$ 134, $^{32}$S/$^{34}$S = 24.4 $\pm$ 5.0, $^{32}$S/$^{33}$S = 153 $\pm$ 40 (see text in Sect.~\ref{sec:isotopic_ratios}).} \label{fig:isotopic_ratios}
\end{figure*}

\subsubsection{$^{12}$C/$^{13}$C in L483} \label{sec:isotopic_ratios_discussion_13c}

The column densities derived in L483 for the isotopologs with $^{13}$C evidence a variety of behaviors. In the top panel of Fig.~\ref{fig:isotopic_ratios} we compare the $^{12}$C/$^{13}$C ratios derived for different molecules with the $^{12}$C/$^{13}$C ratio in the local ISM. Some molecules have $^{12}$C/$^{13}$C ratios in agreement with the local ISM value while others are clearly affected by fractionation.

Carbon monoxide, for example, is slightly enriched in $^{13}$C, as indicated by the C$^{17}$O/$^{13}$C$^{17}$O ratio of 42 $\pm$ 13. We can extrapolate this result to the main $^{16}$O-containing species (something that is correct if $^{17}$O fractionation is not important for CO or if it behaves similarly in $^{12}$CO and $^{13}$CO) because chemical models find that CO should not be particularly affected by oxygen fractionation \citep{Loison2019b}. Carbon isotopic fractionation for CO can be explained as a result of the reaction between $^{13}$C$^+$ and $^{12}$CO which is favored at low temperatures to produce $^{12}$C$^+$ and $^{13}$CO, meaning that CO is enriched in the heavy isotope and C$^+$ is impoverished in $^{13}$C (\citealt{Langer1984}; but see updated exothermicities in \citealt{Mladenovic2014,Mladenovic2017}). This latter fact can lead to a depletion in $^{13}$C for molecules whose synthesis involves C$^+$.

The most dramatic depletion in $^{13}$C is seen for $c$-HC$^{13}$CCH (i.e., $c$-C$_3$H$_2$ with a $^{13}$C substituted in the central carbon atom), which is severely depleted in $^{13}$C, with a $^{12}$C/$^{13}$C of 458 $\pm$ 138. Curiously, the other $^{13}$C isotopolog of $c$-C$_3$H$_2$ substituted in the carbon bonded to hydrogen ($c$-HCC$^{13}$CH) does not suffer such a severe dilution in $^{13}$C, as its $^{12}$C/$^{13}$C ratio is 106 $\pm$ 32, that is, only slightly above the local ISM value. A similar result was recently found in L1527, where the $^{12}$C/$^{13}$C ratios for $c$-HC$^{13}$CCH and $c$-HCC$^{13}$CH were found to be 310 $\pm$ 80 and 122 $\pm$ 22, respectively \citep{Yoshida2015}, or 200 $\pm$ 30 and 82 $\pm$ 16, respectively, according to \cite{Yoshida2019}. It seems clear that $^{13}$C fractionation works differently for these two $^{13}$C isotopologs of $c$-C$_3$H$_2$ and that $^{13}$C dilution is much more favorable for $c$-HC$^{13}$CCH. Results for singly deuterated $c$-C$_3$H$_2$ ($c$-C$_3$HD) indicate little $^{13}$C fractionation when $^{13}$C is substituted in a carbon atom bonded to H or D (see top panel in Fig.~\ref{fig:isotopic_ratios}), in line with what is found for the nondeuterated species ($c$-HCC$^{13}$CH). The strong dilution in $^{13}$C observed for $c$-C$_3$H$_2$ when $^{13}$C is substituted in the central carbon atom ($c$-HC$^{13}$CCH) probably vanishes in the deuterated form (the tentative $c$-C$_3$HD/$c$-HC$^{13}$CCD ratio derived is 89), although a more secure detection of $c$-HC$^{13}$CCD is needed to draw firmer conclusions on this subject.

In the case of C$_2$H, we see an isotopic anomaly as the two $^{13}$C substituted species have significantly different column densities (see Table~\ref{table:column_densities}). The isotopolog C$^{13}$CH is found to be 2.3 times more abundant than $^{13}$CCH. Moreover, C$_2$H is likely to be affected by $^{13}$C dilution, as the C$_2$H/C$^{13}$CH and C$_2$H/$^{13}$CCH ratios are found to be $>$71 and $>$165, respectively. A similar result was found in TMC-1 and L1527, where the C$^{13}$CH/$^{13}$CCH ratio is $\sim$1.6 and the $^{12}$C/$^{13}$C ratios are significantly above the local ISM value \citep{Sakai2010,Yoshida2019}.

For CH$_3$CCH there is little or no fractionation when $^{13}$C is substituted in any of the three carbon atoms. However, its deuterated species CH$_2$DCCH does show a significant enrichment in $^{13}$C when it is substituted in the carbon atom of the methyl group, with a $^{12}$C/$^{13}$C ratio of 22 $\pm$ 7.

Nitriles show a variety of behaviors concerning carbon fractionation. For example, both HCN and HNC are enriched in $^{13}$C by a factor of about two with respect to the local ISM $^{12}$C/$^{13}$C ratio. This result is found for the $^{15}$N substituted species, but it can probably be extrapolated to the main $^{14}$N species as we do not find hints of nitrogen fractionation, at least for HCN (see Sect.~\ref{sec:isotopic_ratios_discussion_rest}). In L1498, \citet{Magalhaes2018} derived a similar enrichment in $^{13}$C for HCN, with a HCN/H$^{13}$CN ratio of 45 $\pm$ 3, while in TMC-1 and L1527, \citet{Liszt2012} found that HNC is only barely enriched in $^{13}$C, with HNC/HN$^{13}$C ratios in the range 33-72. In the case of HC$_3$N, $^{12}$C/$^{13}$C ratios are slightly above, but consistent with, the local ISM value. The three $^{13}$C substituted species of HC$_3$N have similar column densities, with HCC$^{13}$CN being slightly more abundant than the other two isotopologs. This feature has previously been observed in other cold dense clouds. We find H$^{13}$CCCN:HC$^{13}$CCN:HCC$^{13}$CN relative abundances of 1.0:1.0:1.2 in L483, while these ratios are 1.0:1.0:1.4 in TMC-1 and L1527 \citep{Takano1998,Araki2016}, 1.0:1.0:1.5 in L1521B, and 1.5:1.0:2.1 in L134N \citep{Taniguchi2017}. The deuterated species DC$_3$N shows an enrichment in $^{13}$C about three times greater than for HC$_3$N, as indicated by the $^{12}$C/$^{13}$C ratio derived for the isotopolog substituted with $^{13}$C in the middle carbon atom. As in the cases of $c$-C$_3$H$_2$ and CH$_3$CCH, it seems that $^{13}$C fractionation works differently for deuterated and nondeuterated species. The higher $^{13}$C fractionation observed for deuterated species compared to the nondeuterated ones is in agreement with expectations from the larger variation of the zero-point energy. The observation of $^{13}$C isotopologs of the longer cyanopolyynes HC$_5$N and HC$_7$N in TMC-1 indicate that there is little abundance variation among the different $^{13}$C isotopologs, although HC$_5$N seems to be more depleted in $^{13}$C than HC$_3$N and HC$_7$N \citep{Taniguchi2016,Burkhardt2018}. The other nitrogen-bearing molecule for which we have constraints on the $^{12}$C/$^{13}$C ratio is HNCO, which does not show $^{13}$C fractionation.

Regarding sulfur-bearing molecules, CS does not show $^{13}$C fractionation, as measured with the $^{34}$S species. This result can be safely extrapolated to the main $^{32}$S species as we do not see evidence of fractionation on $^{34}$S (see Sect.~\ref{sec:isotopic_ratios_discussion_rest}). It is interesting to note that \citet{Liszt2012} found that neither CS nor H$_2$CS is significantly fractionated in $^{13}$C in TMC-1 and L1527. We find that thioformaldehyde is only slightly diluted in $^{13}$C in L483, with a H$_2$CS/H$_2$$^{13}$CS ratio of 113 $\pm$ 34. On the other hand, C$_2$S shows a significant enrichment in $^{13}$C when $^{13}$C is substituted in the middle carbon atom. The C$_2$S/C$^{13}$CS ratio is 28 $\pm$ 8. \citet{Sakai2007} did not find a significant fractionation for C$^{13}$CS in TMC-1, although they found that C$^{13}$CS is 4.2 times more abundant than $^{13}$CCS, and that this latter isotopolog is heavily diluted in $^{13}$C. We cannot conclude whether or not in L483 $^{13}$CCS is significantly less abundant than C$^{13}$CS because the upper limit derived for the column density of $^{13}$CCS is of the order of the value found for C$^{13}$CS (see Table~\ref{table:isotopic_ratios} and Fig.~\ref{fig:isotopic_ratios}).

\subsubsection{Nitrogen, oxygen, and sulfur isotopic ratios in L483} \label{sec:isotopic_ratios_discussion_rest}

The isotopic ratios involving $^{15}$N, $^{18}$O, $^{17}$O, $^{34}$S, and $^{33}$S derived for different molecules in L483 are shown in the bottom panel of Fig.~\ref{fig:isotopic_ratios}, where they are compared with the corresponding isotopic ratios in the local ISM. Different behaviors can be seen.

In the case of nitrogen, direct $^{14}$N/$^{15}$N ratios could only be obtained for HCN (more specifically for H$^{13}$CN) and HC$_3$N. In both cases the observed ratios are consistent with the local ISM values, and therefore we can safely state that $^{15}$N fractionation is not important for these two molecules in L483. Similar conclusions have been found for these two molecules in other cold dense environments like TMC-1, B1b, L1544, or L1498 (e.g., \citealt{Daniel2013,Taniguchi-Saito2017,Magalhaes2018,Hily-Blant2018}). In the case of N$_2$H$^+$, the two $^{15}$N substituted isotopologs have slightly different abundances, with a N$^{15}$NH$^+$/$^{15}$NNH$^+$ ratio of 1.6, and it is very likely that this molecule is affected by a dilution in $^{15}$N, as the N$_2$H$^+$/$^{15}$NNH$^+$ and N$_2$H$^+$/N$^{15}$NH$^+$ ratios we find are $>$747 and $>$450, respectively. A similar behavior was previously found in B1b \citep{Daniel2013} and L1544 \citep{Bizzocchi2013}. In fact, N$_2$H$^+$ is one of the few molecules which seems to be affected to an significant extent by nitrogen fractionation in cold dense clouds (e.g., \citealt{Redaelli2018}). In general, chemical models predict a low level of nitrogen fractionation for different types of molecules in cold dense clouds \citep{Roueff2015,Wirstrom2018,Loison2019a}, which is in line with our findings for HCN and HC$_3$N but not for N$_2$H$^+$. The observed depletion of $^{15}$N for N$_2$H$^+$ in cold dense clouds remains puzzling. \cite{Furuya2018} suggested that the $^{15}$N depletion is inherited from a diffuse gas component where the gas is depleted in $^{15}$N whereas the ice could be enriched in that same isotope. In that scenario, $^{15}$N atoms are released more efficiently through selective photodissociation of $^{14}$N$^{15}$N around the chemical transition from atomic to molecular nitrogen and are converted efficiently to ammonia ice through hydrogenation on grains. Searching for $^{15}$N in $^{15}$N$^{15}$NH$^+$ and in the various deuterated substitutes of N$_2$H$^+$, $^{14}$N$^{15}$ND$^+$, $^{15}$N$^{14}$ND$^+$ and $^{15}$N$^{15}$ND$^+$ is possible thanks to detailed spectroscopic studies \citep{Dore2009,Dore2017} and could help to further constrain the issue.

Regarding oxygen fractionation, the only constraints we have are for HNCO and SO. Our data show that isocyanic acid is enriched in $^{18}$O by a factor of about two with respect to the local interstellar abundance of this oxygen isotope. The most drastic effect of oxygen fractionation however occurs for SO, which is severely enriched in both $^{18}$O and $^{17}$O (see bottom panel of Fig.~\ref{fig:isotopic_ratios}). This result is in agreement with a recent chemical model in which it is found that in cold dense clouds oxygen fractionation takes place to a greater or lesser degree depending on the molecule, with SO being one of the species that is most affected, and with sizeable enrichments in the heavy isotopes of oxygen \citep{Loison2019b}.

Sulfur isotopic ratios involving $^{34}$S have been obtained for various molecules. It is remarkable that all of them fall in the range of $^{32}$S/$^{34}$S ratios representative of the local interstellar medium (see bottom panel of Fig.~\ref{fig:isotopic_ratios}), thus being consistent with no $^{34}$S fractionation. This finding is in agreement with previous determinations of the $^{32}$S/$^{34}$S ratio in different molecules toward assorted interstellar environments (e.g., \citealt{Wilson1999,Tercero2010,Liszt2012,Vastel2018b}) and with recent chemical model calculations, which predict little $^{34}$S fractionation for SO \citep{Loison2019b}. Regarding the $^{32}$S/$^{33}$S ratio, the only constraint we have is for SO, which shows a ratio that is fully consistent with the  value for the local ISM (see bottom panel of Fig.~\ref{fig:isotopic_ratios}). Our results therefore support the hypothesis that sulfur fractionation is not important in cold dense clouds.

\subsubsection{Deuterated molecules in L483} \label{sec:deuterium}

\begin{table*}
\caption{Deuterated ratios as percentages} \label{table:deuterated_ratios}
\small
\centering
\begin{tabular}{llcccccc}
\hline \hline
Ratio & & L483 & TMC-1 & L1544 & B1-b & L1527 & IRAS\,16293$-$2422 \\
\hline
\multicolumn{8}{c}{Singly deuterated species}  \\
\hline
CH$_2$DOH / CH$_3$OH & /3                          & 0.63 $^a$ & ... & 3.3 $^j$ & ... & $<$1.0 $^r$ & 12 $^u$ \\ 
CH$_3$OD / CH$_3$OH                                    & & 1.4 $^a$ & 2.65 $^d$ & ... & ... & ... & 1.8 $^u$ \\
HDCCO / H$_2$CCO & /2                                  & 4.6 $\pm$ 1.4 & ... & ... & ... & ... & 2.1 $^z$ \\
DCO$_2$$^+$ / HCO$_2$$^+$                          & & 13.3 $\pm$ 4.0 & ... & ... & 13 $^o$ & ... & ... \\
$c$-C$_3$D / $c$-C$_3$H                                 & & 4.4 $\pm$ 1.3 & ... & ... & ... & ... & ... \\
C$_4$D / C$_4$H                                               & & 1.9 $\pm$ 0.6 & 0.43 $^e$ & ... & ... & 1.8 $^r$ & ... \\
$c$-C$_3$HD / $c$-C$_3$H$_2$ & /2               & 5.1 $\pm$ 1.5 & 2.4 $^d$ & 6-8.5 $^k$ & ... & 2.2 $^s$, 3.6 $^r$ & 7 $^v$ \\
$c$-H$^{13}$CCCD / $c$-H$^{13}$CCCH         & & 4.8 $\pm$ 1.4 & ... & ... & ... & ... & ... \\
$c$-HCC$^{13}$CD / $c$-HCC$^{13}$CH         & & 6.2 $\pm$ 1.9 & ... & ... & ... & ... & ... \\
$l$-C$_3$HD / $l$-C$_3$H$_2$ & /2                & 3.8 $\pm$ 1.1 & 2.0 $^f$ & 3.0 $^f$ & ... & ... & ... \\
CH$_2$DCCH / CH$_3$CCH & /3                    & 6.5 $\pm$ 1.9 & 1.8 $^g$ & ... & ... & 4.7 $^s$ & ... \\
$^{13}$CH$_2$DCCH / $^{13}$CH$_3$CCH & /3 & 17.6 $\pm$ 5.3 & ... & ... & ... & ... & ... \\
CH$_3$CCD / CH$_3$CCH                              & & 5.9 $\pm$ 1.8 & 3.95 $^h$ & ... & ... & ... & ... \\
NH$_2$D / NH$_3$ & /3                                   & 1.7 $^b$ & 0.03 $^d$ & 4.3 $^l$ & 4.3 $^p$ & 1.3 $^t$ & 3.3 $^w$\\ 
DC$_3$N / HC$_3$N  &                                   & 2.8 $\pm$ 0.8 & 1.45 $^d$ & ... & ... & 3.9 $^s$ & ... \\
DC$^{13}$CCN / HC$^{13}$CCN  &                & 8.7 $\pm$ 2.6 & ... & ... & ... & ... & ... \\
CH$_2$DCN / CH$_3$CN & /3                        & 4.4 $\pm$ 1.3 & ... & ... & ... & ... & 1.7 $^{aa}$ \\
DNCO / HNCO  &                                             & 3.8 $\pm$ 1.1 & ... & ... & ... & ... & 1 $^x$ \\
HDCS / H$_2$CS &/2~~~~~~~~~~~~~~~~~~~~~~~~~~ & 6.9 $\pm$ 2.1 & 1.0 $^i$ & 11 $^m$ & 15 $^q$ & 14.5 $^s$ & 5 $^{ab}$ \\
\hline
\multicolumn{8}{c}{Doubly deuterated species} \\
\hline
D$_2$CO / H$_2$CO & & 3.2 $^c$ & ... & 4 $^n$ & 5.7 $^q$ & 1.4 $^s$, 44 $^u$ & 5 $^u$ \\ 
CHD$_2$OH / CH$_3$OH & & 0.28 $^a$ & ... &  &  &  & 6 $^y$ \\
$c$-C$_3$D$_2$ / $c$-C$_3$H$_2$        & & 0.97 $\pm$ 0.29 & ... & 1.2-2.1 $^k$ & ... & 0.5 $^s$ & ... \\
D$_2$CS / H$_2$CS & & 4.6 $\pm$ 1.4 & ... & 15 $^m$& 10 $^q$ & ... & ... \\
\hline
\end{tabular}
\tablenoteb{\\
$^a$~CH$_3$OH column density evaluated from $^{13}$CH$_3$OH assuming $^{12}$C/$^{13}$C = 68 \citep{Milam2005}. $^b$~NH$_3$ column density from \cite{Anglada1997}. $^c$~H$_2$CO column density evaluated from H$_2$$^{13}$CO \citep{Tafalla2000} assuming $^{12}$C/$^{13}$C = 68 \citep{Milam2005}. $^d$~\cite{Turner2001}. $^e$~\cite{Turner1989}. $^f$~\cite{Spezzano2016}. $^g$~\cite{Gerin1992a}. $^h$~\cite{Markwick2005}. $^i$~\cite{Minowa1997}. $^j$~\cite{Bizzocchi2014}. $^k$~\cite{Spezzano2013}. $^l$~\cite{Shah2001}. $^m$~\cite{Vastel2018b}. $^n$~\cite{Bacmann2003}. $^o$ \cite{Fuente2016}. $^p$~\cite{Saito2000}. $^q$~\cite{Marcelino2005}. $^r$~\cite{Sakai2009b}. $^s$~\cite{Yoshida2019}. $^t$~\cite{Hatchell2003}. $^u$~\cite{Parise2006}. $^v$~\cite{Majumdar2017}. $^w$~\cite{vanDishoeck1995}. $^x$~\cite{Coutens2016}. $^y$~\cite{Parise2004}. $^z$~\cite{Jorgensen2018}. $^{aa}$~\cite{Calcutt2018b}. $^{ab}$~\cite{Drozdovskaya2018}. \\
}
\end{table*}

This $\lambda$ 3 mm line survey revealed that L483 stands out as a source rich in deuterated molecules, something that was already suggested from the detection of triply deuterated ammonia \citep{Roueff2005}. We detected various singly deuterated molecules and four doubly deuterated species. While some of them are long known in interstellar chemistry, some others have only been discovered recently in interstellar clouds, and a few are detected for the first time in space toward L483. In Table~\ref{table:deuterated_ratios} we list the deuterated species detected, together with the column density ratios between deuterated and nondeuterated species (expressed in \%), where for singly deuterated species with two or more equivalent hydrogen nuclei these ratios are corrected by the number of equivalent hydrogen nuclei to get rid of statistical effects and facilitate the comparison between different molecules. Hereafter, all deuterium isotopic ratios given are corrected for this statistical effect. To put these isotopic ratios in context we also give the values derived in other well-studied low-mass prestellar and protostellar sources: TMC-1, L1544, B1-b, L1527, and IRAS\,16293$-$2422. It is worth noting that the lines of the deuterated species tend to be narrower than those of the nondeuterated form by $\sim$0.1 km s$^{-1}$, which indicates that deuterated molecules are present in less turbulent and colder regions of the cloud. This observational fact is in line with the expectation that deuterium fractionation is more efficient at low temperatures, and leads us to anticipate that the spatial distribution of deuterated molecules in L483 should differentiate from that of the nondeuterated counterpart, with the former being preferentially present in colder regions.

We detect the two singly deuterated versions of methanol (CH$_2$DOH and CH$_3$OD), with isotopic ratios of $\sim$1 \%, which are comparable to the ratios derived in similar sources. The CH$_2$DOH/CH$_3$OH ratio is slightly higher (by a factor of $\sim$5) in L1544 and significantly higher (by a factor of $\sim$20) in IRAS\,16293$-$2422. It is interesting to note that while in this latter source deuteration on the methyl group is favored over deuteration on the hydroxyl group, the contrary is found in L483. This is in line with the fact that we detected methanol deuterated twice on the methyl group, with a CHD$_2$OH/CH$_3$OH ratio substantially lower than in IRAS\,16293$-$2422. This result may be related to the different temperatures in each source. The $\lambda$ 3 mm line survey covered the 2$_{1,2}$-1$_{1,1}$ line of doubly deuterated formaldehyde, which was detected, leading to a D$_2$CO/H$_2$CO ratio of 3.2 \%, which is comparable to that found in L1544, B1-b, and IRAS\,16293$-$2422. This ratio has been reported to be as high as 44 \% in L1527 \citep{Parise2006}, although more recently \cite{Yoshida2019} derived a much lower value of 1.4 \%, which is more in line with the D$_2$CO/H$_2$CO ratios found in the other sources. The singly deuterated form of formaldehyde, HDCO, does not have favorable lines in the 80-116 GHz range and therefore could not be observed. We also detected the deuterated form of protonated carbon dioxide (DCO$_2$$^+$), for which the rotational spectrum is well known \citep{Bogey1988,Bizzocchi2017}. The tentative detection of DCO$_2$$^+$ in L1544 claimed by \cite{Vastel2016} is unlikely because of a difference of 0.29 MHz between the observed and laboratory frequency of the 5$_{0,5}$-4$_{0,4}$ transition. More recently, \cite{Fuente2016} reported on the first secure detection of this ion toward B1-b. In our line survey of L483, we detect clearly the 4$_{0,4}$-3$_{0,3}$ line lying at 80288.759 MHz at the correct position (the $V_{\rm LSR}$ derived is 5.30 $\pm$ 0.03 km s$^{-1}$, in perfect agreement with the systemic velocity of the source). Moreover, the 5$_{0,5}$-4$_{0,4}$ transition at 100359.521 MHz is marginally detected, although again at the correct position ($V_{\rm LSR}$ is 5.25 $\pm$ 0.11 km s$^{-1}$), which adds support to the identification of DCO$_2$$^+$. We derive a DCO$_2$$^+$/HCO$_2$$^+$ ratio of 13.3 $\pm$ 4.0 \%, which is in the high end of deuterium enrichment ratios determined in L483, and fully consistent with the value derived in B1-b ($\sim$13 \%; \citealt{Fuente2016}).

Hydrocarbons also show a variety of deuterated molecules in L483. We detected C$_4$D, with an isotopic ratio of 1.9 $\pm$ 0.6 \%, which is slightly above that found in TMC-1 (0.43; \citealt{Turner1989}) and similar to that derived in L1527 (1.8; \citealt{Sakai2009b}). We also detected the singly and doubly deuterated forms of $c$-C$_3$H$_2$, with isotopic ratios of 5.1 $\pm$ 1.5 \% and 0.97 $\pm$ 0.29 \%, respectively, which are similar to the values found in L1544 \citep{Spezzano2013} and L1527 \citep{Sakai2009b,Yoshida2019}. Moreover, $c$-C$_3$HD and $c$-C$_3$D$_2$ have been surveyed in a sample of low-mass prestellar and protostellar cores \citep{Chantzos2018}, finding that the corresponding isotopic ratios are relatively uniform, within a factor of a few, and similar to those in L1544 and L483. Thanks to the high sensitivity of our line survey, we detected the different deuterated forms of the two $^{13}$C substituted isotopologs of $c$-C$_3$H$_2$, that is, $c$-H$^{13}$CCCD, $c$-HCC$^{13}$CD, and $c$-HC$^{13}$CCD; the latter only tentatively. This is to our knowledge the first time these species have been detected in space. The deuterium ratios derived for $c$-H$^{13}$CCCD and $c$-HCC$^{13}$CD are in line with that found for $c$-C$_3$HD. We also detected the deuterated form of the linear isomer of C$_3$H$_2$, $l$-C$_3$HD, which was recently observed for the first time toward TMC-1 and L1544 \citep{Spezzano2016}. The isotopic ratio we find in L483, 3.8 $\pm$ 1.1 \%, is similar to the values derived in TMC-1 and L1544. Moreover, it seems that the linear isomer of C$_3$H$_2$ shows very similar levels of deuterium fractionation to the cyclic isomer in both low-mass prestellar and protostellar cores. The two deuterated forms of methyl acetylene, CH$_2$DCCH and CH$_3$CCD, which have previously been observed in TMC-1 with isotopic ratios of a few percent \citep{Gerin1992a,Markwick2005}, are also detected in L483 with slightly higher deuterium ratios ($\sim$6 \%). The deuterium ratio found for CH$_2$DCCH in L483, 6.5 $\pm$ 1.9 \%, is similar to that derived in L1527 (4.7 \%; \citealt{Yoshida2019}).

There are also several N-bearing species among the deuterated molecules. Singly deuterated ammonia (NH$_2$D) is detected with an isotopic ratio of 1.7 \%, which is much higher than the value derived in TMC-1 but of the order of the ratios found in L1544, B1-b, L1527, and IRAS\,16293$-$2422. We also detected the deuterated form of HC$_3$N, with an isotopic ratio of 2.8 $\pm$ 0.8 \%, similar to those found in TMC-1 and L1527. Moreover, the deuterated form of one of the $^{13}$C isotopologs (DC$^{13}$CCN) was also detected for the first time in interstellar space. The isotopic ratio derived is about three times higher than that found for the main species, which suggests that deuterium fractionation is more efficient for HC$^{13}$CCN than for HC$_3$N. The singly deuterated form of methyl cyanide (CH$_2$DCN) has been observed previously in high-mass star-forming regions, concretely in the hot core Orion KL and tentatively in G34.3 \citep{Gerin1992b}, toward Sgr\,B2 \citep{Belloche2016}, and more recently in a hot core associated with the infrared dark cloud G34.43\,+00.24 MM3 \citep{Sakai2018}, and in the hot corino IRAS\,16293$-$2422 \citep{Calcutt2018b}. We detected this species in L483, with an isotopic ratio of 4.4 $\pm$ 1.3 \%, which is at least ten times higher than those found in Orion\,KL and Sgr\,B2. Another deuterated N-bearing molecule not very widely observed before is DNCO, which has only been recently detected toward IRAS\,16293$-$2422 \citep{Coutens2016}. The deuterium fractionation of HNCO derived in L483 (3.8 $\pm$ 1.1 \%) is slightly higher than, but comparable to, that found in the hot corino source (1 \%).

We also detected HDCS and D$_2$CS with isotopic ratios of 6.9 $\pm$ 2.1 \% and 4.6 $\pm$ 1.4 \%, respectively. These values are slightly lower than, although of the same order as, those found in L1544 and B1-b \citep{Marcelino2005,Vastel2018b}. The deuterium ratio of HDCS in L483 is also only slightly lower than that found in L1527 \citep{Yoshida2019}. In TMC-1, deuterium fractionation of H$_2$CS seems to be around ten times less efficient than in L1544, B1-b, L1527, and L483, as inferred from observations of HDCS \citep{Minowa1997}.

We also report deuterium ratios for HDCCO and $c$-C$_3$D. Detection of singly deuterated ketene (HDCCO) was possible thanks to the laboratory characterization of its rotational spectrum at millimeter wavelengths \citep{Nemes1976,Guarnieri2005}. This species was recently observed for the first time in space toward IRAS\,16293$-$2422 using ALMA \citep{Jorgensen2018}. We derive a deuterium enrichment of 4.6 $\pm$ 1.4 \% for HDCCO, which is in line with the deuterated ratios of a few percent derived for other molecules. Recently, \citep{Yoshida2019} reported on the detection of the 5$_{0,5}$-4$_{0,4}$ line of HDCCO in L1527, although these authors did not derive a column density. The deuterated form of cyclic C$_3$H is detected for the first time in space toward L483 in this study. The rotational spectrum of $c$-C$_3$D has been measured in the laboratory by \cite{Yamamoto1990} and \cite{Lovas1992}. The isotopic ratio $c$-C$_3$D/$c$-C$_3$H is found to be 4.4 $\pm$ 1.3 \%, which is similar to the deuterium fractionation level in $c$-C$_3$H$_2$ and $l$-C$_3$H$_2$. The deuterated form of the linear isomer of C$_3$H has not yet been observed in space, although a similar deuterium enrichment of a few percent could be expected for $l$-C$_3$D based on the behavior of $c$-C$_3$D, $c$-C$_3$HD, and $l$-C$_3$HD. 

In general, isotopic ratios in singly and doubly deuterated species in L483 are in the range 1-10 \%. When comparison with other low-mass prestellar and protostellar sources is possible, deuterium fractionation ratios in L483 tend to be systematically higher than in TMC-1 (except for CH$_3$OD, which is slightly more abundant in TMC-1), slightly smaller than in L1544 and B1-b, similar to those in L1527 (except for D$_2$CO, which is substantially more abundant in L1527), and smaller than in IRAS\,16293$-$2422. There are some molecules which show a higher degree of variability in the deuterium fractionation among different sources. For example, deuteration in ammonia proceeds much less efficiently in the starless cloud TMC-1 than in other sources, while deuteration in methanol on the methyl group seems to be especially favored in the hot corino IRAS\,16293$-$2422.

\section{Conclusions}

We used the IRAM 30m telescope to perform a $\lambda$ 3 mm line survey in the 80-116 GHz frequency range of the dense core L483, which hosts a low-mass Class\,0 protostar. We detected 631 lines (all of them in emission with the exception of one methanol line which was observed in absorption), 613 of which were assigned to rotational transitions of different molecules, while 18 remain unidentified. Most lines are narrow (FWHM $\sim$ 0.5 km s$^{-1}$) and derived rotational temperatures are low (4.1-10.5 K, except for HC$_5$N which has 28 $\pm$ 3 K), indicating that their emission arises exclusively from the ambient cold and quiescent cloud, rather than from the bipolar outflow, or the surroundings of the protostar.

We detected 71 molecules (140 if different isotopologs are taken into account), which comprise O-bearing molecules (including the $cis$ conformer of HCOOH and the three complex organic molecules HCOOCH$_3$, CH$_3$OCH$_3$, and C$_2$H$_5$OH), hydrocarbons (most of them being carbon chains), a wide variety of N-containing molecules ranging from carbon chains to oxides like N$_2$O, several S-bearing molecules (including carbon chains and saturated species like CH$_3$SH), and one Si-containing molecule (SiO). Of particular interest is the detection of several new interstellar molecules (HCCO, HCS, HSC, NCCNH$^+$, CNCN, NCO, H$_2$NCO$^+$, and NS$^+$), which have recently been  reported toward L483, most of them thanks to observations carried out in the context of this line survey.

In general, fractional molecular abundances in L483 are systematically lower than in TMC-1 (especially in the case of carbon chains, which are significantly more abundant in TMC-1), while they tend to be higher than in L1544 and B1-b (probably as a consequence of a more important depletion on dust grains in these latter sources), and are similar to those in L1527. The use of chemical indicators such as the C$_4$H/CH$_3$OH abundance ratio indicates that the chemical composition of dense cores around Class\,0 protostars like B1-b, L1527, and L483 resemble more that of starless/prestellar cores like TMC-1 and L1544 than that of Class\,0 hot corino sources like IRAS\,16293$-$2422. This fact suggests that the chemical composition of the ambient cloud of some Class\,0 sources could be largely inherited from the dark cloud starless/prestellar phase. Potential chemical evolutionary indicators to trace the prestellar/protostellar transition, such as the HNCO/C$_3$S, SO$_2$/C$_2$S, and CH$_3$SH/C$_2$S ratios, are explored.

We also derived isotopic ratios for a variety of molecules for which we detected minor isotopologs containing $^{13}$C, $^{15}$N, $^{18}$O, $^{17}$O, $^{34}$S, and $^{33}$S, as well as deuterium. Many of the molecular isotopic ratios in L483 are close to the values of the local ISM, many of those involving $^{13}$C and $^{15}$N, and remarkably all those involving $^{34}$S and $^{33}$S. There are however several isotopic anomalies like an extreme depletion in $^{13}$C for one of the two isotopologs of $c$-C$_3$H$_2$, an important enrichment in $^{13}$C in one of the two isotopologs of C$_2$S, a drastic enrichment in $^{18}$O for SO and HNCO (SO being also largely enriched in $^{17}$O), and different abundances for the two $^{13}$C substituted species of C$_2$H and the two $^{15}$N substituted species of N$_2$H$^+$. We reported the first detection in space of several doubly isotopically substituted species with $^{13}$C and deuterium and evaluated for the first time the deuterium fractionation for $c$-C$_3$D. We find that deuterium fractionation in L483 tends to be higher than in TMC-1, similar to in L1544, B1-b, and L1527, and lower than in IRAS\,16293$-$2422.

The detailed chemical characterization of the L483 dense core presented here provides information that, together with similar exhaustive characterizations of other low-mass prestellar and protostellar sources foreseen for the near future, should help constrain the main factors that affect the chemical evolution of cores along the process of formation of low-mass protostars.

\begin{acknowledgements}

We thank the IRAM 30m staff for their help during the observations and the GILDAS team, in particular S\'ebastien Bardeau, for assistance with CLASS. We acknowledge the referee for a report that helped to improve the article. We also thank B. Parise for the help with the partition function of CHD$_2$OH and H. S. P. M\"uller for informing us about recent literature on IRAS\,16293$-$2422. M.A., N.M., and J.C. acknowledge funding support from the European Research Council (ERC Grant 610256: NANOCOSMOS) and from Spanish MICINN through grant AYA2016-75066-C2-1-P. M.A. also thanks funding support from the Ram\'on y Cajal programme of Spanish MICINN (RyC-2014-16277). M.T. acknowledges support from Spanish MICINN under grant AYA2016-79006-P. E.R. acknowledges the support of the Programme National "Physique et Chimie du Milieu Interstellaire" (PCMI) of CNRS/INSU with INC/INP co-funded by CEA and CNES.

\end{acknowledgements}

\appendix
\onecolumn

\section{Supplementary tables and figure}

\small
\begin{longtable}{llrrrrrl}
\caption{\label{table:lines} Line parameters of assigned lines in L483}\\\hline\hline
Molecule & Transition & \multicolumn{1}{c}{Frequency} & \multicolumn{1}{c}{$V_{\rm LSR}$} & \multicolumn{1}{c}{$\Delta v$}      & \multicolumn{1}{c}{$T_A^*$ peak} & \multicolumn{1}{c}{$\int T_A^* dv$}  & Notes \\
               &                  & \multicolumn{1}{c}{(MHz)}        & \multicolumn{1}{c}{(km s$^{-1}$)}    & \multicolumn{1}{c}{(km s$^{-1}$)} & \multicolumn{1}{c}{(K)}                   & \multicolumn{1}{c}{(K km s$^{-1}$)} & \\
\hline
\endfirsthead
\caption{continued.}\\
\hline\hline
Molecule & Transition & \multicolumn{1}{c}{Frequency} & \multicolumn{1}{c}{$V_{\rm LSR}$} & \multicolumn{1}{c}{$\Delta v$}      & \multicolumn{1}{c}{$T_A^*$ peak} & \multicolumn{1}{c}{$\int T_A^* dv$}  & Notes \\
               &                  & \multicolumn{1}{c}{(MHz)}        & \multicolumn{1}{c}{(km s$^{-1}$)}    & \multicolumn{1}{c}{(km s$^{-1}$)} & \multicolumn{1}{c}{(K)}                   & \multicolumn{1}{c}{(K km s$^{-1}$)} & \\
\hline
\endhead
\hline
\endfoot
H$_2$C$_4$                     & 9$_{1,9}$-8$_{1,8}$                                                              &  80046.678 &  5.29( 3) & 0.46( 3) & 0.046 & 0.023(  2)   &           \\
$c$-HCC$^{13}$CH               & 2$_{0,2}$-1$_{1,1}$                                                              &  80047.537 &  5.26( 3) & 0.40( 3) & 0.295 & 0.126( 12)   &           \\
H$_2$CCO                       & 4$_{1,4}$-3$_{1,3}$                                                              &  80076.652 &  5.30( 3) & 0.47( 3) & 0.139 & 0.069(  6)   &           \\
H$_2$NCO$^+$                   & 4$_{1,4}$-3$_{1,3}$                                                              &  80246.484 &  5.16( 3) & 0.41(11) & 0.005 & 0.0022(   2) &           \\
$c$-C$_3$HD                    & 3$_{2,2}$-3$_{0,3}$                                                              &  80255.666 &  5.25( 3) & 0.36( 7) & 0.006 & 0.0023(   2) & 5         \\
HSCN                           & 7$_{0,7}$-6$_{0,6}$                                                              &  80283.167 &  5.30( 3) & 0.49( 9) & 0.013 & 0.007(  1)   &           \\
DCO$_2$$^+$                    & 4$_{0,4}$-3$_{0,3}$                                                              &  80288.759 &  5.30( 3) & 0.37( 3) & 0.014 & 0.0053(   5) &           \\
H$_2$C$_4$                     & 9$_{0,9}$-8$_{0,8}$                                                              &  80383.887 &  5.17( 3) & 0.42( 3) & 0.023 & 0.010(  1)   &           \\
\end{longtable}
\tablenoteb{Notes:\\
(1) Complex line profile. Impossible to fit to Gaussian function.\\
(2) Line affected by frequency switching negative artifact of other line.\\
(3) Line blended. Only one component could be fitted.\\
(4) Line blended. Several components could be fitted.\\
(5) Marginal detection.\\
(6) Hyperfine components resolved in astronomical spectrum but not in the laboratory.}

\clearpage

\small
\begin{longtable}{lrrrrl}
\caption{\label{table:ulines} Line parameters of unidentified lines in L483}\\
\hline\hline
\multicolumn{1}{l}{Label} & \multicolumn{1}{c}{Frequency} & \multicolumn{1}{c}{$\Delta v$}      & \multicolumn{1}{c}{$T_A^*$ peak} & \multicolumn{1}{c}{$\int T_A^* dv$} & \multicolumn{1}{l}{Comment} \\
          & \multicolumn{1}{c}{(MHz)}        & \multicolumn{1}{c}{(km s$^{-1}$)} & \multicolumn{1}{c}{(K)}                   & \multicolumn{1}{c}{(K km s$^{-1}$)} & \\
\hline
\endfirsthead
\caption{continued.}\\
\hline\hline
\multicolumn{1}{l}{Label} & \multicolumn{1}{c}{Frequency} & \multicolumn{1}{c}{$\Delta v$}      & \multicolumn{1}{c}{$T_A^*$ peak} & \multicolumn{1}{c}{$\int T_A^* dv$} & \multicolumn{1}{l}{Comment} \\
          & \multicolumn{1}{c}{(MHz)}        & \multicolumn{1}{c}{(km s$^{-1}$)} & \multicolumn{1}{c}{(K)}                   & \multicolumn{1}{c}{(K km s$^{-1}$)} & \\
\hline
\endhead
\hline
\endfoot
U82619     &  82618.59(2) & 0.33(8) & 0.010 & 0.004(1) & \\
U82667     &  82666.72(2) & 0.47(7) & 0.014 & 0.007(1) & \\
U87998     &  87998.01(2) & 0.58(9) & 0.009 & 0.006(1) & \\
U88306     &  88306.51(2) & 0.43(9) & 0.012 & 0.005(1) & \\
U91341     &  91340.84(2) & 0.50(13) & 0.006 & 0.003(1) & | U91341, U91343, and U91344 are probably hyperfine components of the \\
U91343     &  91342.70(2) & 0.53(5) & 0.016 & 0.009(1) & | same rotational transition. \\
U91344     &  91344.01(2) & 0.66(10) & 0.009 & 0.006(1) & | Seen in B1-b at 91344.0 MHz \citep{Lefloch2018}. \\
U92373     &  92372.79(2) & 0.27(2) & 0.031 & 0.009(1) & Nearby U line seen in L1448-R2 at 92374.0 MHz \citep{Lefloch2018}. \\
\end{longtable}

\begin{figure*}
\centering
\includegraphics[width=\textwidth]{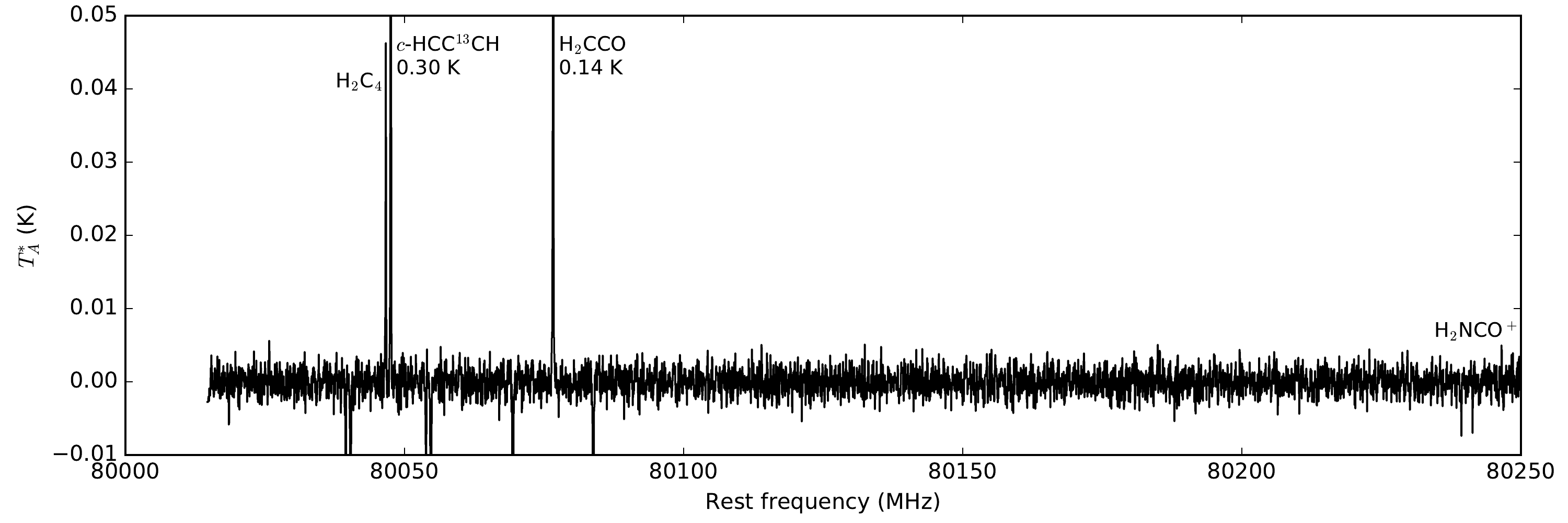}
\includegraphics[width=\textwidth]{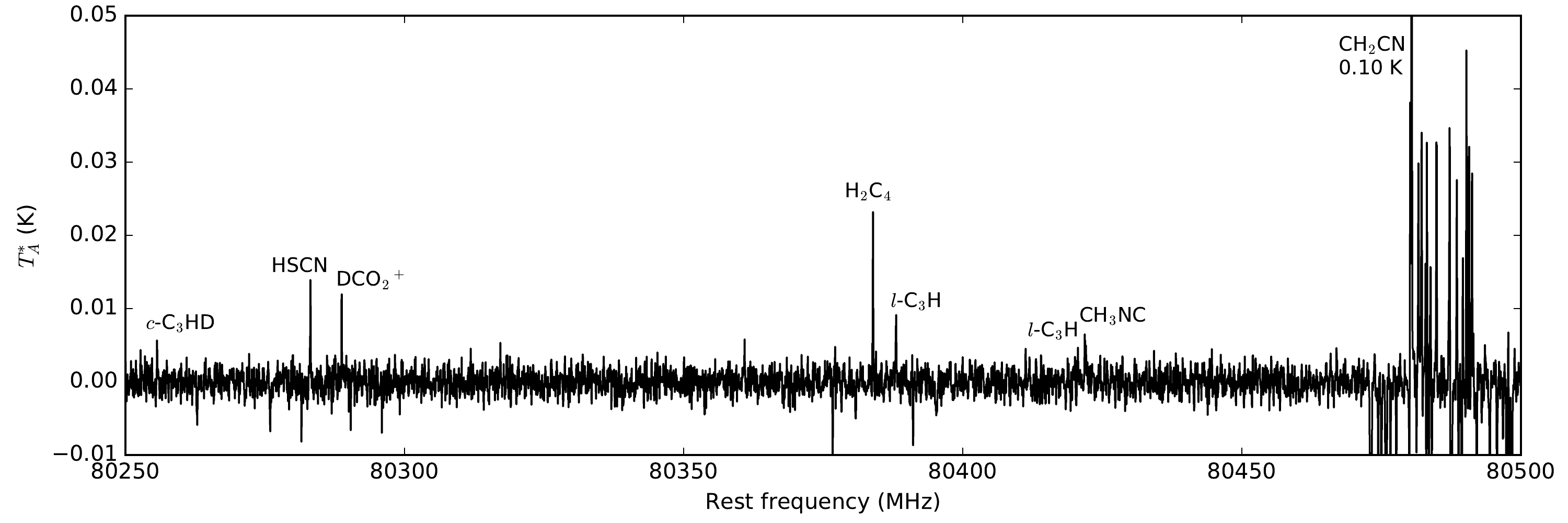}
\includegraphics[width=\textwidth]{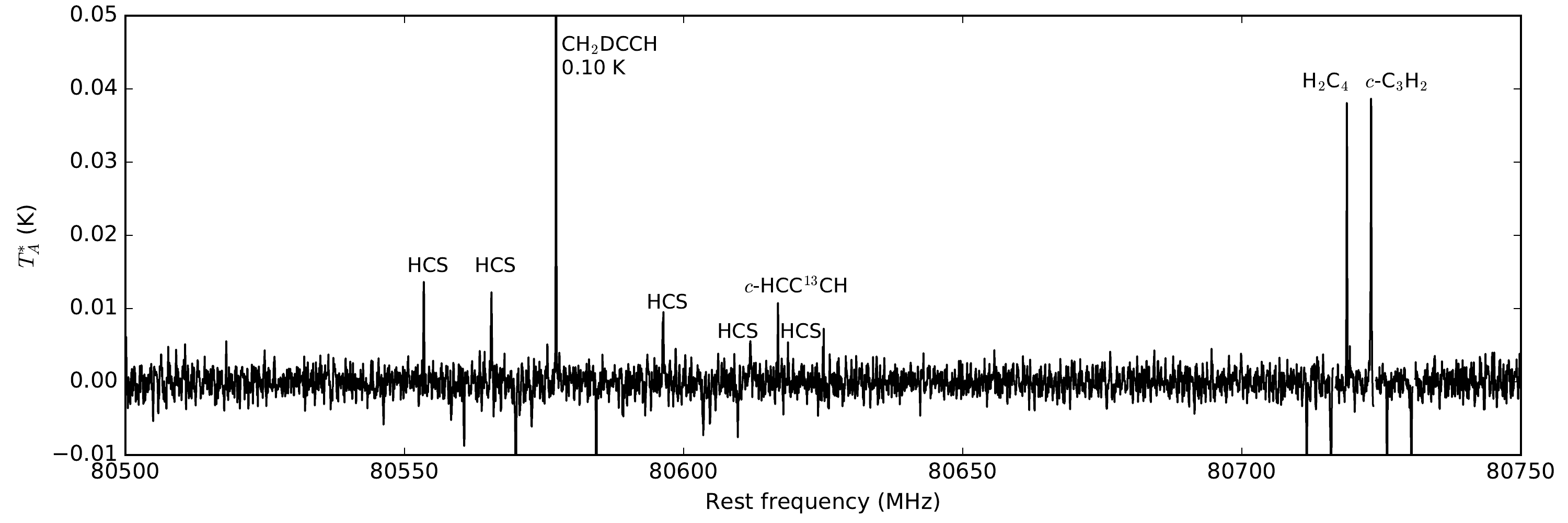}
\includegraphics[width=\textwidth]{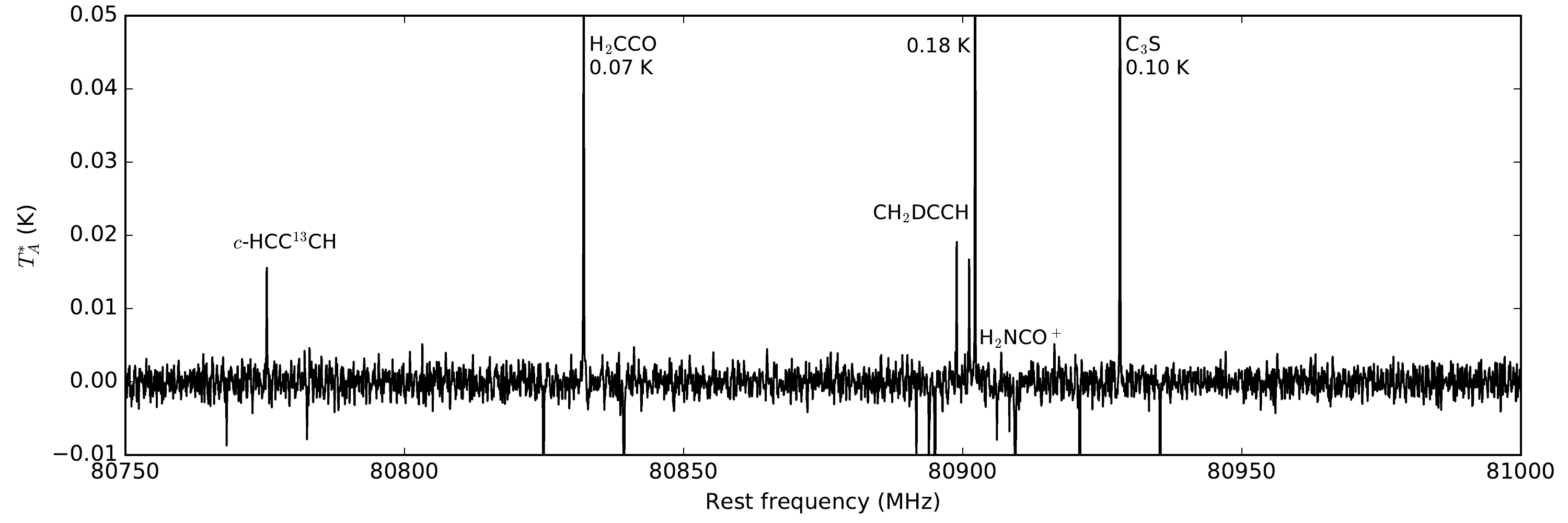}
\caption{The 80-116 GHz spectrum of L483.}
\label{fig:spectrum}
\end{figure*}

\setcounter{figure}{0}
\begin{figure*}
\centering
\includegraphics[width=\textwidth]{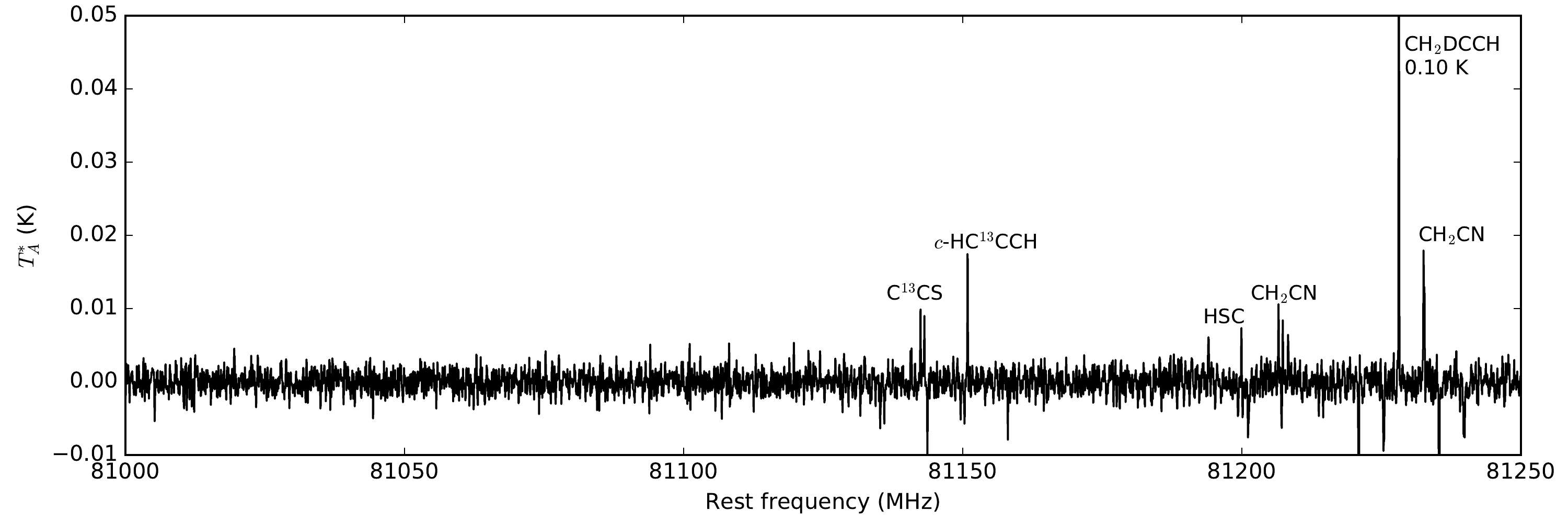}
\includegraphics[width=\textwidth]{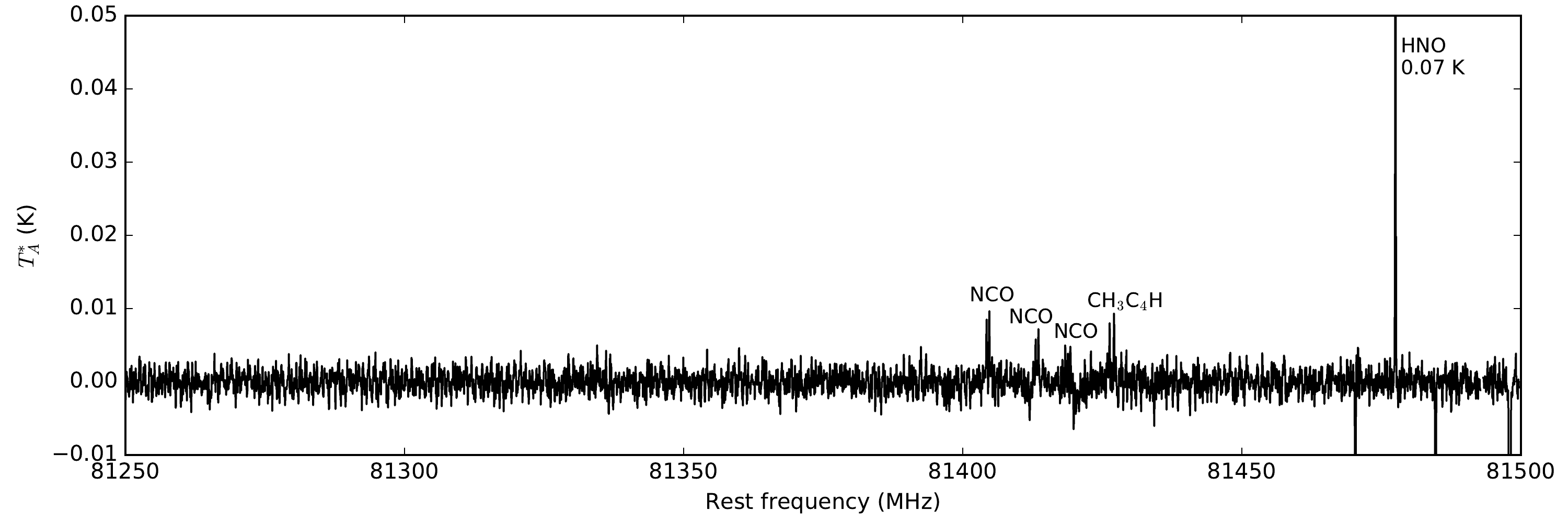}
\includegraphics[width=\textwidth]{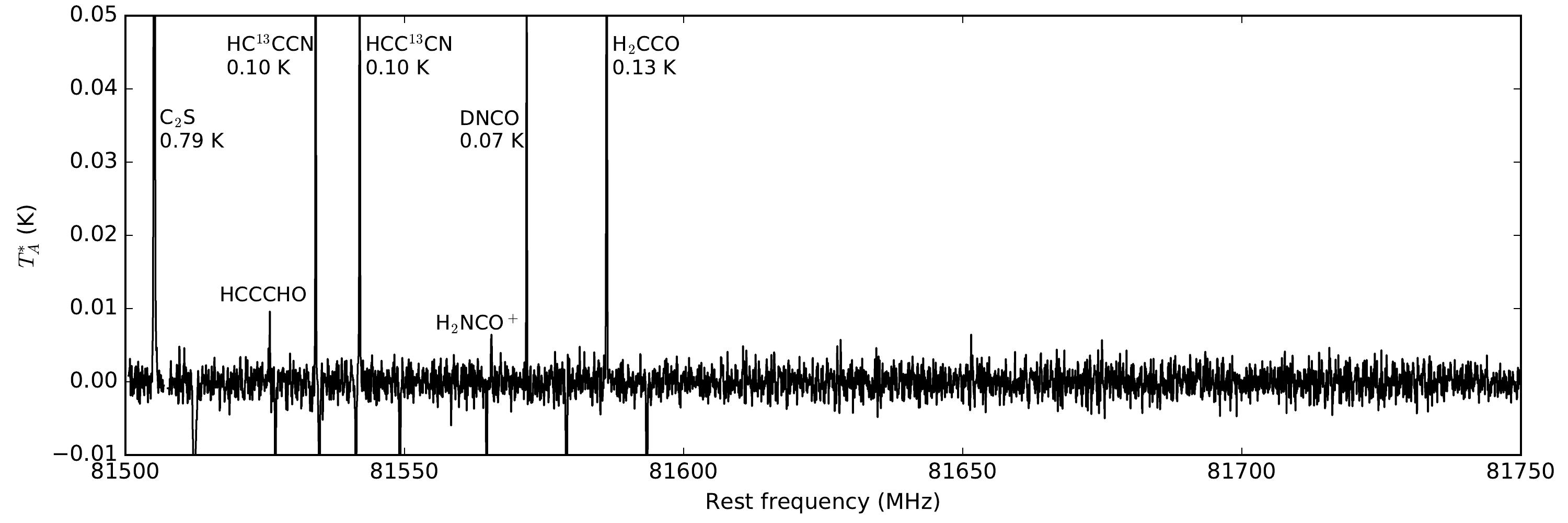}
\includegraphics[width=\textwidth]{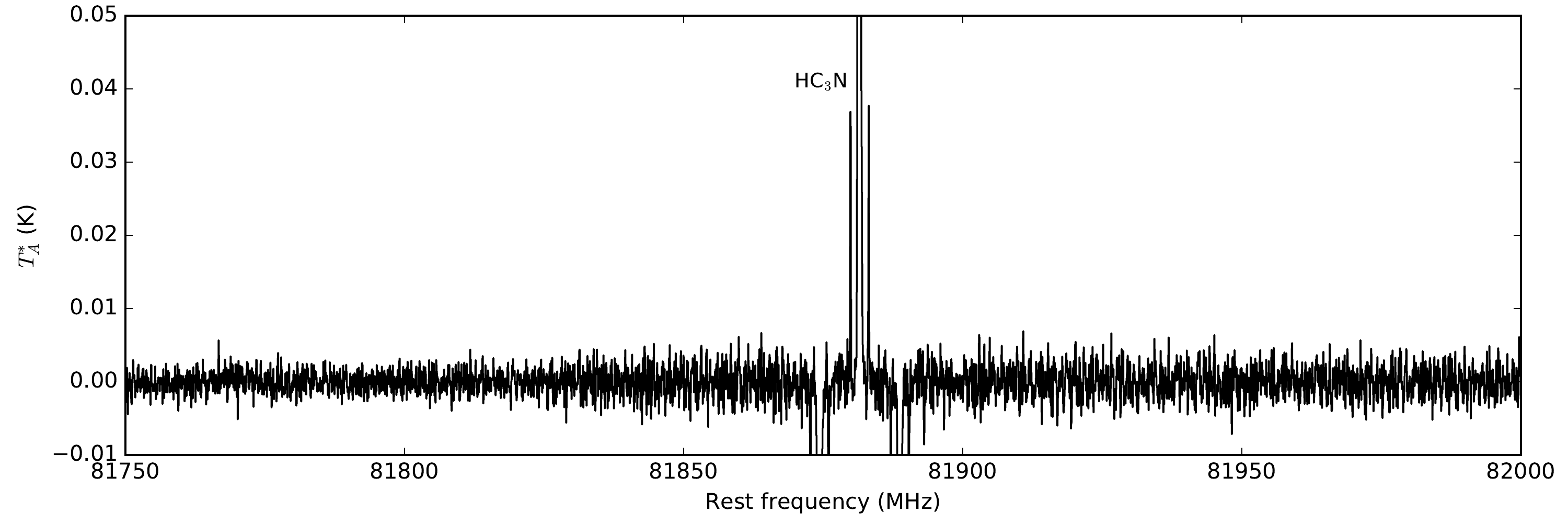}
\caption{Continued}
\end{figure*}

\setcounter{figure}{0}
\begin{figure*}
\centering
\includegraphics[width=\textwidth]{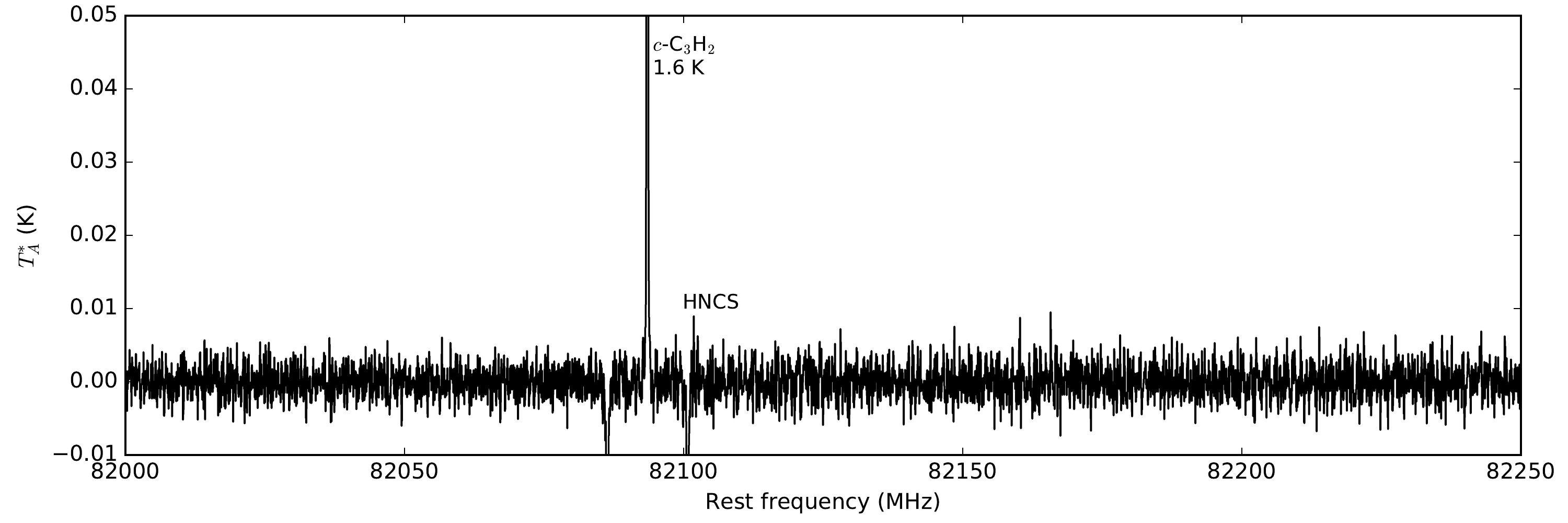}
\includegraphics[width=\textwidth]{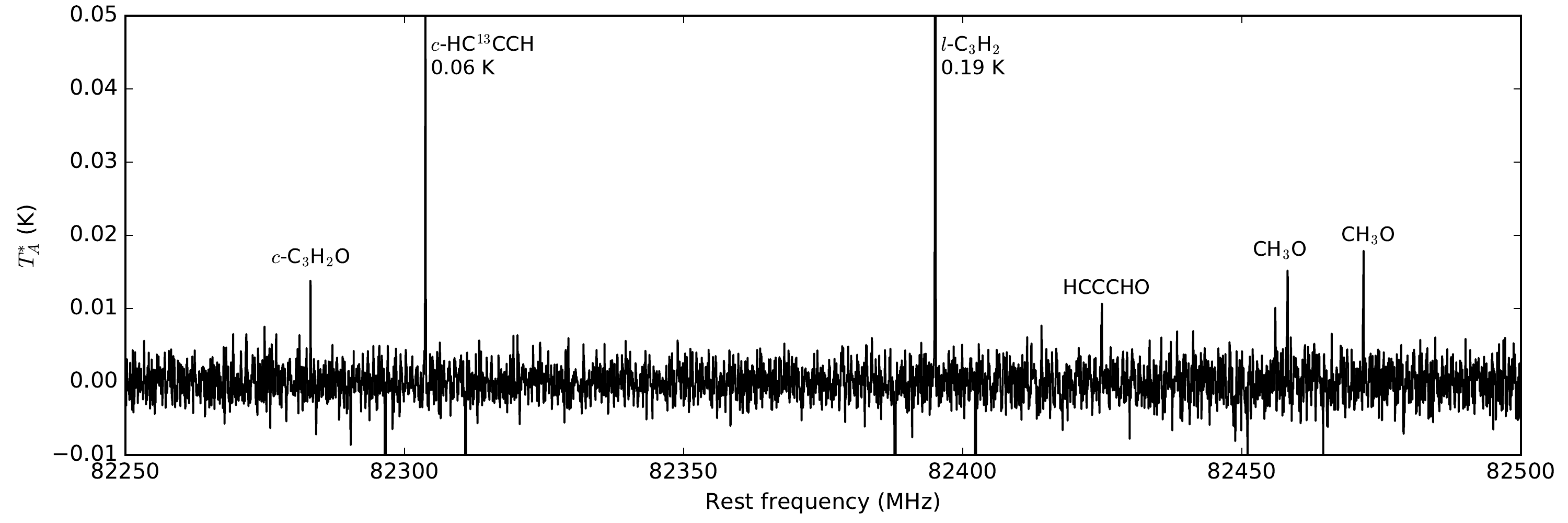}
\includegraphics[width=\textwidth]{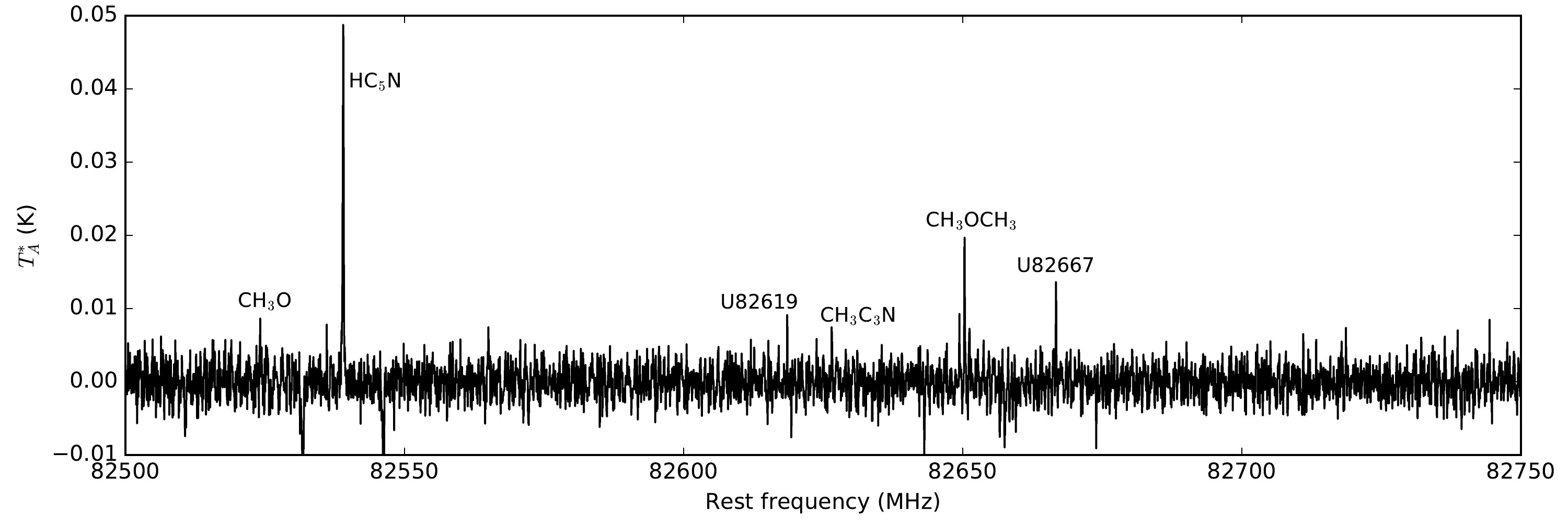}
\includegraphics[width=\textwidth]{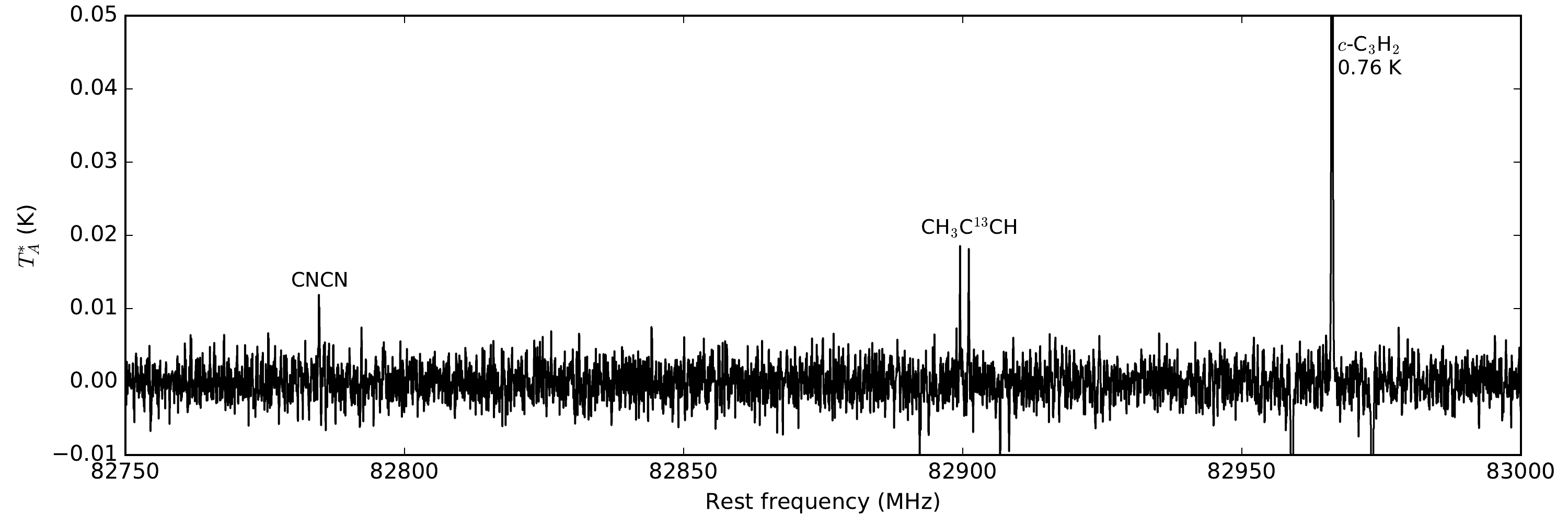}
\caption{Continued}
\end{figure*}

\setcounter{figure}{0}
\begin{figure*}
\centering
\includegraphics[width=\textwidth]{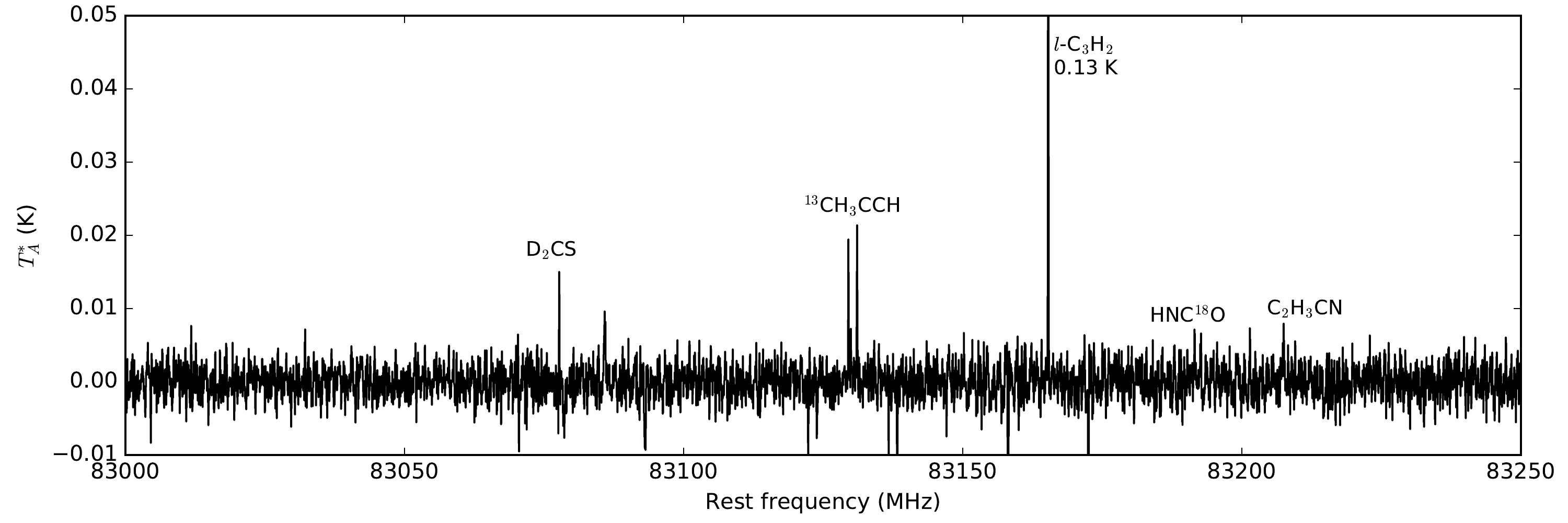}
\includegraphics[width=\textwidth]{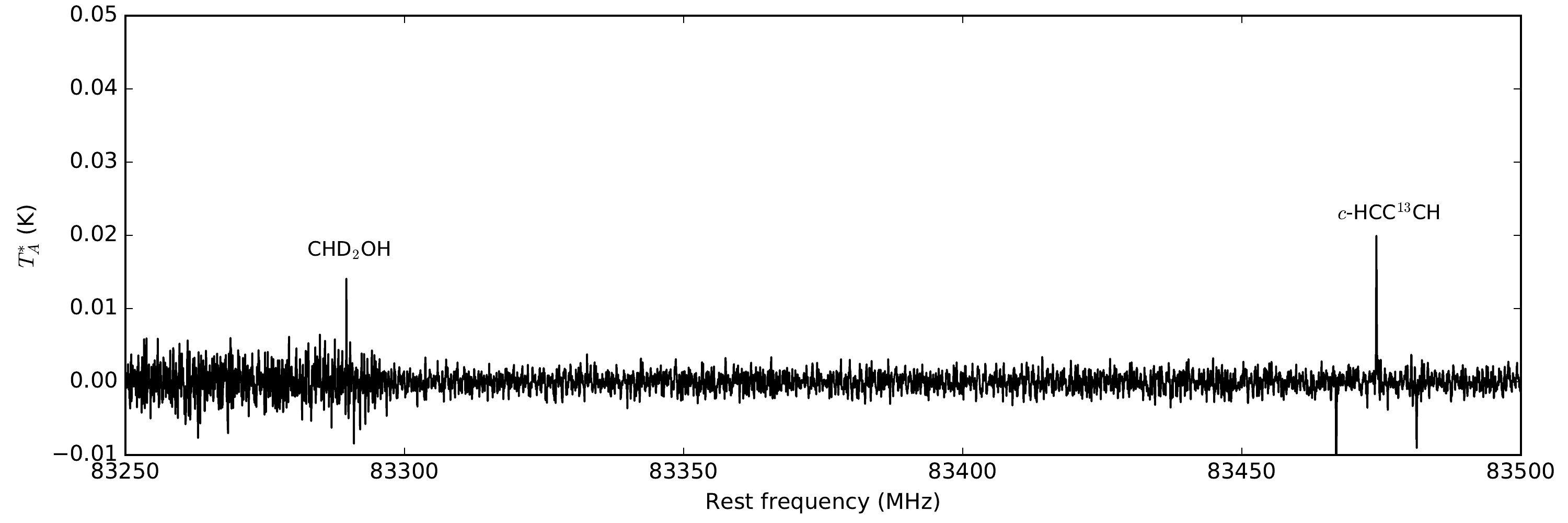}
\includegraphics[width=\textwidth]{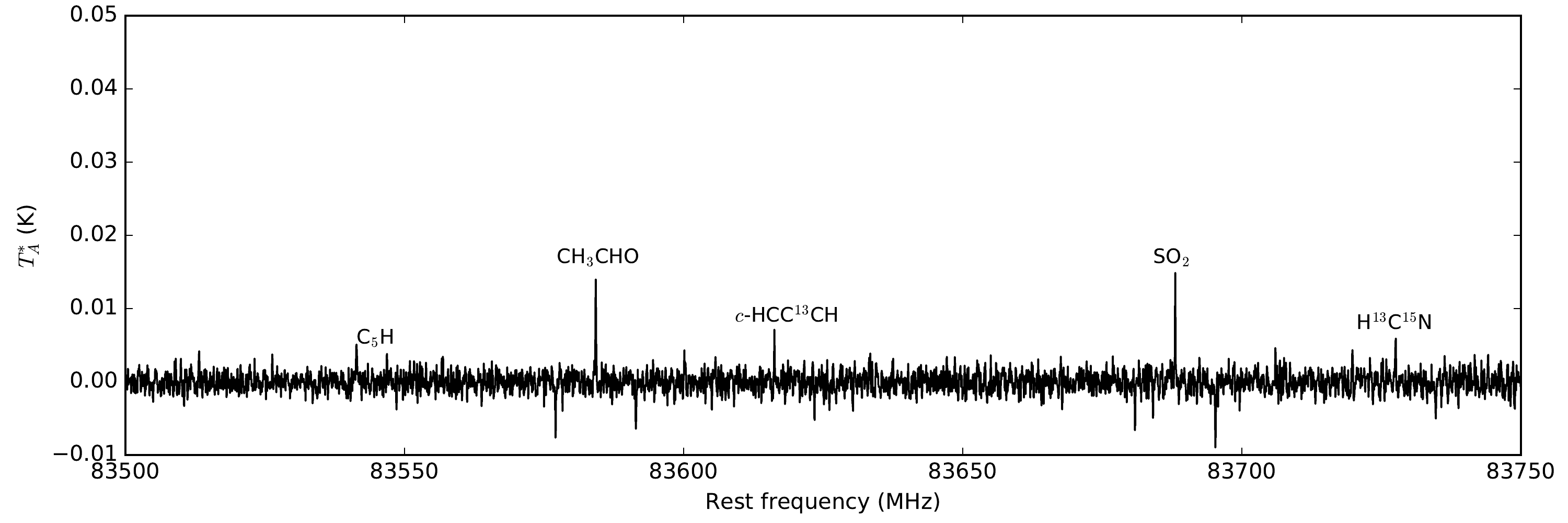}
\includegraphics[width=\textwidth]{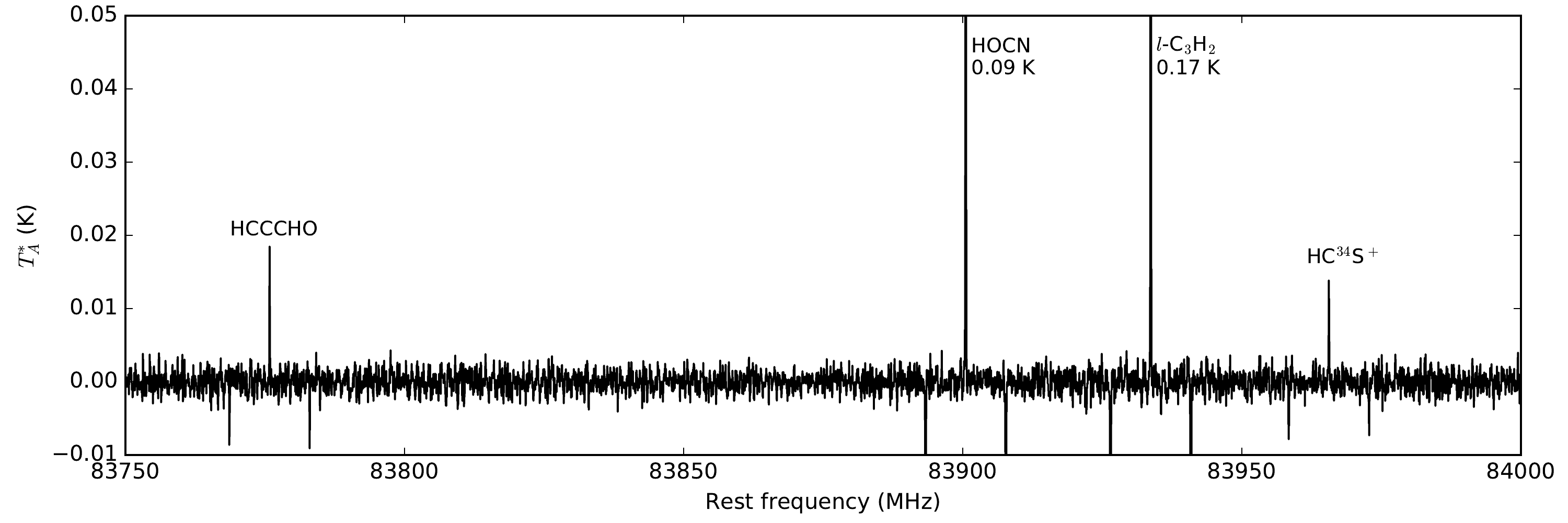}
\caption{Continued}
\end{figure*}

\setcounter{figure}{0}
\begin{figure*}
\centering
\includegraphics[width=\textwidth]{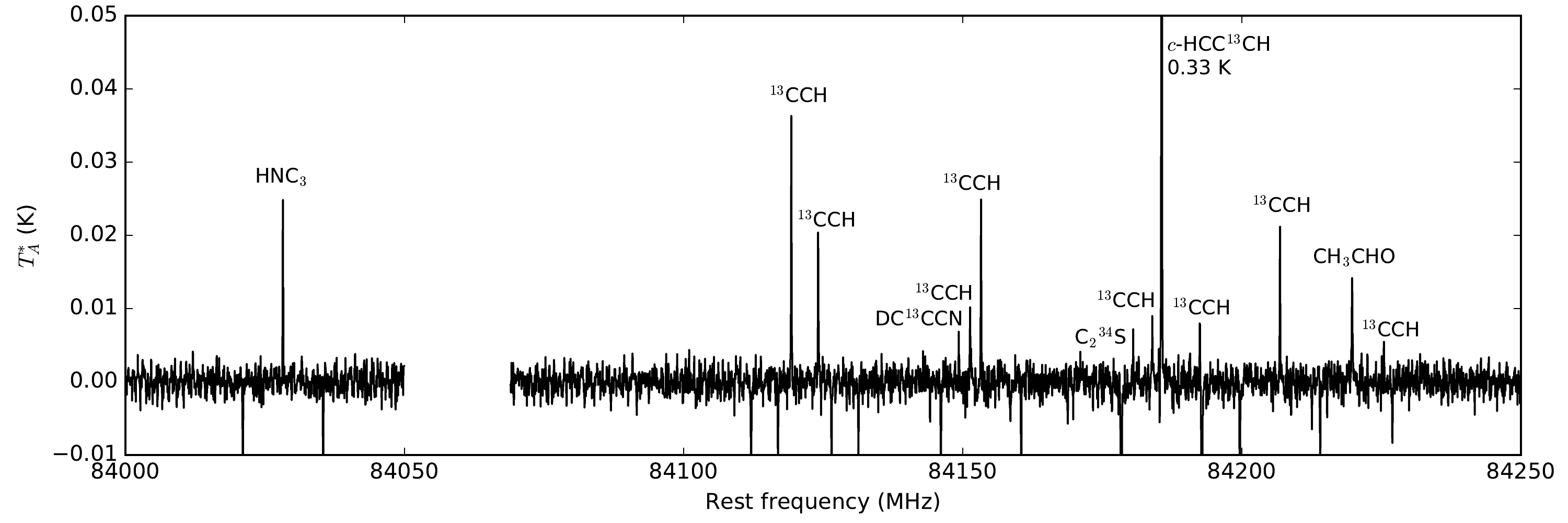}
\includegraphics[width=\textwidth]{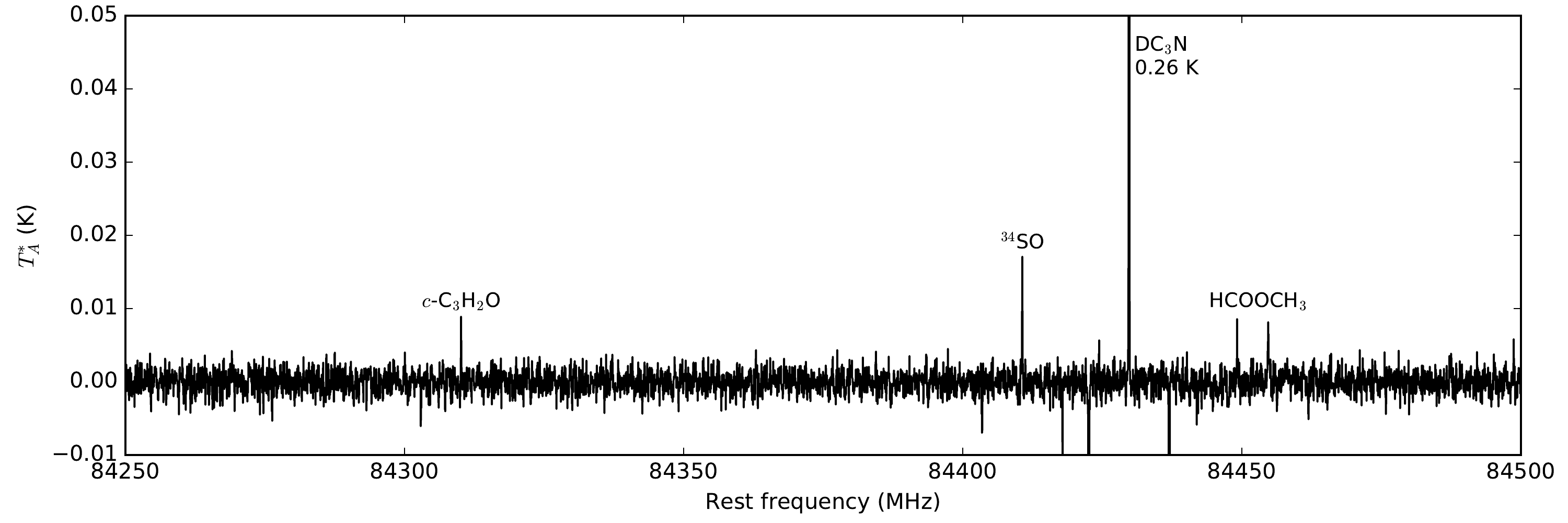}
\includegraphics[width=\textwidth]{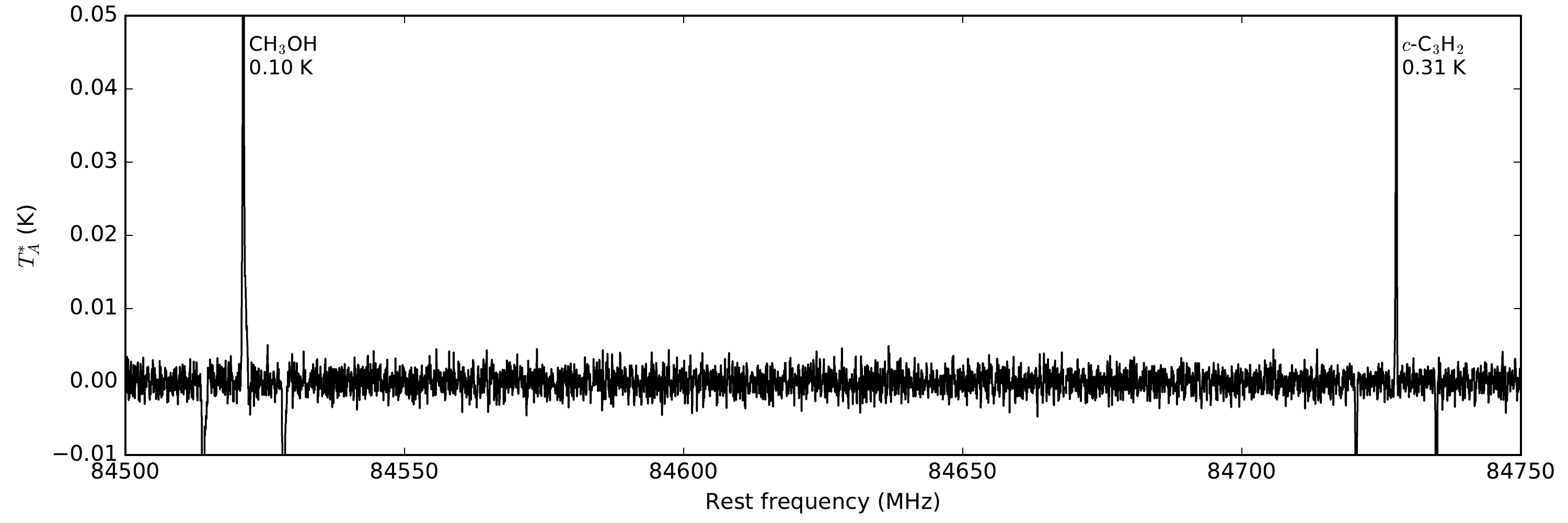}
\includegraphics[width=\textwidth]{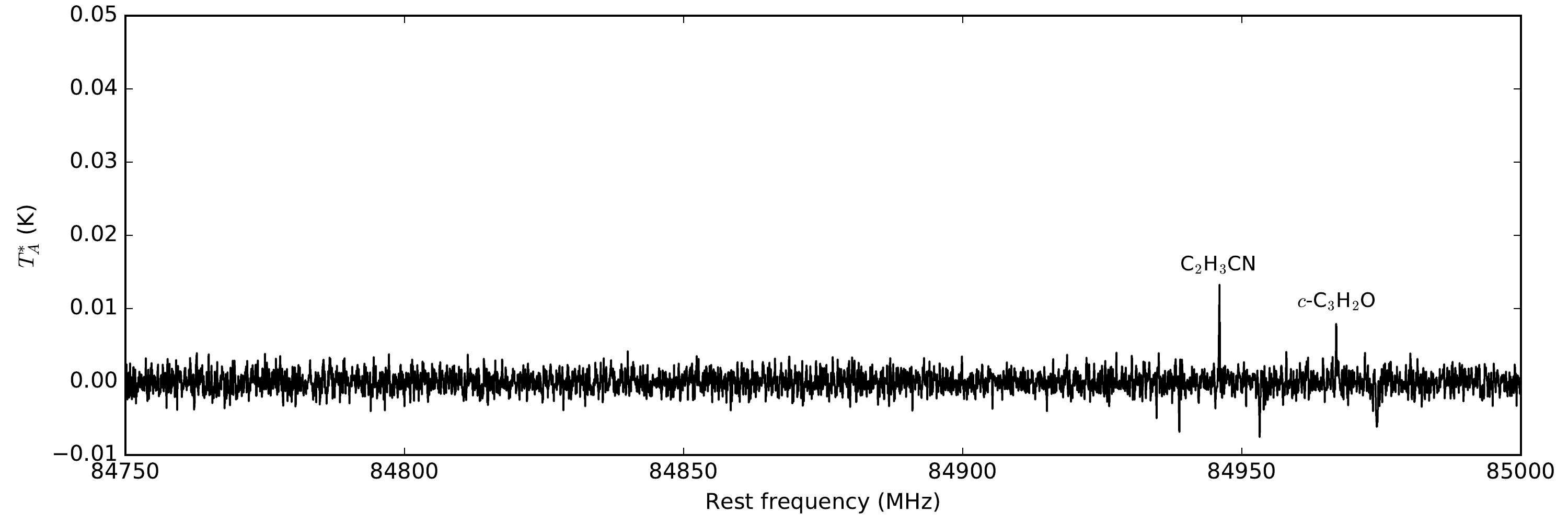}
\caption{Continued}
\end{figure*}

\setcounter{figure}{0}
\begin{figure*}
\centering
\includegraphics[width=\textwidth]{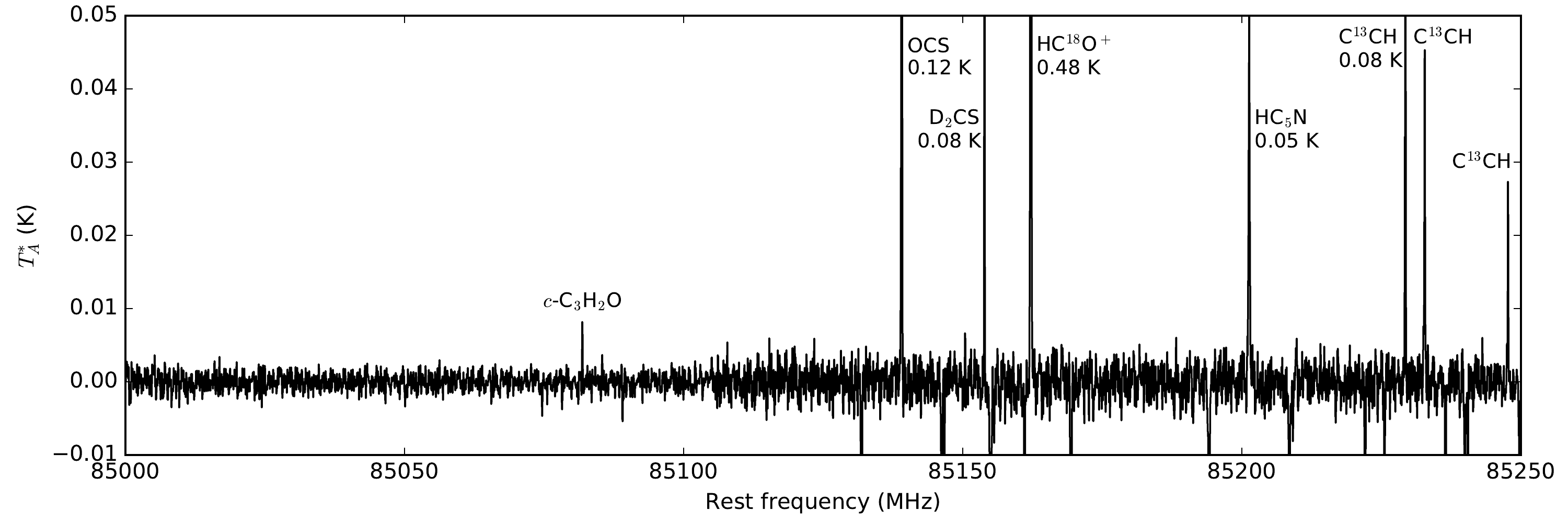}
\includegraphics[width=\textwidth]{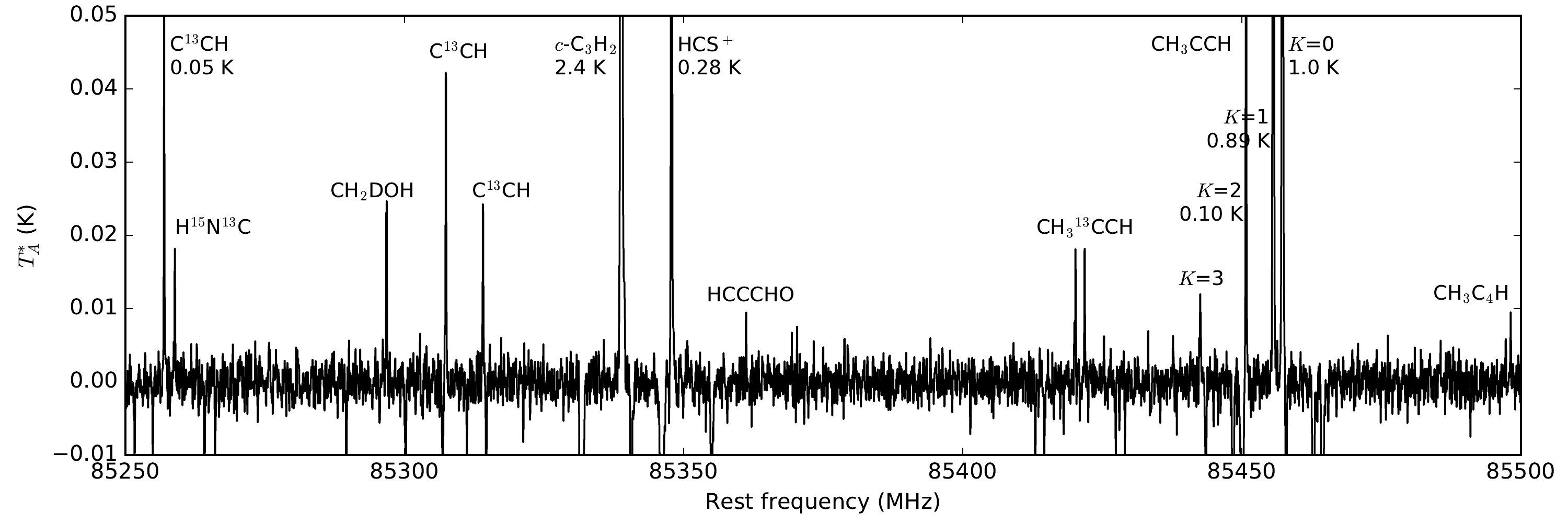}
\includegraphics[width=\textwidth]{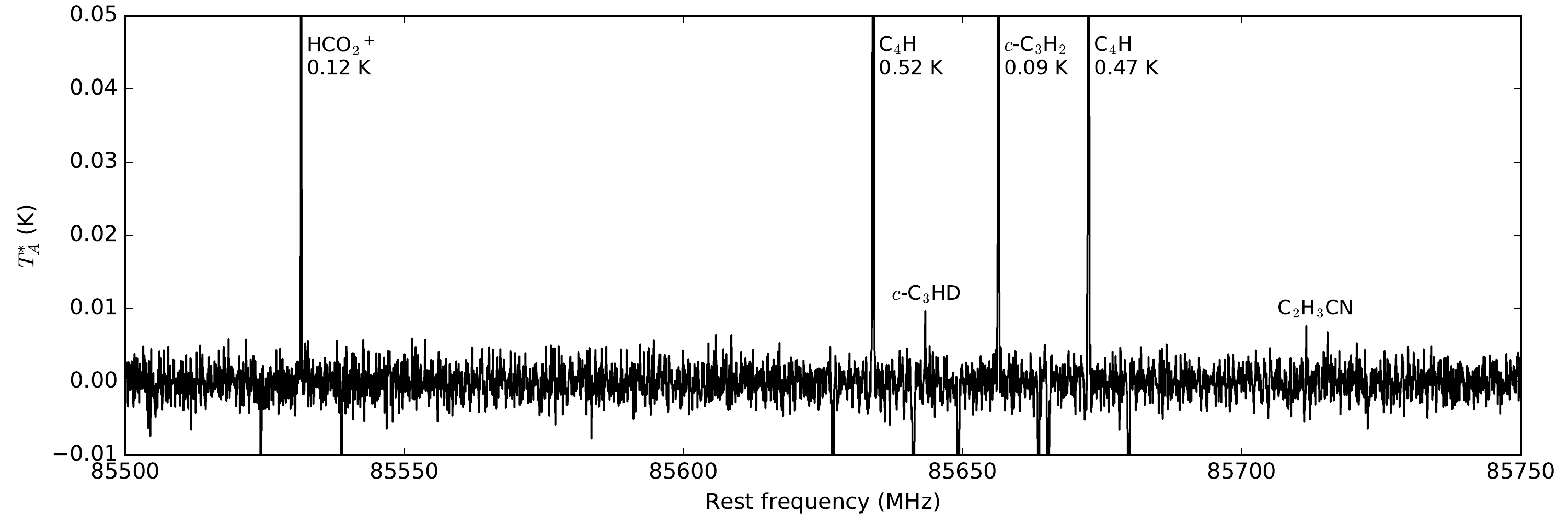}
\includegraphics[width=\textwidth]{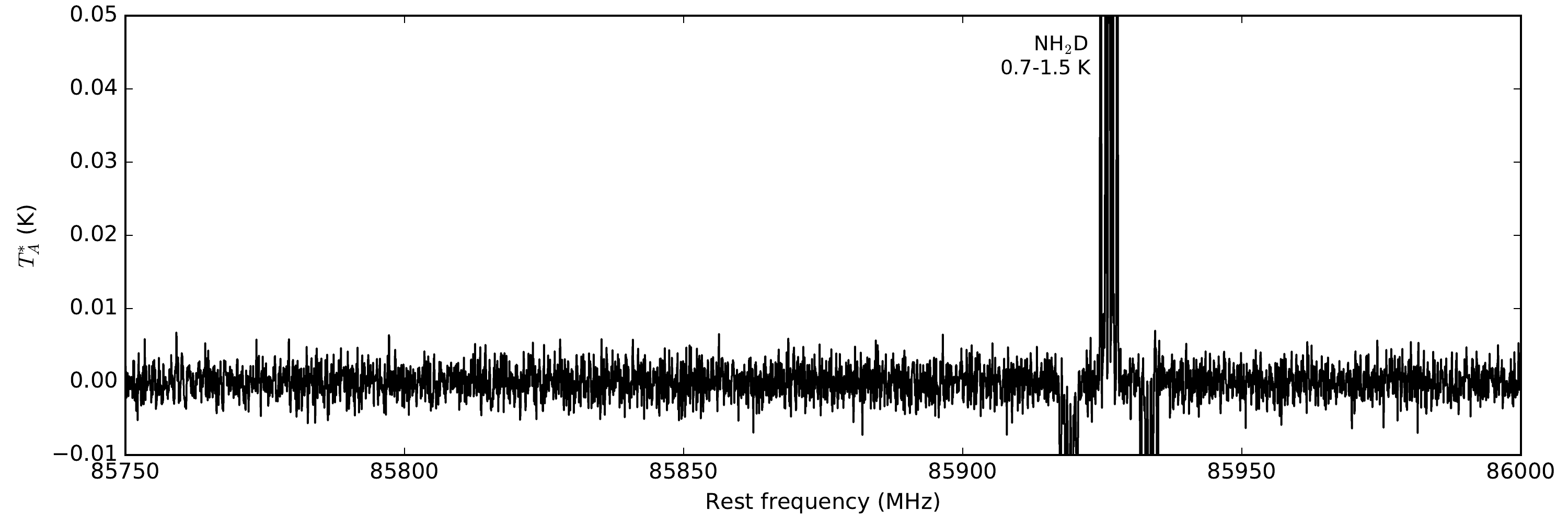}
\caption{Continued}
\end{figure*}

\setcounter{figure}{0}
\begin{figure*}
\centering
\includegraphics[width=\textwidth]{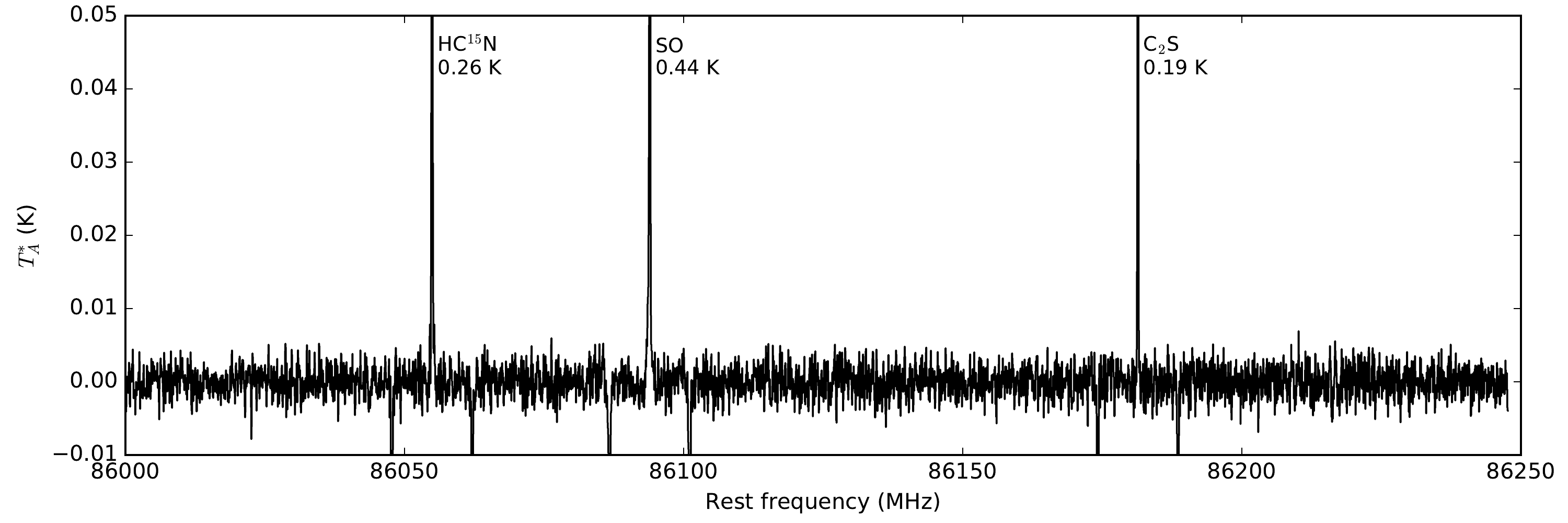}
\includegraphics[width=\textwidth]{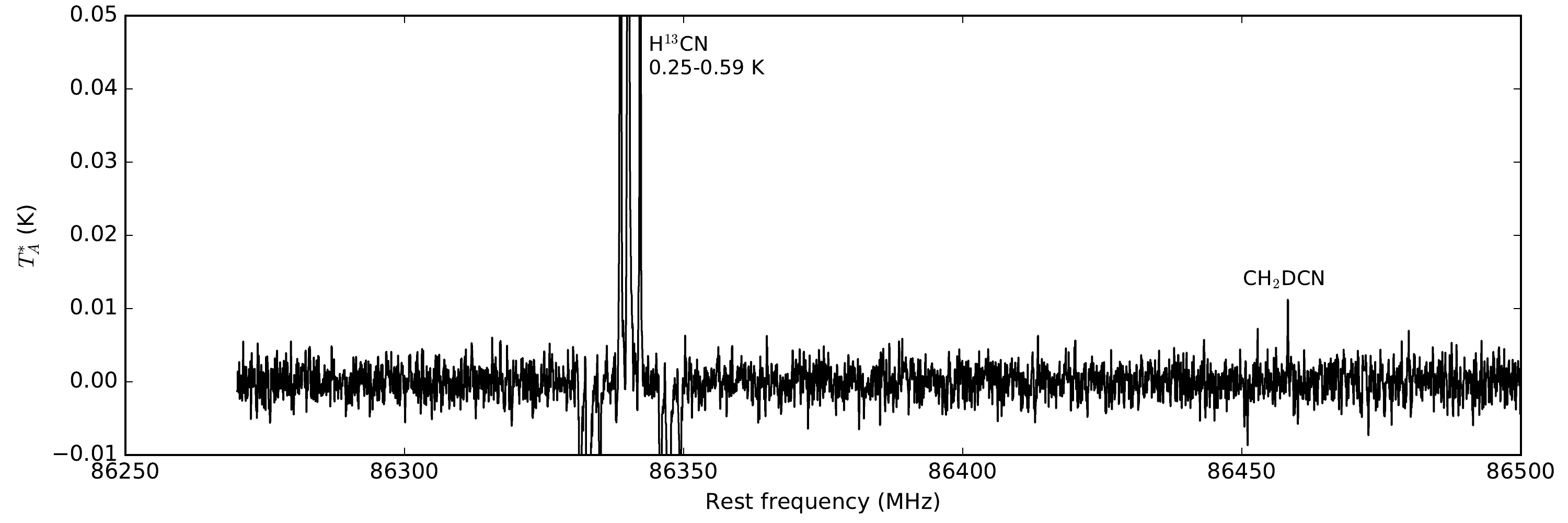}
\includegraphics[width=\textwidth]{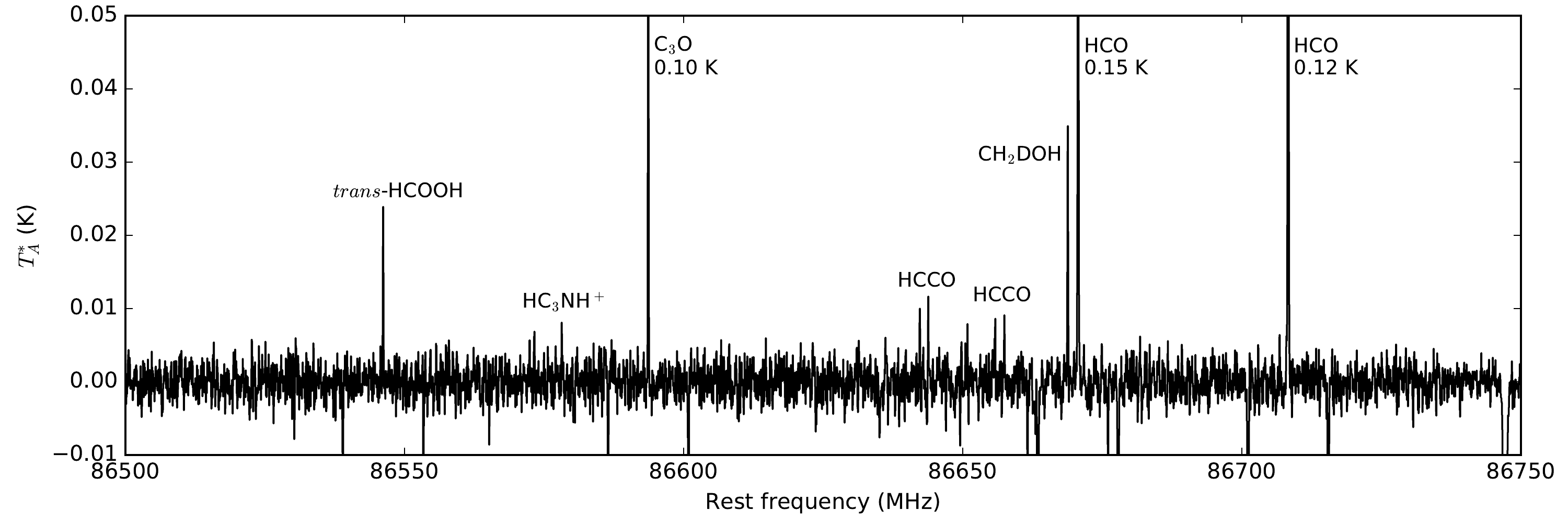}
\includegraphics[width=\textwidth]{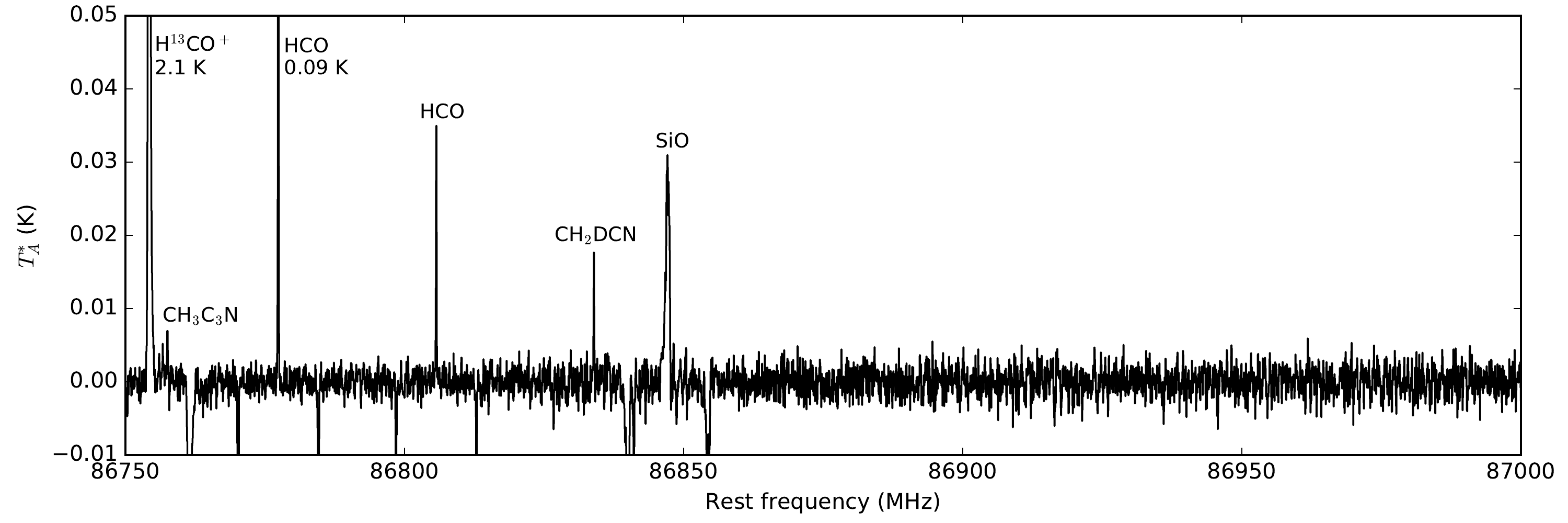}
\caption{Continued}
\end{figure*}

\setcounter{figure}{0}
\begin{figure*}
\centering
\includegraphics[width=\textwidth]{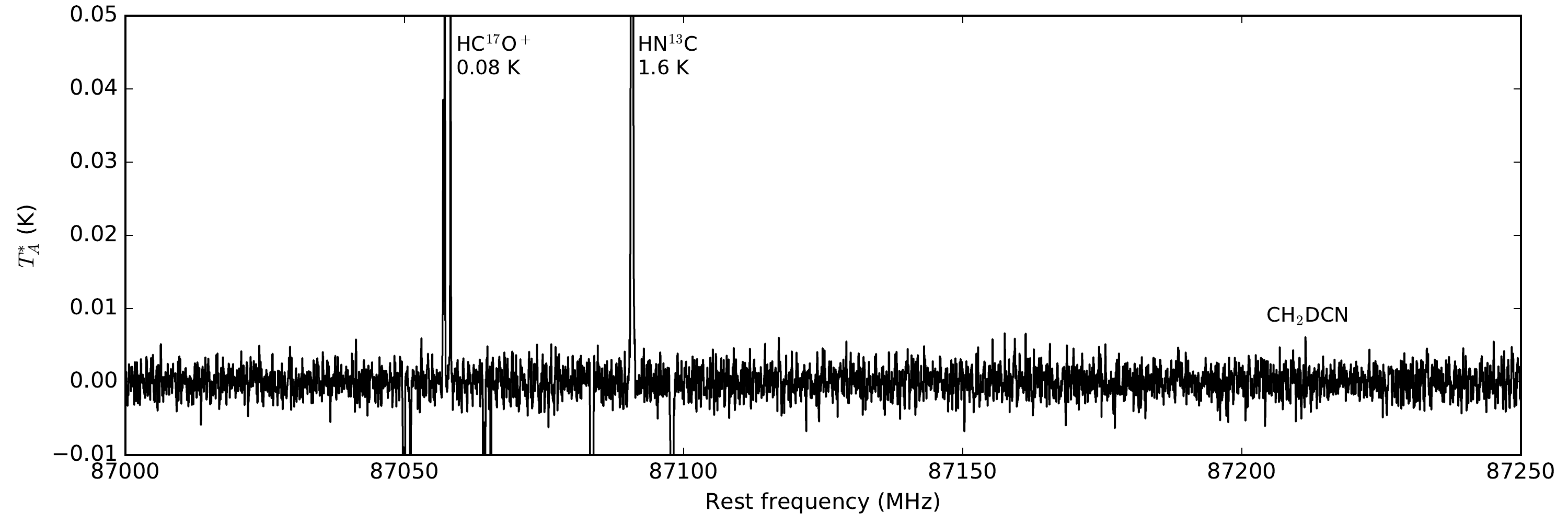}
\includegraphics[width=\textwidth]{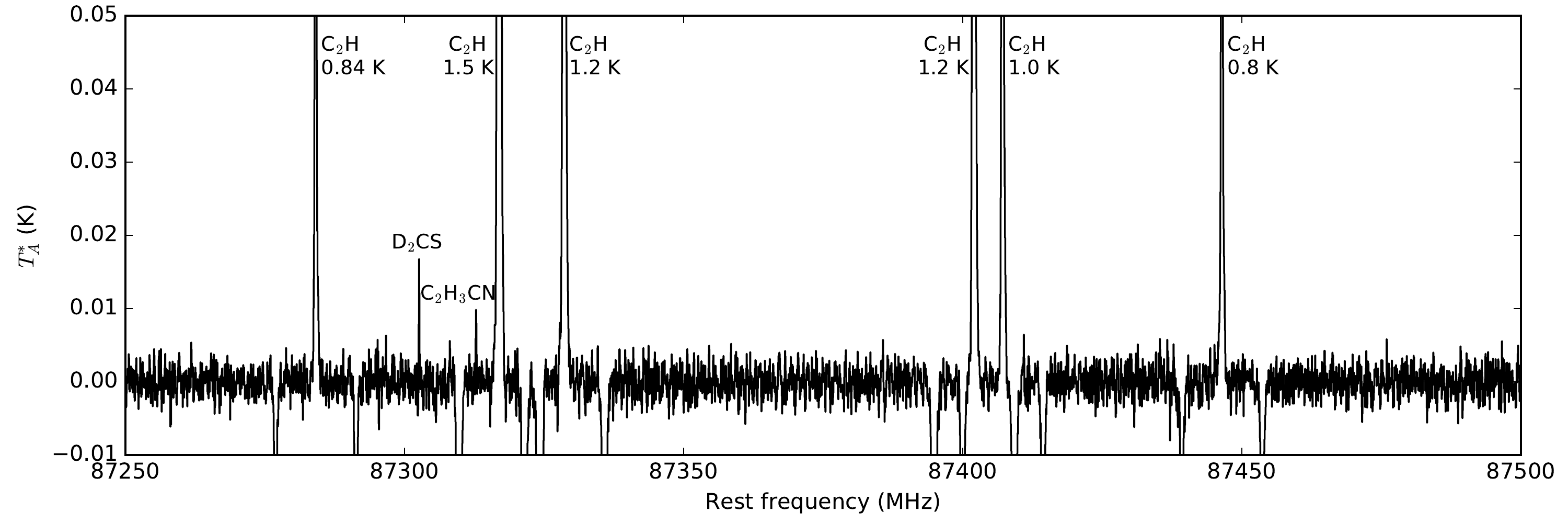}
\includegraphics[width=\textwidth]{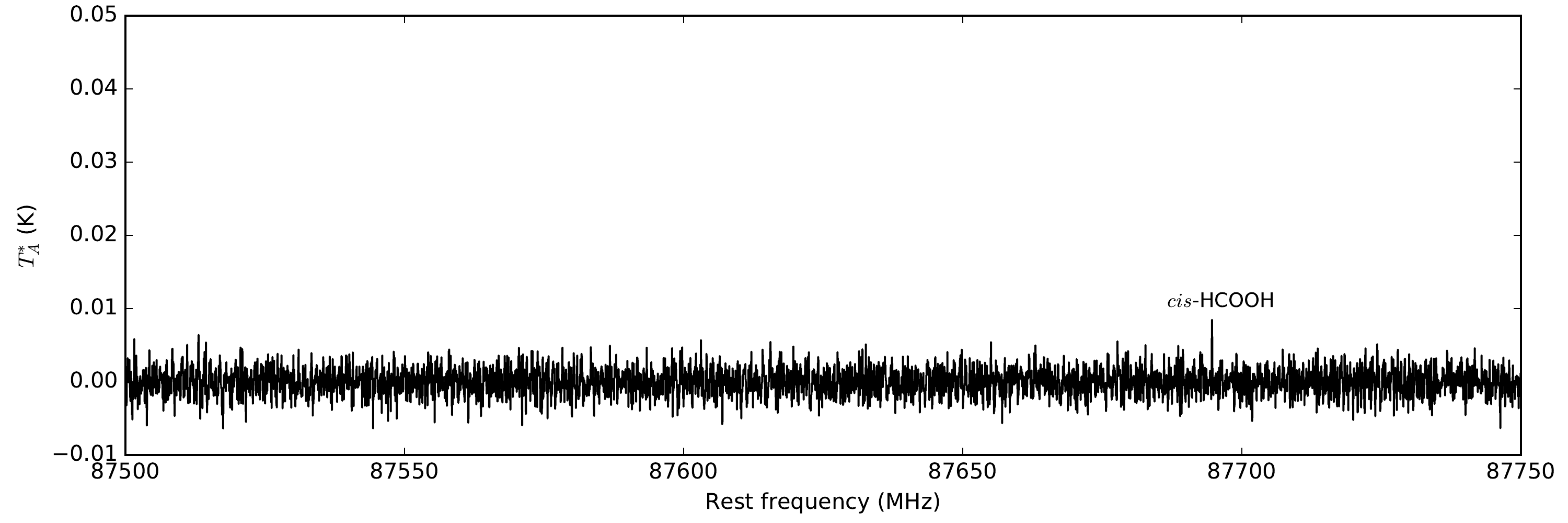}
\includegraphics[width=\textwidth]{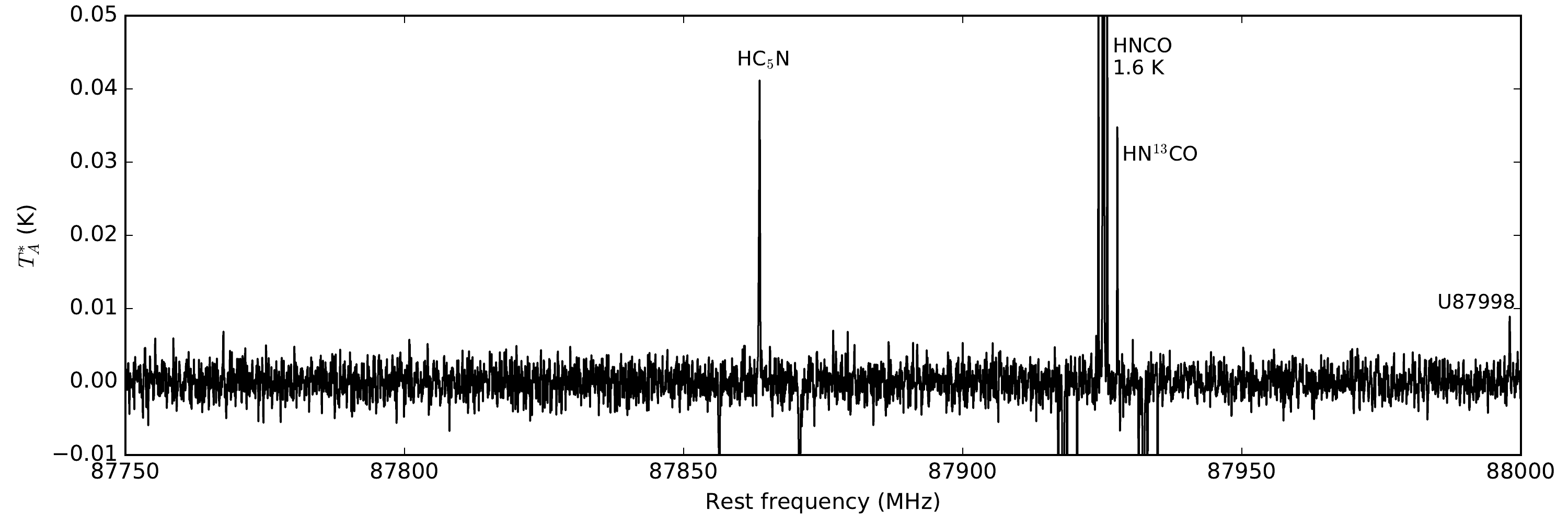}
\caption{Continued}
\end{figure*}

\setcounter{figure}{0}
\begin{figure*}
\centering
\includegraphics[width=\textwidth]{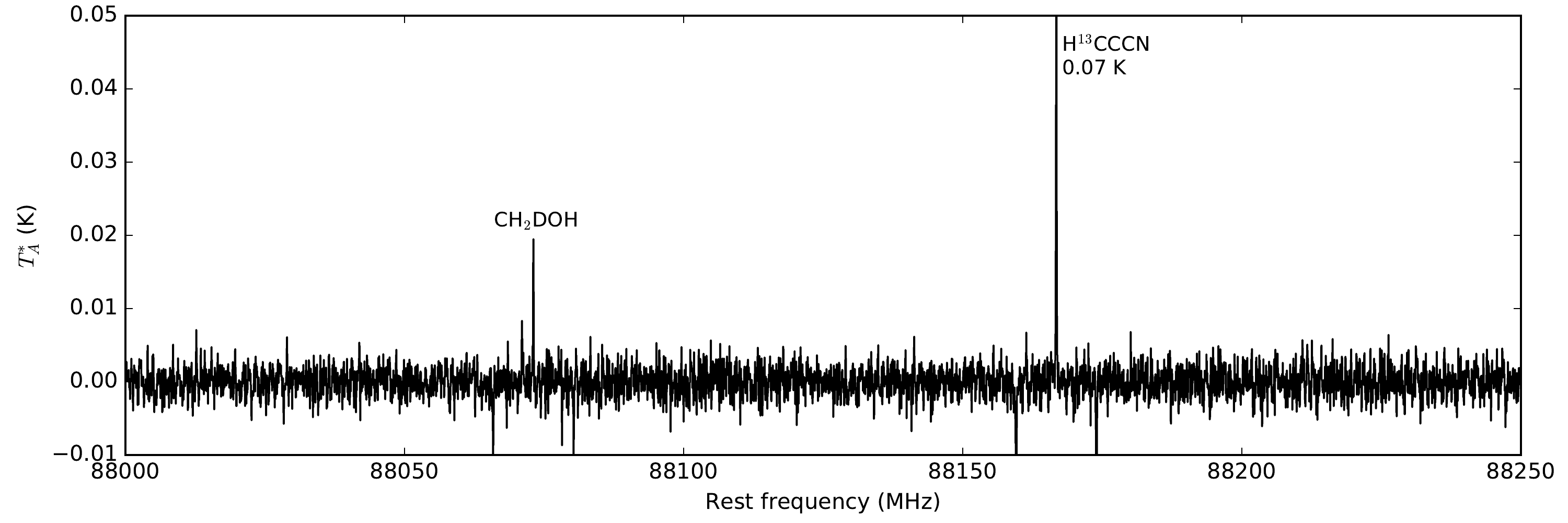}
\includegraphics[width=\textwidth]{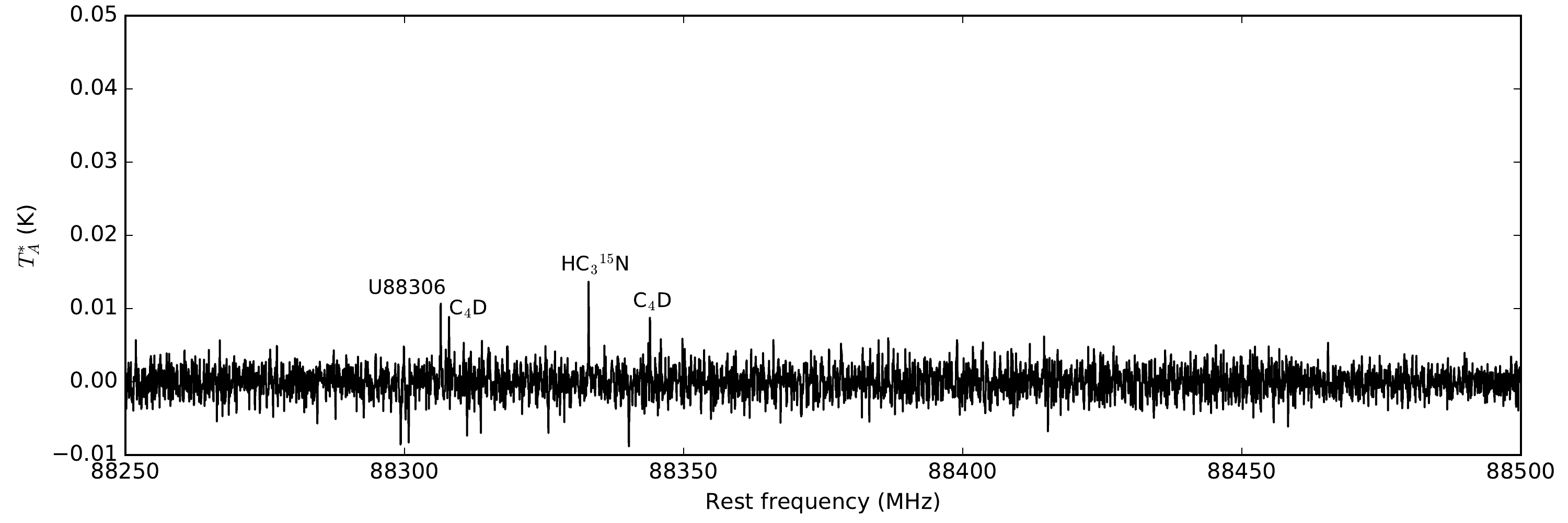}
\includegraphics[width=\textwidth]{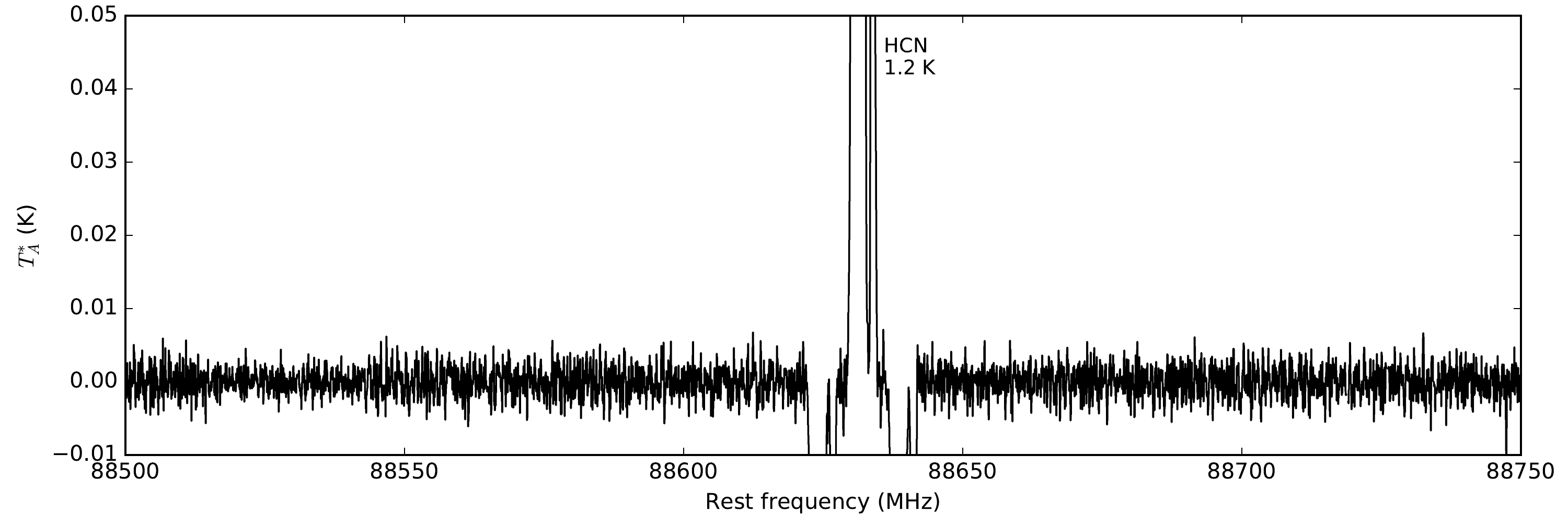}
\includegraphics[width=\textwidth]{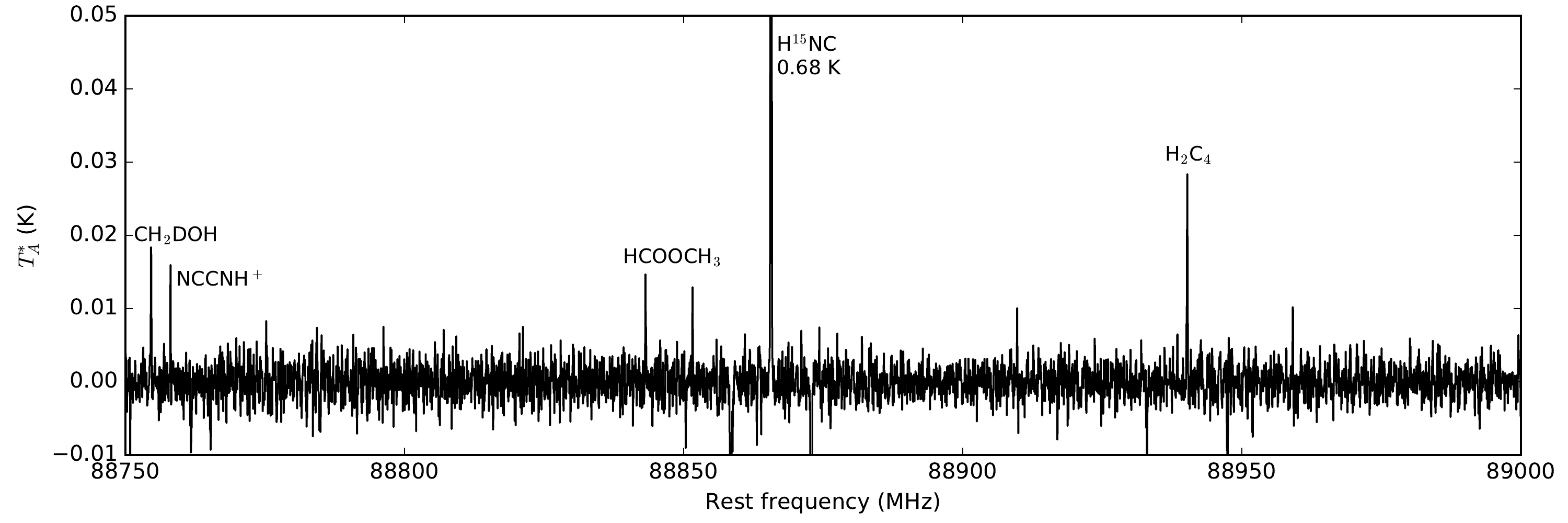}
\caption{Continued}
\end{figure*}

\setcounter{figure}{0}
\begin{figure*}
\centering
\includegraphics[width=\textwidth]{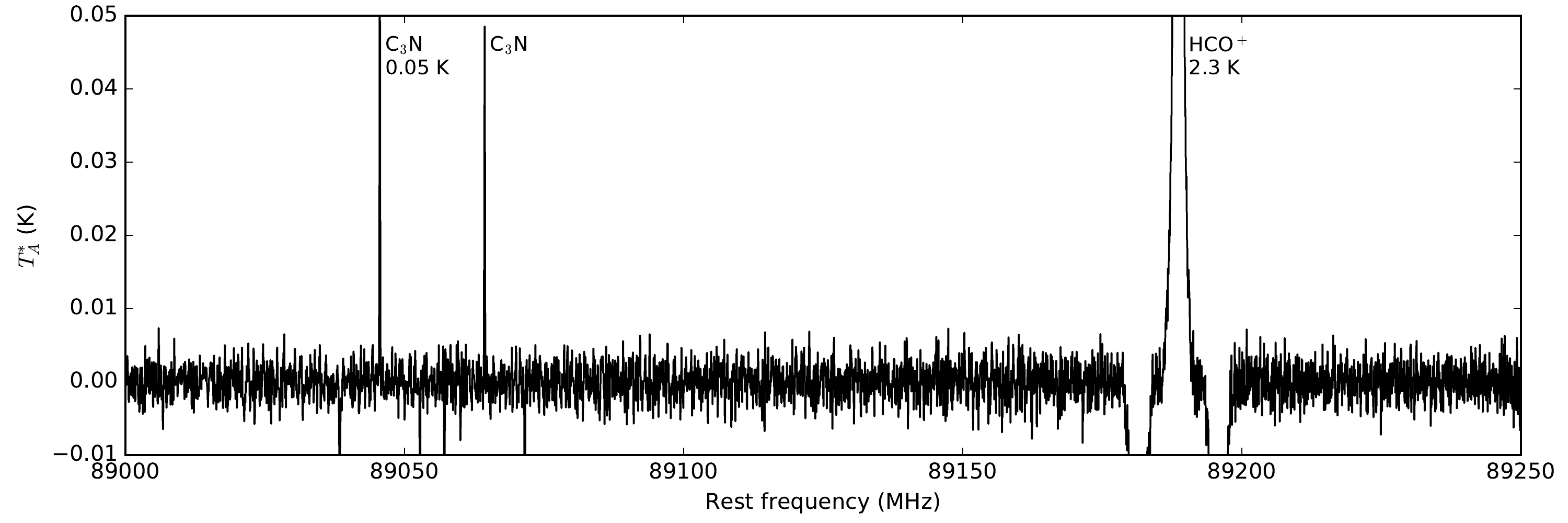}
\includegraphics[width=\textwidth]{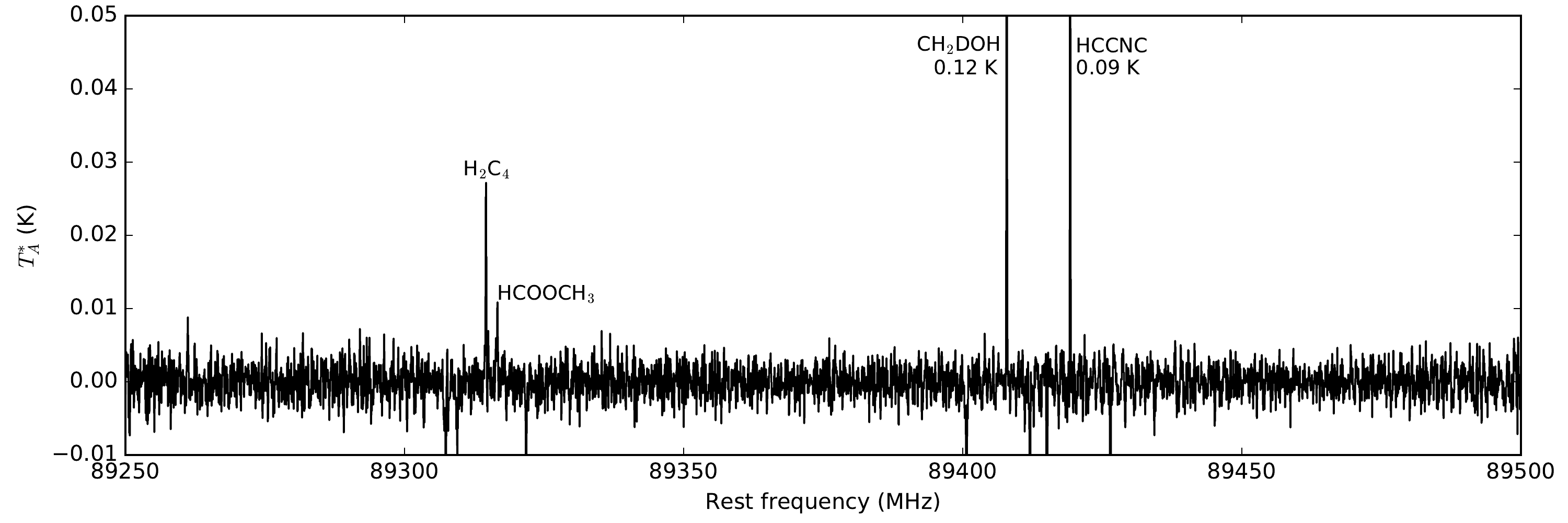}
\includegraphics[width=\textwidth]{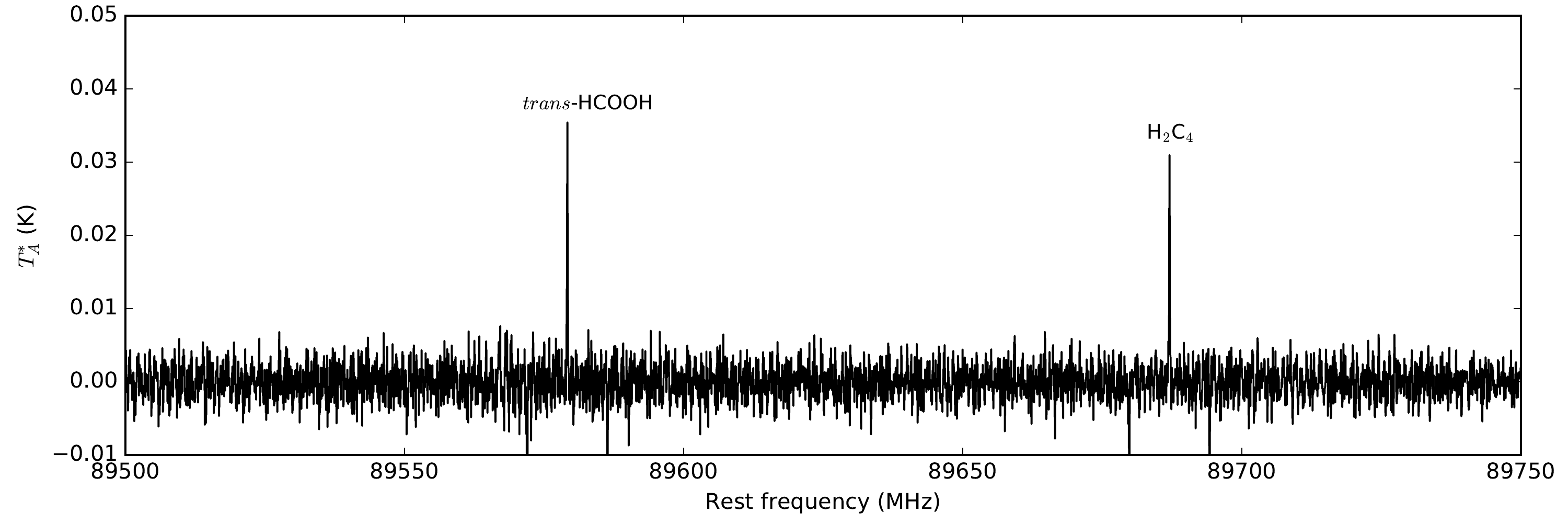}
\includegraphics[width=\textwidth]{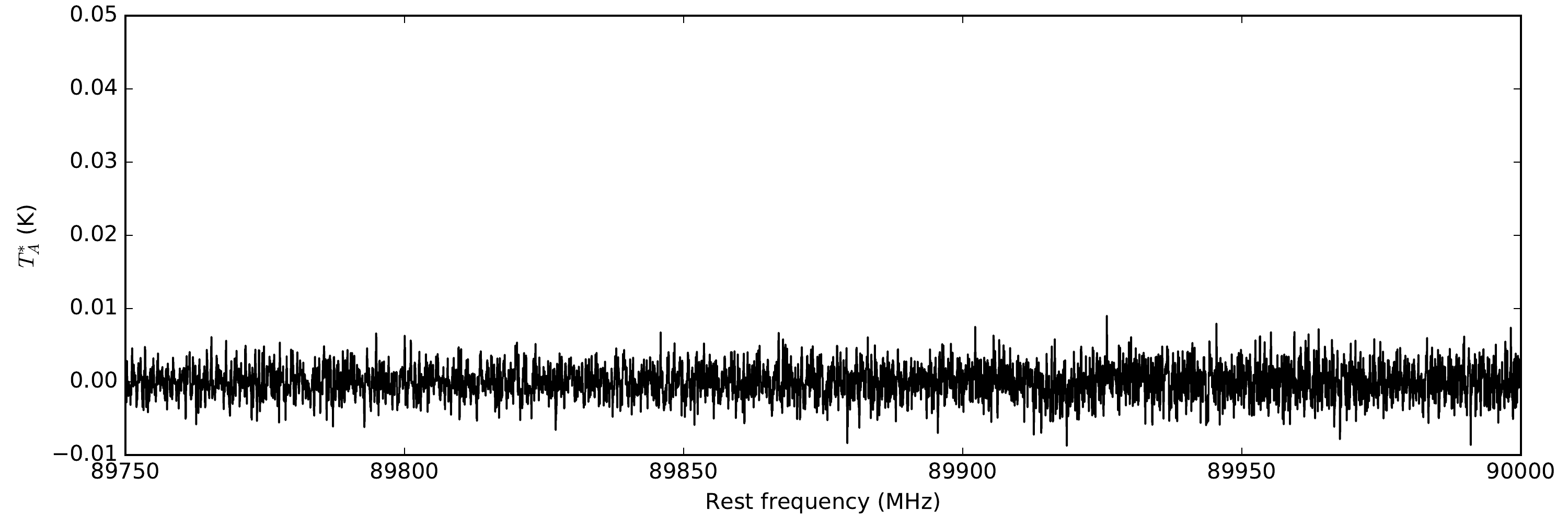}
\caption{Continued}
\end{figure*}

\setcounter{figure}{0}
\begin{figure*}
\centering
\includegraphics[width=\textwidth]{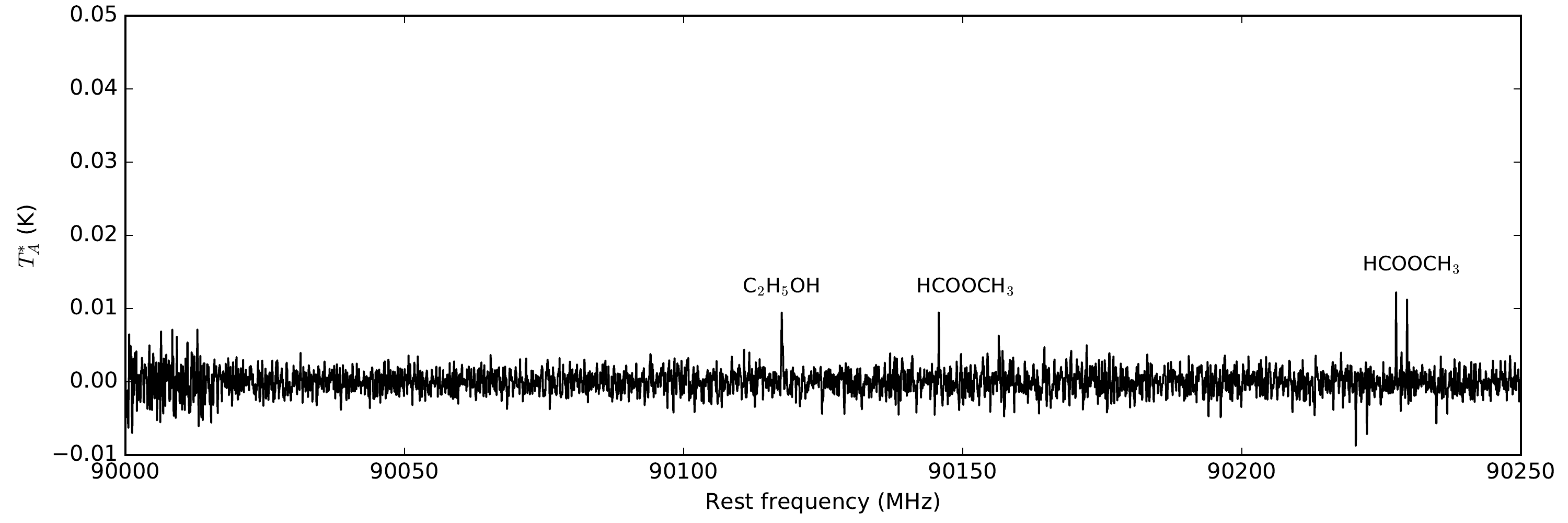}
\includegraphics[width=\textwidth]{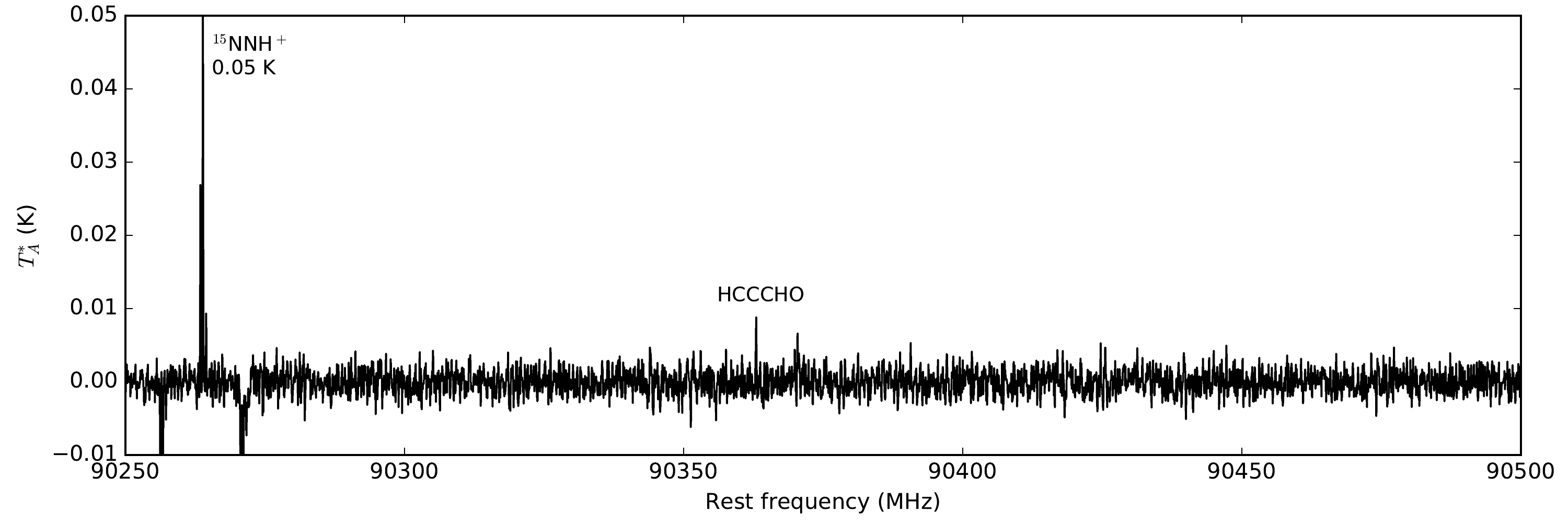}
\includegraphics[width=\textwidth]{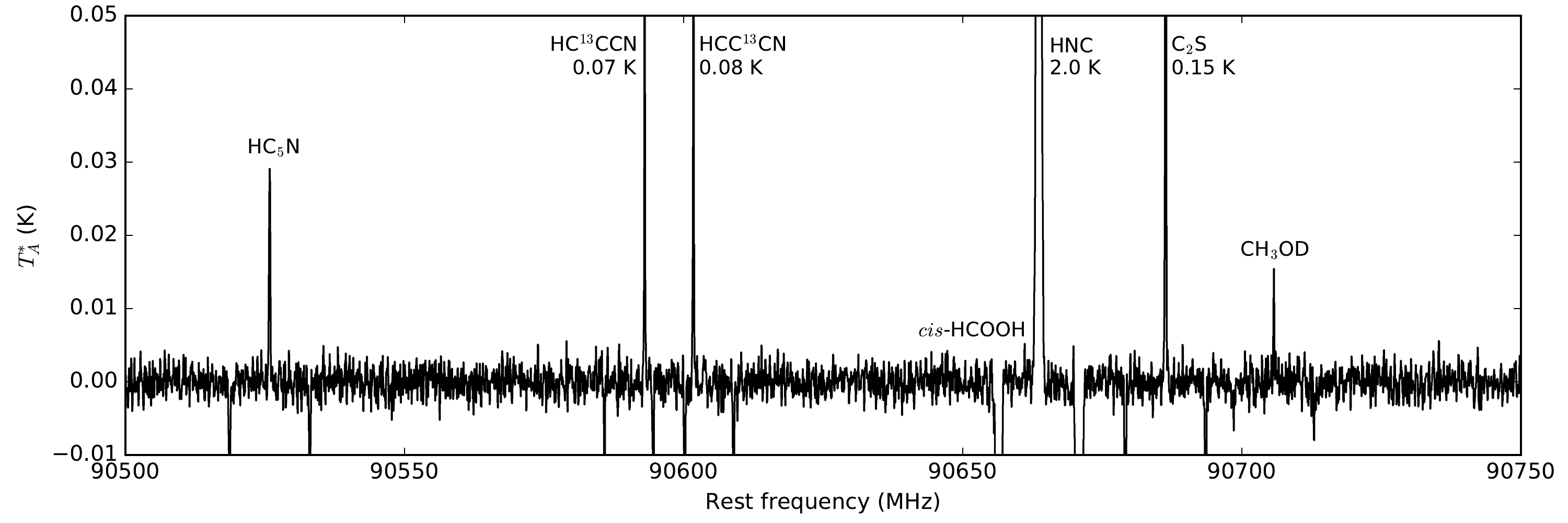}
\includegraphics[width=\textwidth]{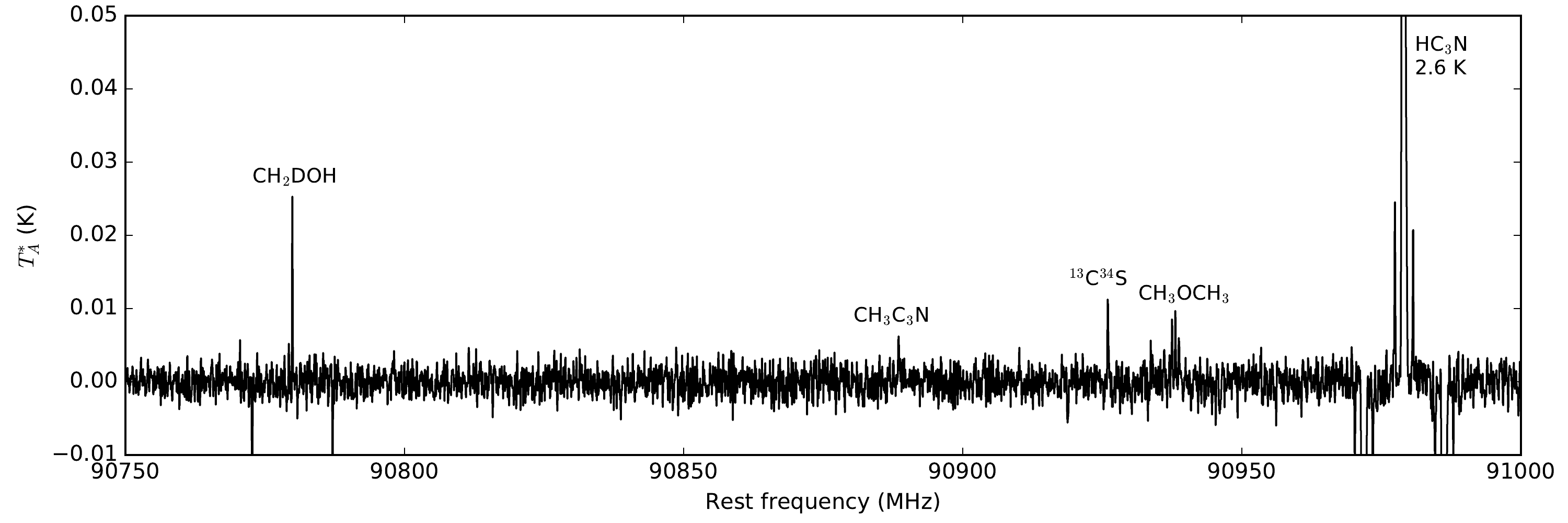}
\caption{Continued}
\end{figure*}

\setcounter{figure}{0}
\begin{figure*}
\centering
\includegraphics[width=\textwidth]{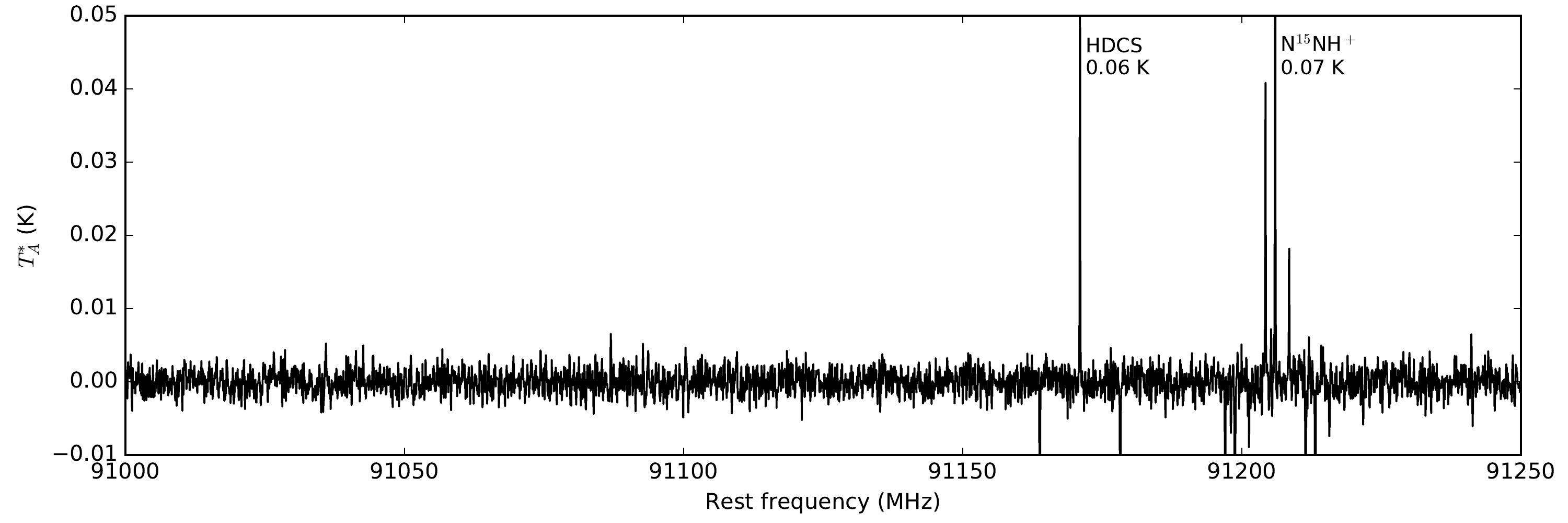}
\includegraphics[width=\textwidth]{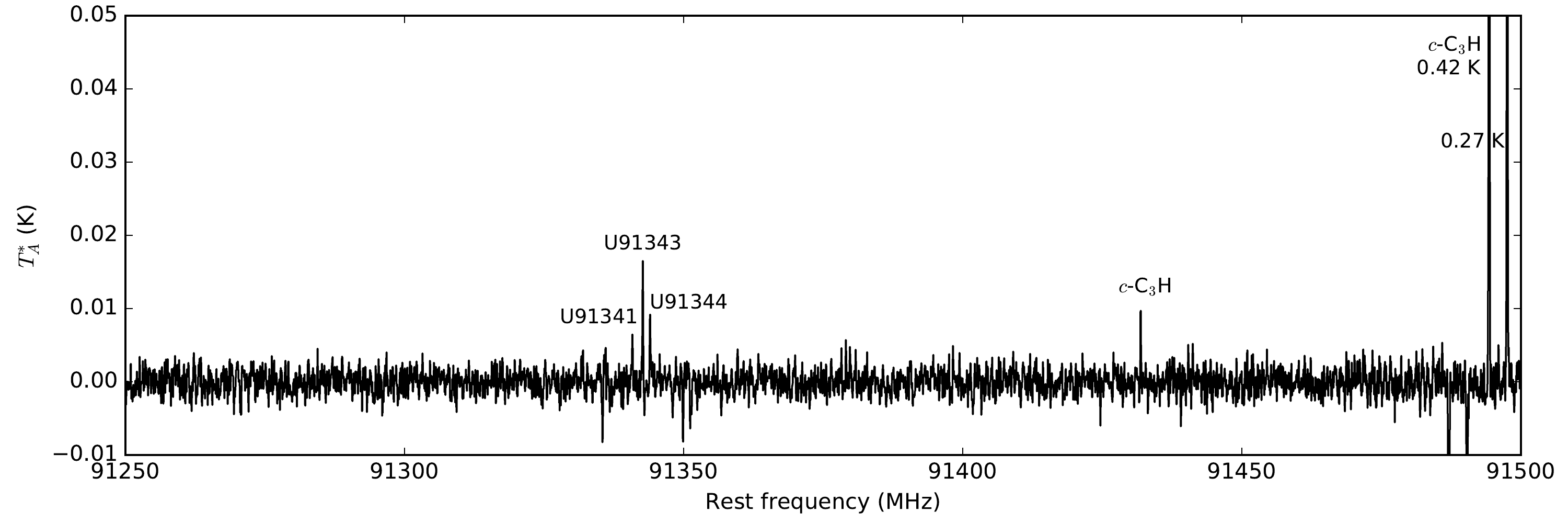}
\includegraphics[width=\textwidth]{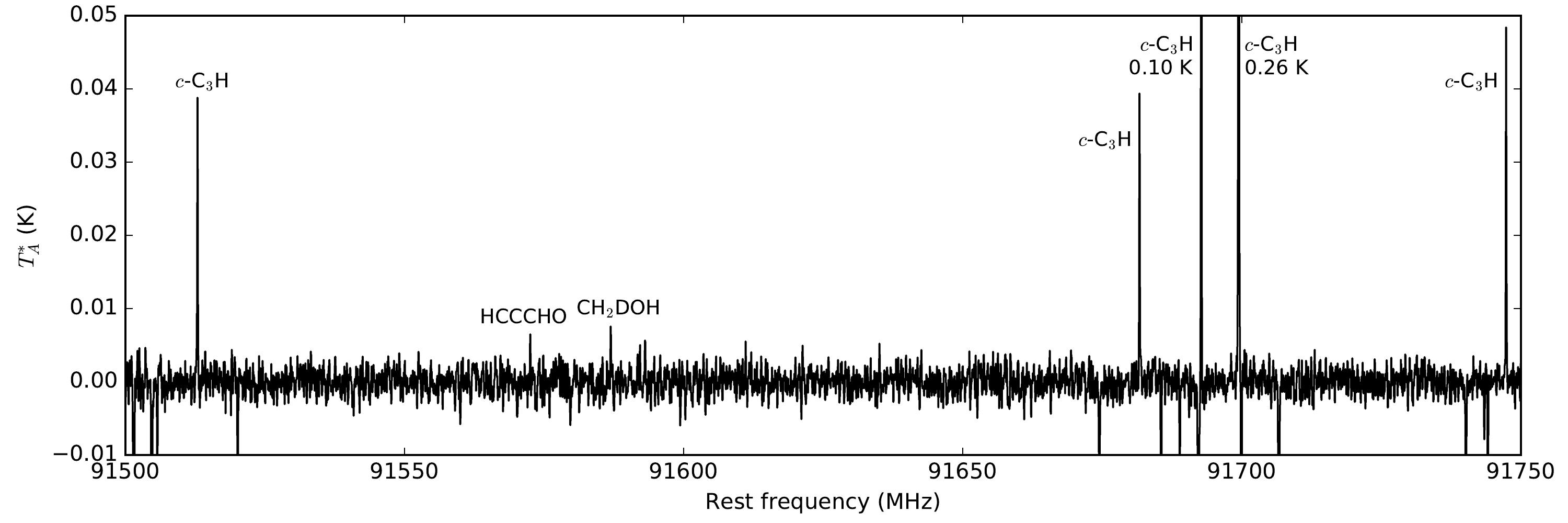}
\includegraphics[width=\textwidth]{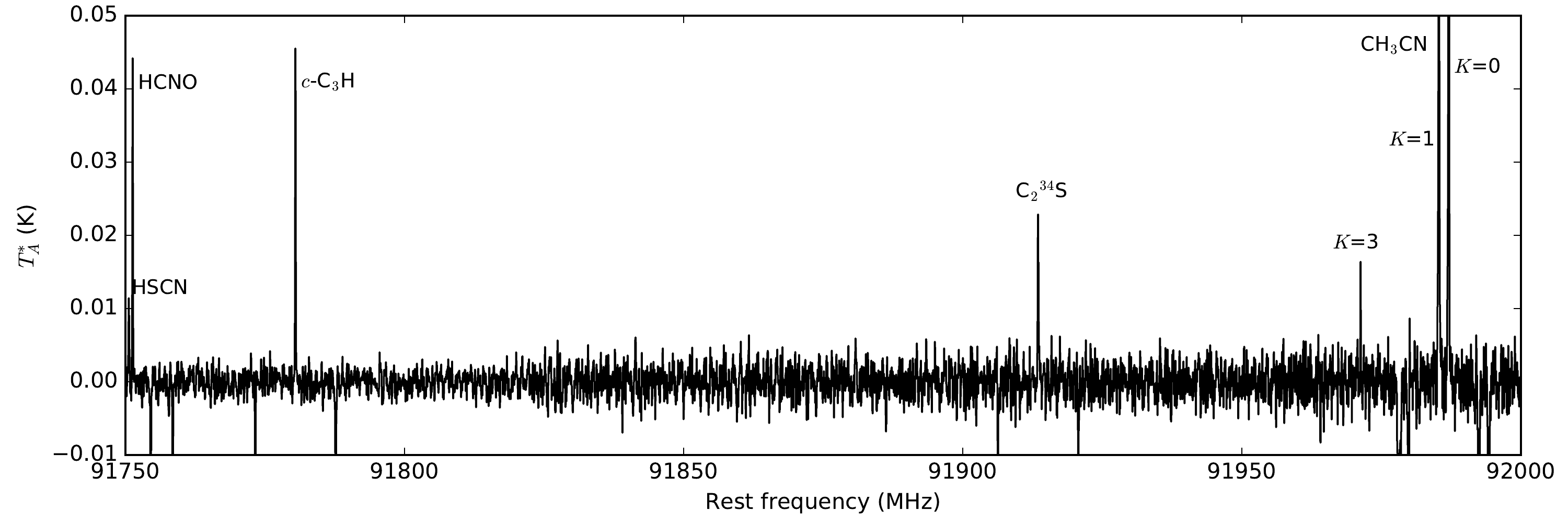}
\caption{Continued}
\end{figure*}

\setcounter{figure}{0}
\begin{figure*}
\centering
\includegraphics[width=\textwidth]{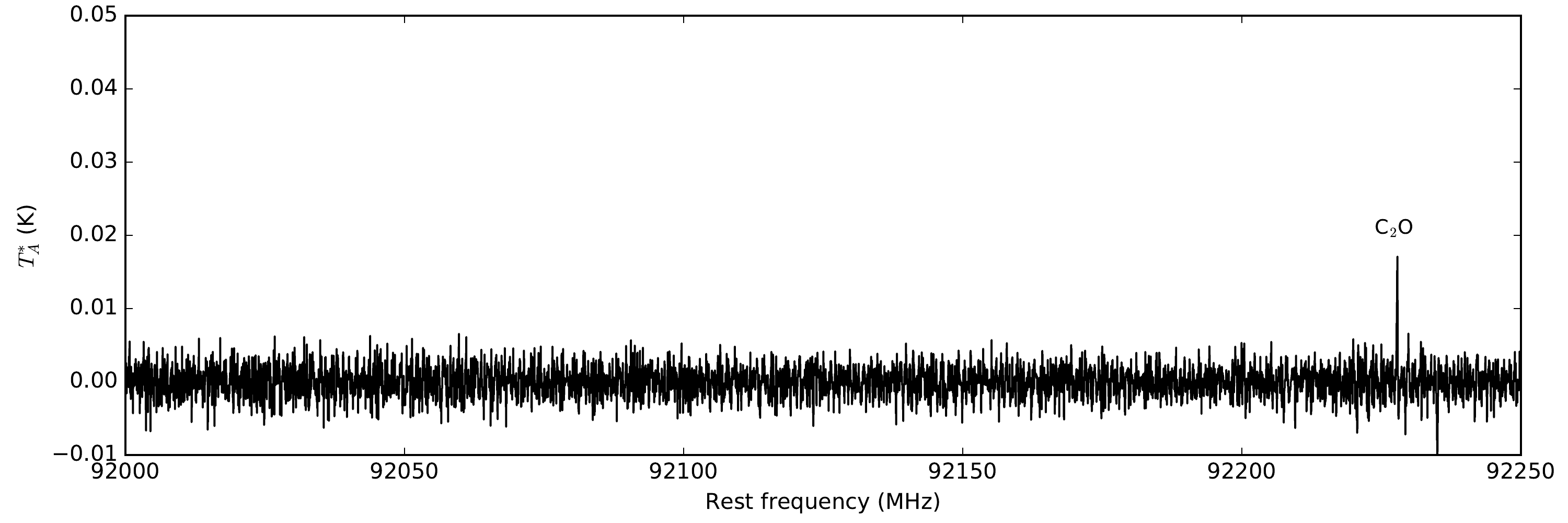}
\includegraphics[width=\textwidth]{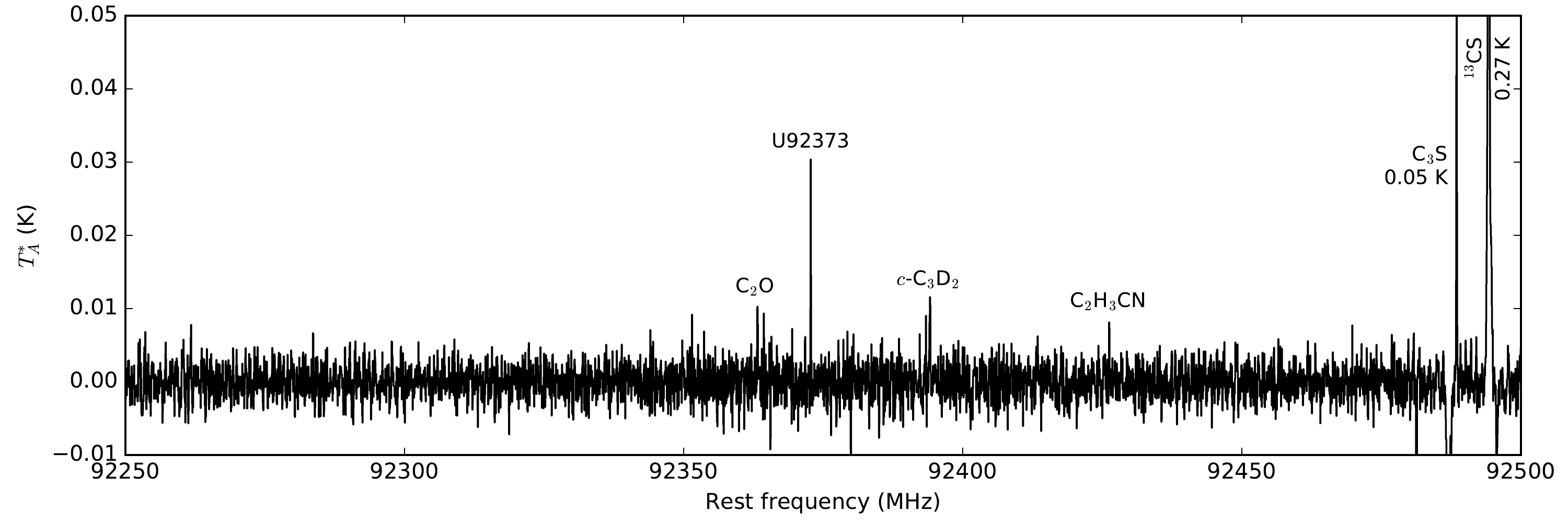}
\includegraphics[width=\textwidth]{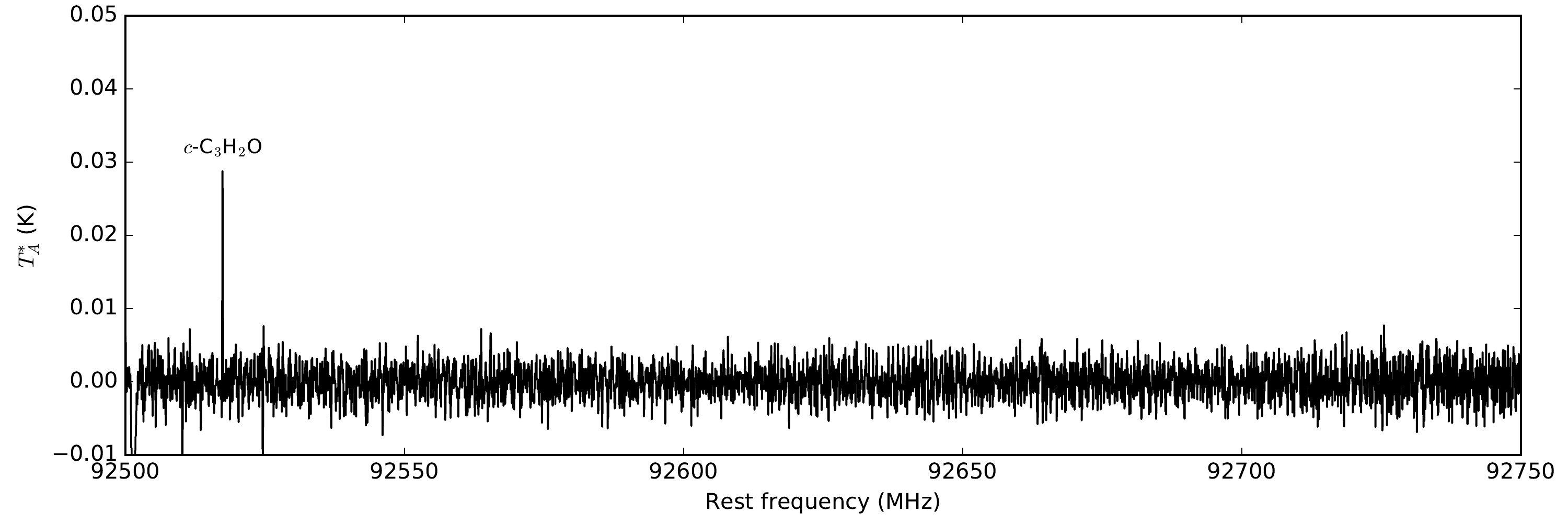}
\includegraphics[width=\textwidth]{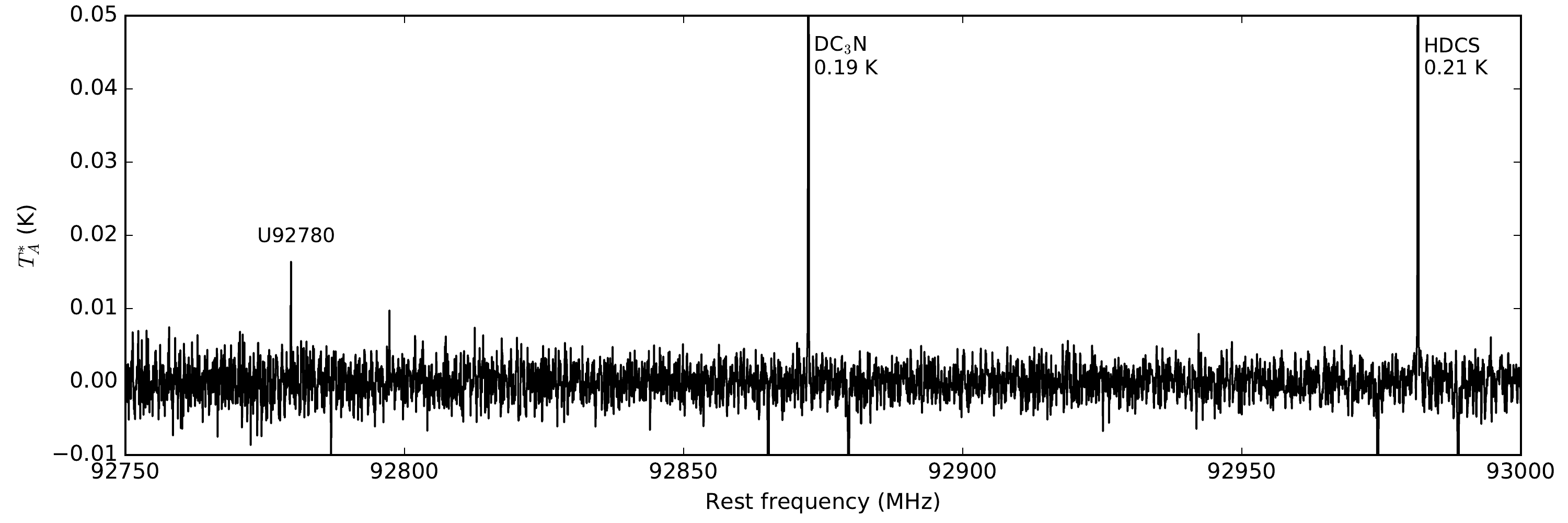}
\caption{Continued}
\end{figure*}

\setcounter{figure}{0}
\begin{figure*}
\centering
\includegraphics[width=\textwidth]{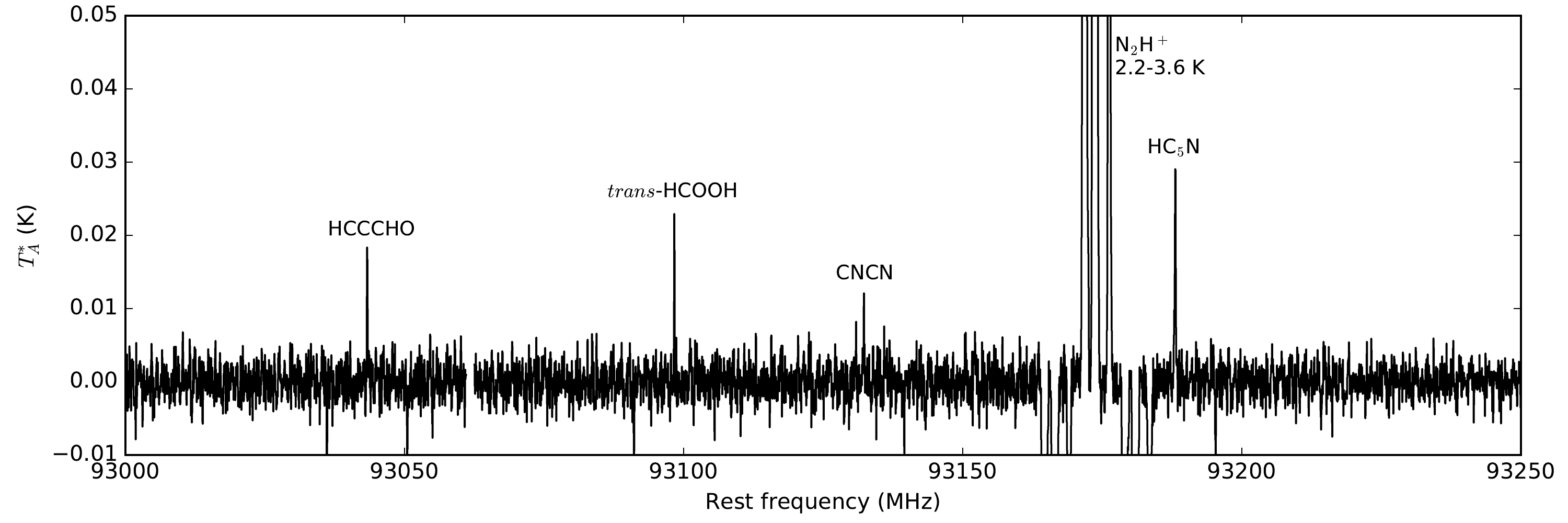}
\includegraphics[width=\textwidth]{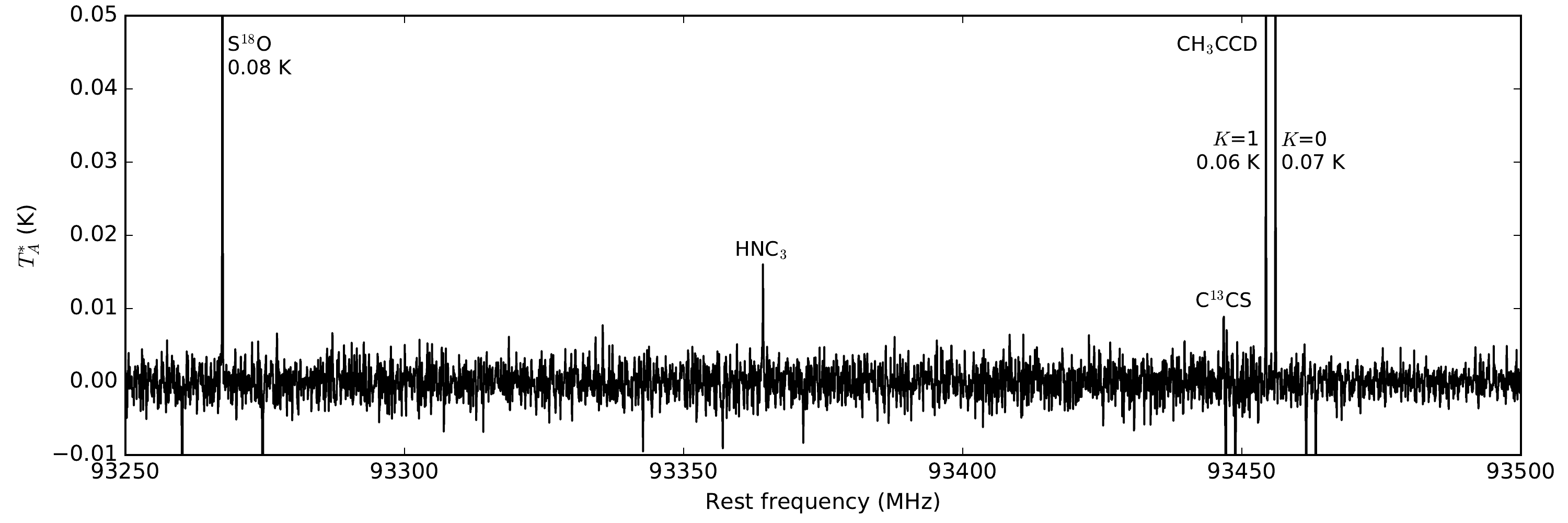}
\includegraphics[width=\textwidth]{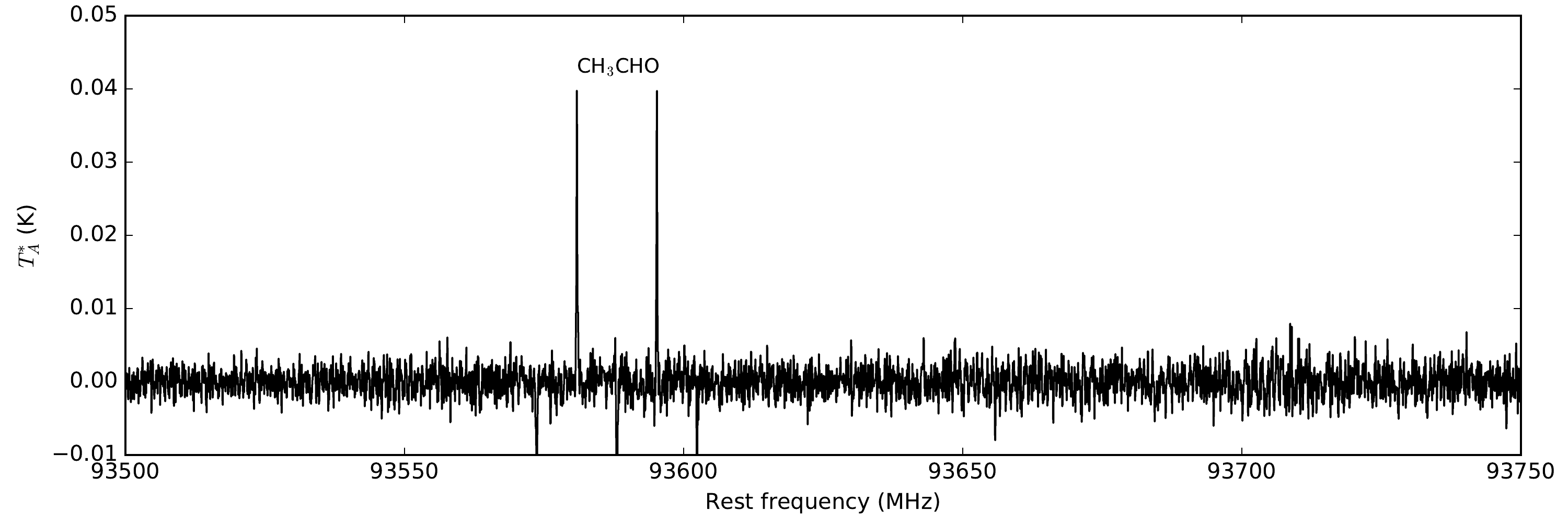}
\includegraphics[width=\textwidth]{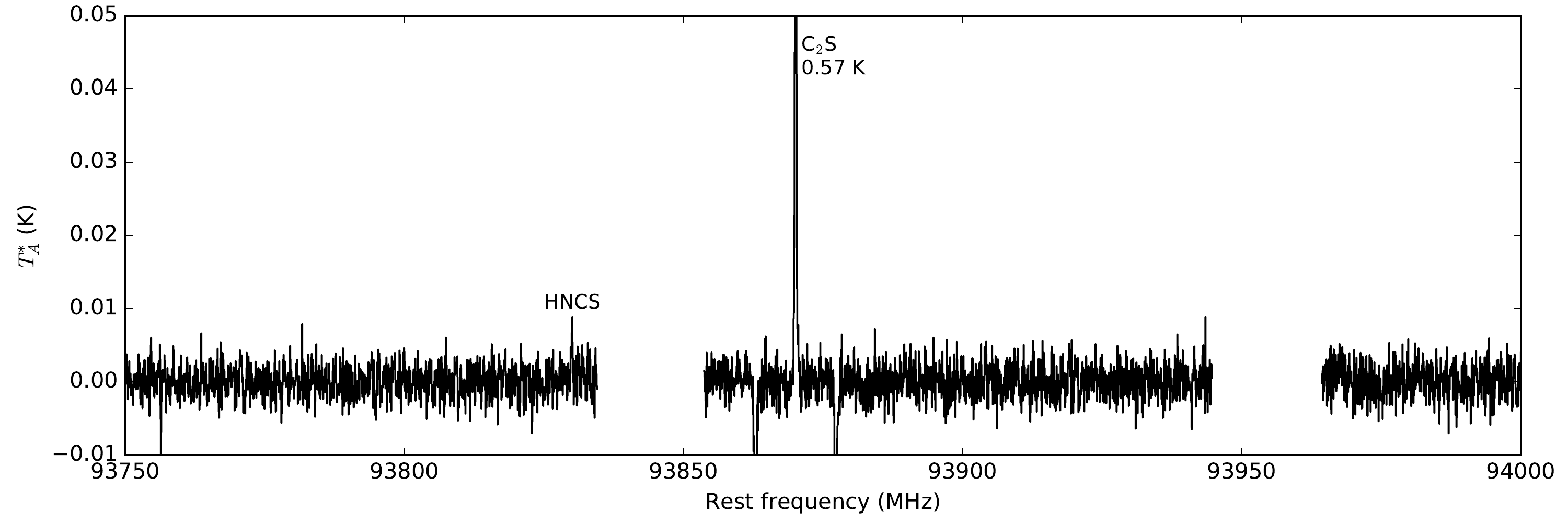}
\caption{Continued}
\end{figure*}

\setcounter{figure}{0}
\begin{figure*}
\centering
\includegraphics[width=\textwidth]{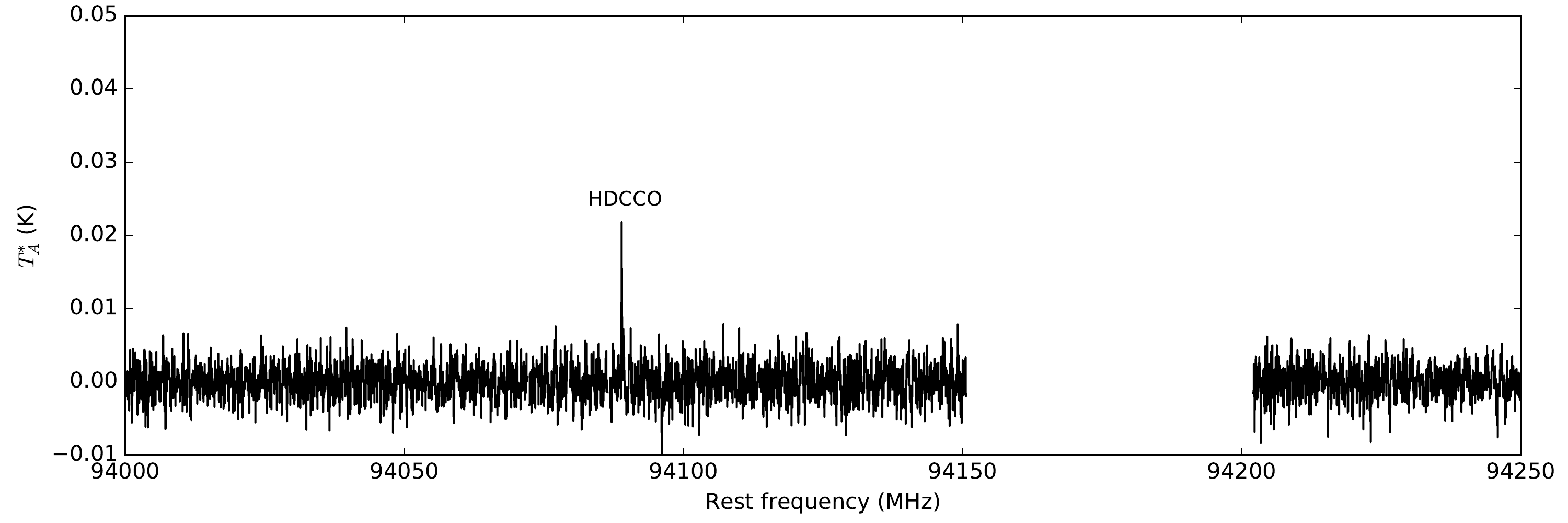}
\includegraphics[width=\textwidth]{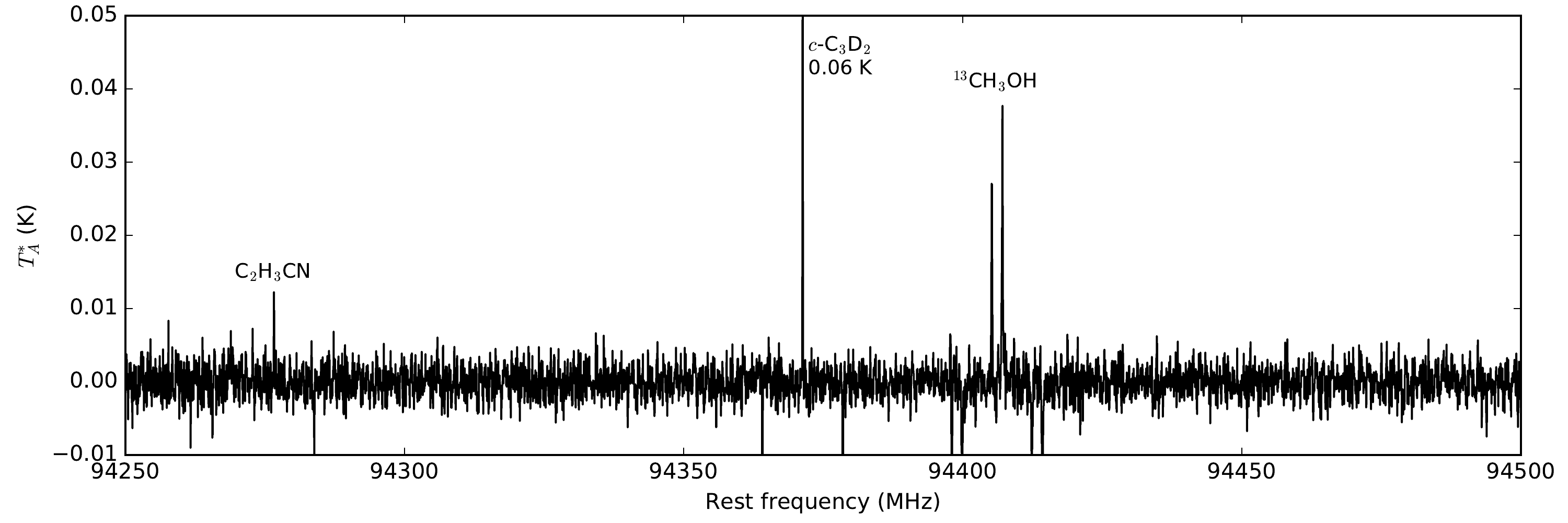}
\includegraphics[width=\textwidth]{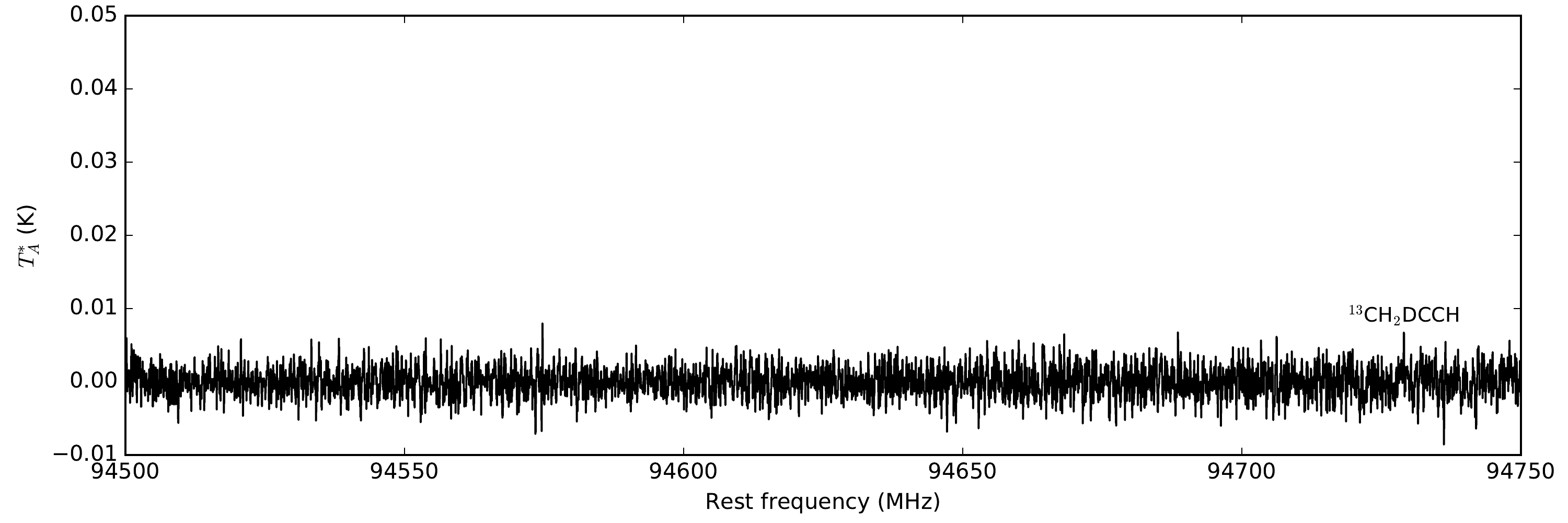}
\includegraphics[width=\textwidth]{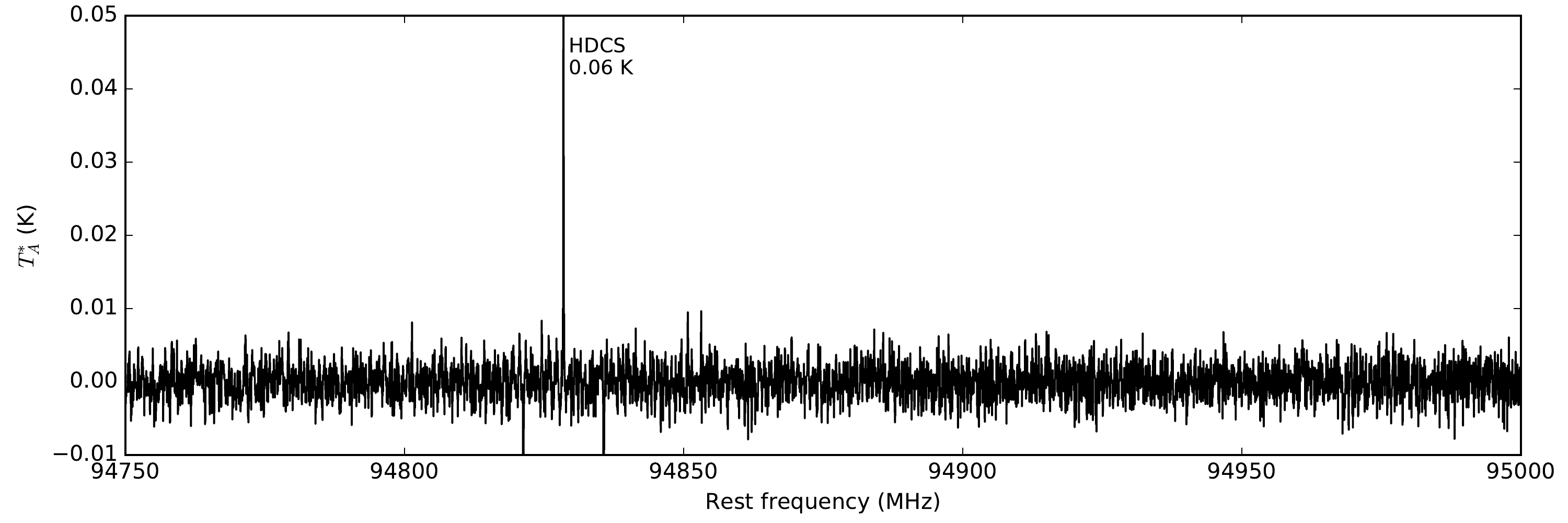}
\caption{Continued}
\end{figure*}

\setcounter{figure}{0}
\begin{figure*}
\centering
\includegraphics[width=\textwidth]{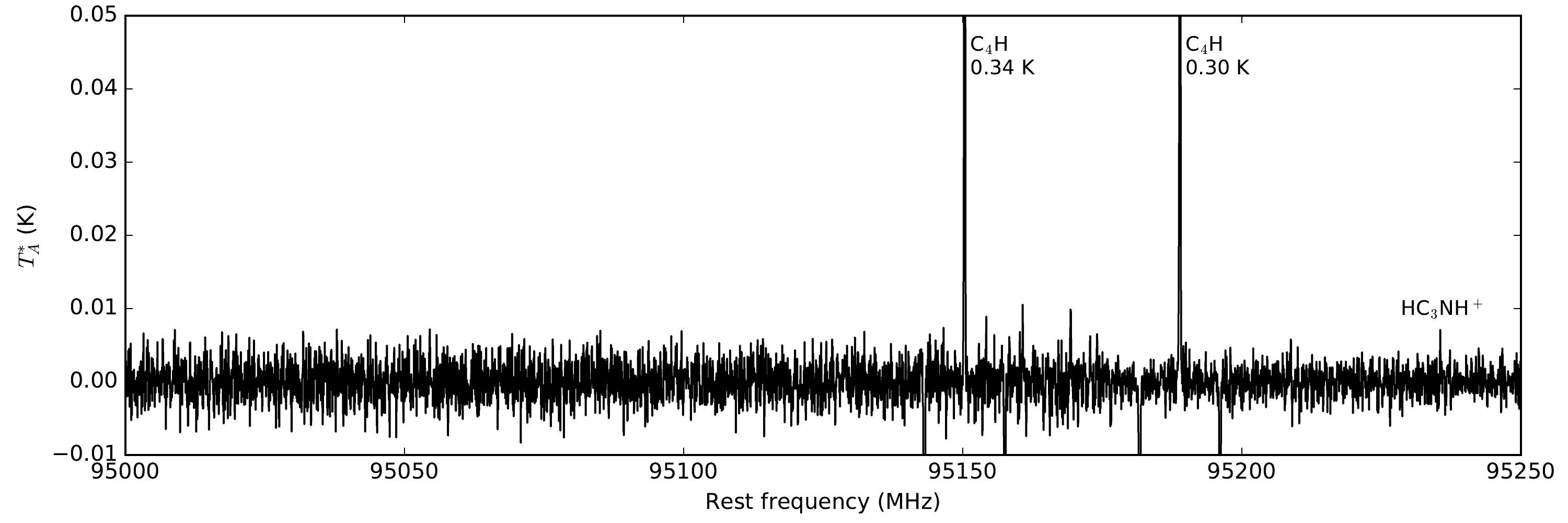}
\includegraphics[width=\textwidth]{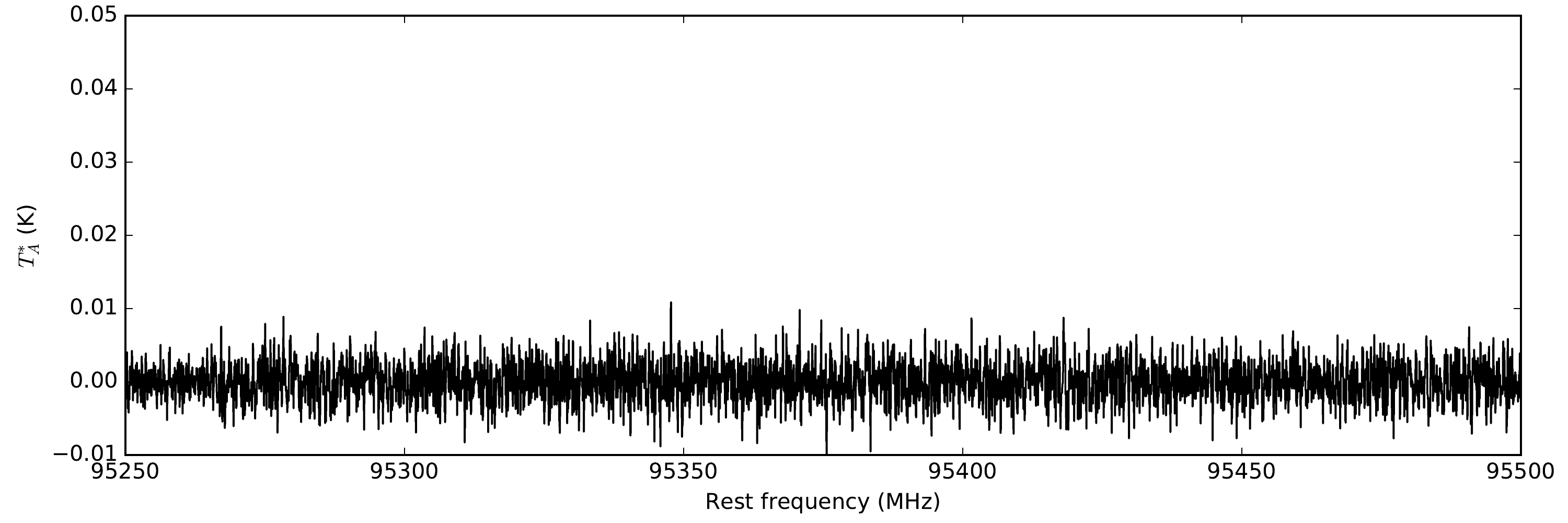}
\includegraphics[width=\textwidth]{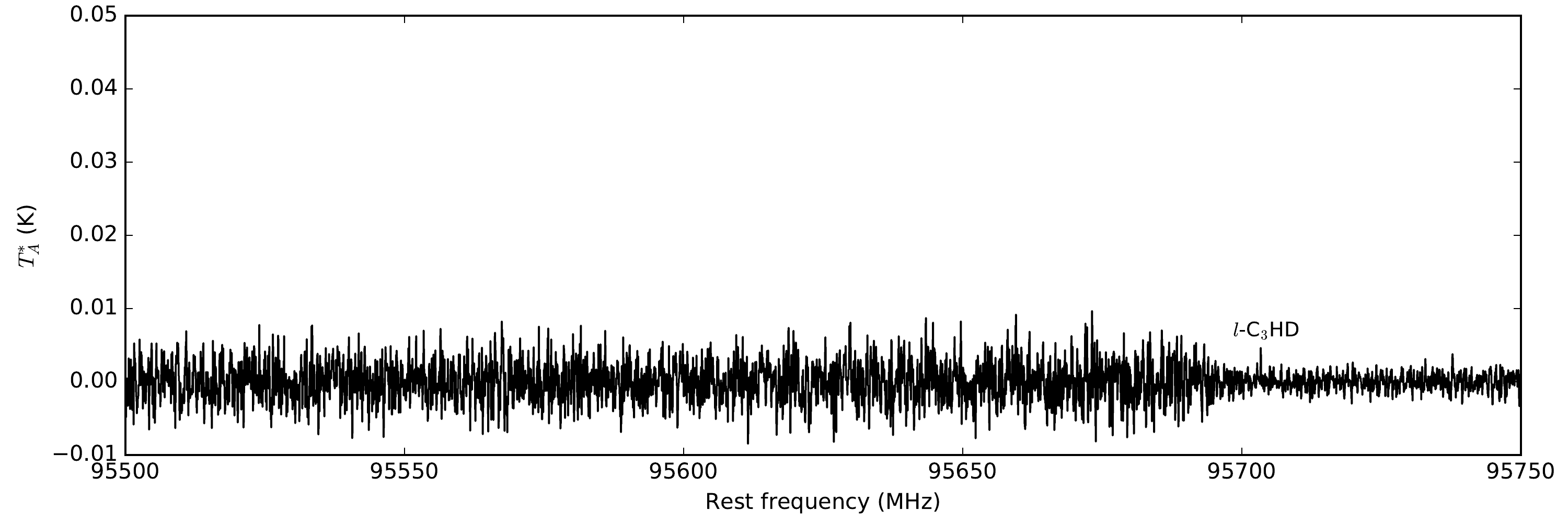}
\includegraphics[width=\textwidth]{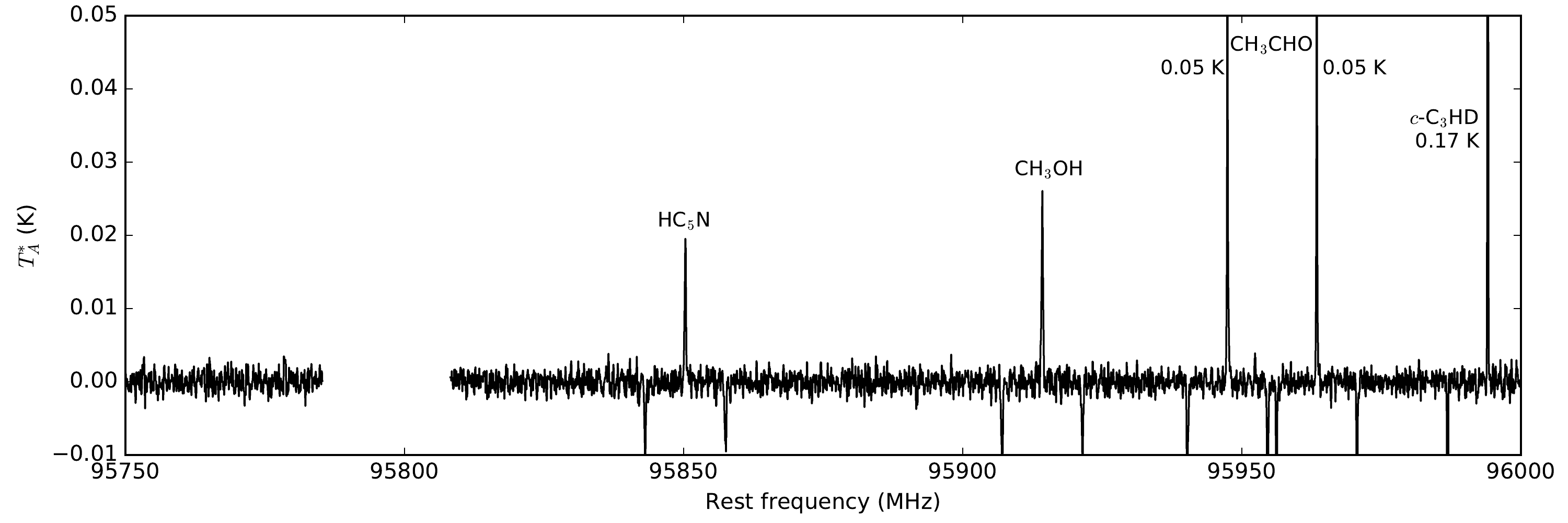}
\caption{Continued}
\end{figure*}

\setcounter{figure}{0}
\begin{figure*}
\centering
\includegraphics[width=\textwidth]{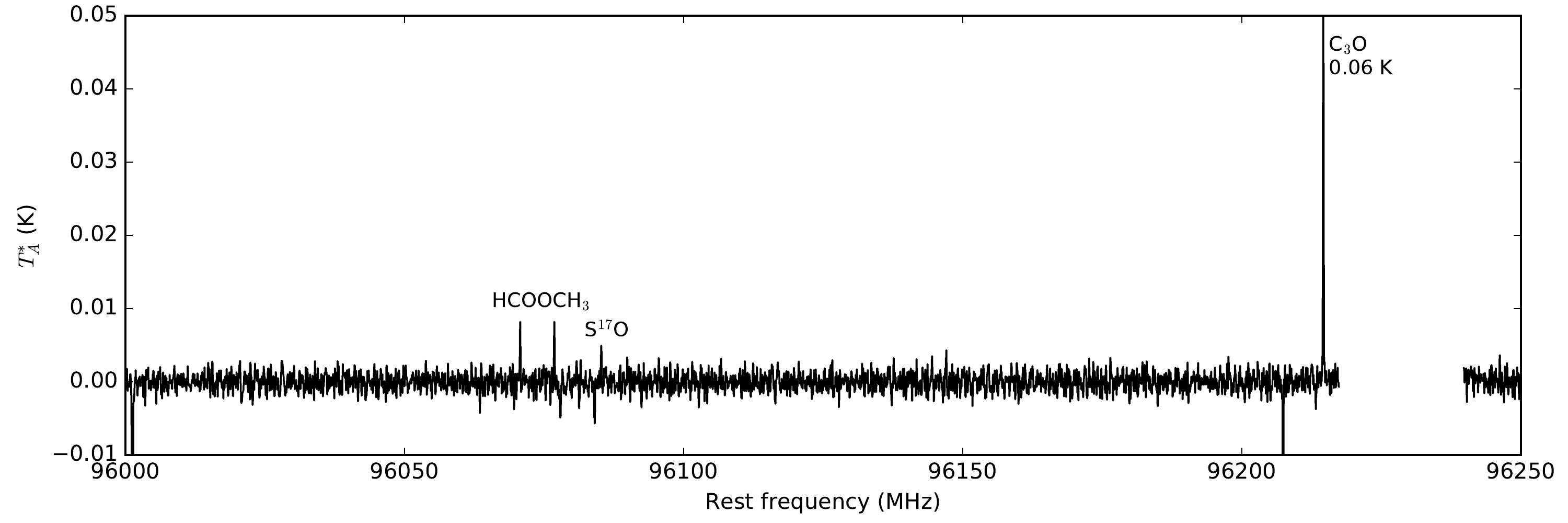}
\includegraphics[width=\textwidth]{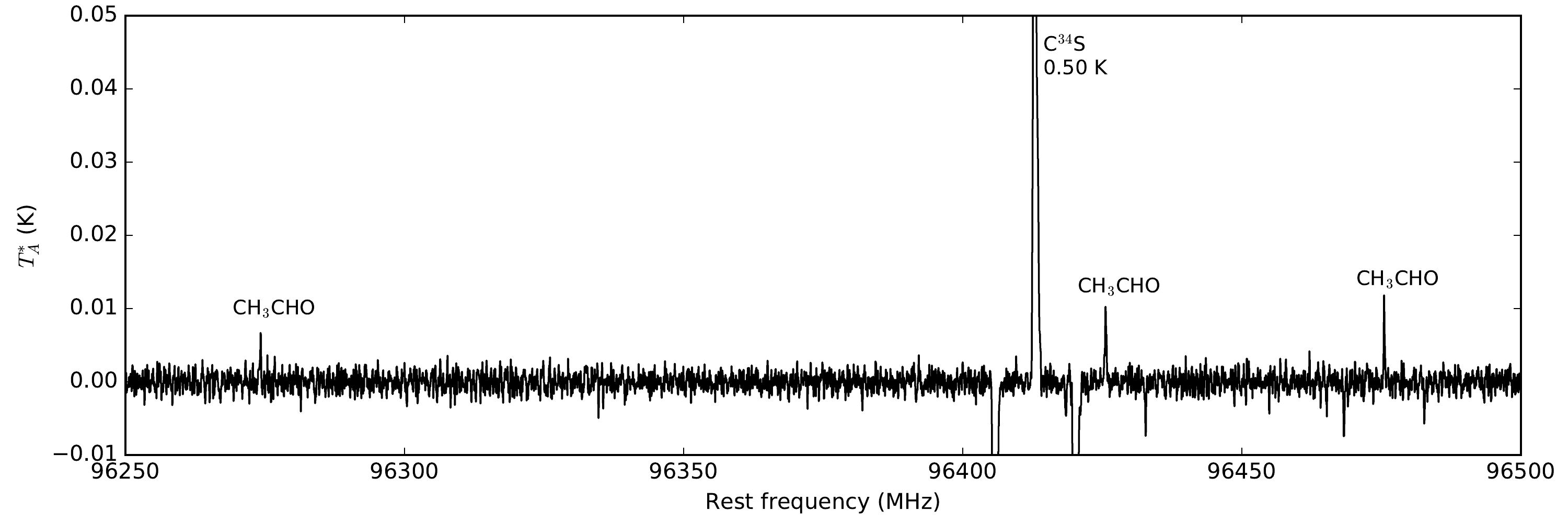}
\includegraphics[width=\textwidth]{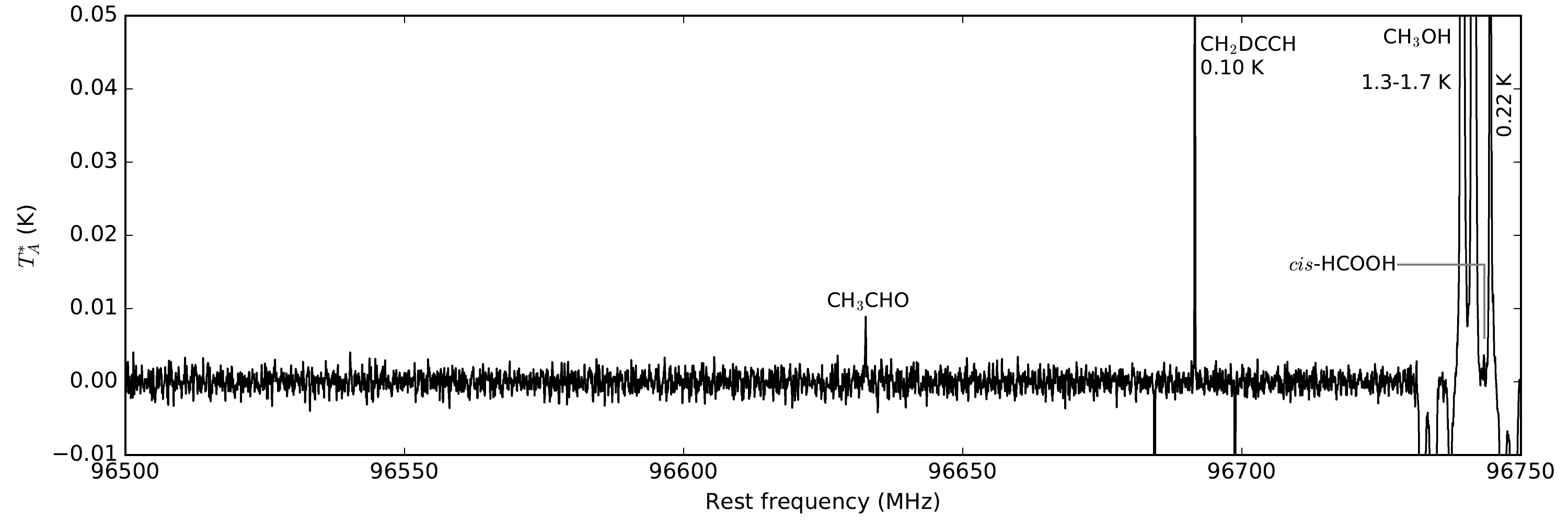}
\includegraphics[width=\textwidth]{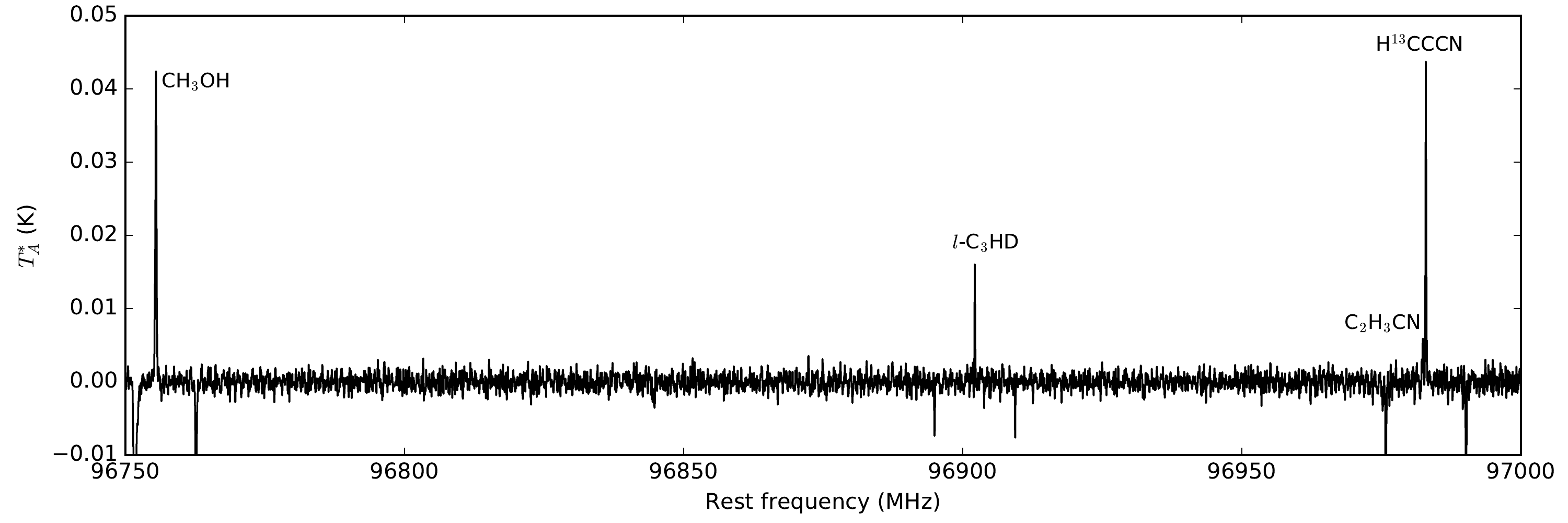}
\caption{Continued}
\end{figure*}

\setcounter{figure}{0}
\begin{figure*}
\centering
\includegraphics[width=\textwidth]{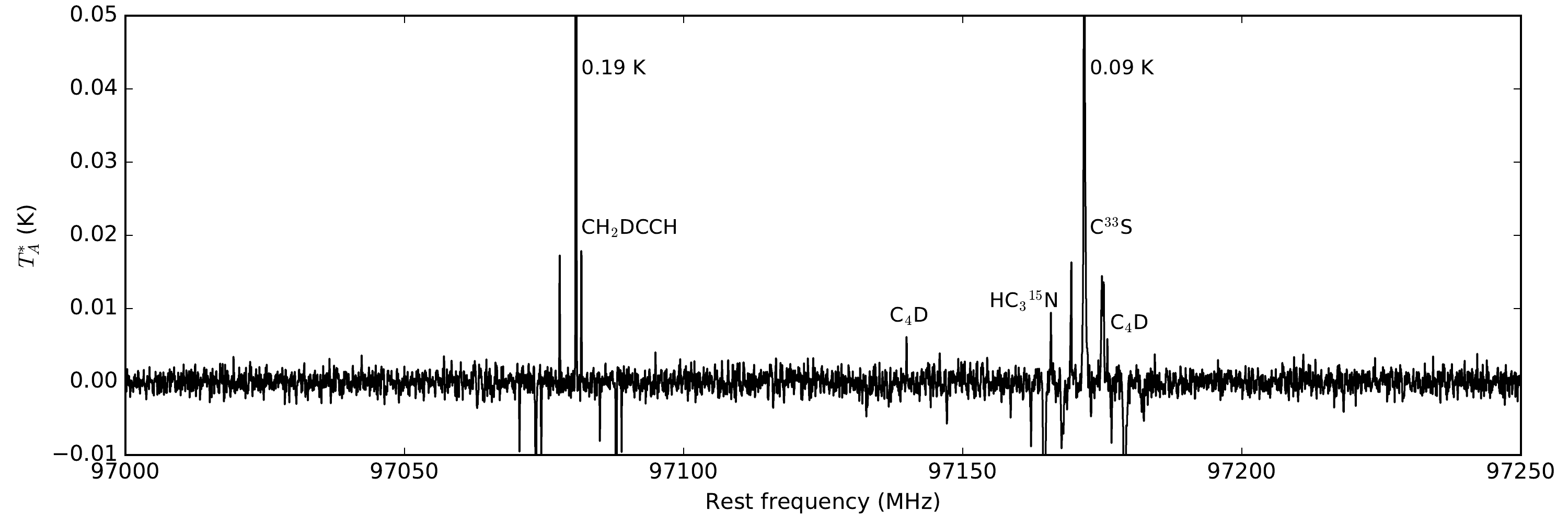}
\includegraphics[width=\textwidth]{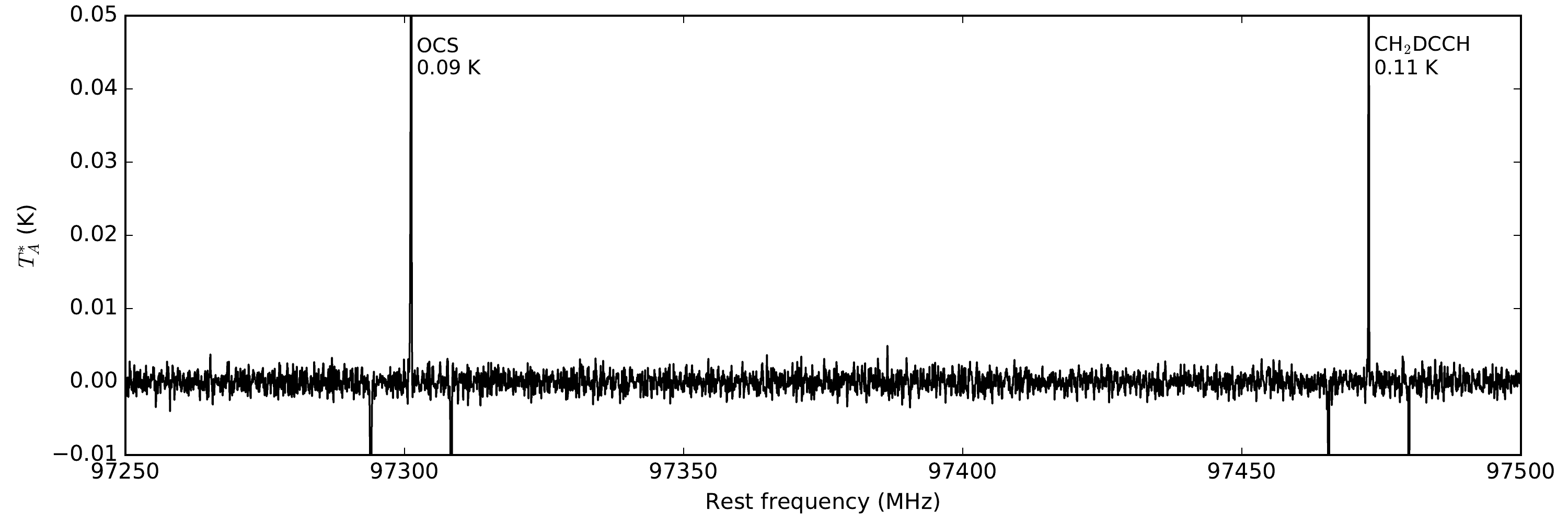}
\includegraphics[width=\textwidth]{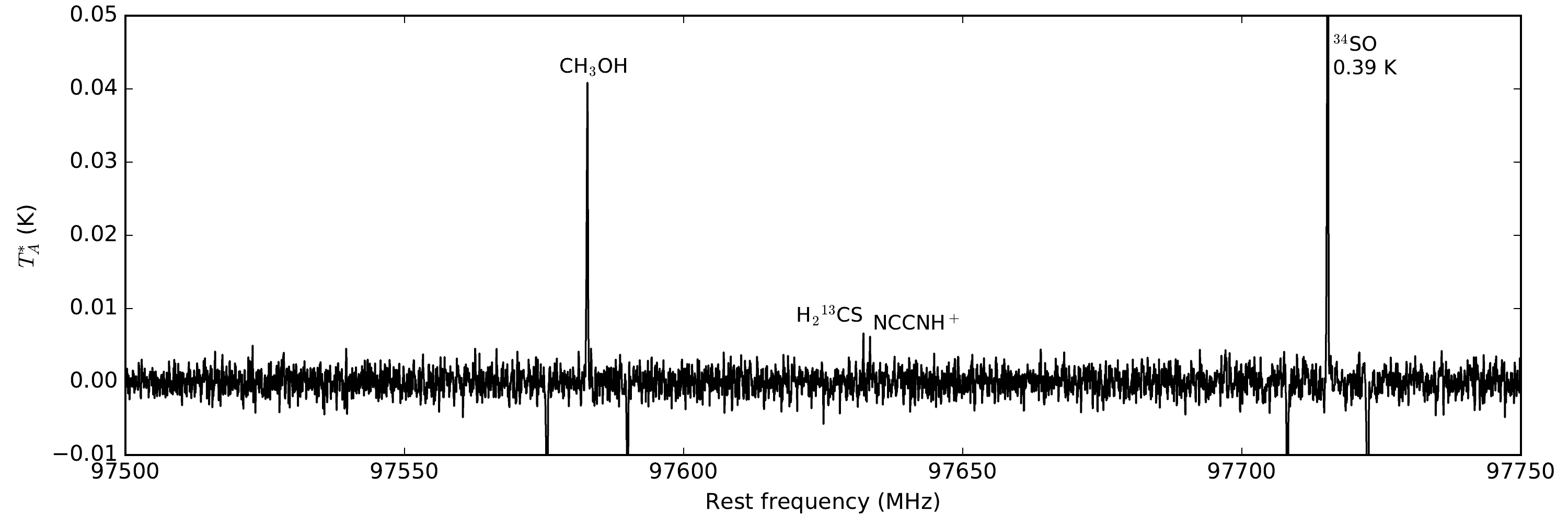}
\includegraphics[width=\textwidth]{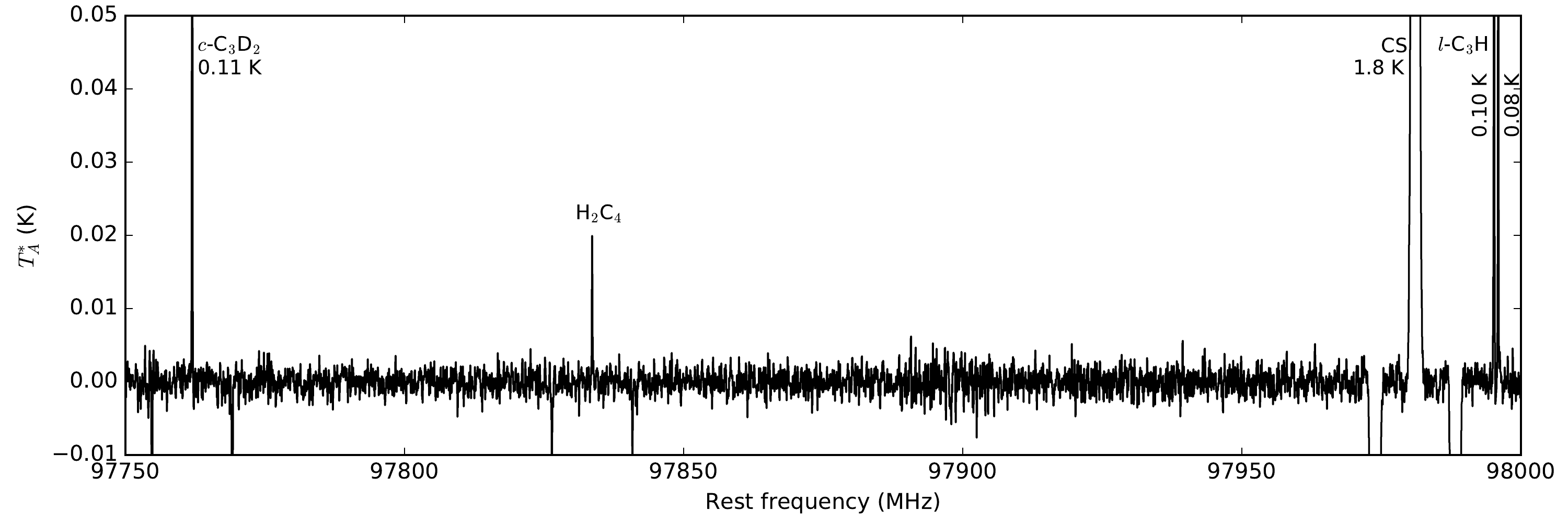}
\caption{Continued}
\end{figure*}

\setcounter{figure}{0}
\begin{figure*}
\centering
\includegraphics[width=\textwidth]{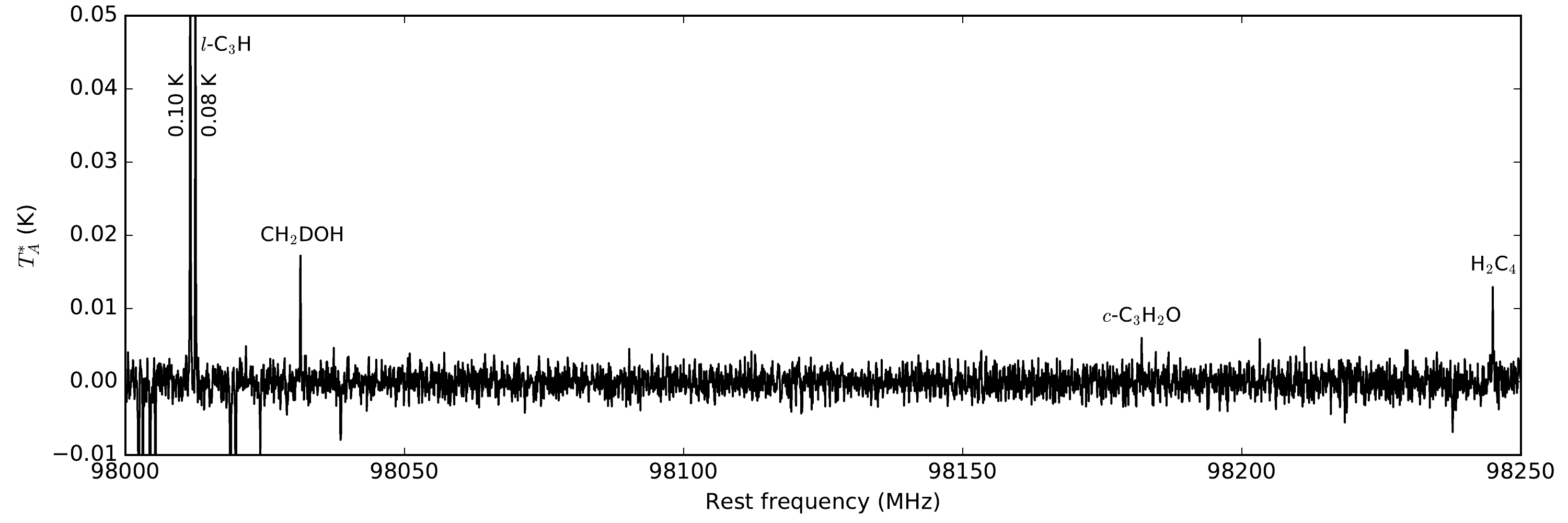}
\includegraphics[width=\textwidth]{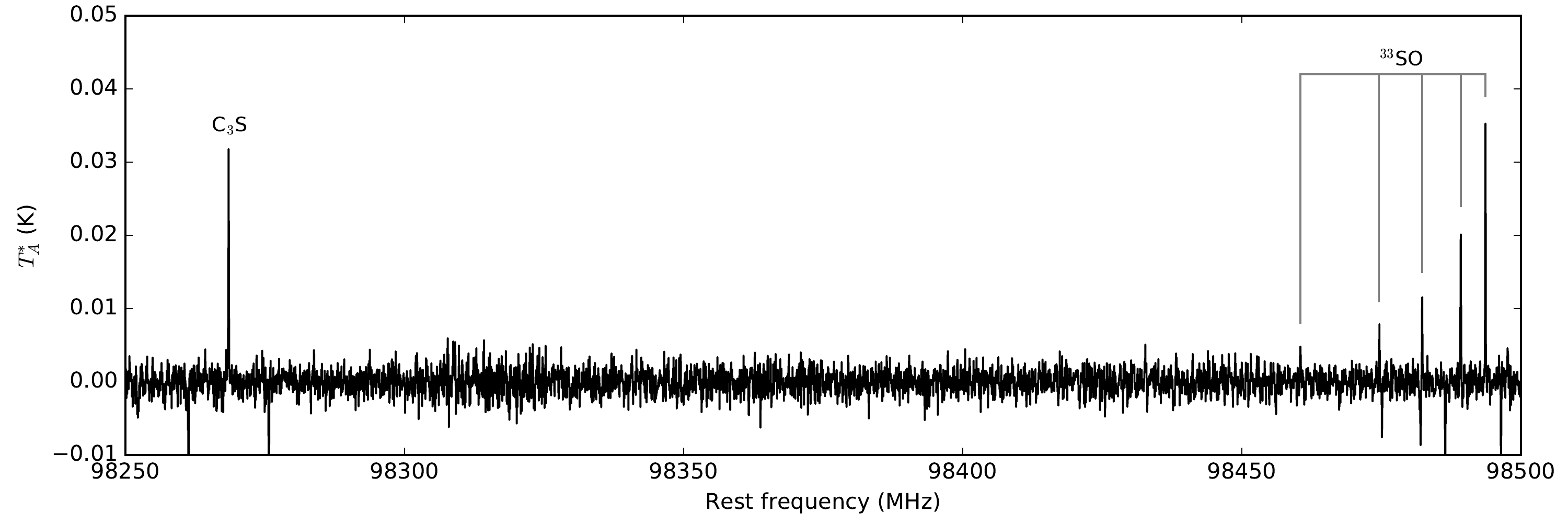}
\includegraphics[width=\textwidth]{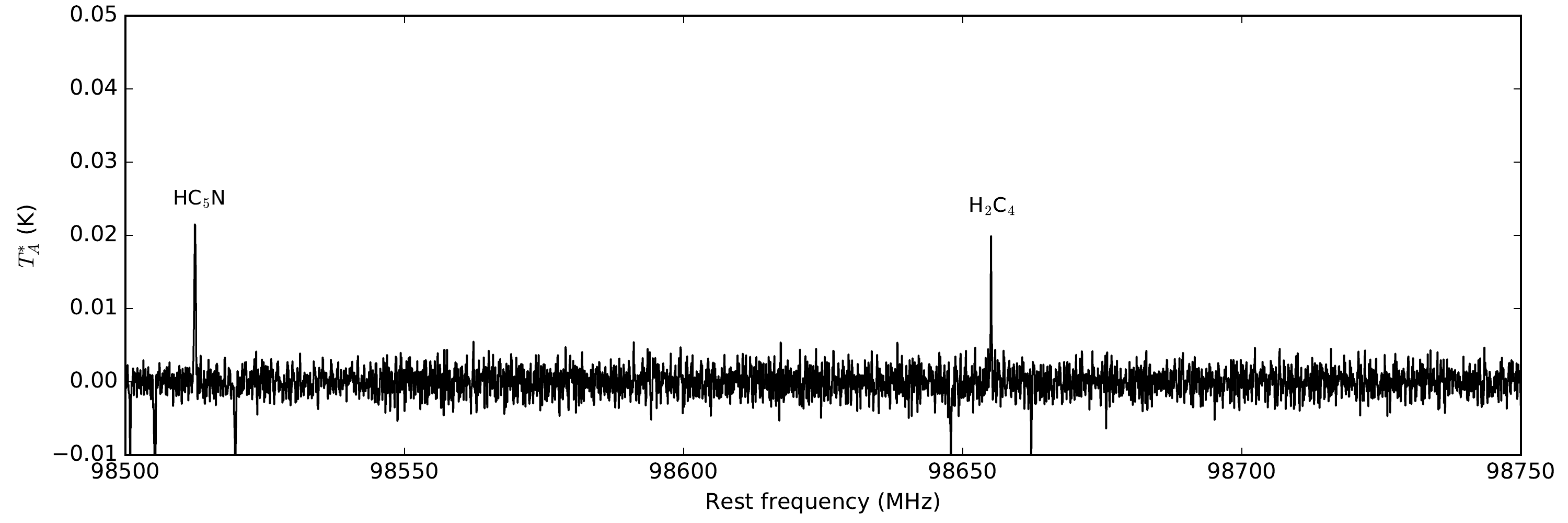}
\includegraphics[width=\textwidth]{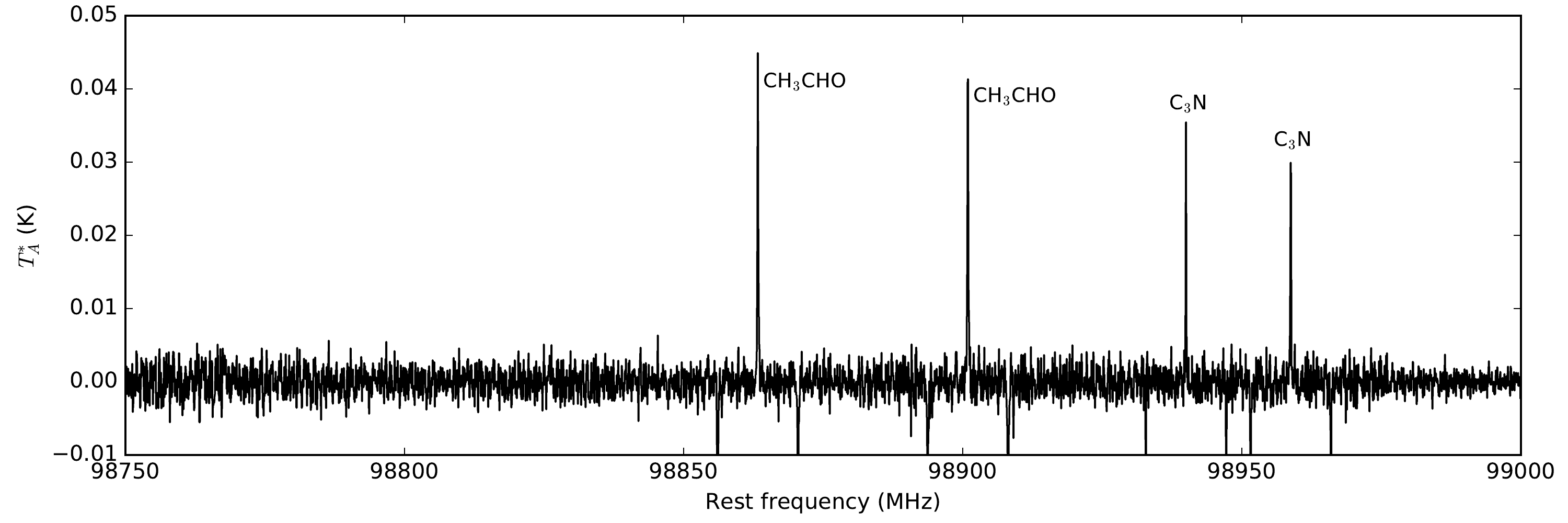}
\caption{Continued}
\end{figure*}

\setcounter{figure}{0}
\begin{figure*}
\centering
\includegraphics[width=\textwidth]{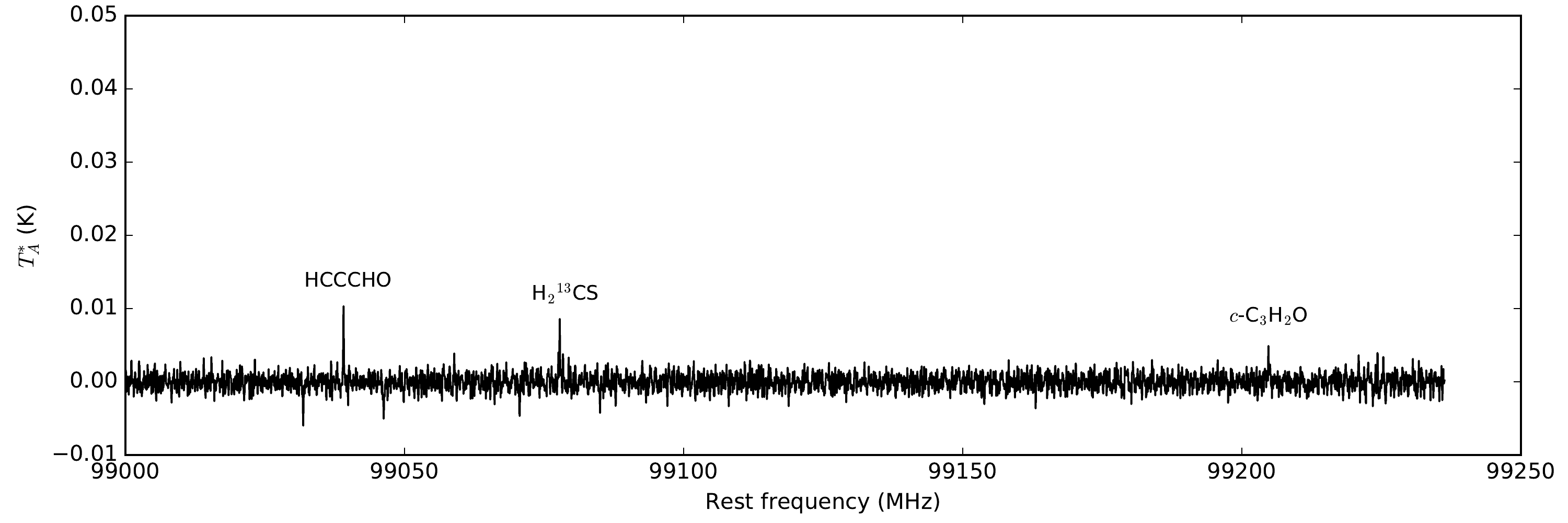}
\includegraphics[width=\textwidth]{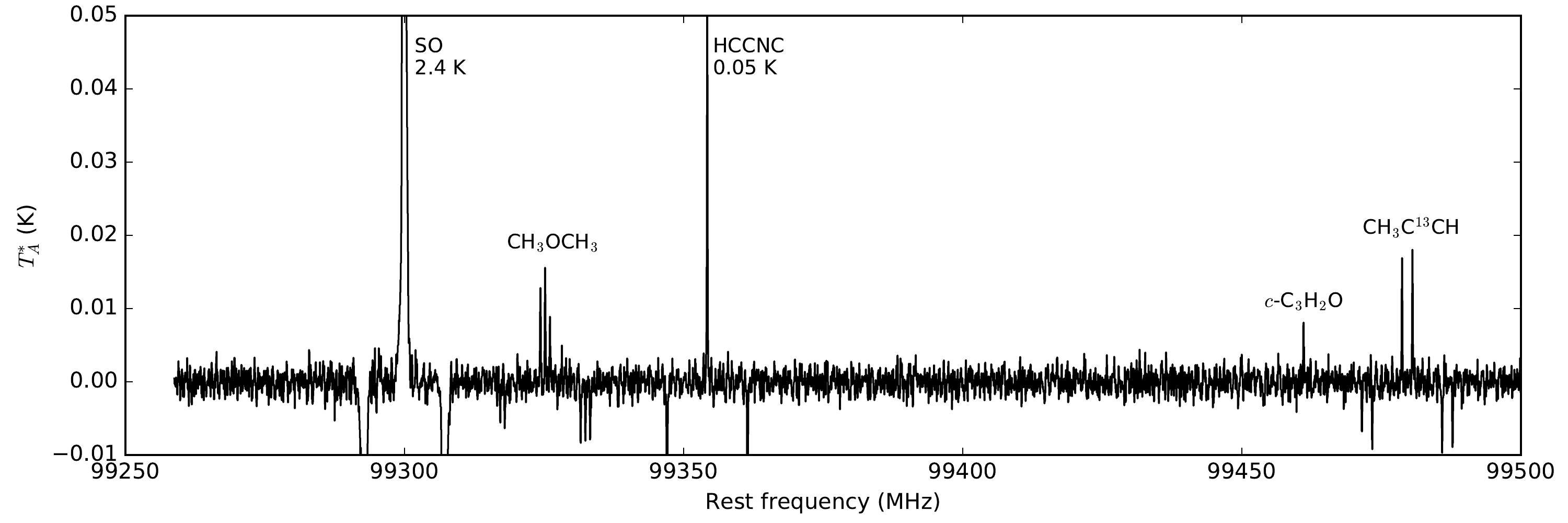}
\includegraphics[width=\textwidth]{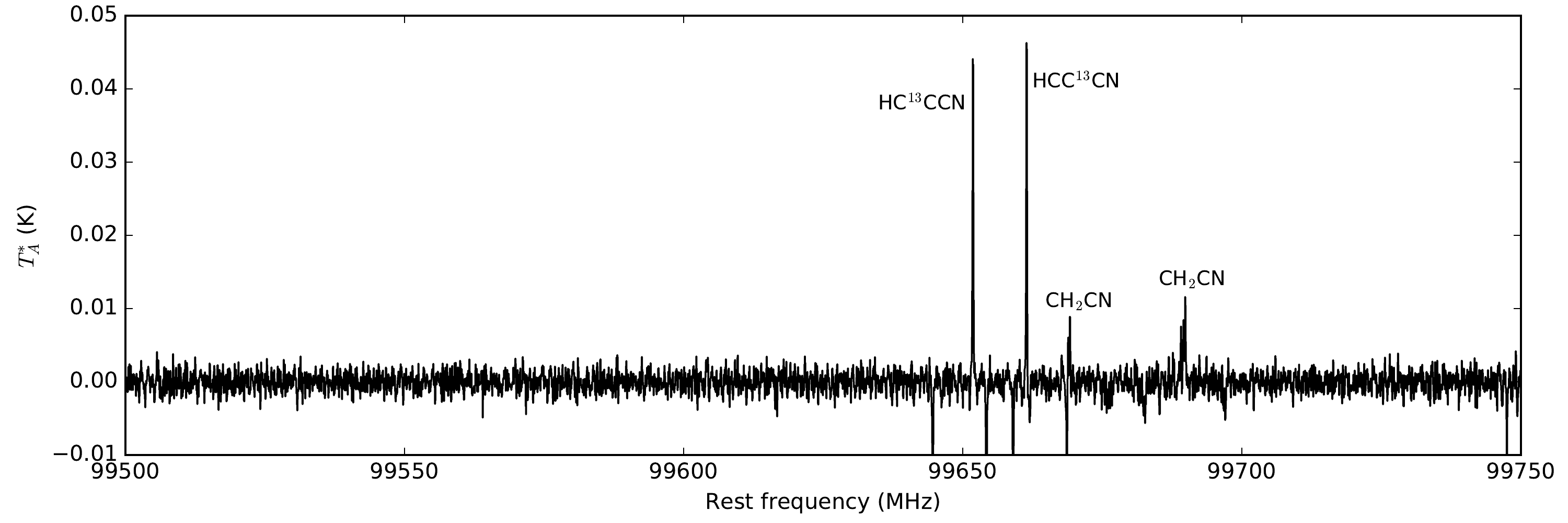}
\includegraphics[width=\textwidth]{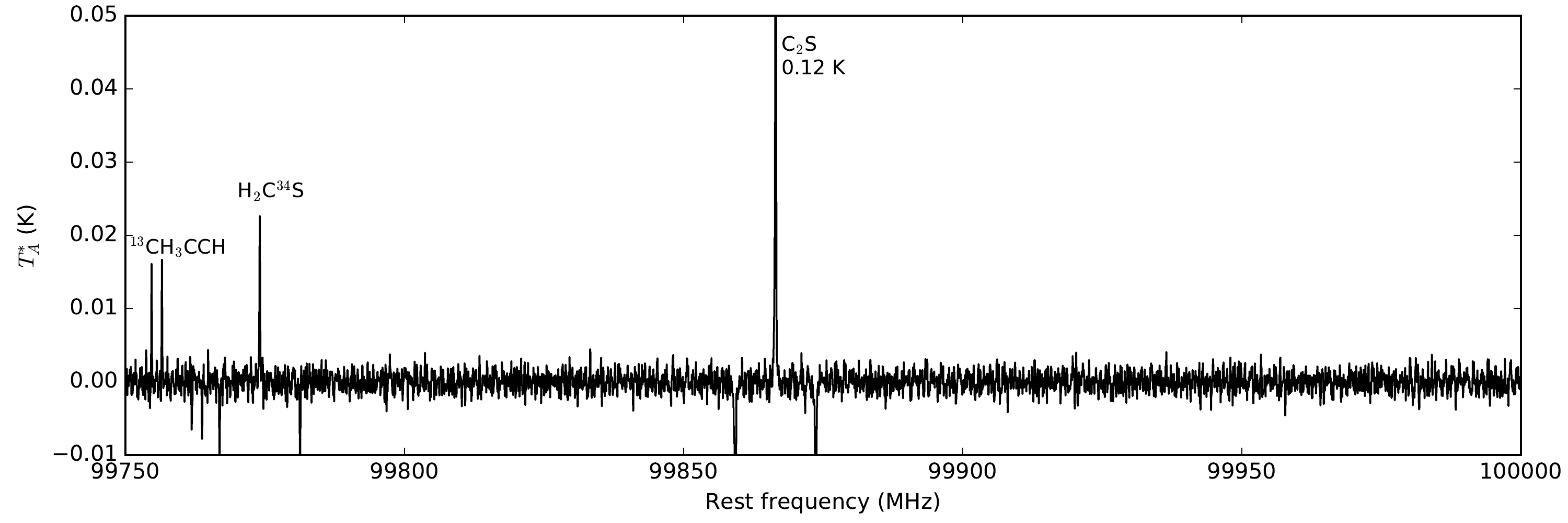}
\caption{Continued}
\end{figure*}

\setcounter{figure}{0}
\begin{figure*}
\centering
\includegraphics[width=\textwidth]{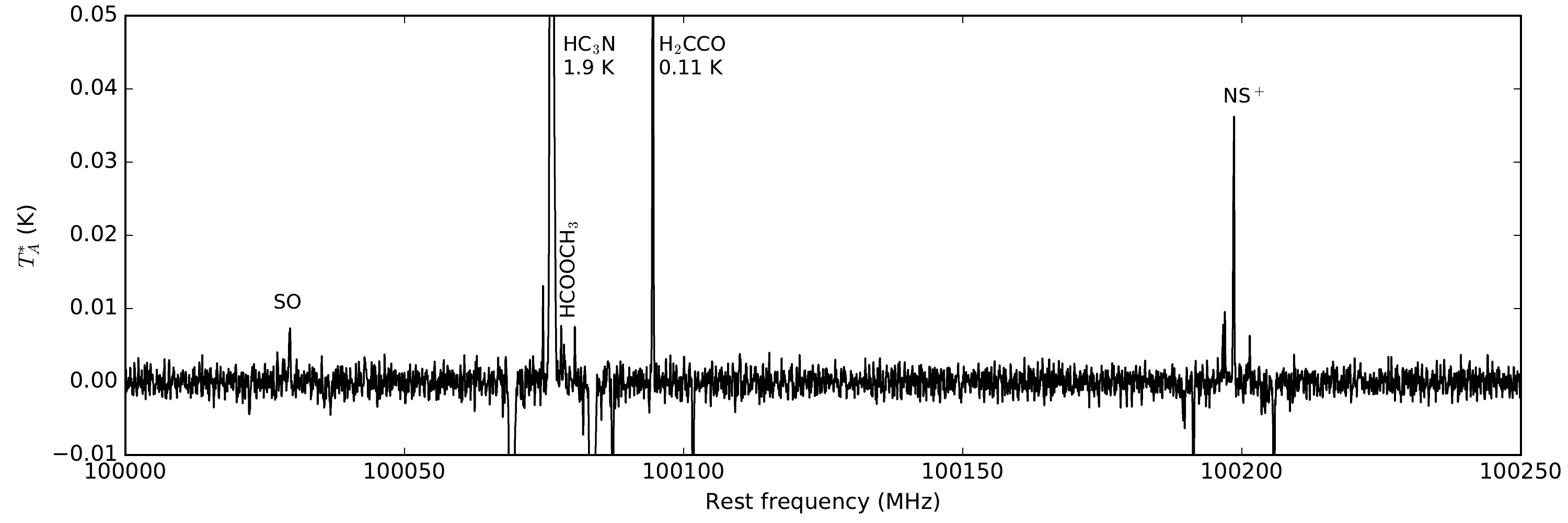}
\includegraphics[width=\textwidth]{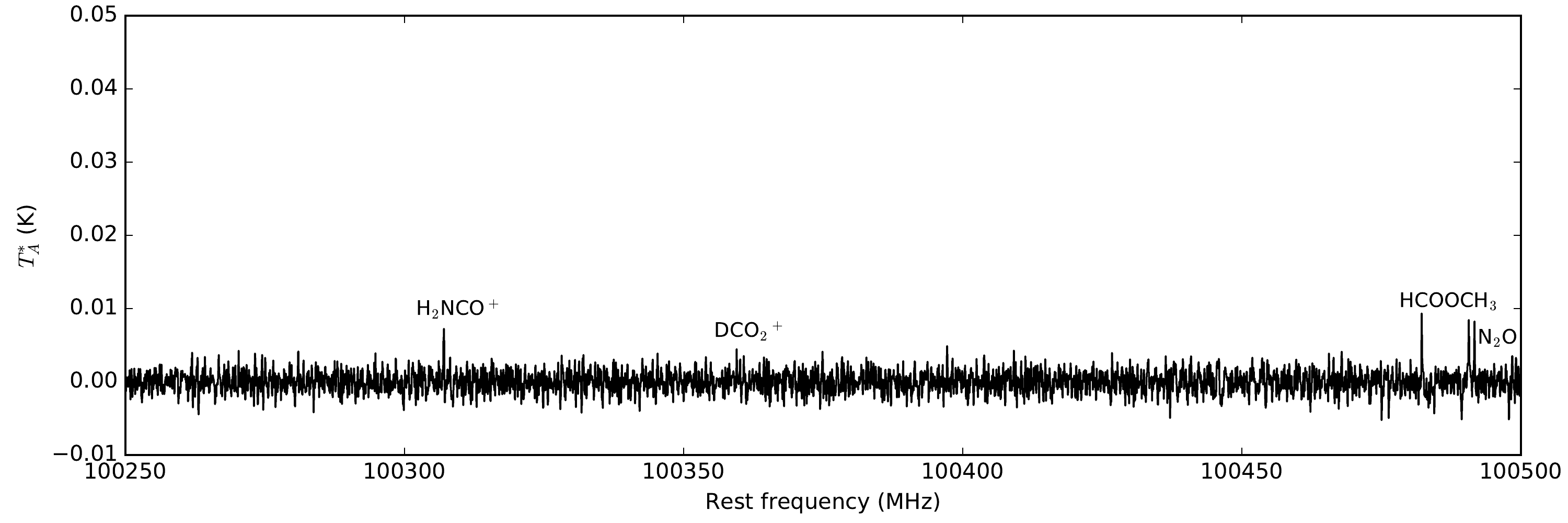}
\includegraphics[width=\textwidth]{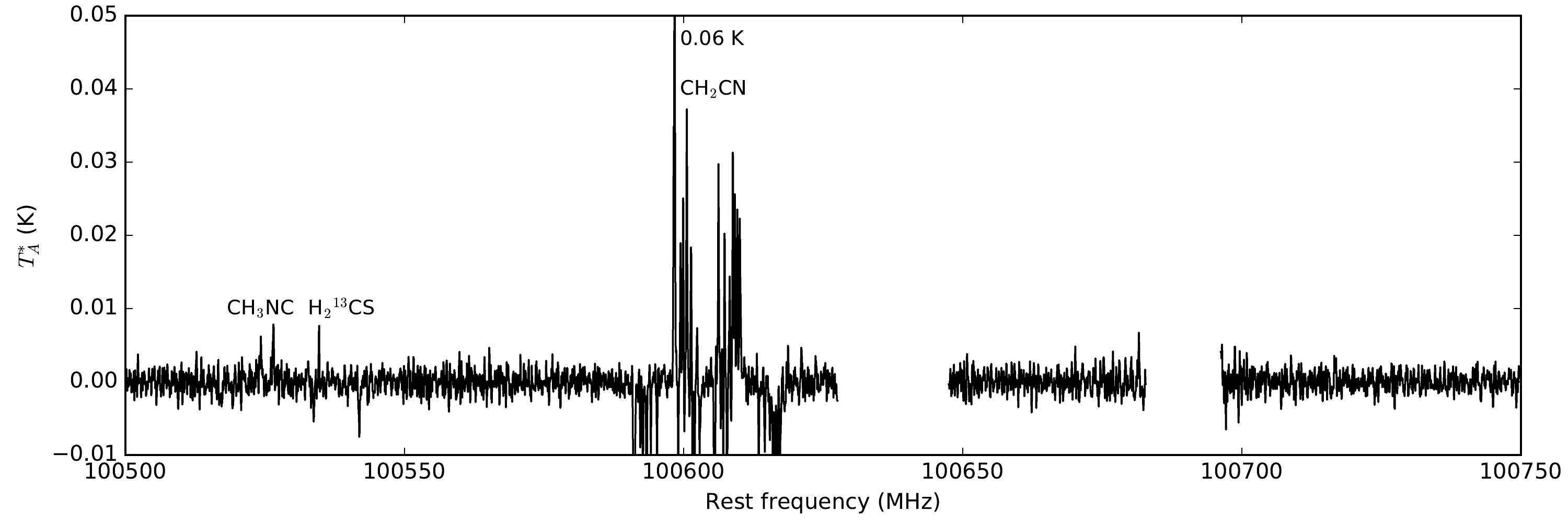}
\includegraphics[width=\textwidth]{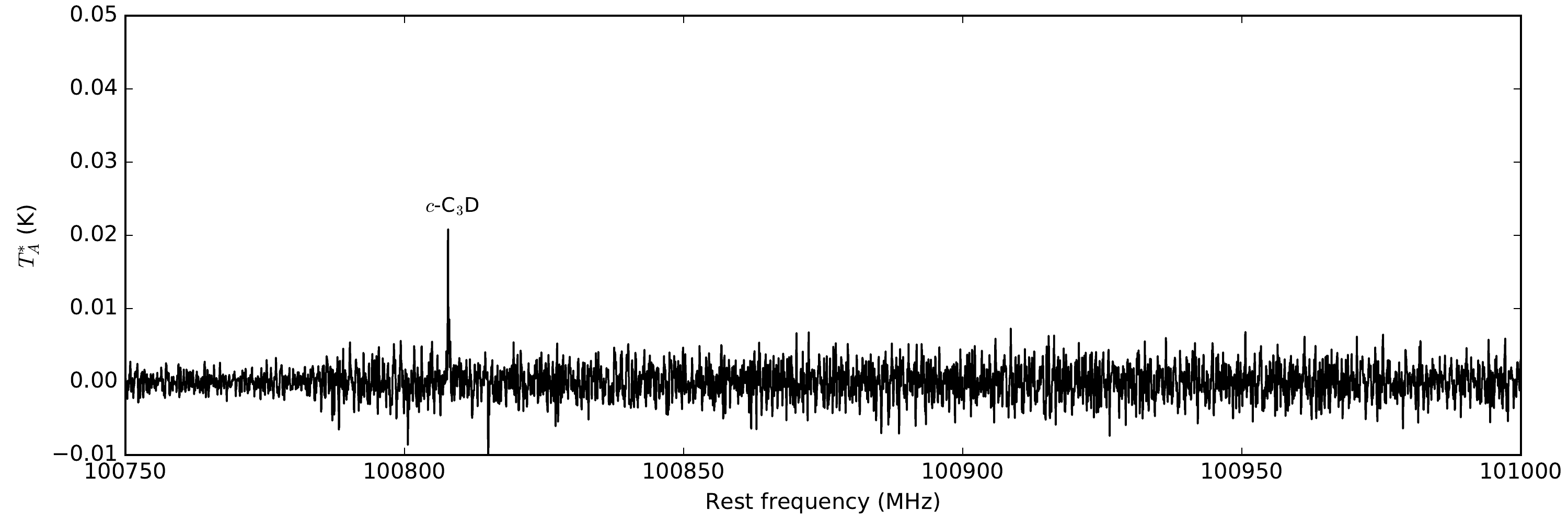}
\caption{Continued}
\end{figure*}

\setcounter{figure}{0}
\begin{figure*}
\centering
\includegraphics[width=\textwidth]{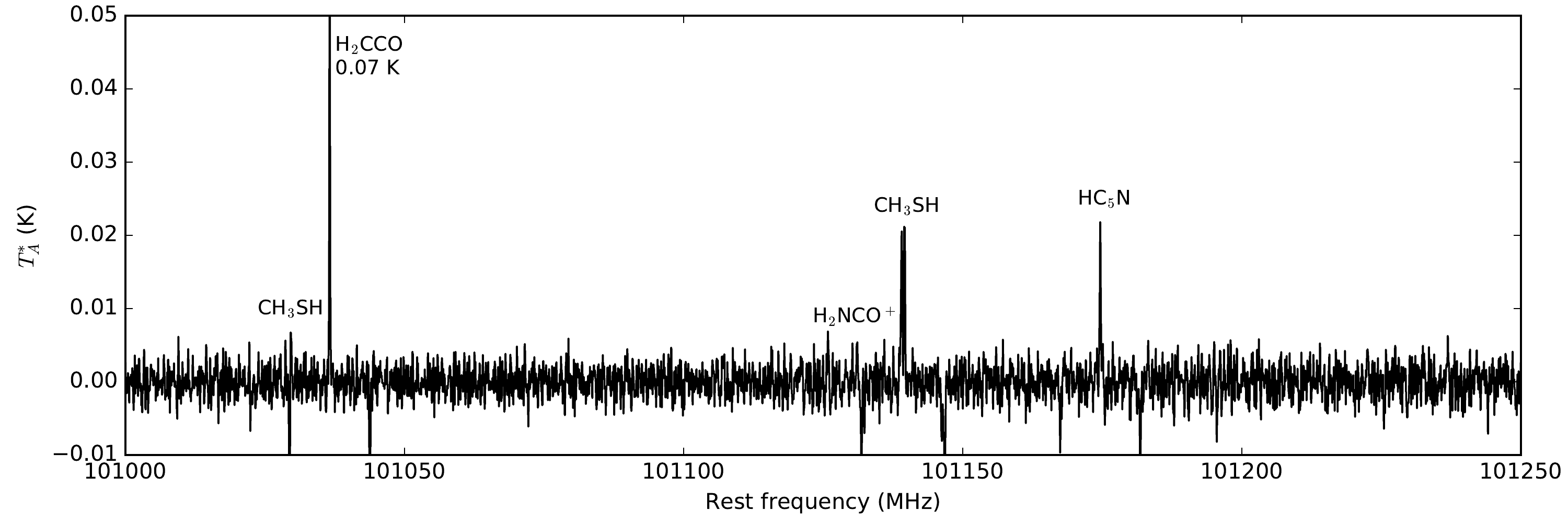}
\includegraphics[width=\textwidth]{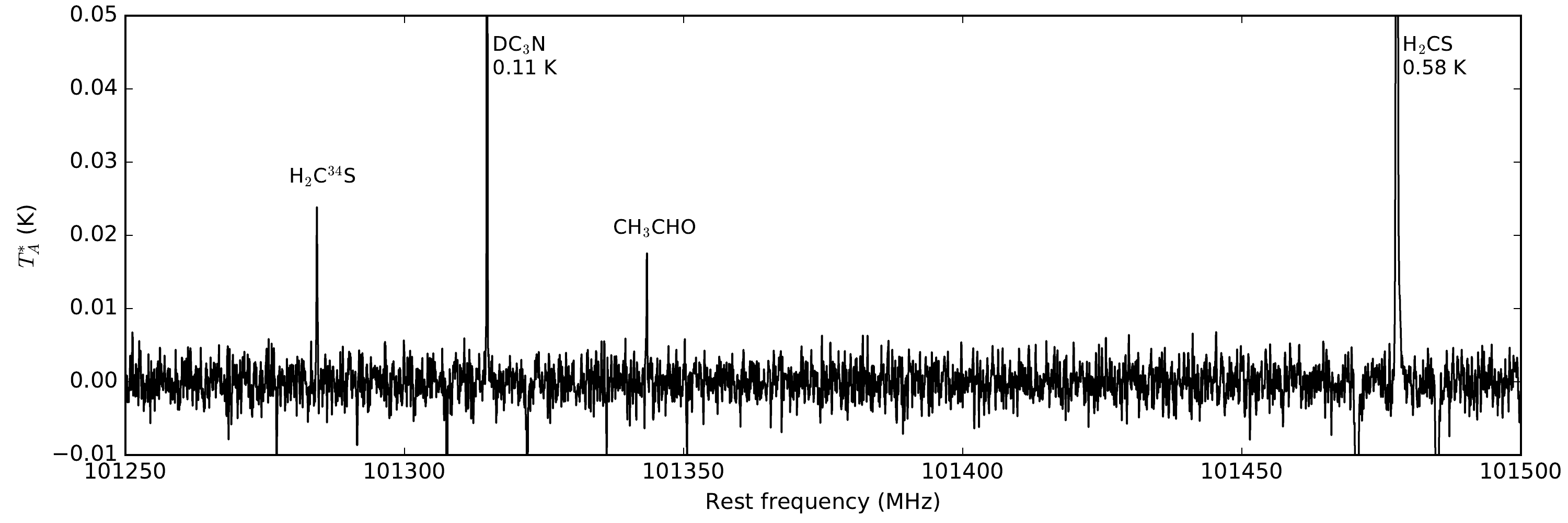}
\includegraphics[width=\textwidth]{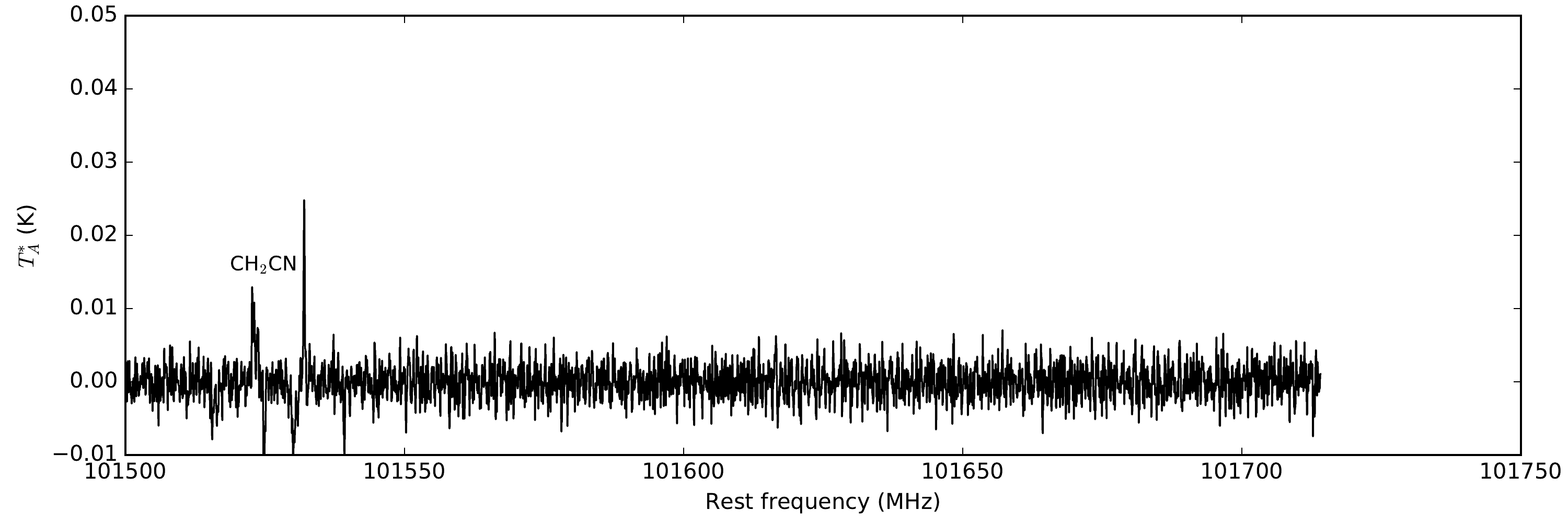}
\includegraphics[width=\textwidth]{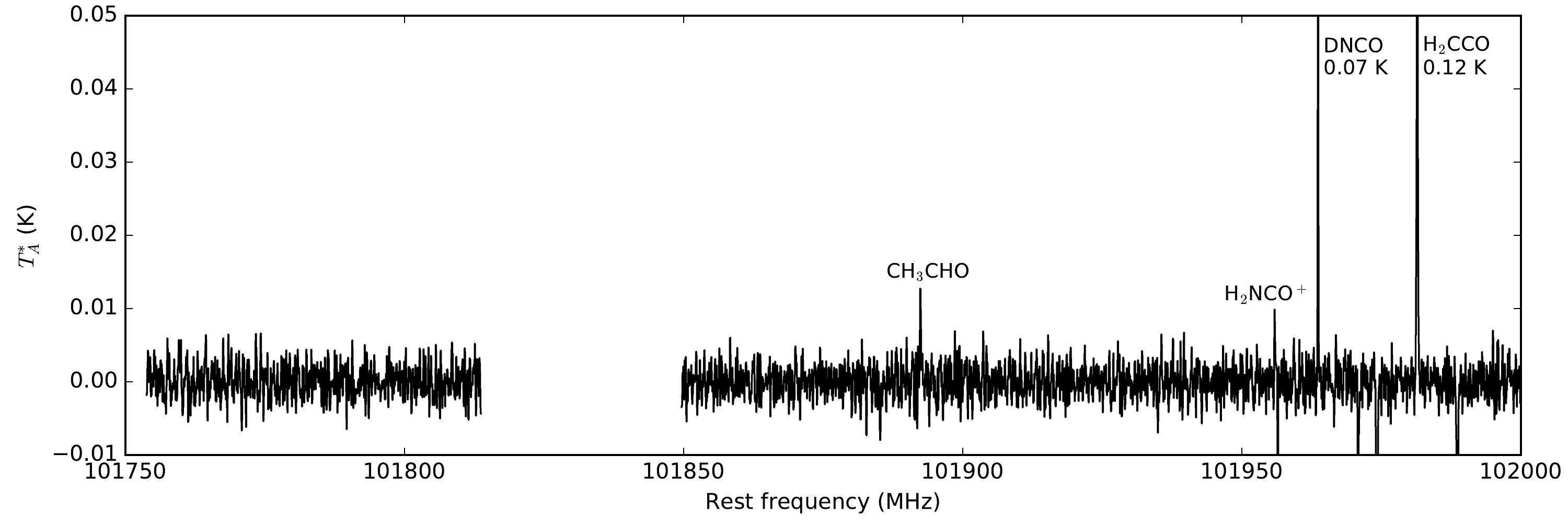}
\caption{Continued}
\end{figure*}

\setcounter{figure}{0}
\begin{figure*}
\centering
\includegraphics[width=\textwidth]{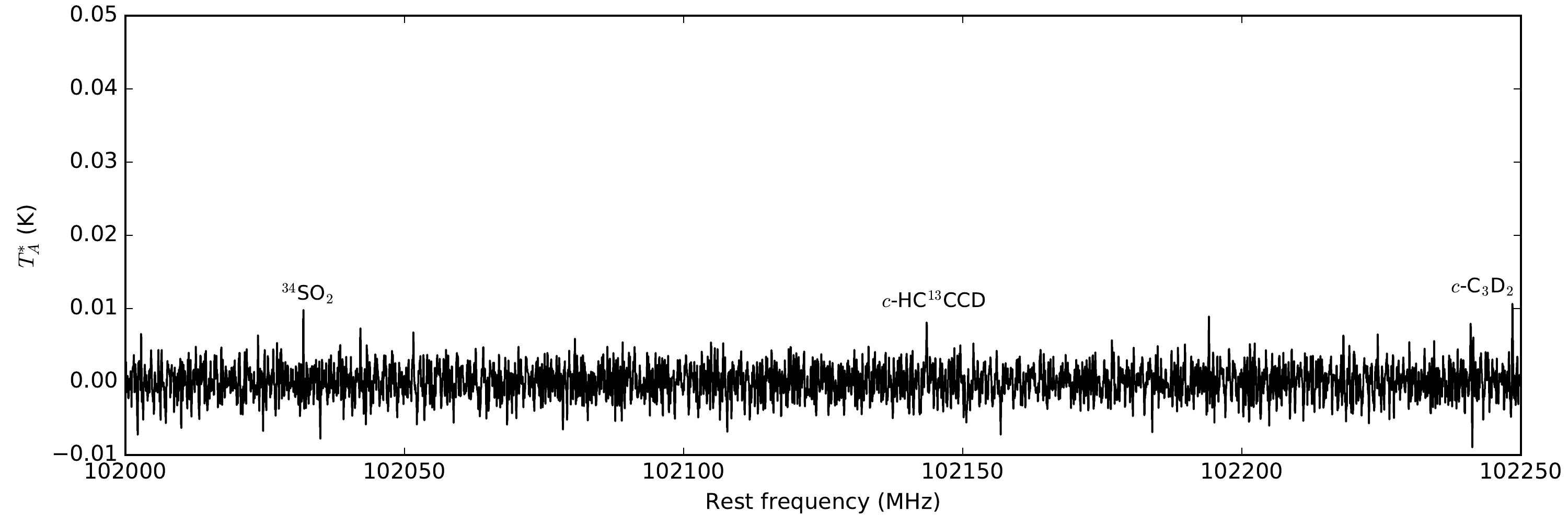}
\includegraphics[width=\textwidth]{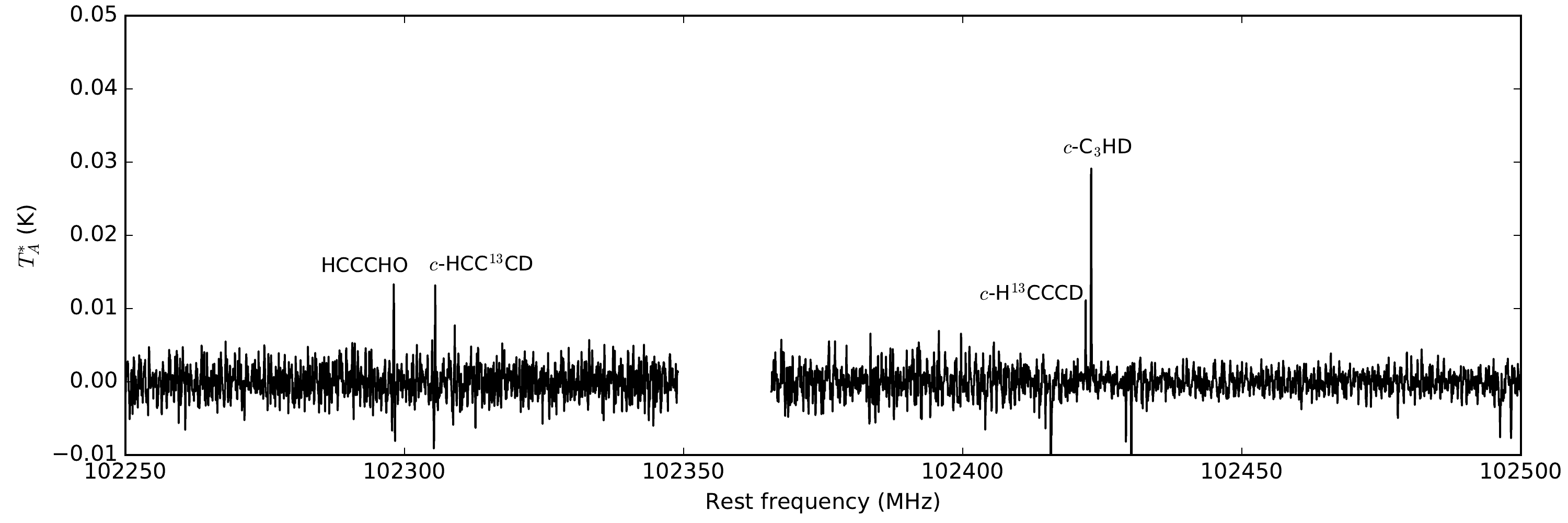}
\includegraphics[width=\textwidth]{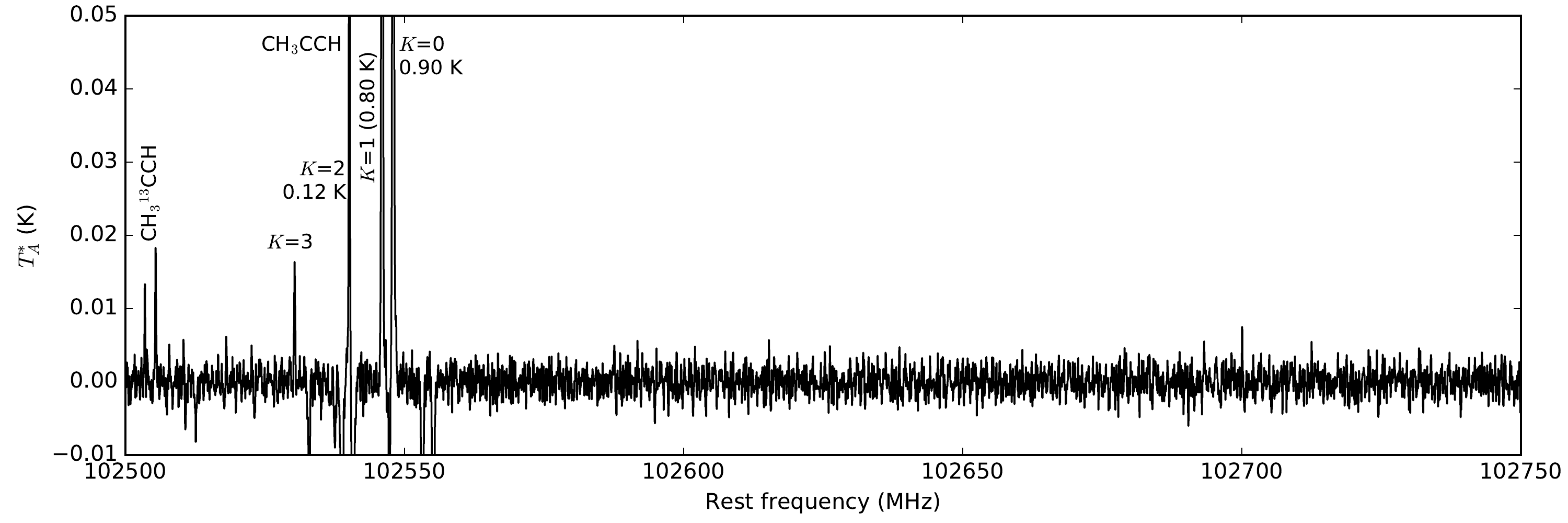}
\includegraphics[width=\textwidth]{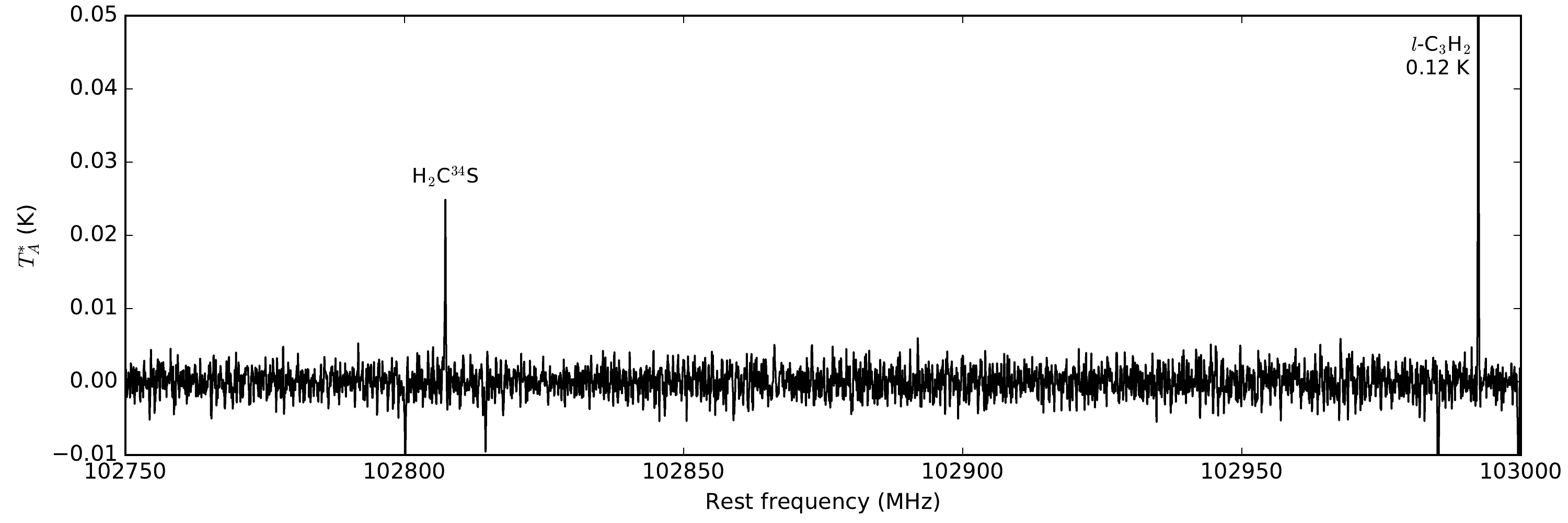}
\caption{Continued}
\end{figure*}

\setcounter{figure}{0}
\begin{figure*}
\centering
\includegraphics[width=\textwidth]{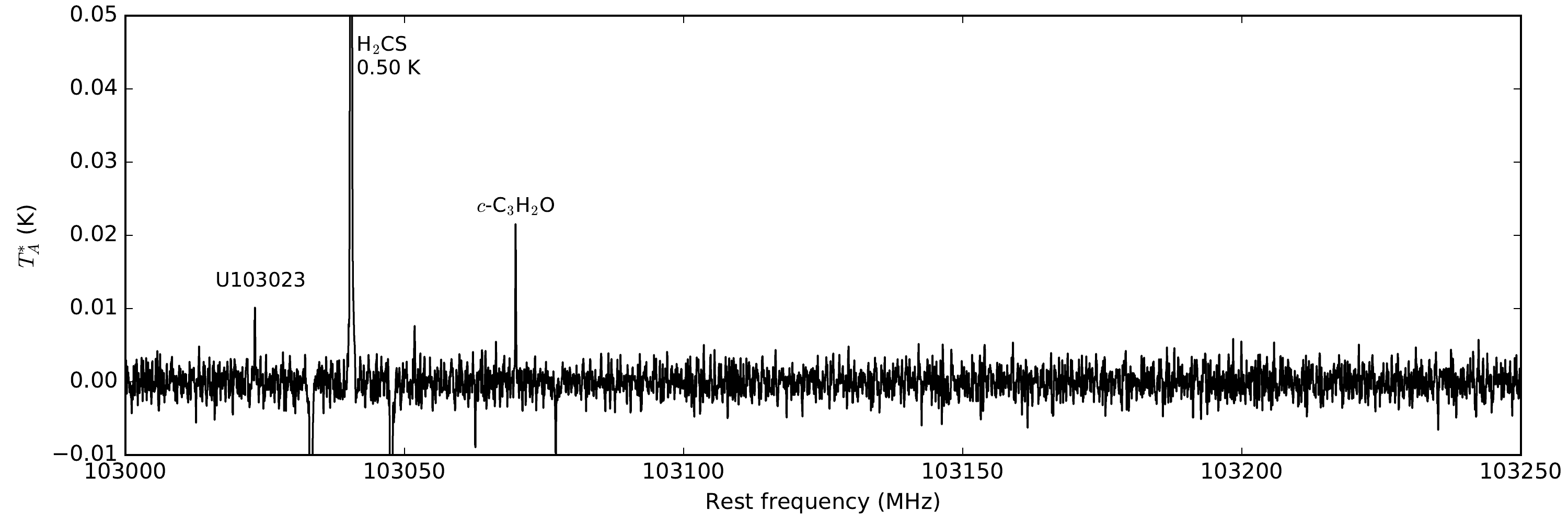}
\includegraphics[width=\textwidth]{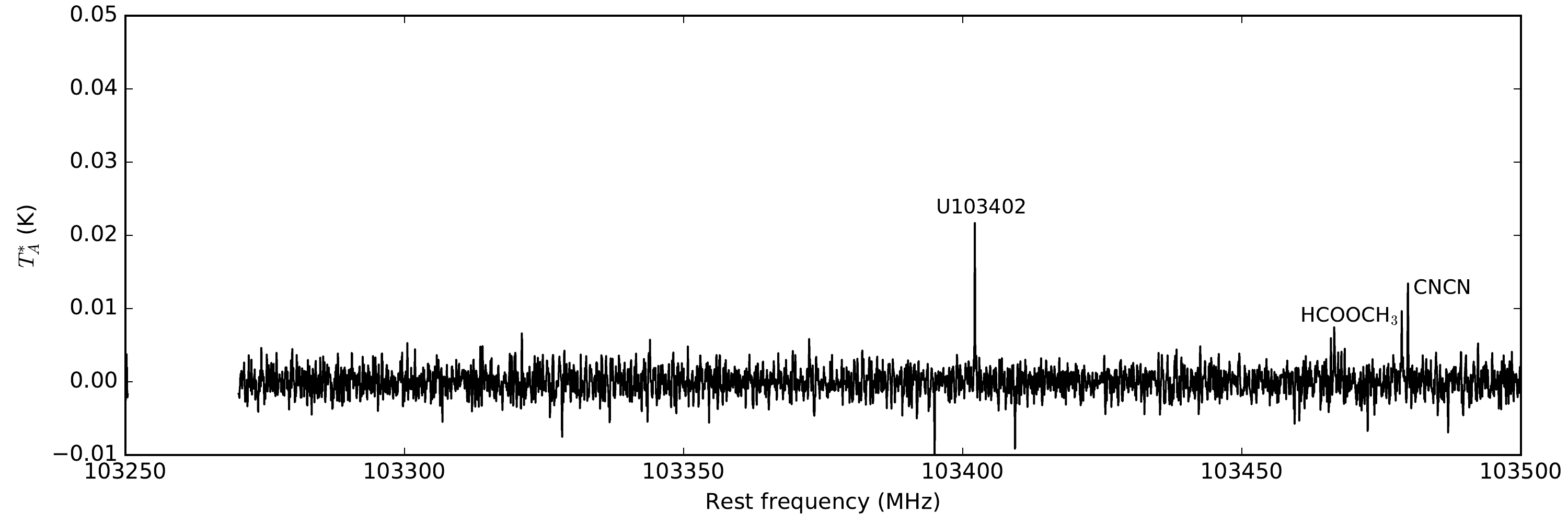}
\includegraphics[width=\textwidth]{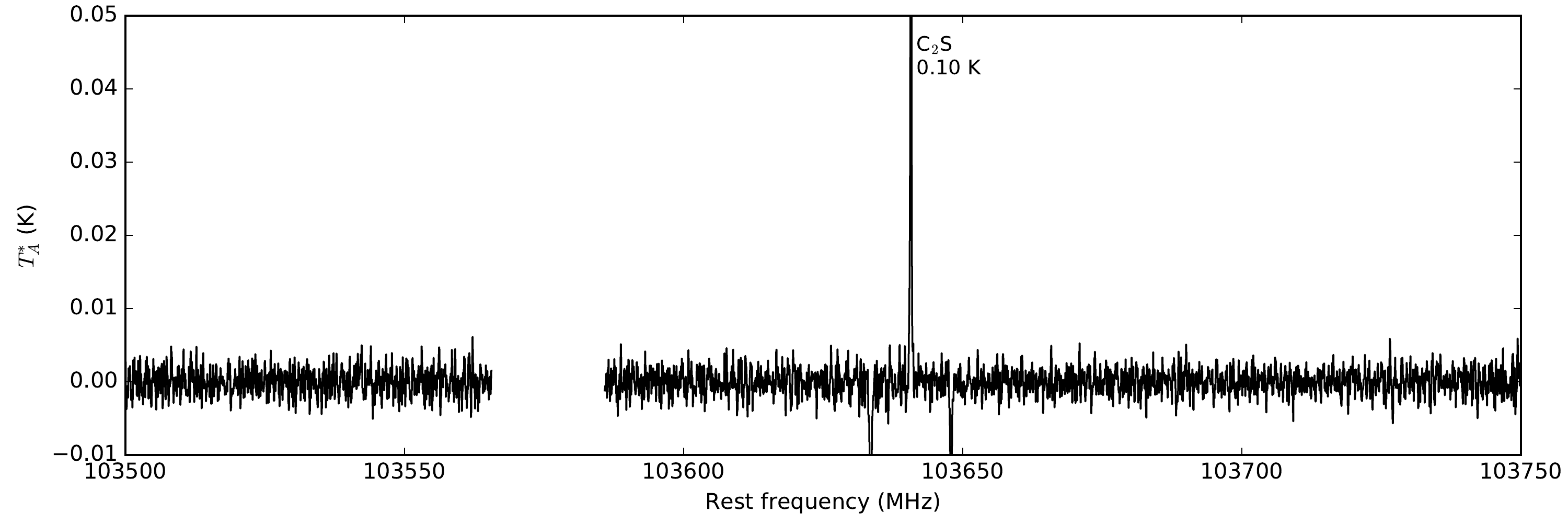}
\includegraphics[width=\textwidth]{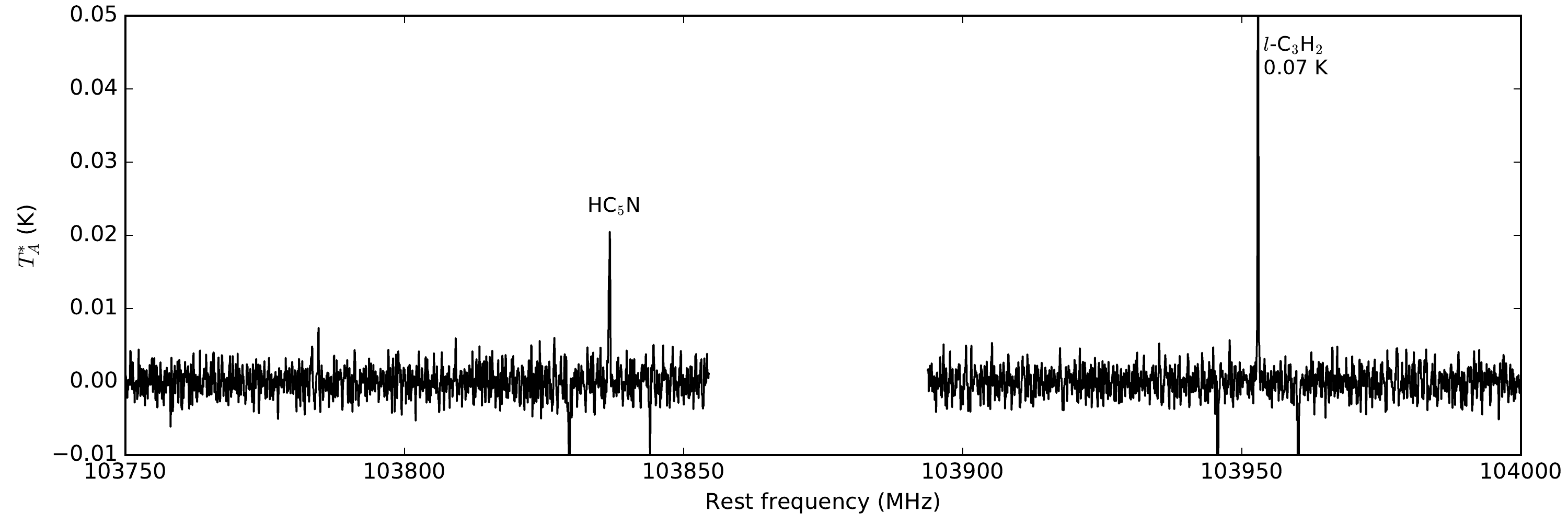}
\caption{Continued}
\end{figure*}

\setcounter{figure}{0}
\begin{figure*}
\centering
\includegraphics[width=\textwidth]{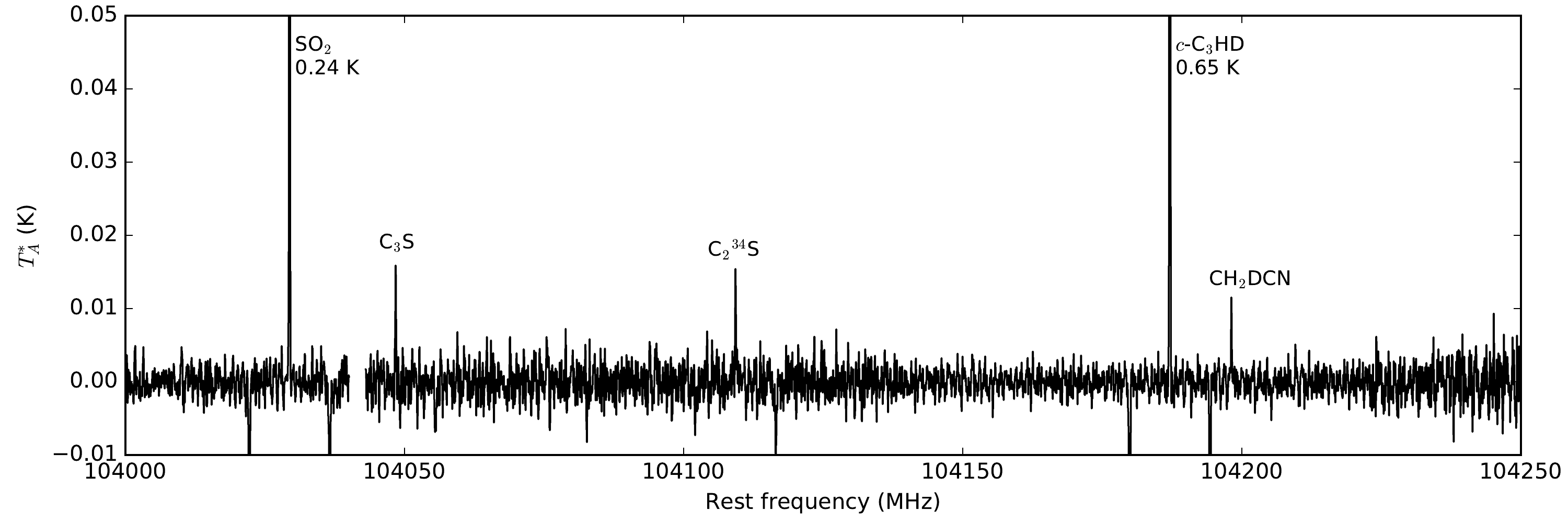}
\includegraphics[width=\textwidth]{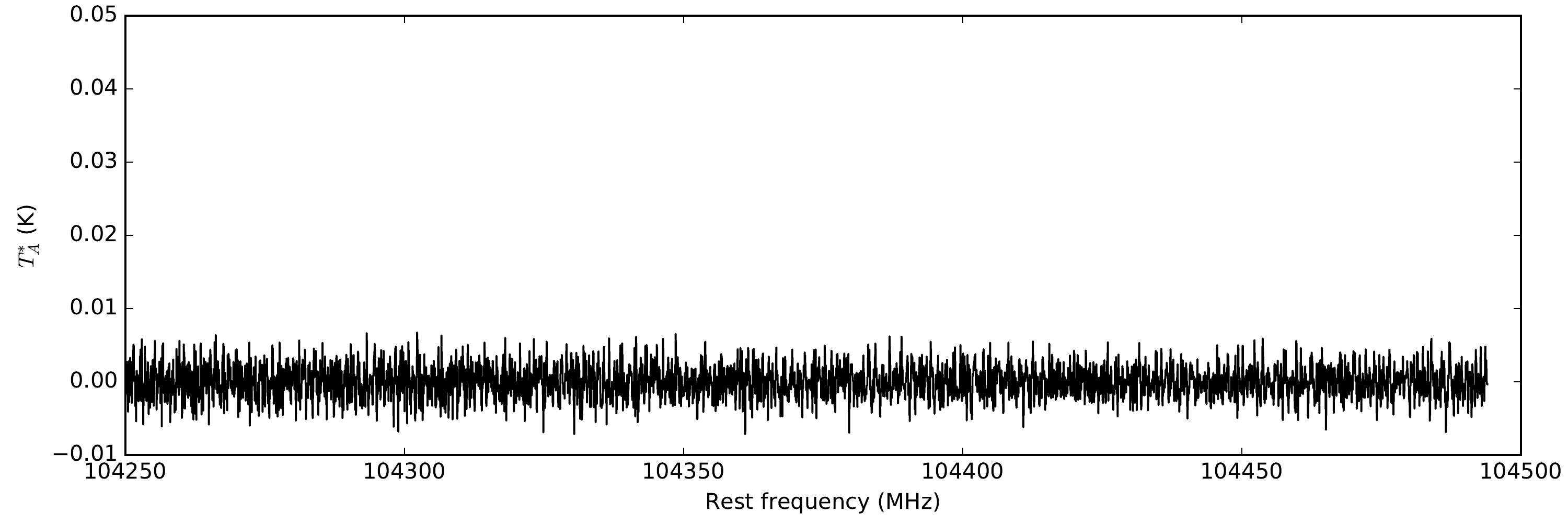}
\includegraphics[width=\textwidth]{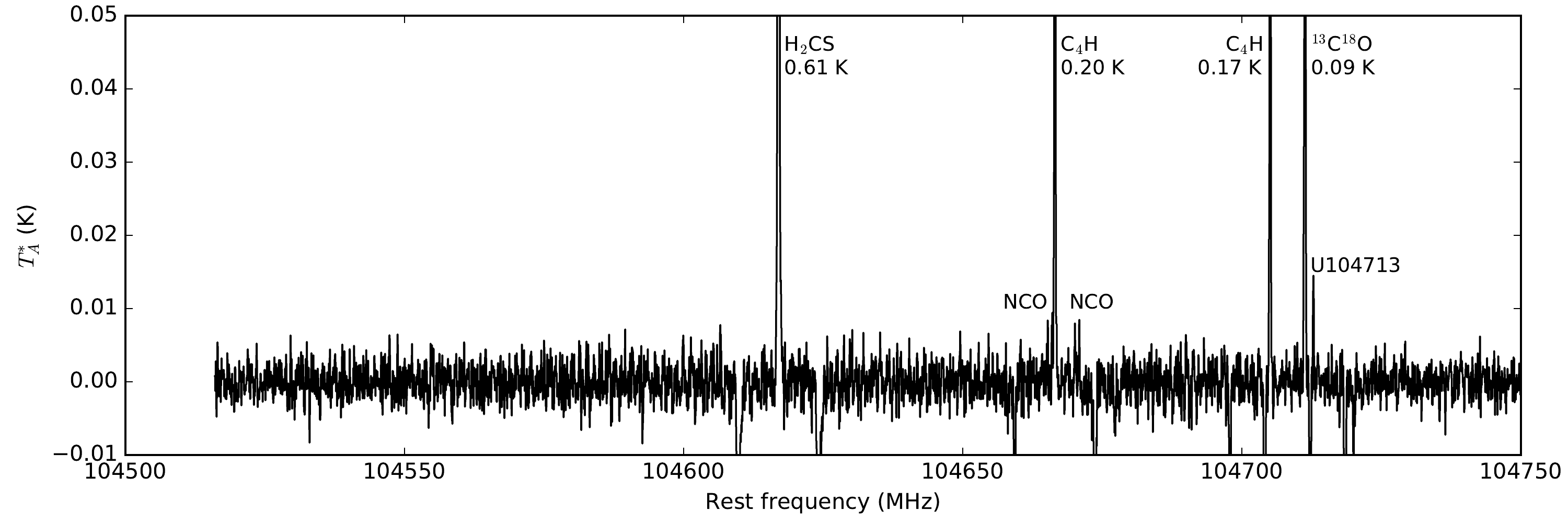}
\includegraphics[width=\textwidth]{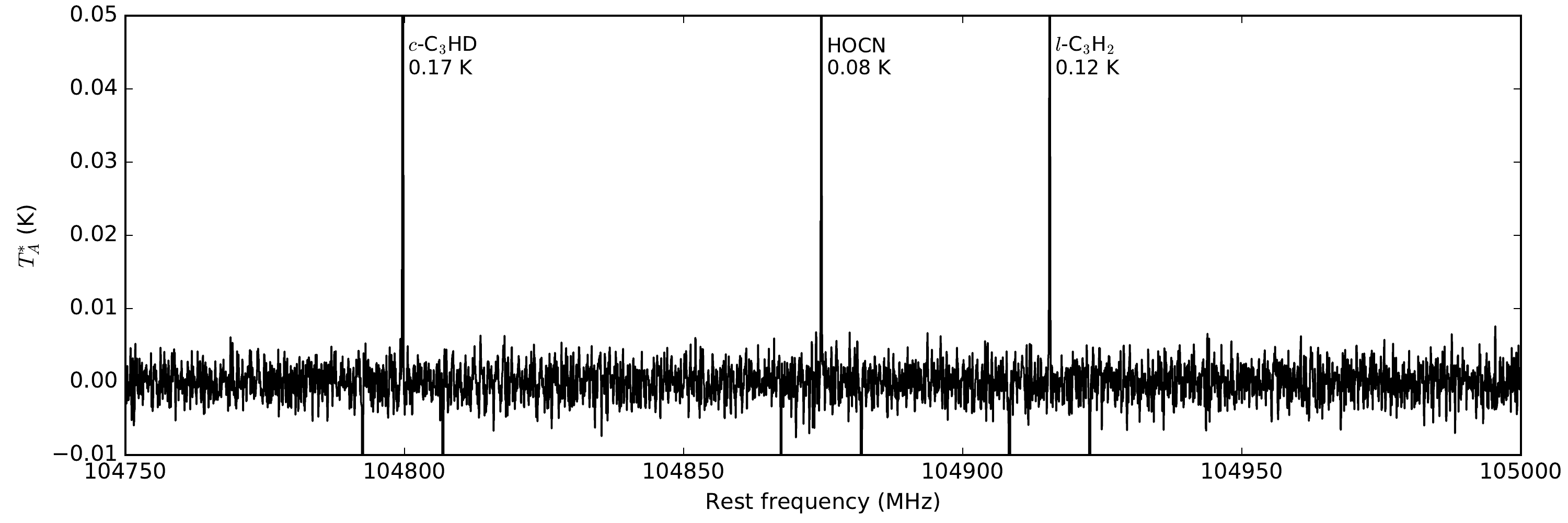}
\caption{Continued}
\end{figure*}

\setcounter{figure}{0}
\begin{figure*}
\centering
\includegraphics[width=\textwidth]{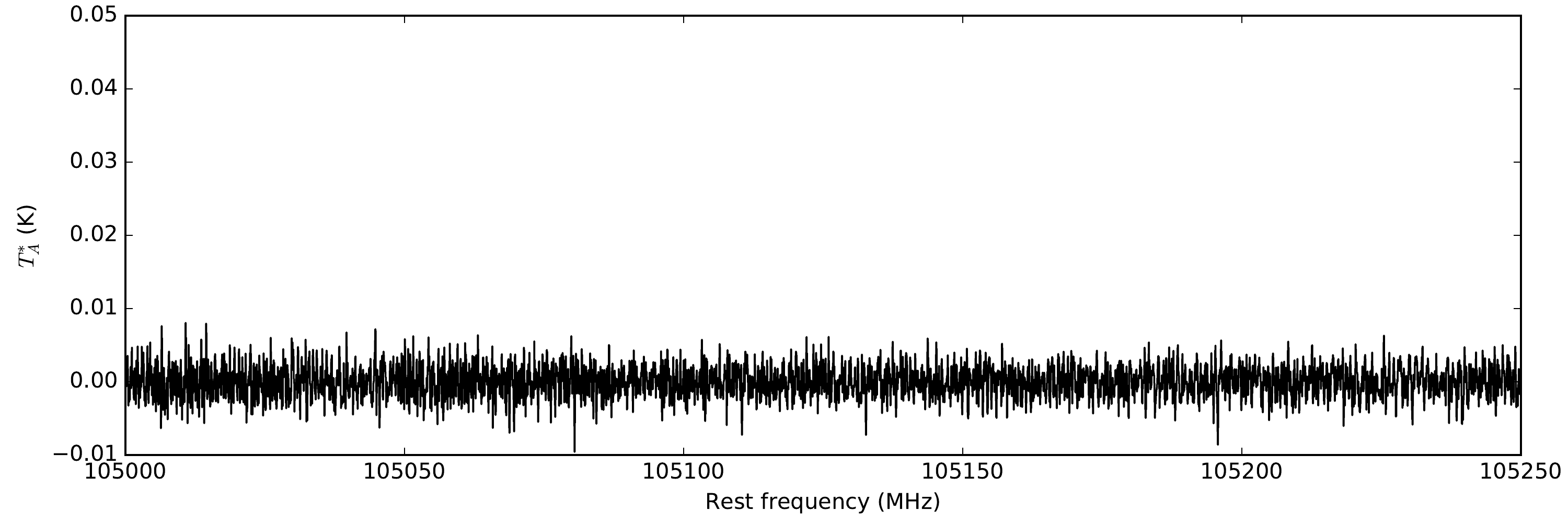}
\includegraphics[width=\textwidth]{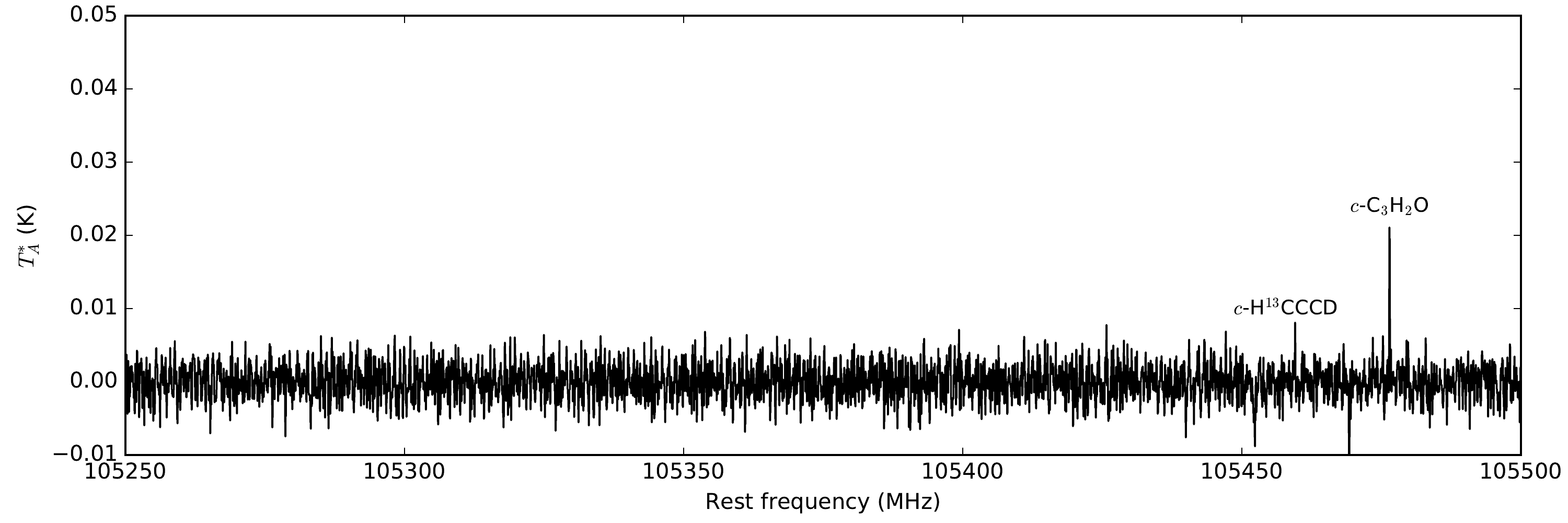}
\includegraphics[width=\textwidth]{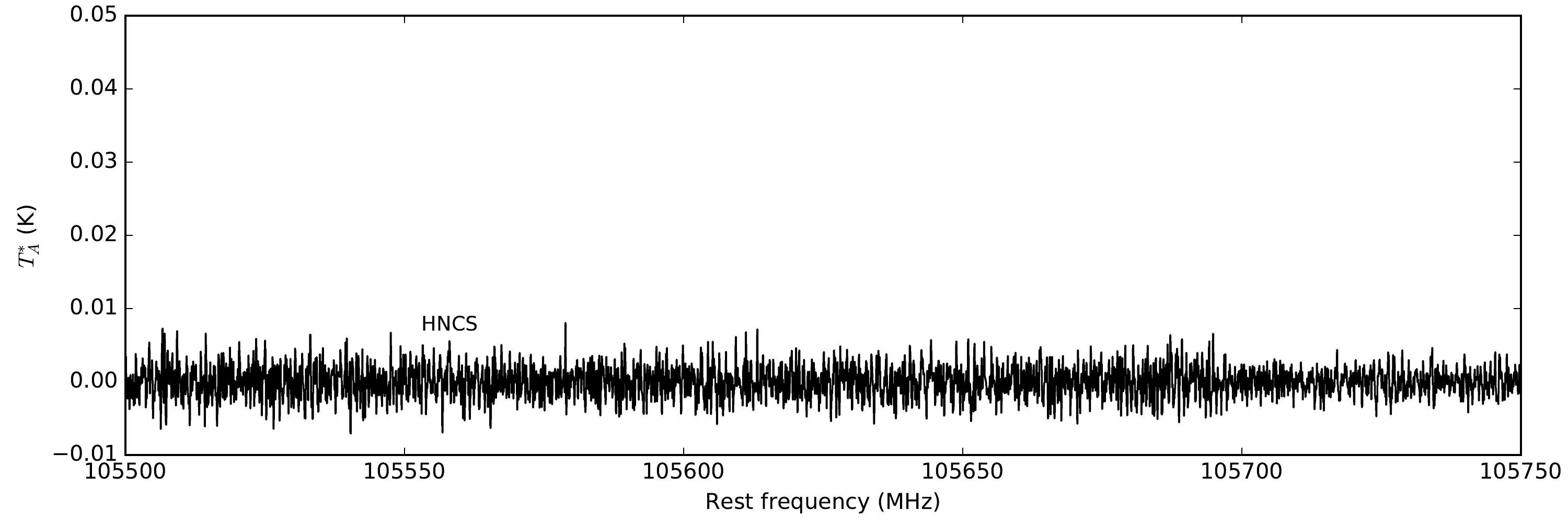}
\includegraphics[width=\textwidth]{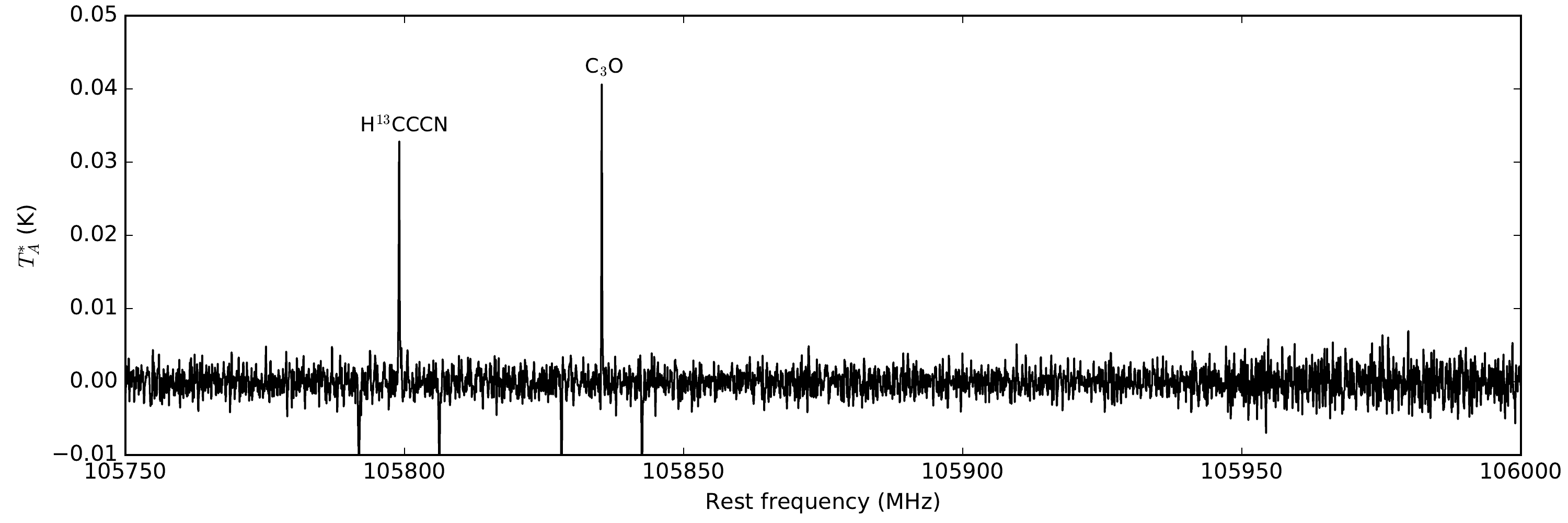}
\caption{Continued}
\end{figure*}

\setcounter{figure}{0}
\begin{figure*}
\centering
\includegraphics[width=\textwidth]{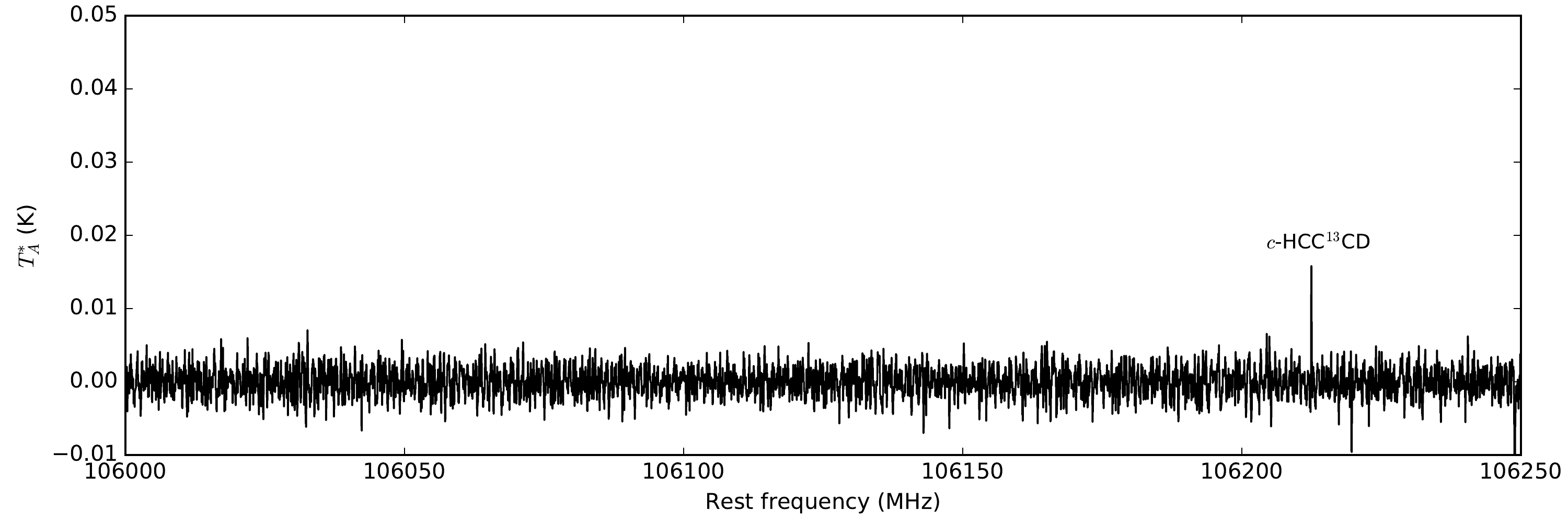}
\includegraphics[width=\textwidth]{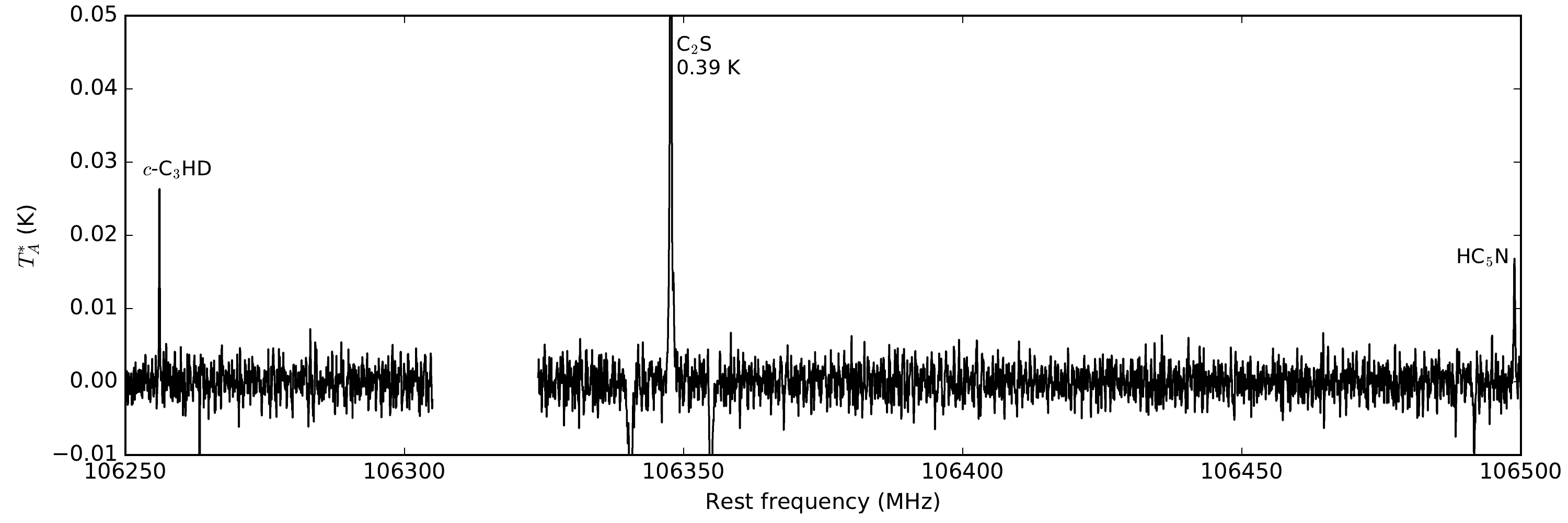}
\includegraphics[width=\textwidth]{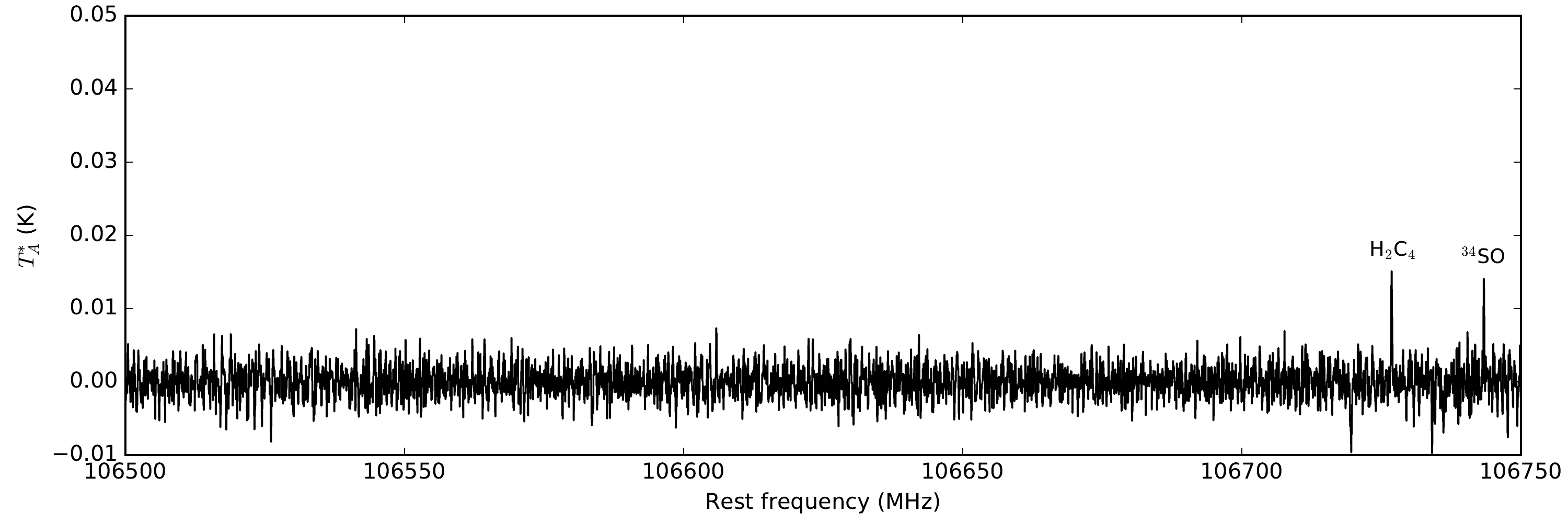}
\includegraphics[width=\textwidth]{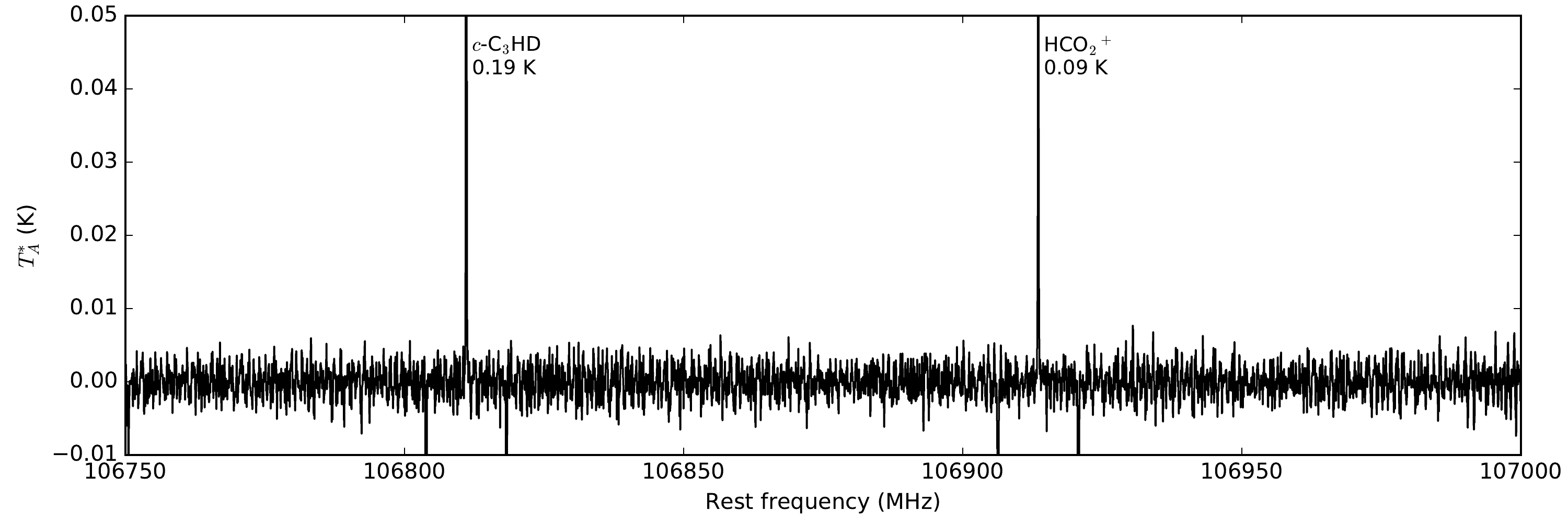}
\caption{Continued}
\end{figure*}

\setcounter{figure}{0}
\begin{figure*}
\centering
\includegraphics[width=\textwidth]{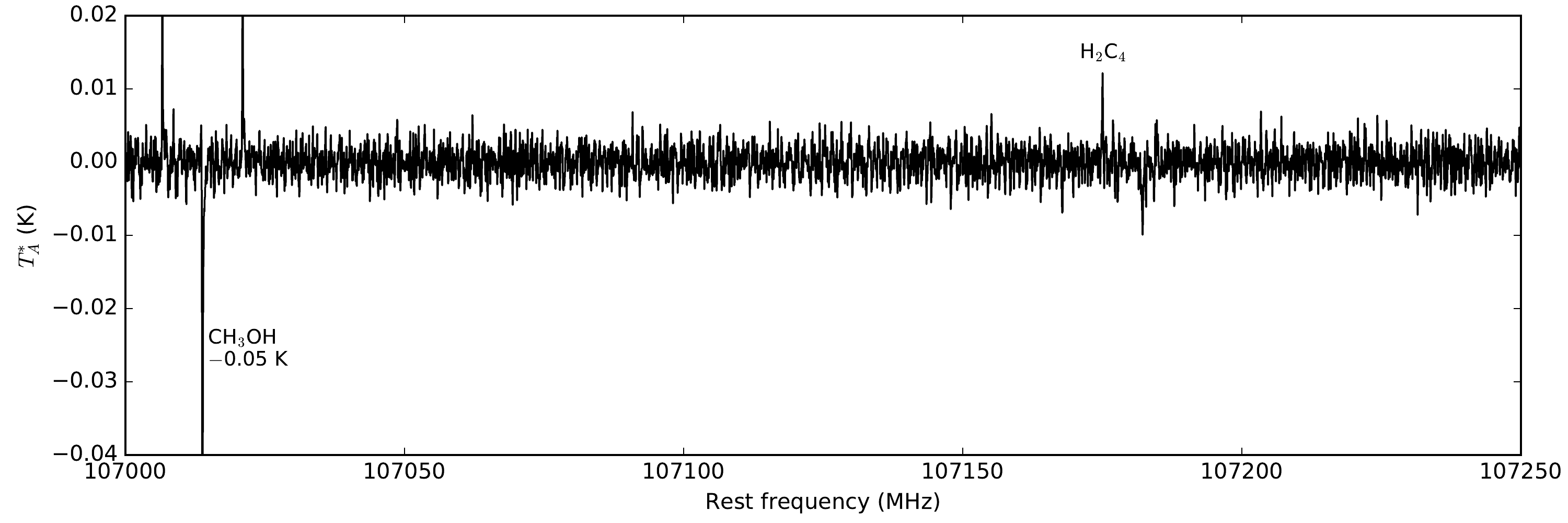}
\includegraphics[width=\textwidth]{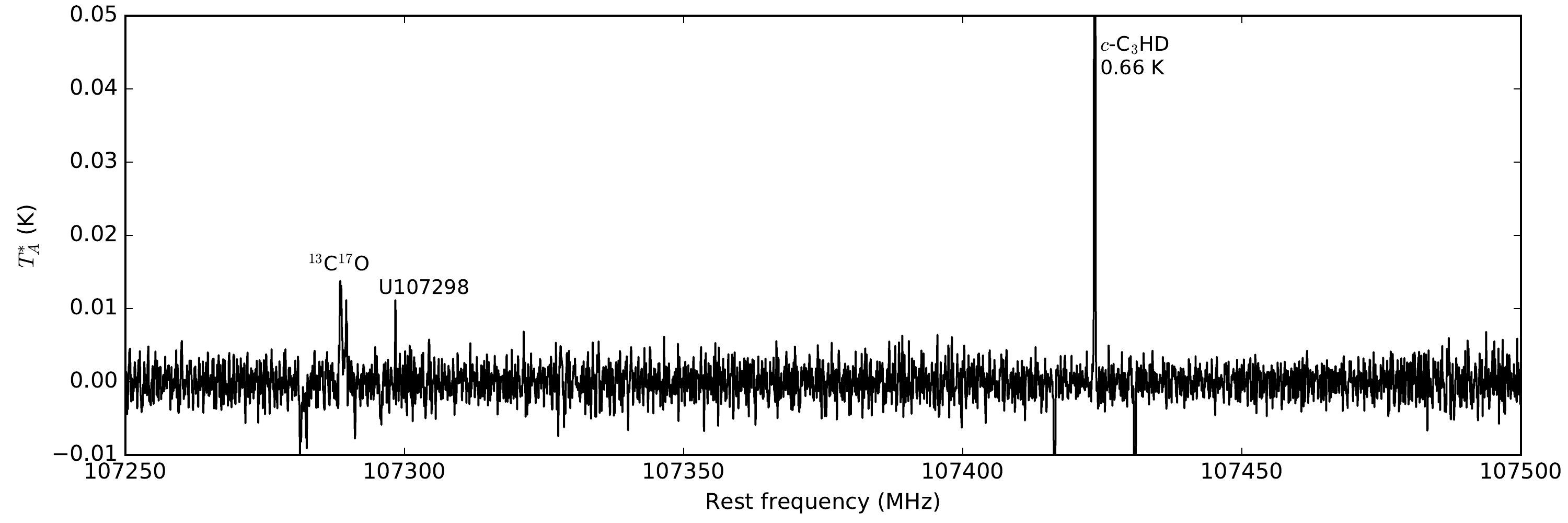}
\includegraphics[width=\textwidth]{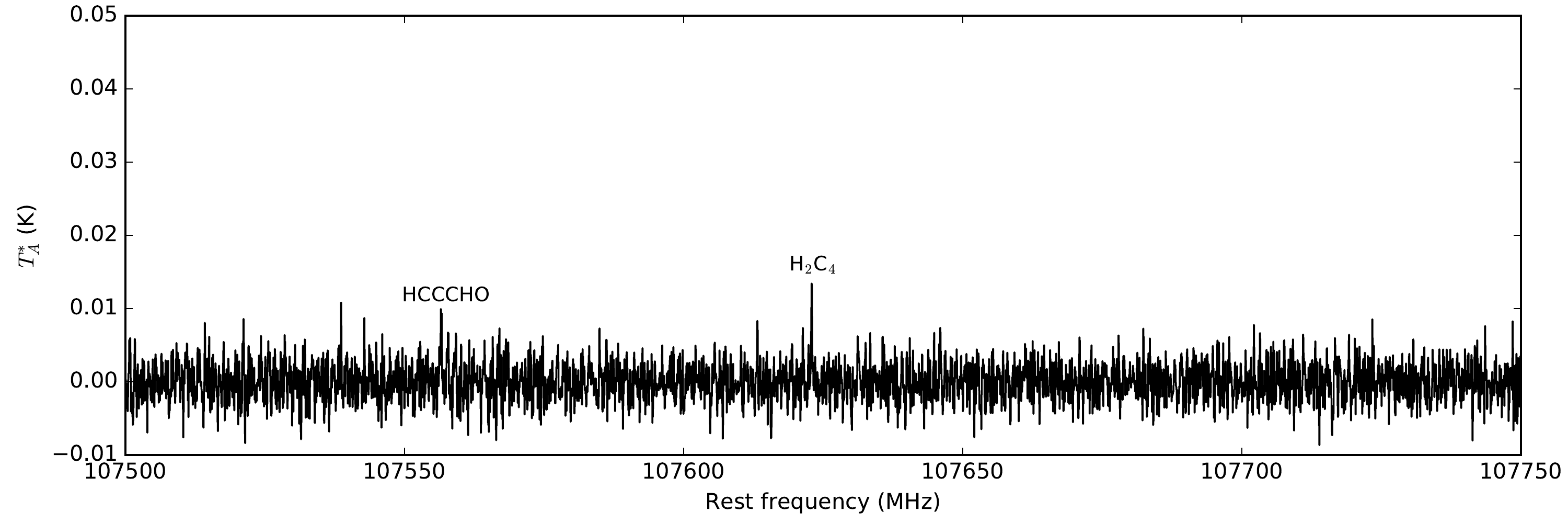}
\includegraphics[width=\textwidth]{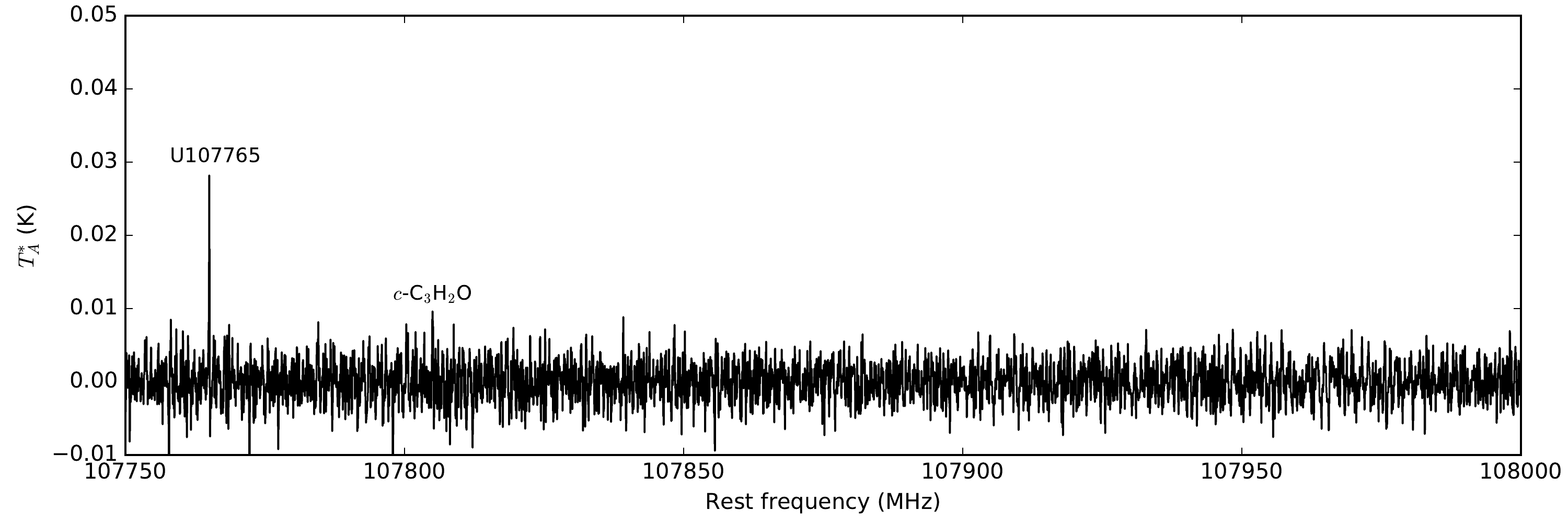}
\caption{Continued}
\end{figure*}

\setcounter{figure}{0}
\begin{figure*}
\centering
\includegraphics[width=\textwidth]{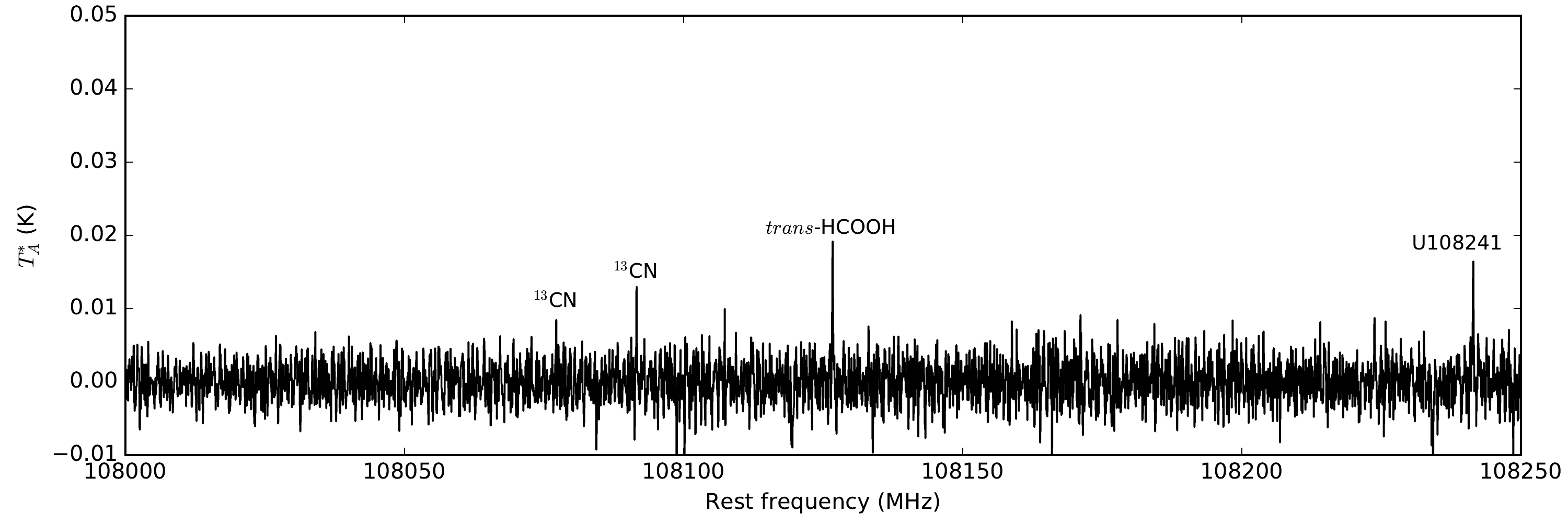}
\includegraphics[width=\textwidth]{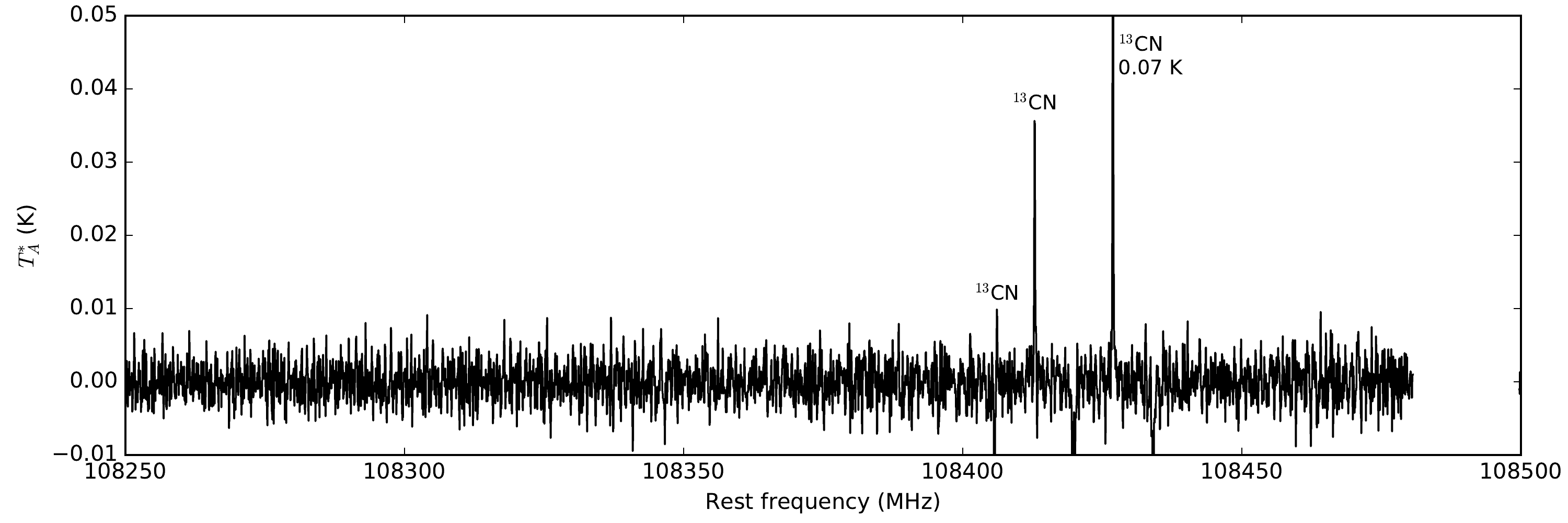}
\includegraphics[width=\textwidth]{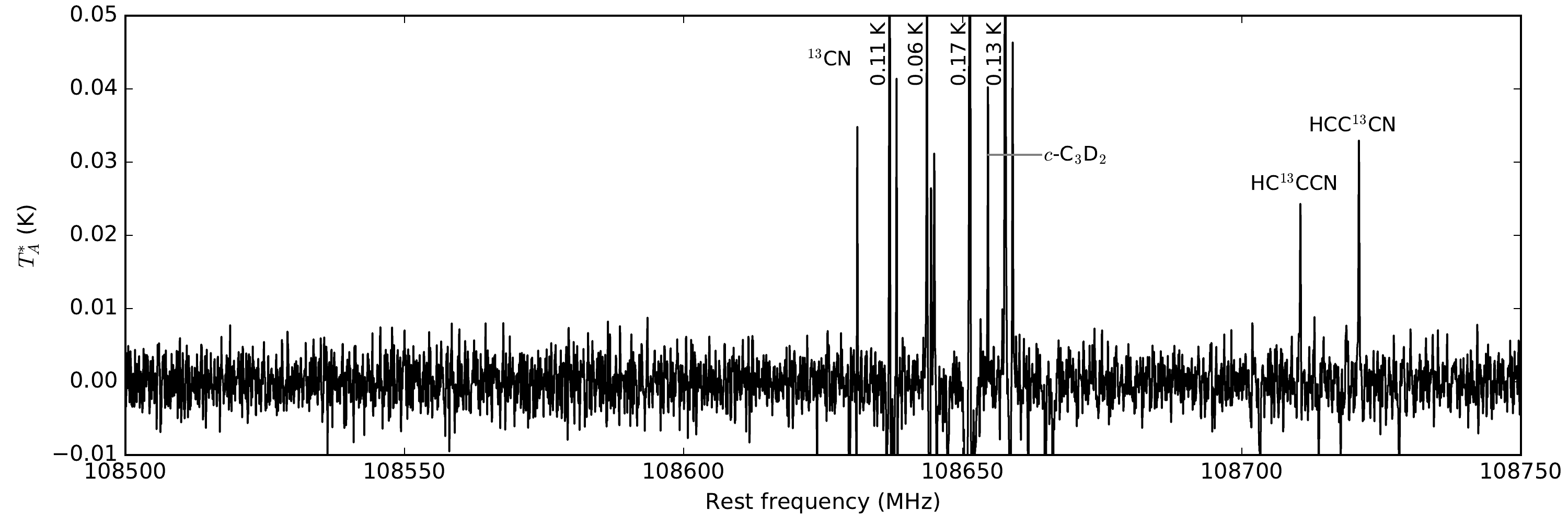}
\includegraphics[width=\textwidth]{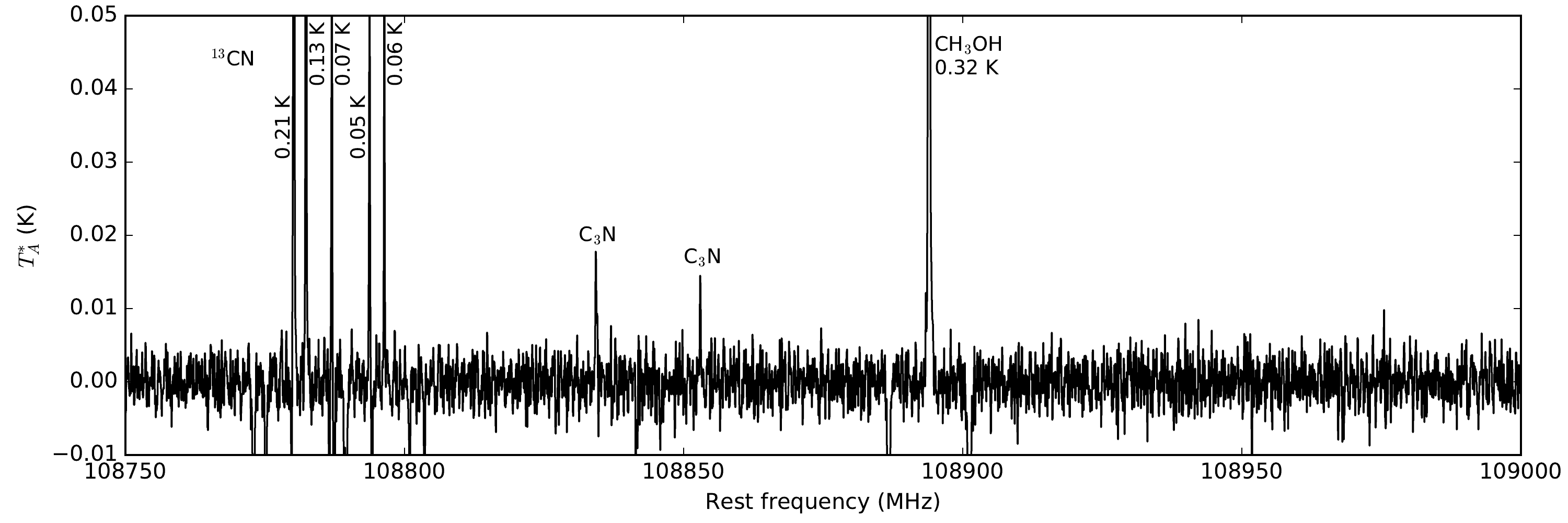}
\caption{Continued}
\end{figure*}

\setcounter{figure}{0}
\begin{figure*}
\centering
\includegraphics[width=\textwidth]{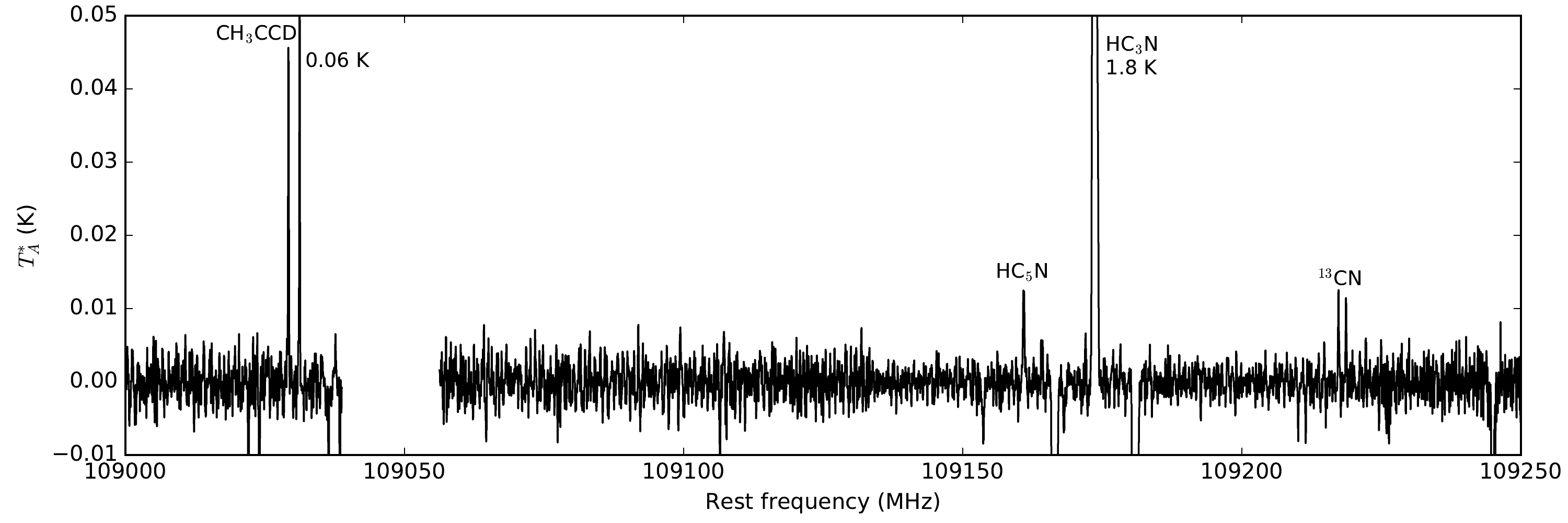}
\includegraphics[width=\textwidth]{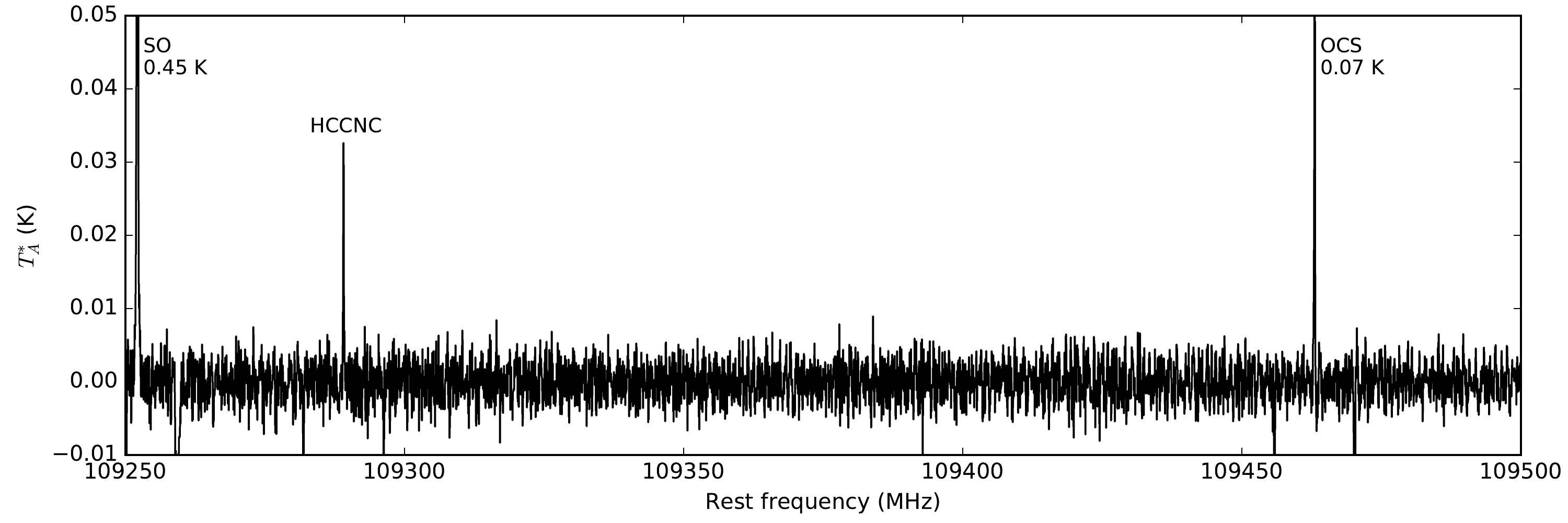}
\includegraphics[width=\textwidth]{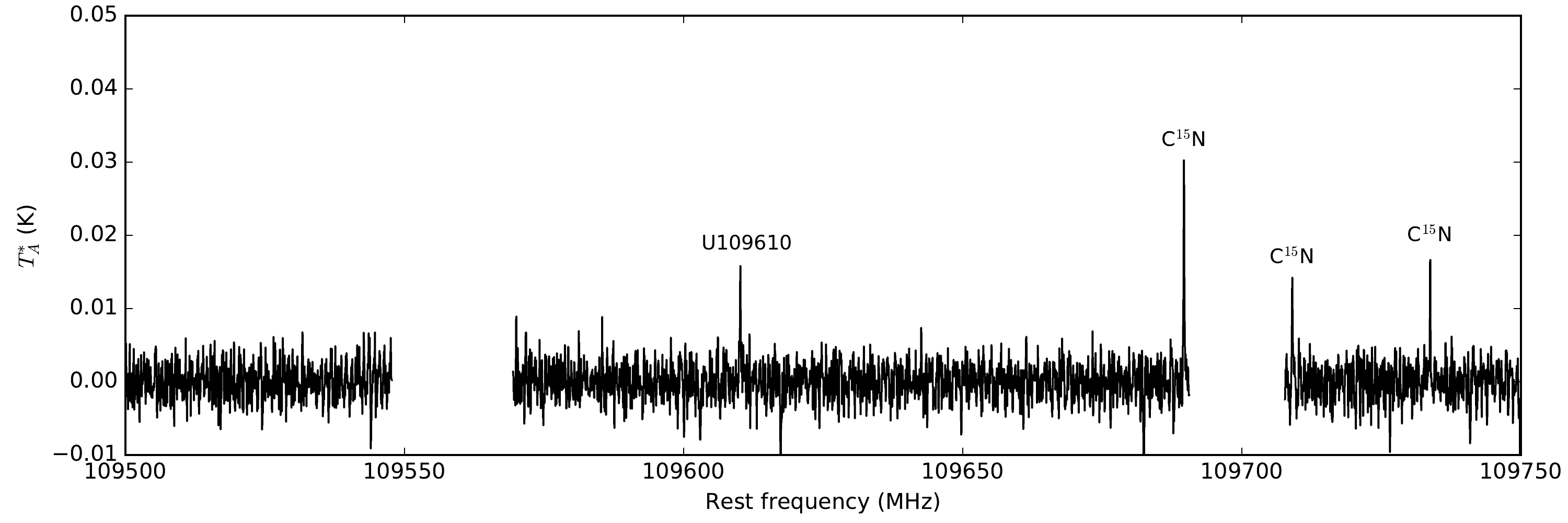}
\includegraphics[width=\textwidth]{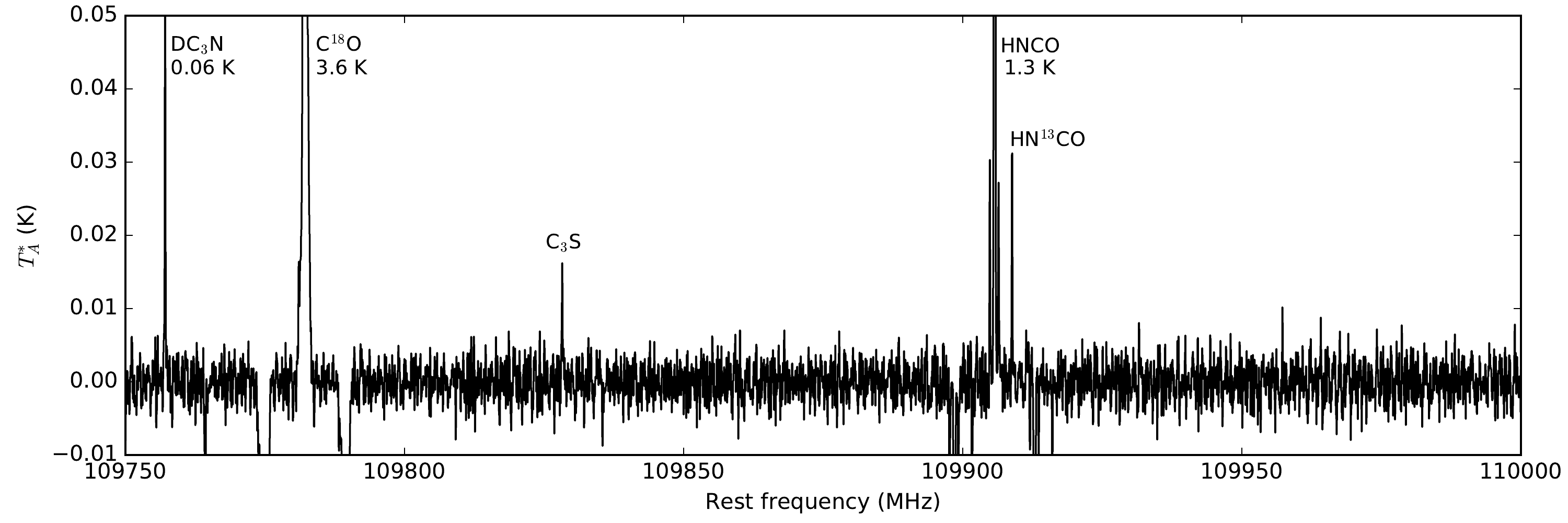}
\caption{Continued}
\end{figure*}

\setcounter{figure}{0}
\begin{figure*}
\centering
\includegraphics[width=\textwidth]{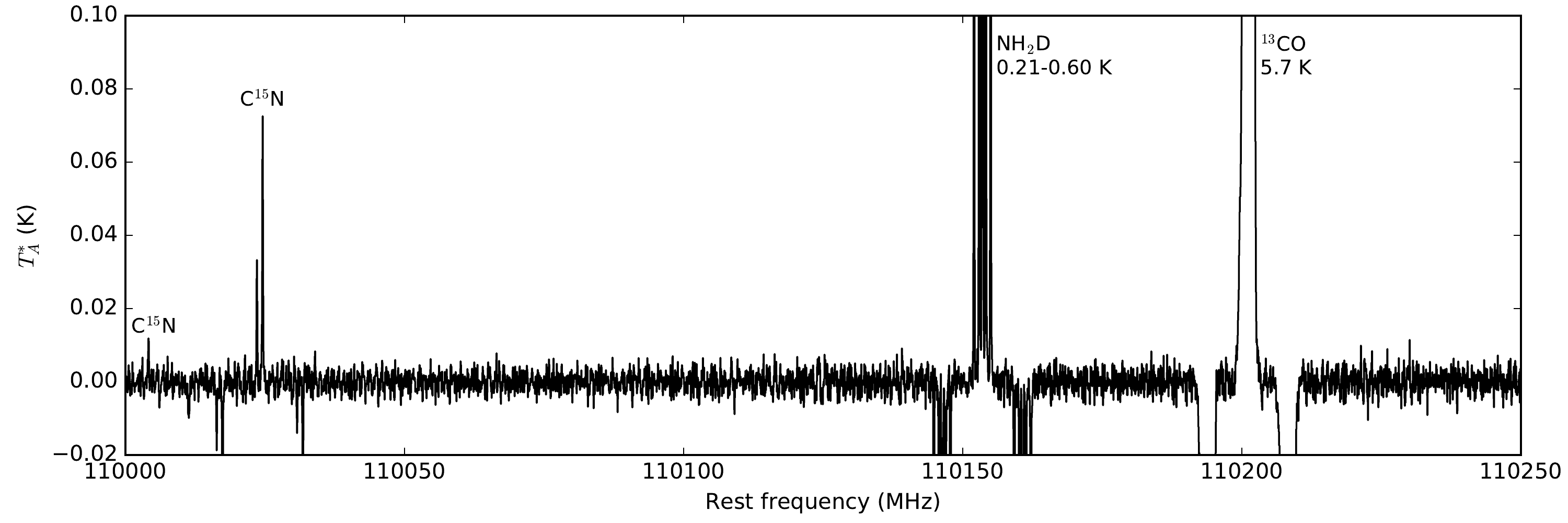}
\includegraphics[width=\textwidth]{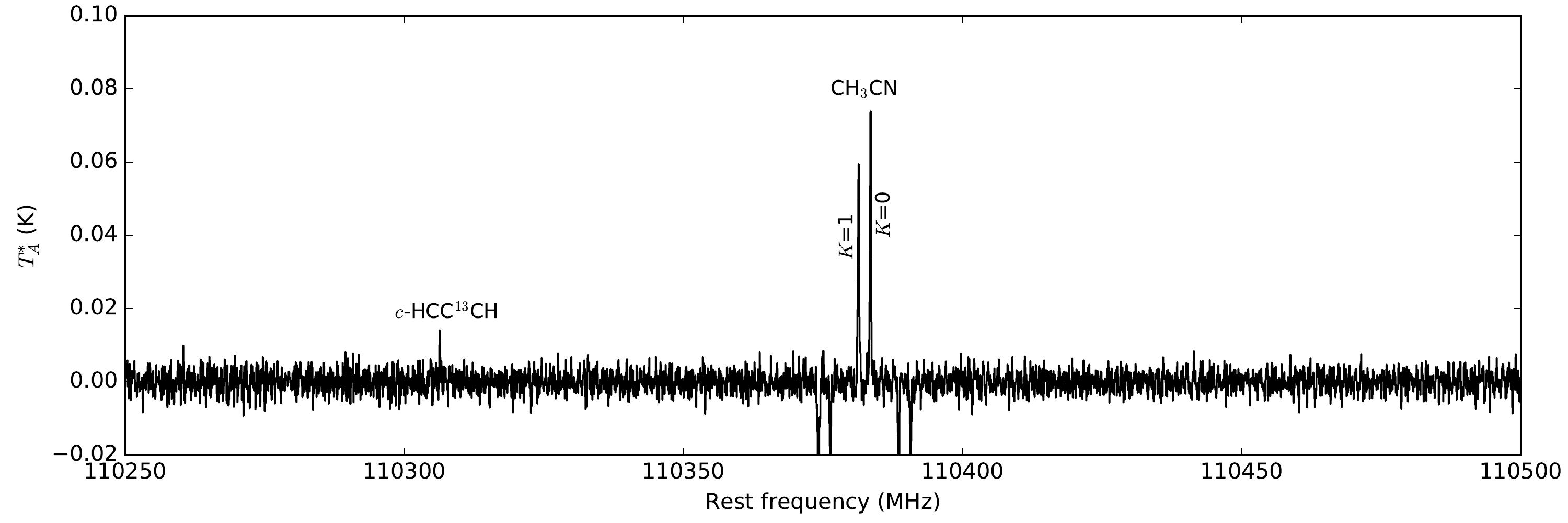}
\includegraphics[width=\textwidth]{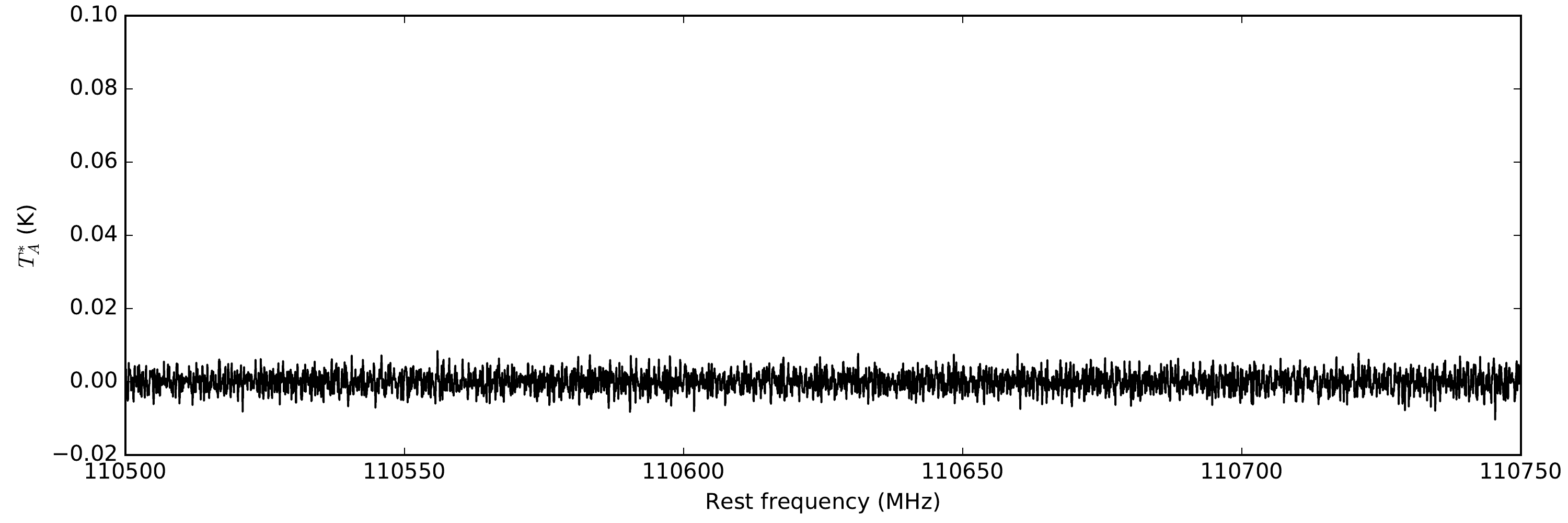}
\includegraphics[width=\textwidth]{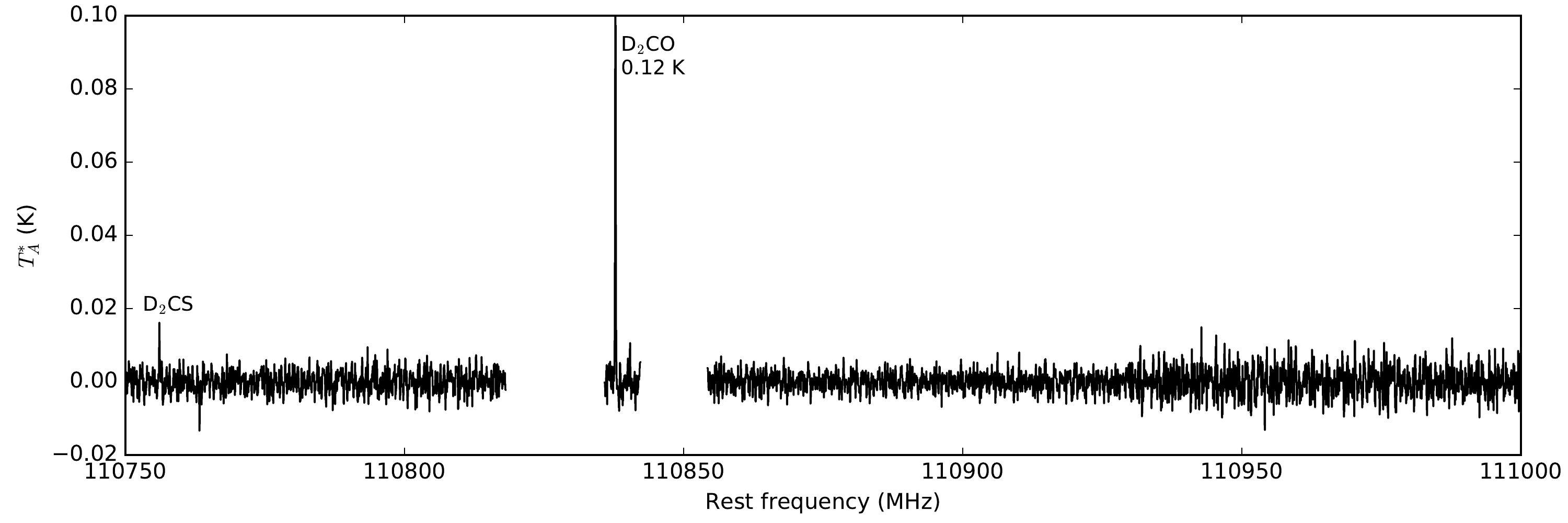}
\caption{Continued}
\end{figure*}

\setcounter{figure}{0}
\begin{figure*}
\centering
\includegraphics[width=\textwidth]{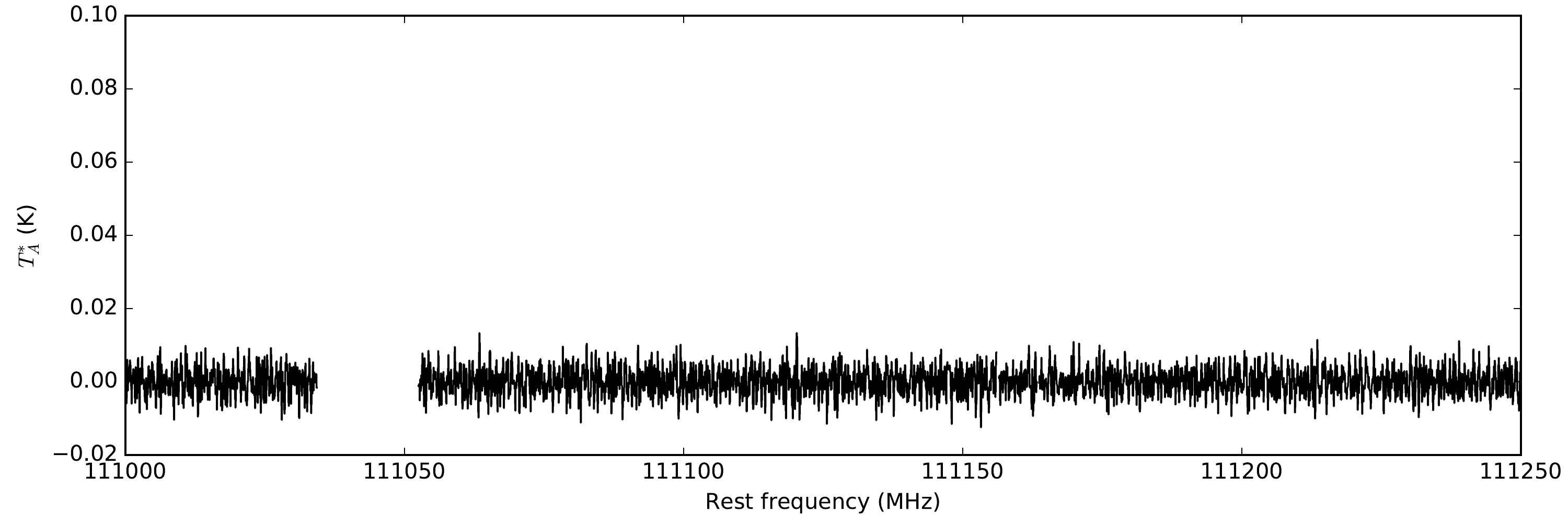}
\includegraphics[width=\textwidth]{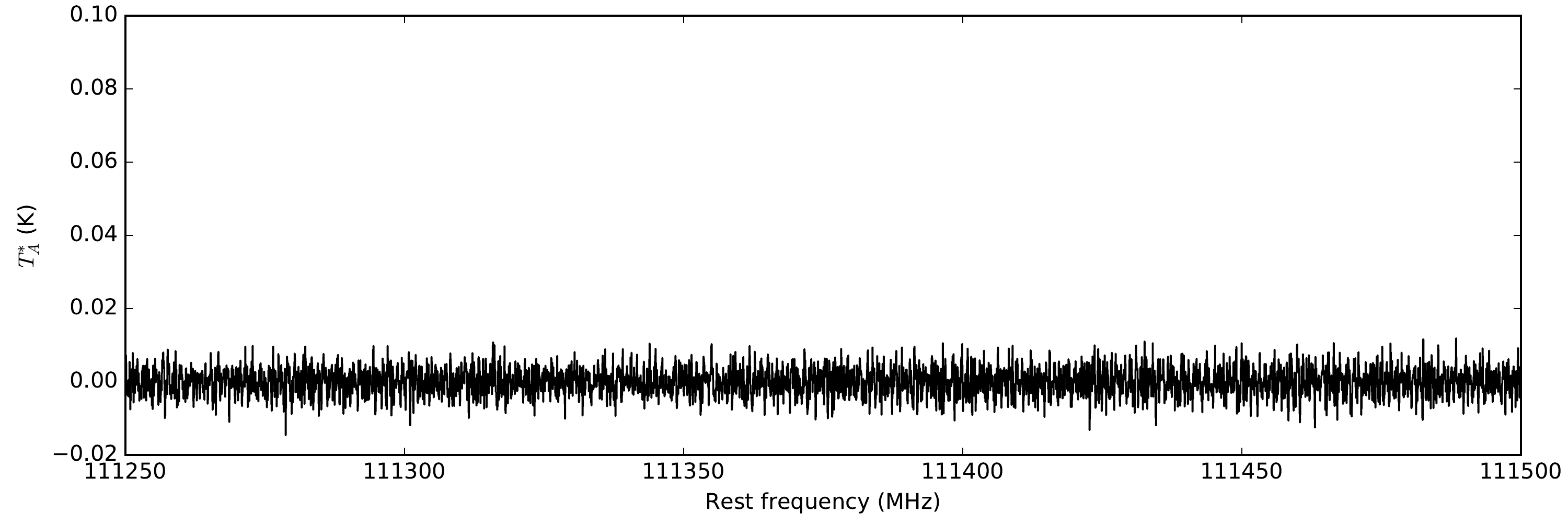}
\includegraphics[width=\textwidth]{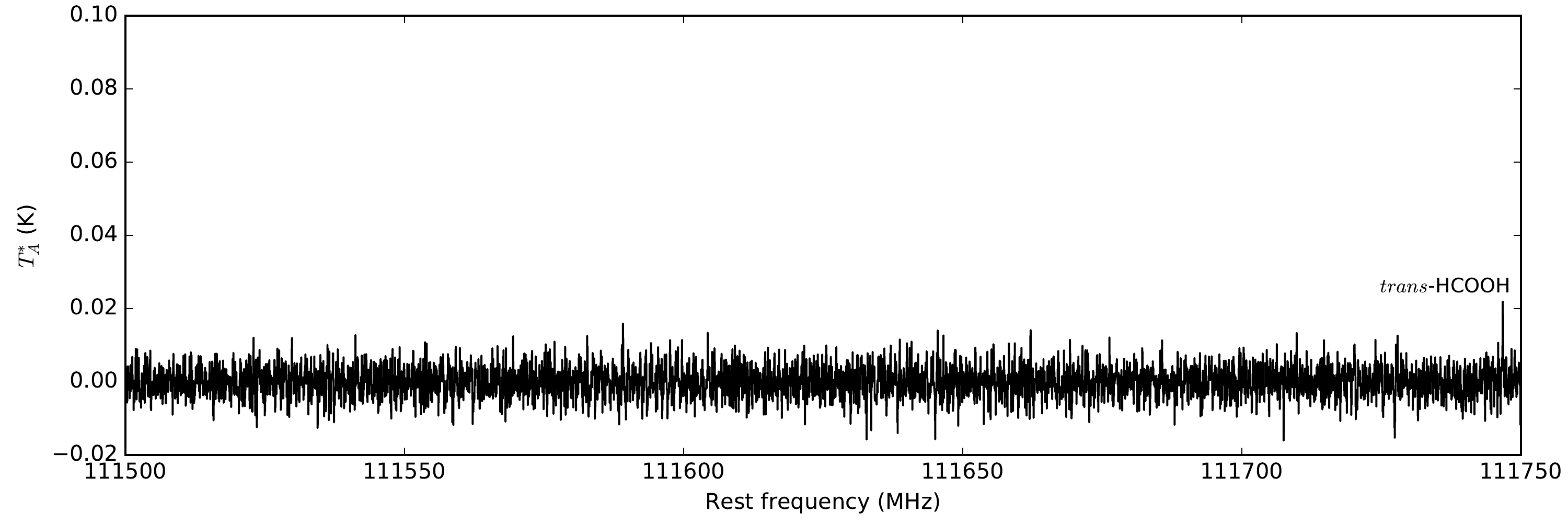}
\includegraphics[width=\textwidth]{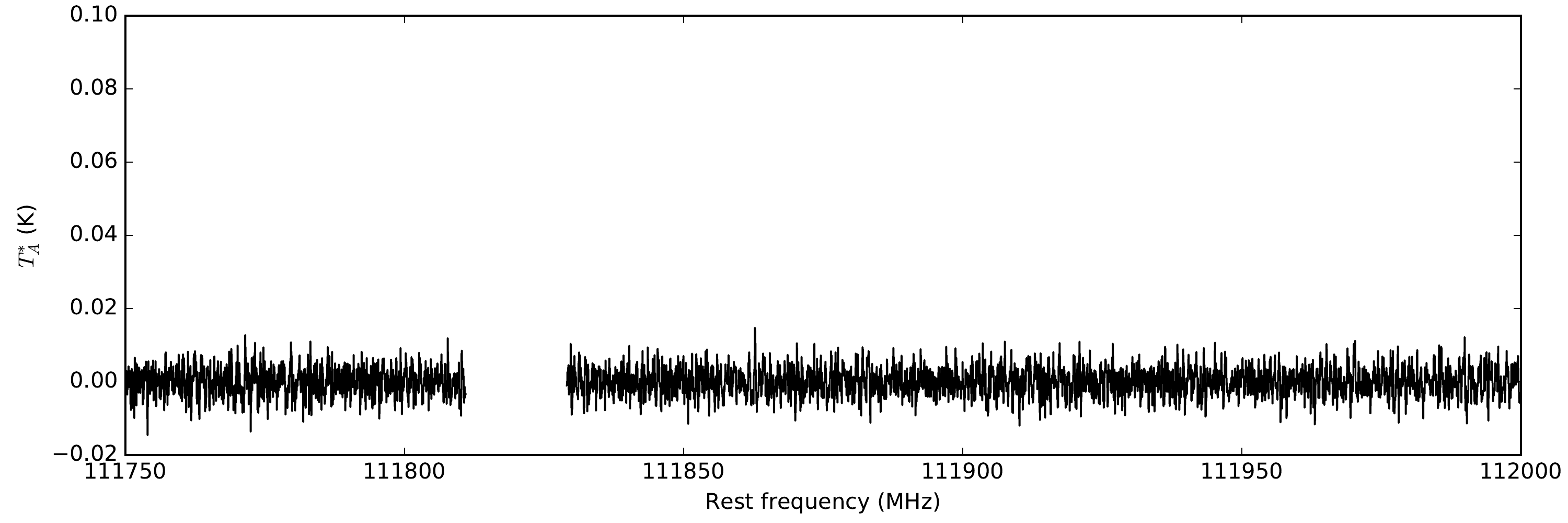}
\caption{Continued}
\end{figure*}

\setcounter{figure}{0}
\begin{figure*}
\centering
\includegraphics[width=\textwidth]{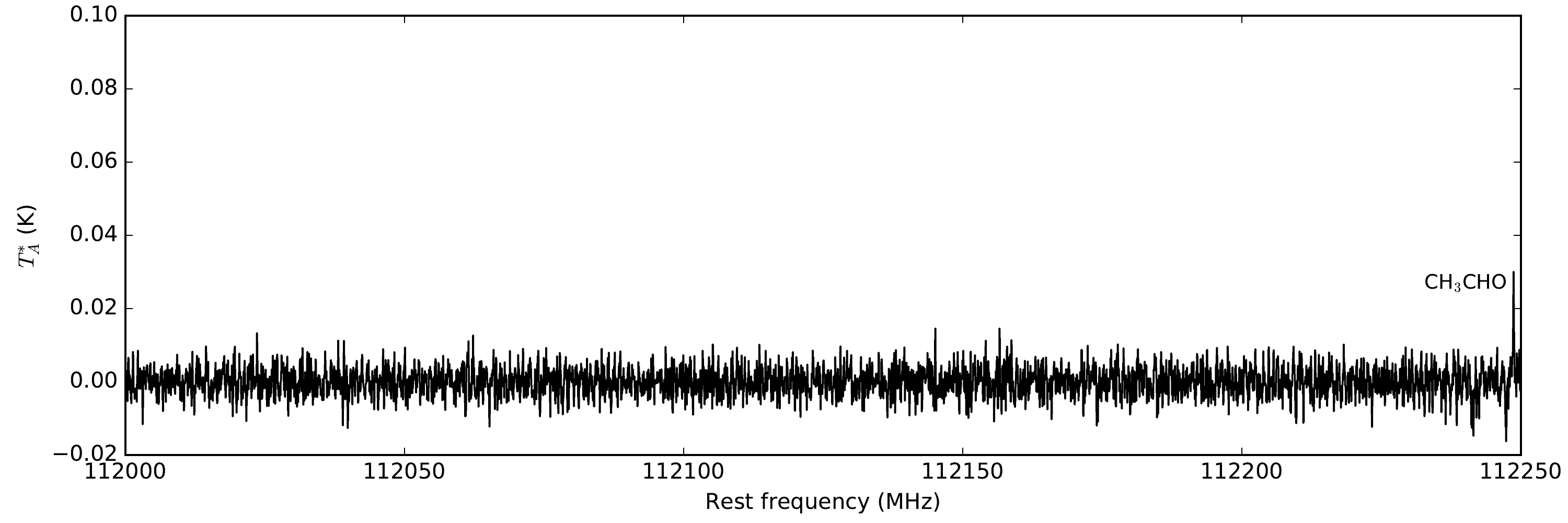}
\includegraphics[width=\textwidth]{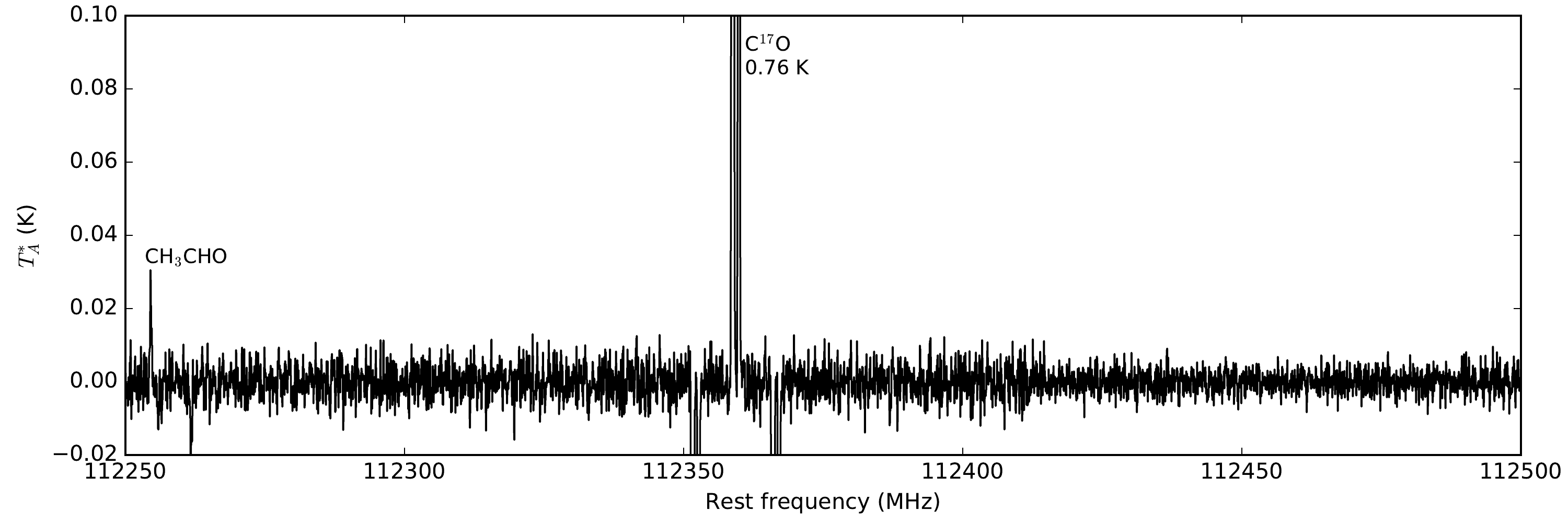}
\includegraphics[width=\textwidth]{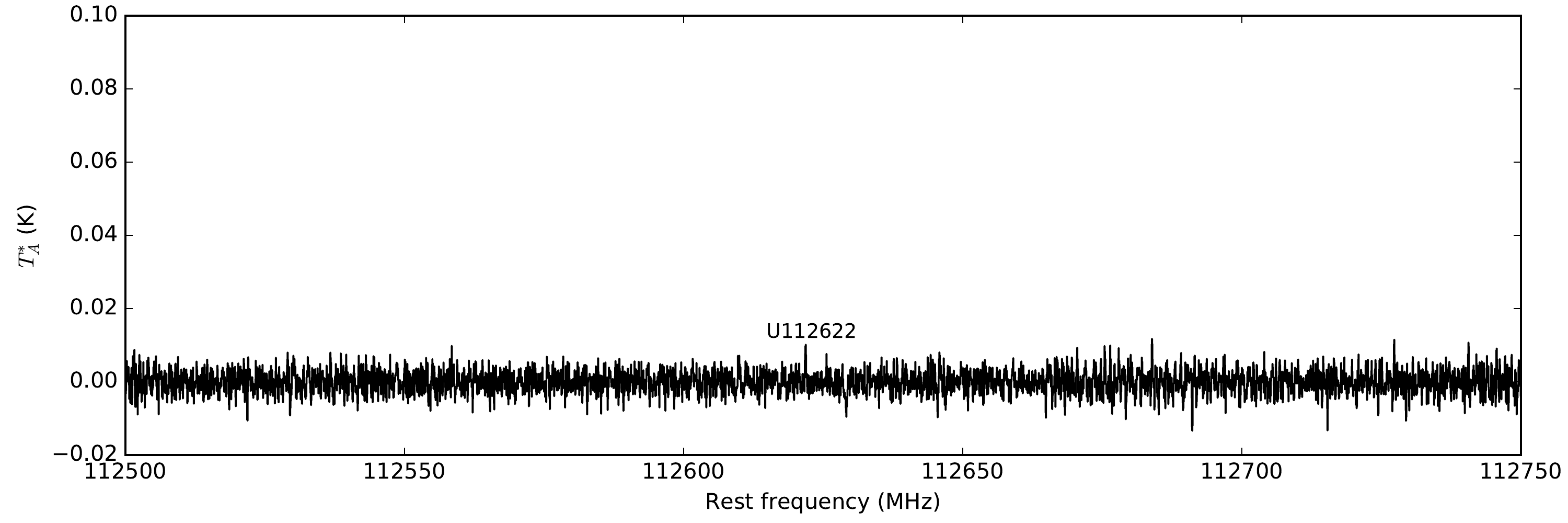}
\includegraphics[width=\textwidth]{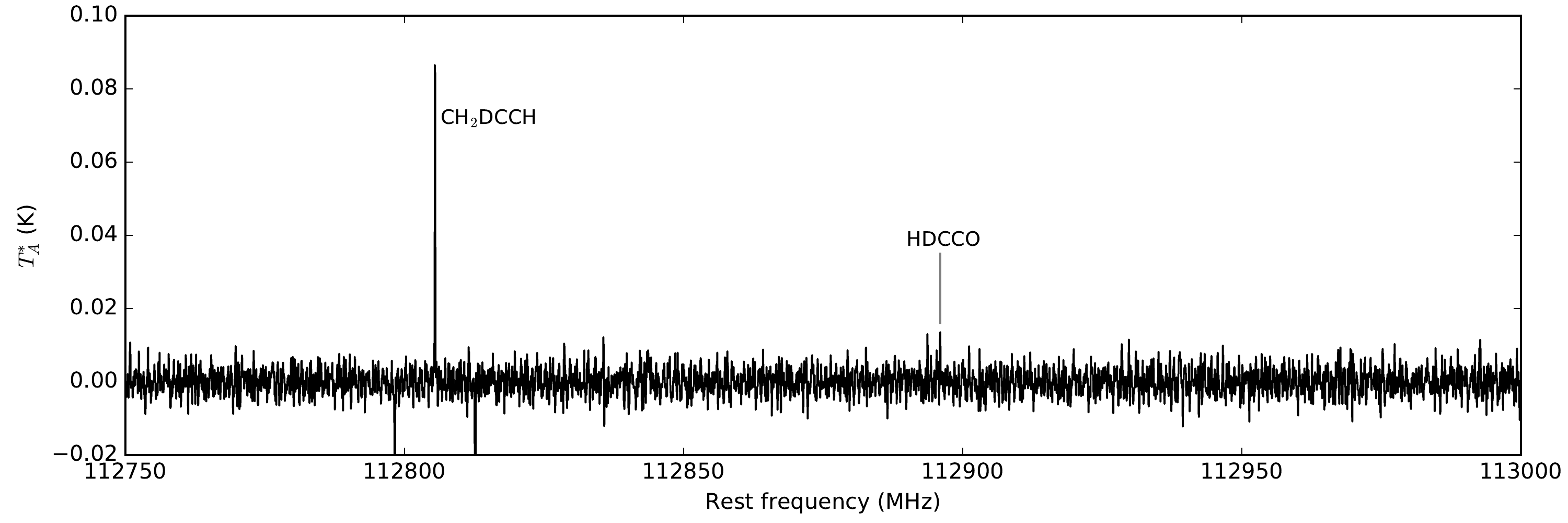}
\caption{Continued}
\end{figure*}

\setcounter{figure}{0}
\begin{figure*}
\centering
\includegraphics[width=\textwidth]{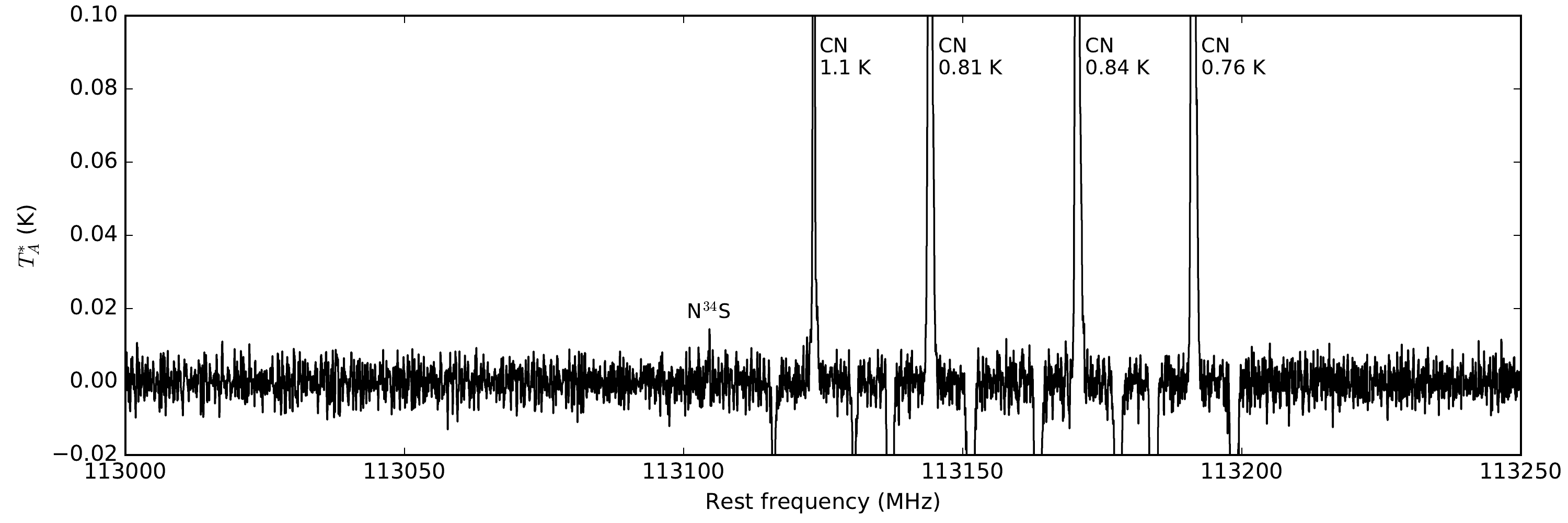}
\includegraphics[width=\textwidth]{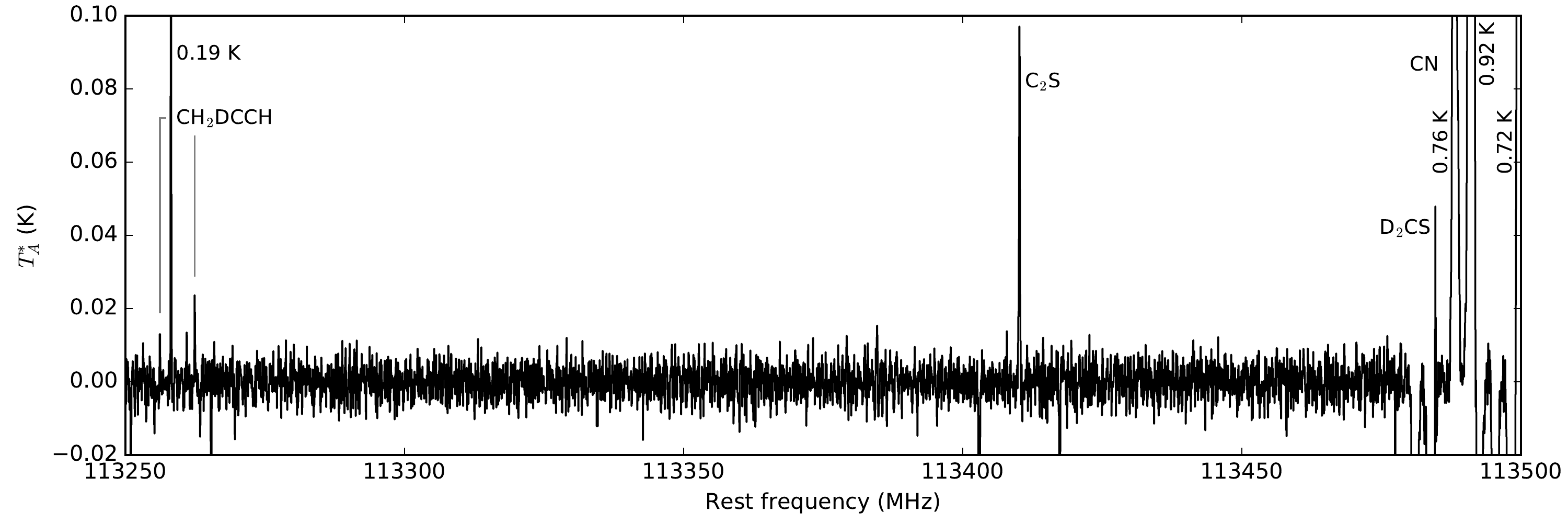}
\includegraphics[width=\textwidth]{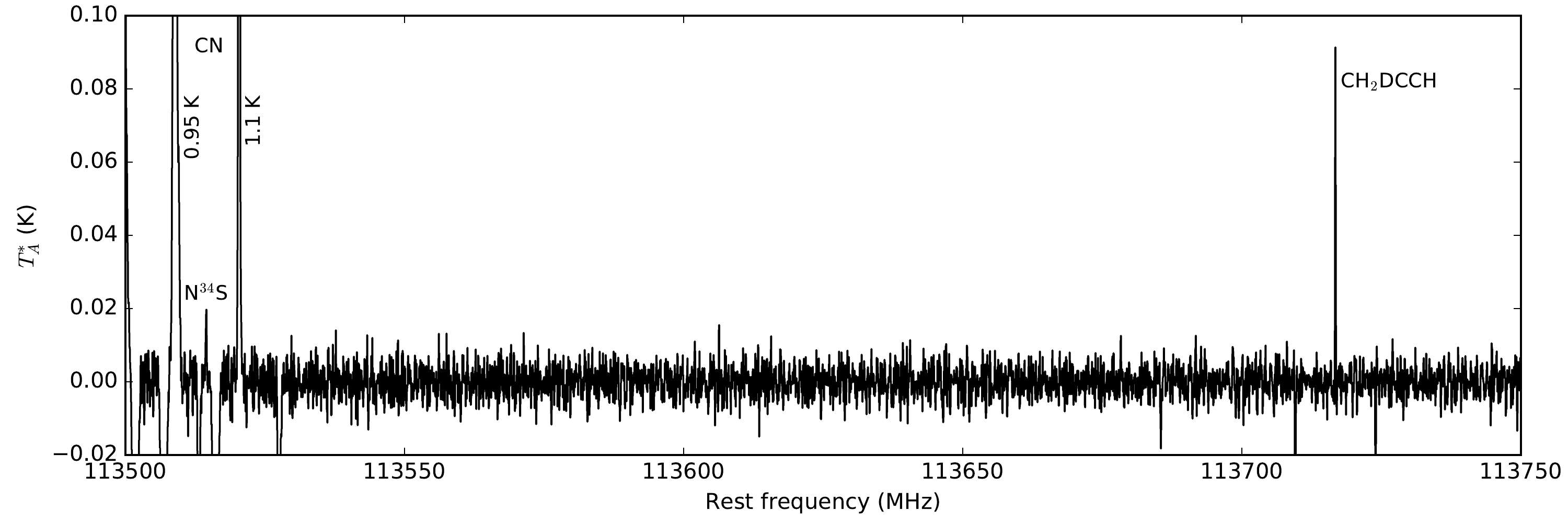}
\includegraphics[width=\textwidth]{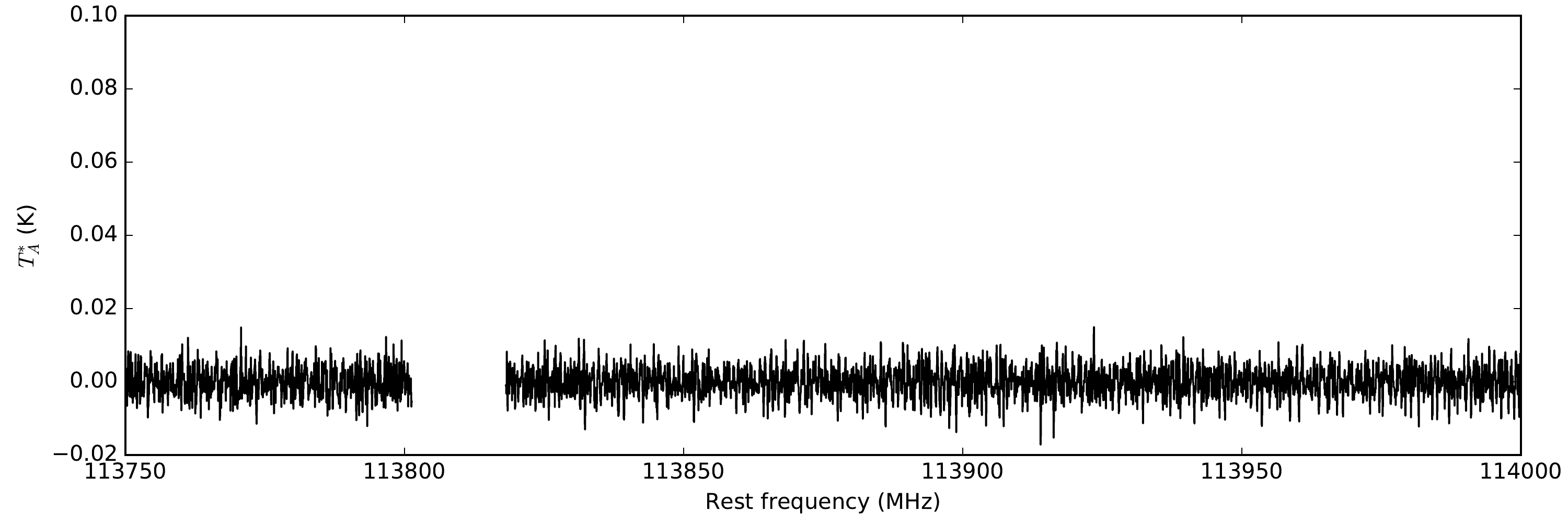}
\caption{Continued}
\end{figure*}

\setcounter{figure}{0}
\begin{figure*}
\centering
\includegraphics[width=\textwidth]{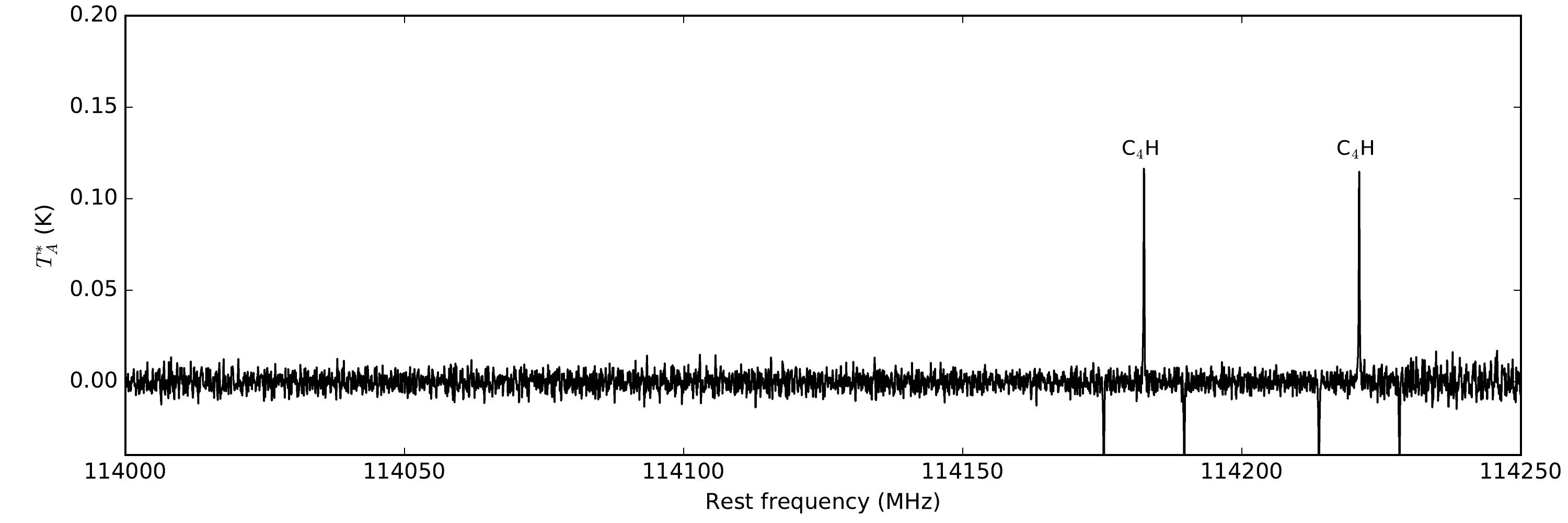}
\includegraphics[width=\textwidth]{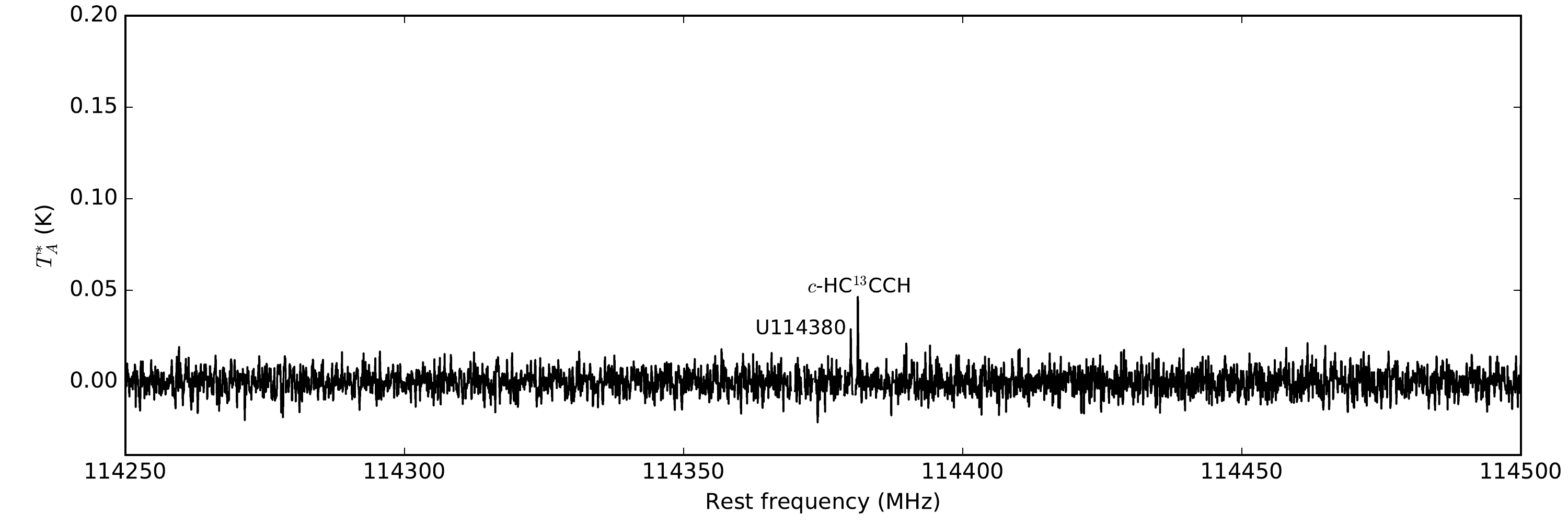}
\includegraphics[width=\textwidth]{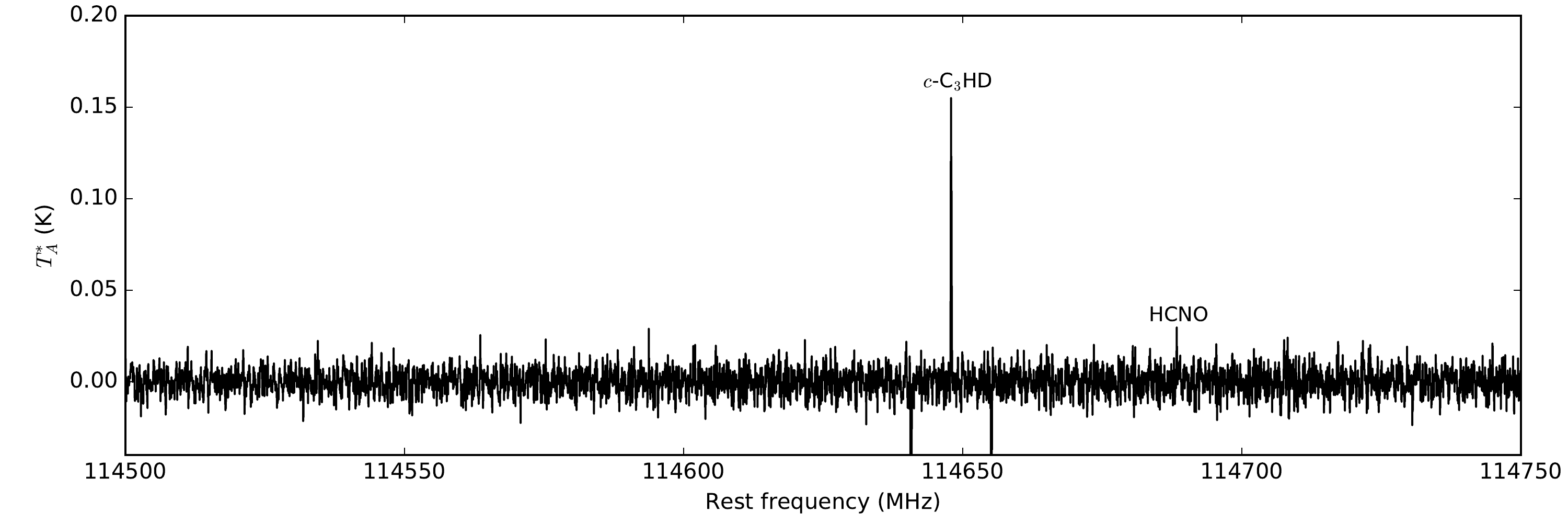}
\includegraphics[width=\textwidth]{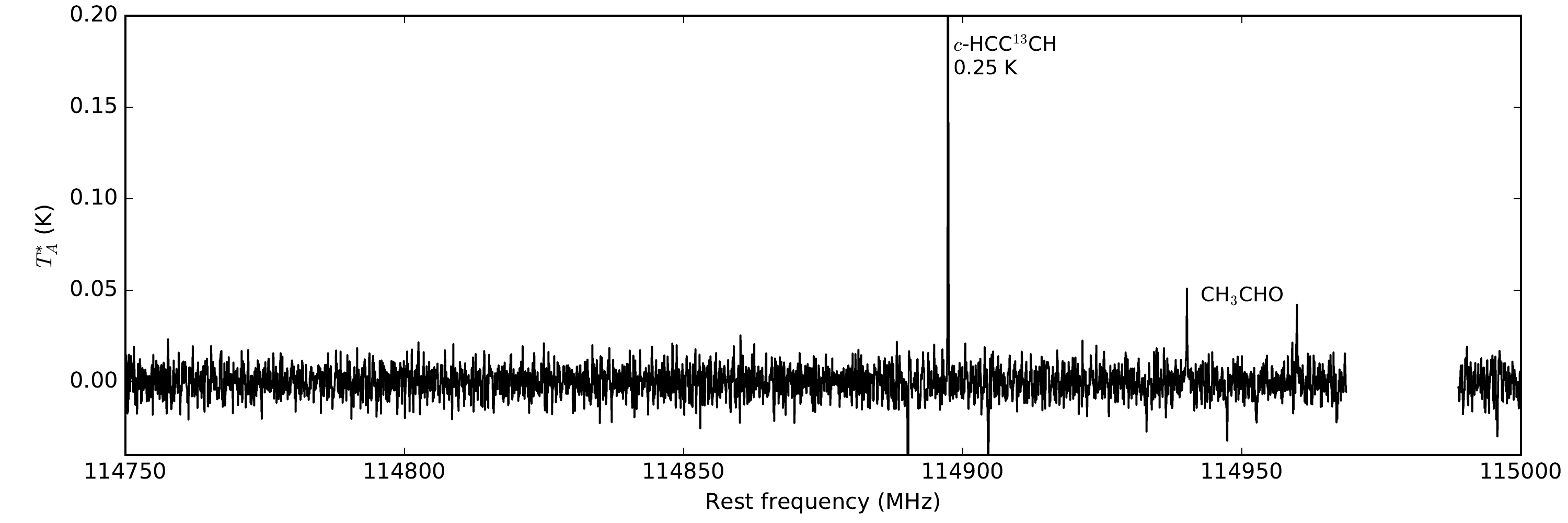}
\caption{Continued}
\end{figure*}

\setcounter{figure}{0}
\begin{figure*}
\centering
\includegraphics[width=\textwidth]{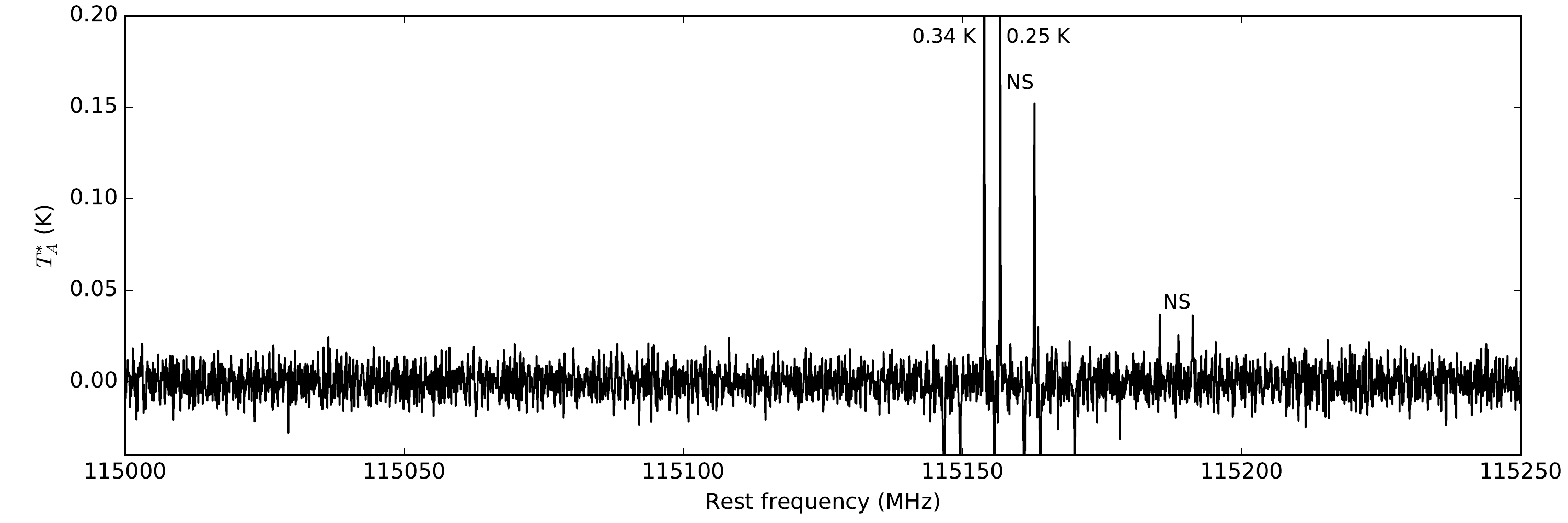}
\includegraphics[width=\textwidth]{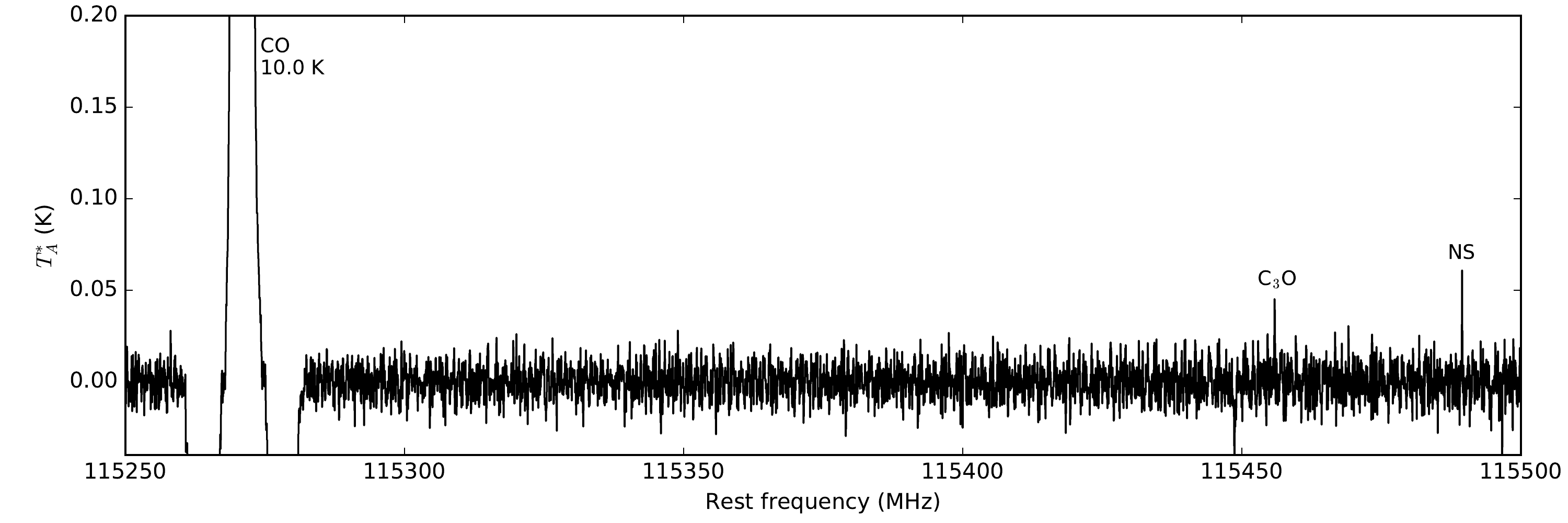}
\includegraphics[width=\textwidth]{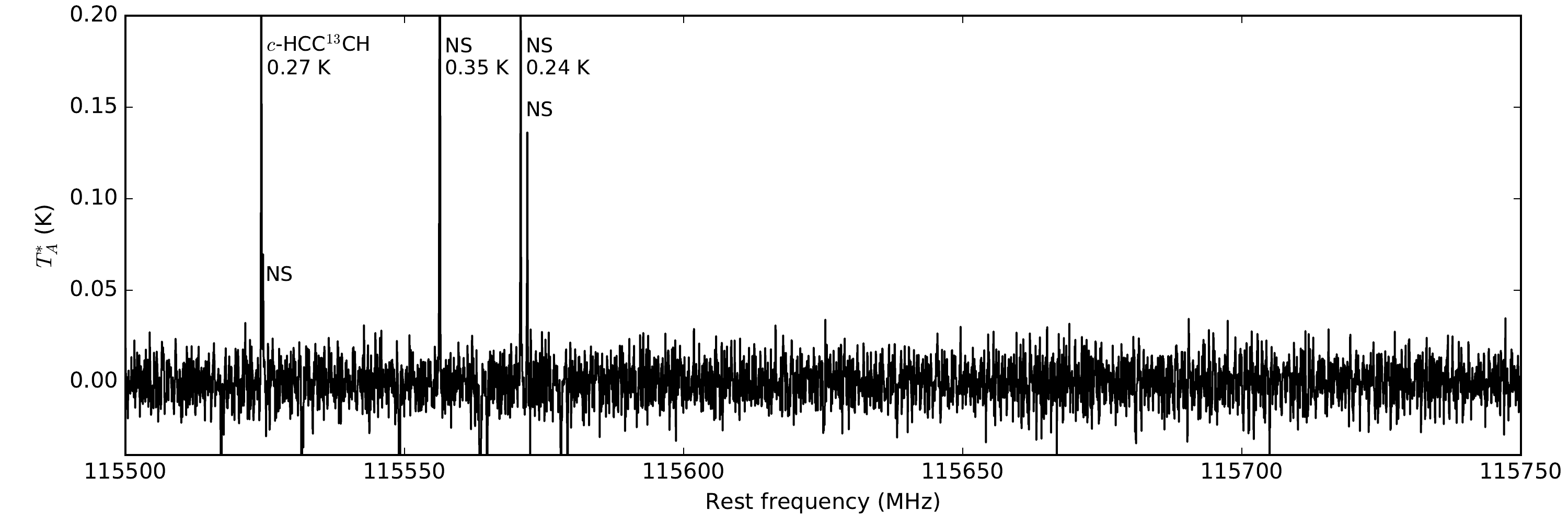}
\includegraphics[width=\textwidth]{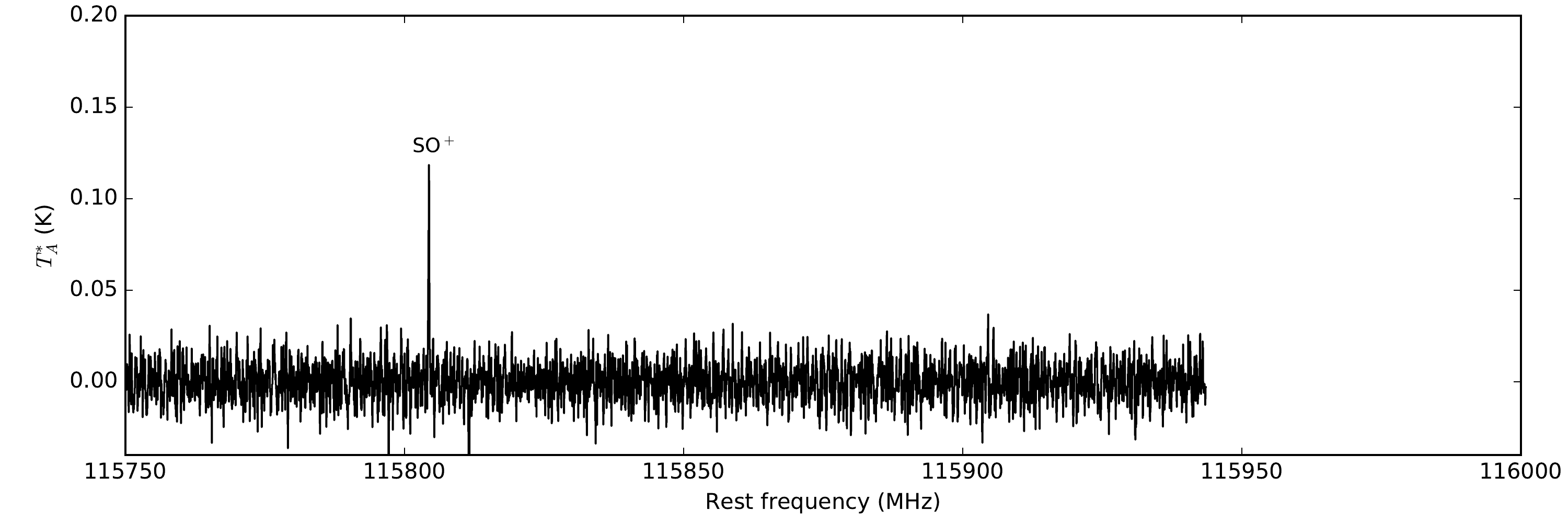}
\caption{Continued}
\end{figure*}


\begin{thebibliography}{}

\bibitem[Adande et al.(2010)]{Adande2010} Adande, G. R., Halfen, D. T., Ziurys, L. M., et al. 2010, \apj, 725, 561
\bibitem[Adande et al.(2012)]{Adande2012} Adande, G. R. \& Ziurys, L. M. 2012, \apj, 744, 194
\bibitem[Ag\'undez et al.(2008)]{Agundez2008} Ag\'undez, M., Cernicharo, J., Gu\'elin, M., et al. 2008, \aap, 478, L19
\bibitem[Ag\'undez \& Wakelam(2013)]{Agundez2013} Ag\'undez, M. \& Wakelam, V. 2013, \chemrev, 113, 8710
\bibitem[Ag\'undez et al.(2015a)]{Agundez2015a} Ag\'undez, M., Cernicharo, J., \& Gu\'elin, M. 2015a, \aap, 577, L5
\bibitem[Ag\'undez et al.(2015b)]{Agundez2015b} Ag\'undez, M., Cernicharo, J., de Vicente, et al. 2015b, \aap, 579, L10
\bibitem[Ag\'undez et al.(2018a)]{Agundez2018a} Ag\'undez, M., Marcelino, N., Cernicharo, J., \& Tafalla, M. 2018a, \aap, 611, L1
\bibitem[Ag\'undez et al.(2018b)]{Agundez2018b} Ag\'undez, M., Marcelino, N., \& Cernicharo, J. 2018b, \apj, 861, L22
\bibitem[Aikawa et al.(2008)]{Aikawa2008} Aikawa, Y., Wakelam, V., Garrod, R. T., \& Herbst, E. 2008, \apj, 674, 993
\bibitem[Aikawa et al.(2012)]{Aikawa2012} Aikawa, Y., Wakelam, V., Hersant, F., et al. 2012, \apj, 760, 40
\bibitem[Aikawa(2013)]{Aikawa2013} Aikawa, Y. 2013, \chemrev, 113, 8961
\bibitem[Albertsson et al.(2013)]{Albertsson2013} Albertsson, T., Semenov, D. A., Vasyunin, A. I., et al. 2013, \apjs, 207, 27
\bibitem[Andr\'e et al.(2000)]{Andre2000} Andr\'e, P., Ward-Thompson, D., \& Barsony, M. 2000, in Protostars and Planets IV, ed. V. Manning, A. P. Boss, \& S. S. Russell (Tucson, AZ: Univ. Arizona Press), 59
\bibitem[Anglada et al.(1997)]{Anglada1997} Anglada, G., Sep\'ulveda, I., \& G\'omez, J. F. 1997, \aaps, 121, 255
\bibitem[Anti\~nolo et al.(2016)]{Antinolo2016} Anti\~nolo, M., Ag\'undez, M., Jim\'enez, E., et al. 2016, \apj, 823, 25
\bibitem[Araki et al.(2016)]{Araki2016} Araki, M., Takano, S., Sakai, N., et al. 2016, \apj, 833, 291
\bibitem[Araki et al.(2017)]{Araki2017} Araki, M., Takano, S., Sakai, N., et al. 2017, \apj, 847, 51
\bibitem[Bachiller \& P\'erez Guti\'errez(1997)]{Bachiller1997} Bachiller, R. \& P\'erez Guti\'errez, M. 1997, \apj, 487, L93
\bibitem[Bacmann et al.(2003)]{Bacmann2003} Bacmann, A., Lefloch, B., Ceccarelli, C., et al. 2003, \apj, 585, L55
\bibitem[Bacmann et al.(2012)]{Bacmann2012} Bacmann, A., Taquet, V., Faure, A., et al. 2012, \aap, 541, L12
\bibitem[Bacmann \& Faure(2016)]{Bacmann2016} Bacmann, A. \& Faure, A. 2016, \aap, 587, A130
\bibitem[Balucani et al.(2015)]{Balucani2015} Balucani, N., Ceccarelli, C., \& Taquet, V. 2015, \mnras, 449, L16
\bibitem[Belloche et al.(2016)]{Belloche2016} Belloche, A., M\"uller, H. S. P., Garrod, R. T., \& Menten, K. M. 2016, \aap, 587, A91
\bibitem[Bianchi et al.(2019)]{Bianchi2019} Bianchi, E., Codella, C., Ceccarelli, C., et al. 2019, \mnras, 483, 1850
\bibitem[Bizzocchi et al.(2013)]{Bizzocchi2013} Bizzocchi, L., Caselli, P., Leonardo, E., \& Dore, L. 2013, \aap, 555, A109
\bibitem[Bizzocchi et al.(2014)]{Bizzocchi2014} Bizzocchi, L., Caselli, P., Spezzano, S., \& Leonardo, E. 2014, \aap, 569, A27
\bibitem[Bizzocchi et al.(2017)]{Bizzocchi2017} Bizzocchi, L., Lattanzi, V., Laas, J., et al. 2017, \aap, 602, A34
\bibitem[Bogey et al.(1988)]{Bogey1988} Bogey, M., Demuynck, C., Destombes, J. L., \& Krupnov, A. 1988, \jmst, 190, 465
\bibitem[Bottinelli et al.(2004)]{Bottinelli2004} Bottinelli, S., Ceccarelli, C., Lefloch, B., et al. 2004, \apj, 615, 354
\bibitem[Bottinelli et al.(2007)]{Bottinelli2007} Bottinelli, S., Ceccarelli, C., Williams, J. P., \& Lefloch, B. 2007, \aap, 463, 601
\bibitem[Broten et al.(1984)]{Broten1984} Broten, N. W., MacLeod, J. M., Avery, L. W., et al. 1984, \apj, 276, L25
\bibitem[Brown et al.(1975)]{Brown1975} Brown, J. M., Hougen, J. T., Huber, K.-P., et al. 1975, \jms, 55, 500
\bibitem[Burkhardt et al.(2018)]{Burkhardt2018} Burkhardt, A. M., Herbst, E., Kalenskii, S. V., et al. 2018, \mnras, 474, 5068
\bibitem[Calcutt et al.(2018a)]{Calcutt2018a} Calcutt, H., Fietcher, M. R., Willis, E. R., et al. 2018a, \aap, 617, A95
\bibitem[Calcutt et al.(2018b)]{Calcutt2018b} Calcutt, H., J{\o}rgensen, J. K., M\"uller, H. S. P., et al. 2018b, \aap, 616, A90
\bibitem[Caselli et al.(1999)]{Caselli1999} Caselli, P., Walmsley, C. M., Tafalla, M., et al. 1999, \apj, 523, L165
\bibitem[Caselli et al.(2002)]{Caselli2002} Caselli, P., Walmsley, C. M., Zucconi, A., et al. 2002, \apj, 565, 344
\bibitem[Caselli et al.(2012)]{Caselli2012} Caselli, P., Keto, E., Bergin, E. A., et al. 2012, \apj, 759, L37
\bibitem[Cazaux et al.(2003)]{Cazaux2003} Cazaux, S., Tielens, A. G. G. M., Ceccarelli, C., et al. 2003, \apj, 593, L51
\bibitem[Ceccarelli et al.(2017)]{Ceccarelli2017} Ceccarelli, C., Caselli, P., Fontani, F., et al. 2017, \apj, 850, 176
\bibitem[Cernicharo(1985)]{Cernicharo1985} Cernicharo, J. 1985, IRAM Internal Report 52
\bibitem[Cernicharo \& Gu\'elin(1987)]{Cernicharo1987} Cernicharo, J. \& Gu\'elin, M. 1987, \aap, 176, 299
\bibitem[Cernicharo et al.(1988)]{Cernicharo1988} Cernicharo, J., Kahane, C., Gu\'elin, M., \& G\'omez-Gonz\'alez, J. 1988, \aap, 189, L1
\bibitem[Cernicharo et al.(2011)]{Cernicharo2011} Cernicharo, J., Spielfiedel, A., Balan{\c c}a, C., et al. 2011, \aap, 531, A103
\bibitem[Cernicharo et al.(2012)]{Cernicharo2012} Cernicharo, J., Marcelino, N., Roueff, E., et al. 2012, \apj, 759, L43
\bibitem[Cernicharo(2012)]{Cernicharo2012:madex} Cernicharo, J. 2012, in EAS Publications Series, Vol. 58, ed. C. Stehl, C. Joblin, \& L. d'Hendecourt (Cambridge: Cambridge Univ. Press), 251
\bibitem[Cernicharo et al.(2018)]{Cernicharo2018} Cernicharo, J., Lefloch, B., Ag\'undez, M., et al. 2018, \apj, 853, L22
\bibitem[Chantzos et al.(2018)]{Chantzos2018} Chantzos, J., Spezzano, S., Caselli, P., et al. 2018, \apj, 863, 126
\bibitem[Chin et al.(1996)]{Chin1996} Chin, Y.-N., Henkel, C., Whiteoak, J. B., et al. 1996, \aap, 305, 960
\bibitem[Cordiner et al.(2013)]{Cordiner2013} Cordiner, M. A., Buckle, J. V., Wirstr\"om, E. S., et al. 2013, \apj, 770, 48
\bibitem[Coutens et al.(2016)]{Coutens2016} Coutens, A., J{\o}rgensen, J. K., van der Wiel, M. H. D., et al. 2016, \aap, 590, L6
\bibitem[Cuadrado et al.(2016)]{Cuadrado2016} Cuadrado, S., Goicoechea, J. R., Roncero, O., et al. 2016, \aap, 596, L1
\bibitem[Dahmen et al.(1995)]{Dahmen1995} Dahmen, G., Wilson, T. L., \& Matteucci, F. 1995, \aap, 295, 194
\bibitem[Dame \& Thaddeus(1985)]{Dame1985} Dame, T. M. \& Thaddeus, P. 1985, \apj, 297, 751
\bibitem[Daniel et al.(2013)]{Daniel2013} Daniel, F., G\'erin, M., Roueff, E., et al. 2013, \aap, 560, A3
\bibitem[Dore et al.(2009)]{Dore2009} Dore, L., Bizzocchi, L., Degli Esposti, C., \& Tinti, F. 2009, \aap, 496, 275
\bibitem[Dore et al.(2017)]{Dore2017} Dore, L., Bizzocchi, L., Wirstr\"om, E. S., et al. 2017, \aap, 604, A26
\bibitem[Drozdovskaya et al.(2018)]{Drozdovskaya2018} Drozdovskaya, M. N., van Dishoeck, E. F., J{\o}rgensen, J. K., et al. 2018, \mnras, 476, 4949
\bibitem[Endo et al.(1986)]{Endo1986} Endo, Y., Saito, S., \& Hirota, E. 1986, \jcp, 85, 1770
\bibitem[Esplugues et al.(2013)]{Esplugues2013} Esplugues, G. B., Tercero, B., Cernicharo, J., et al. 2013, \aap, 556, A143
\bibitem[Foss\'e et al.(2001)]{Fosse2001} Foss\'e, D., Cernicharo, J., Gerin, M., \& Cox, P. 2001, \apj, 552, 168
\bibitem[Frerking et al.(1982)]{Frerking1982} Frerking, M. A., Langer, W. D., \& Wilson, R. W. 1982, \apj, 262, 590
\bibitem[Fuente et al.(2016)]{Fuente2016} Fuente, A., Cernicharo, J., Roueff, E., et al. 2016, \aap, 593, A94
\bibitem[Fuller \& Myers(1993)]{Fuller1993} Fuller, G. A. \& Myers, P. C. 1993, \apj, 418, 273
\bibitem[Fuller et al.(1995)]{Fuller1995} Fuller, G. A., Lada, E. A., \& Masson, C. R., \& Myers, P. C. 1995, \apj, 453, 754
\bibitem[Furuya \& Aikawa(2018)]{Furuya2018} Furuya, K. \& Aikawa, Y. 2018, \apj, 857, 105
\bibitem[Gerin et al.(1987)]{Gerin1987} Gerin, M., Wootten, H. A., Combes, F., et al. 1987, \aap, 173, L1
\bibitem[Gerin et al.(1992a)]{Gerin1992a} Gerin, M., Combes, F., Wlodarczak, G., et al. 1992a, \aap, 253, L29
\bibitem[Gerin et al.(1992b)]{Gerin1992b} Gerin, M., Combes, F., Wlodarczak, G., et al. 1992b, \aap, 259, L35
\bibitem[Graninger et al.(2016a)]{Graninger2016a} Graninger, D. M., Wilkins, O. H., \& \"Oberg, K. I. 2016a, \apj, 819, 140
\bibitem[Graninger et al.(2016b)]{Graninger2016b} Graninger, D. M., Wilkins, O. H., \& \"Oberg, K. I. 2016b, \apj, 833, 125
\bibitem[Gratier et al.(2013)]{Gratier2013} Gratier, P., Pety, J., Guzm\'an, V., et al. 2013, \aap, 557, A101
\bibitem[Gratier et al.(2016)]{Gratier2016} Gratier, P., Majumdar, L., Ohishi, M., et al. 2016, \apjs, 225, 25
\bibitem[Guarnieri(2005)]{Guarnieri2005} Guarnieri, A. 2005, \znata, 60, 619
\bibitem[Gupta et al.(2009)]{Gupta2009} Gupta, H., Gottlieb, C. A., McCarthy, M. C., \& Thaddeus, P. 2009, \apj, 691, 1494
\bibitem[Halfen et al.(2001)]{Halfen2001} Halfen, D. T., Apponi, A. J., \& Ziurys, L. M. 2001, \apj, 561, 244
\bibitem[Halfen et al.(2006)]{Halfen2006} Halfen, D. T., Apponi, A. J., Woolf, N., et al. 2006, \apj, 639, 237
\bibitem[Hassel et al.(2008)]{Hassel2008} Hassel, G. E., Herbst, E., \& Garrod, R. T. 2008, \apj, 681, 1385
\bibitem[Hassel et al.(2011)]{Hassel2011} Hassel, G. E., Harada, N., \& Herbst, E. 2011, \apj, 743, 182
\bibitem[Hatchell et al.(1999)]{Hatchell1999} Hatchell, J., Fuller, G. A., \& Ladd, E. F. 1999, \aap, 344, 687
\bibitem[Hatchell(2003)]{Hatchell2003} Hatchell, J. 2003, \aap, 403, L25
\bibitem[Herbst \& van Dishoeck(2009)]{Herbst2009} Herbst, E. \& van Dishoeck, E. F. 2009, \araa, 47, 427
\bibitem[Higuchi et al.(2018)]{Higuchi2018} Higuchi, A. E., Sakai, N., Watanabe, Y., et al. 2018, \apj, 236, 52
\bibitem[Hily-Blant et al.(2018)]{Hily-Blant2018} Hily-Blant, P., Faure, A., Vastel, C., et al. 2018, \mnras, 480, 1174
\bibitem[Hincelin et al.(2016)]{Hincelin2016} Hincelin, U., Commer{\c c}on, B., Wakelam, V., et al. 2016, \apj, 822, 12
\bibitem[Hirano et al.(1999)]{Hirano1999} Hirano, N., Kamazaki, T., Mikami, H., et al. 1999, in Star Formation 1999, ed. T. Nakamoto, 181
\bibitem[Hirota et al.(1998)]{Hirota1998} Hirota, T., Yamamoto, S., Mikami, H., \& Ohishi, M. 1998, \apj, 503, 717
\bibitem[Hirota et al.(2009)]{Hirota2009} Hirota, T., Ohishi, M., \& Yamamoto, S. 2009, \apj, 699, 585
\bibitem[Imai et al.(2016)]{Imai2016} Imai, M., Sakai, N., Oya, Y., et al. 2016, \apj, 830, L37
\bibitem[Ioppolo et al.(2011)]{Ioppolo2011} Ioppolo, S., Cuppen, H. M., van Dishoeck, E. F., \& Linnartz, H. 2011, \mnras, 410, 1089
\bibitem[Irvine et al.(1988)]{Irvine1988} Irvine, W. M., Friberg, P., Hjalmarson, \AA., et al. 1988, \apj, 334, L107
\bibitem[Jacobsen et al.(2018)]{Jacobsen2018} Jacobsen, S. K., J{\o}rgensen, J. K., Di Francesco, J., et al. 2018, arXiv:1809.00390
\bibitem[Jim\'enez-Serra et al.(2005)]{Jimenez-Serra2005} Jim\'enez-Serra, I., Mart\'in-Pintado, J., Rodr\'iguez-Franco, A., \& Mart\'in, S. 2005, \apj, 627, L121
\bibitem[Jim\'enez-Serra et al.(2016)]{Jimenez-Serra2016} Jim\'enez-Serra, I., Vasyunin, A., Caselli, P., et al. 2016, \apj, 830, L6
\bibitem[J{\o}rgensen et al.(2002)]{Jorgensen2002} J{\o}rgensen, J. K., Sch\"oier, F. L., \& van Dishoeck, E. F. 2002, \aap, 389, 908
\bibitem[J{\o}rgensen(2004)]{Jorgensen2004} J{\o}rgensen, J. K. 2004, \aap, 424, 589
\bibitem[J{\o}rgensen et al.(2018)]{Jorgensen2018} J{\o}rgensen, J. K., M\"uller, H. S. P., Calcutt, H., et al. 2018, \aap, 620, A170
\bibitem[Kalv{\= a}ns(2015)]{Kalvans2015} Kalv{\= a}ns, J. 2015, \apj, 803, 52
\bibitem[Langer et al.(1984)]{Langer1984} Langer, W. D., Graedel, T. E., Frerking, M. A., \& Armentrout, P. B. 1984, \apj, 277, 581
\bibitem[Law et al.(2018)]{Law2018} Law, C. J., \"Oberg, K. I., Bergner, J. B., \& Graninger, D. 2018, \apj, 863, 88
\bibitem[Lee \& Myers(1999)]{Lee1999} Lee, C. W. \& Myers, P. C. 1999, \apjs, 123, 233
\bibitem[Lee et al.(2004)]{Lee2004} Lee, C. W., Myers, P. C., \& Plume, R. 2004, \apjs, 153, 523
\bibitem[Lefloch et al.(2018)]{Lefloch2018} Lefloch, B., Bachiller, R., Ceccarelli, C., et al. 2018, \mnras, 477, 4792
\bibitem[Leung et al.(2016)]{Leung2016} Leung, G. Y. C., Lim, J., \& Takakuwa, S. 2016, \apj, 833, 55
\bibitem[Ligterink et al.(2018)]{Ligterink2018} Ligterink, N. F. W., Calcutt, H., Coutens, A., et al. 2018, \aap, 619, A28
\bibitem[Lindberg et al.(2016)]{Lindberg2016} Lindberg, J. E., Charnley, S. B., \& Cordiner, M. A. 2016, \apj, 833, L14
\bibitem[Lique et al.(2006)]{Lique2006} Lique, F., Cernicharo, J., \& Cox, P. 2006, \apj, 653, 1342
\bibitem[Liszt \& Ziurys(2012)]{Liszt2012} Liszt, H. S. \& Ziurys, L. M. 2012, \apj, 747, 55
\bibitem[Loison et al.(2016)]{Loison2016} Loison, J.-C., Ag\'undez, M., Marcelino, N., et al. 2016, \mnras, 456, 4101
\bibitem[Loison et al.(2017)]{Loison2017} Loison, J.-C., Ag\'undez, M., Wakelam, V., et al. 2017, \mnras, 470, 4075
\bibitem[Loison et al.(2019a)]{Loison2019a} Loison, J.-C., Wakelam, V., Gratier, P., \& Hickson, K. H. 2019a, \mnras, 484, 2747
\bibitem[Loison et al.(2019b)]{Loison2019b} Loison, J.-C., Wakelam, V., Gratier, P., et al. 2019b, \mnras, 485, 5777
\bibitem[L\'opez et al.(2014)]{Lopez2014} L\'opez, A., Tercero, B., Kisiel, Z., et al. 2014, \aap, 572, A44
\bibitem[Lovas et al.(1992)]{Lovas1992} Lovas, F. J., Suenram, R. D., Ogata, T., \& Yamamoto, S. 1992, \apj, 399, 325
\bibitem[Lucas \& Liszt(1998)]{Lucas1998} Lucas, R. \& Liszt, H. 1998, \aap, 337, 246
\bibitem[Magalh\~aes et al.(2018)]{Magalhaes2018} Magalh\~aes, V. S., Hily-Blant, P., Faure, A., et al. 2018, \aap, 615, A52
\bibitem[Majumdar et al.(2017)]{Majumdar2017} Majumdar, L., Gratier, P., Andron, I., et al. 2017, \mnras, 467, 3525
\bibitem[Marcelino et al.(2005)]{Marcelino2005} Marcelino, N., Cernicharo, J., Roueff, E., et al. 2005, \apj, 620, 308
\bibitem[Marcelino et al.(2009)]{Marcelino2009} Marcelino, N., Cernicharo, J., Tercero, B., \& Roueff, E. 2009, \apj, 690, L27
\bibitem[Marcelino et al.(2010)]{Marcelino2010} Marcelino, N., Br\"unken, S., Cernicharo, J., et al. 2010, \aap, 516, A105
\bibitem[Marcelino et al.(2018a)]{Marcelino2018a} Marcelino, N., Ag\'undez, M., Cernicharo, J., et al. 2018a, \aap, 612, L10
\bibitem[Marcelino et al.(2018b)]{Marcelino2018b} Marcelino, N., Gerin, M., Cernicharo, J., et al. 2018b, \aap, 620, A80
\bibitem[Mardones et al.(1997)]{Mardones1997} Mardones, D., Myers, P. C., Tafalla, M., et al. 1997, \apj, 489, 719
\bibitem[Maret et al.(2005)]{Maret2005} Maret, S., Ceccarelli, C., Tielens, A. G. G. M., et al. 2005, \aap, 442, 527
\bibitem[Mart\'in-Pintado et al.(1992)]{Martin-Pintado1992} Mart\'in-Pintado, J., Bachiller, R., \& Fuente, A. 1992, \aap, 254, 315
\bibitem[Markwick et al.(2005)]{Markwick2005} Markwick, A. J., Charnley, S. B., Butner, H. M., \& Millar, T. J. 2005, \apj, 627, L117
\bibitem[Matthews et al.(1984)]{Matthews1984} Matthews, H. E., Irvine, W. M., Friberg, P., et al. 1984, \nature, 310, 125
\bibitem[Matthews et al.(1987)]{Matthews1987} Matthews, H. E., MacLeod, J. M., Broten, N. W., et al. 1987, \apj, 315, 646
\bibitem[Milam et al.(2005)]{Milam2005} Milam, S. N., Savage, C., Brewster, M. A., \& Ziurys, L. M. 2005, \apj, 634, 1126
\bibitem[Minissale et al.(2016)]{Minissale2016} Minissale, M., Moudens, A., Baouche, S., et al. 2016, \mnras, 458, 2953
\bibitem[Minowa et al.(1997)]{Minowa1997} Minowa, H., Satake, M., Hirota, T., et al. 1997, \apj, 491, L63
\bibitem[Mladenovi\'c \& Roueff(2014)]{Mladenovic2014} Mladenovi\'c, M. \& Roueff, E. 2014, \aap, 566, A144
\bibitem[Mladenovi\'c \& Roueff(2017)]{Mladenovic2017} Mladenovi\'c, M. \& Roueff, E. 2017, \aap, 605, A22
\bibitem[Mottram et al.(2014)]{Mottram2014} Mottram, J. C., Kristensen, L. E., van Dishoeck, E. F., et al. 2014, \aap, 572, A21
\bibitem[M\"uller et al.(2005)]{Muller2005} M\"uller, H. S. P., Schl\"oder, F., Stutzki, J., \& Winnewisser, G. 2005, \jmst, 742, 215
\bibitem[Myers et al.(1995)]{Myers1995} Myers, P. C., Bachiller, R., Caselli, P., et al. 1995, \apj, 449, L65
\bibitem[Nemes \& Winnewisser(1976)]{Nemes1976} Nemes, L. \& Winnewisser, M. 1976, \znata, 31, 272
\bibitem[\"Oberg et al.(2010)]{Oberg2010} \"Oberg, K. I., Bottinelli, S., J{\o}rgensen, J. K., \& van Dishoeck, E. F. 2010, \apj, 716, 825
\bibitem[Oca\~na et al.(2017)]{Ocana2017} Oca\~na, A. J., Jim\'enez, E., Ballesteros, B., et al. 2017, \apj, 850, 28
\bibitem[Ohishi et al.(1991)]{Ohishi1991} Ohishi, M., Suzuki, H., Ishikawa, S.-I., et al. 1991, \apj, 380, L39
\bibitem[Ohishi \& Kaifu(1998)]{Ohishi1998} Ohishi, M. \& Kaifu, N. 1998, \fdis, 109, 205
\bibitem[Oya et al.(2017)]{Oya2017} Oya, Y., Sakai, N., Watanabe, Y., et al. 2017, \apj, 837, 174
\bibitem[Oya et al.(2018)]{Oya2018} Oya, Y., Sakai, N., Watanabe, Y., et al. 2018, \apj, 863, 72
\bibitem[Palumbo et al.(2008)]{Palumbo2008} Palumbo, M. E., Leto, P., Siringo, C., \& Trigilio, C. 2008, \apj, 685, 1033
\bibitem[Pardo et al.(2001)]{Pardo2001} Pardo, J. R., Cernicharo, J., \& Serabyn, E. 2001, \ieee, 49, 1683
\bibitem[Parise et al.(2004)]{Parise2004} Parise, B., Castets, A., Herbst, E., et al. 2004, \aap, 416, 159
\bibitem[Parise et al.(2006)]{Parise2006} Parise, B., Ceccarelli, C., Tielens, A. G. G. M., et al. 2006, \aap, 453, 949
\bibitem[Park et al.(2000)]{Park2000} Park, Y.-S., Panis, J.-F., Ohashi, N., et al. 2000, \apj, 542, 344
\bibitem[Pickett et al.(1998)]{Pickett1998} Pickett, H. M., Poynter, R. L., Cohen, E. A., et al. 1998, \jqsrt, 60, 883
\bibitem[Pratap et al.(1997)]{Pratap1997} Pratap, P., Dickens, J. E., Snell, R. L., et al. 1997, \apj, 486, 862
\bibitem[Qu\'enard et al.(2017)]{Quenard2017} Qu\'enard, D., Vastel, C., Ceccarelli, C., et al. 2017, \mnras, 470, 3194
\bibitem[Redaelli et al.(2018)]{Redaelli2018} Redaelli, E., Bizzocchi, L., Caselli, P., et al. 2018, \aap, 617, A7
\bibitem[Remijan et al.(2005)]{Remijan2005} Remijan, A. J., Hollis, J. M., Lovas, F. J., et al. 2005, \apj, 632, 333
\bibitem[Ritchey et al.(2011)]{Ritchey2011} Ritchey, A. M., Federman, S. R., \& Lambert, D. L. 2011, \apj, 728, 36
\bibitem[Ritchey et al.(2015)]{Ritchey2015} Ritchey, A. M., Federman, S. R., \& Lambert, D. L. 2015, \apj, 804, L3
\bibitem[Roueff \& Gerin(2003)]{Roueff2003} Roueff, E. \& Gerin, M. 2003, \ssr, 106, 61
\bibitem[Roueff et al.(2005)]{Roueff2005} Roueff, E., Lis, D. C., van der Tak, F. F. S., et al. 2005, \aap, 438, 585
\bibitem[Roueff et al.(2015)]{Roueff2015} Roueff, E., Loison, J.-C., \& Hickson, K. M. 2015, \aap, 576, A99
\bibitem[Ruaud et al.(2015)]{Ruaud2015} Ruaud, M., Loison, J. C., Hickson, K. M., et al. 2015, \mnras, 447, 4004
\bibitem[Saito et al.(2000)]{Saito2000} Saito, S., Ozeki, H., Ohishi, M., \& Yamamoto, S. 2000, \apj, 535, 227
\bibitem[Sakai et al.(2007)]{Sakai2007} Sakai, N., Ikeda, M., Morita, M., et al. 2007, \apj, 663, 1174
\bibitem[Sakai et al.(2008a)]{Sakai2008a} Sakai, N., Sakai, T., Hirota, T., \& Yamamoto, S. 2008a, \apj, 672, 371
\bibitem[Sakai et al.(2008b)]{Sakai2008b} Sakai, N., Sakai, T., Aikawa, Y., \& Yamamoto, S. 2008b, \apj, 675, L89
\bibitem[Sakai et al.(2009a)]{Sakai2009a} Sakai, N., Sakai, T., Hirota, T., et al. 2009a, \apj, 697, 769
\bibitem[Sakai et al.(2009b)]{Sakai2009b} Sakai, N., Sakai, T., Hirota, T., \& Yamamoto 2009b, \apj, 702, 1025
\bibitem[Sakai et al.(2010)]{Sakai2010} Sakai, N., Saruwatari, O., Sakai, T., et al. 2010, \aap, 512, A31
\bibitem[Sakai \& Yamamoto(2013)]{Sakai2013} Sakai, N. \& Yamamoto, S. 2013, \chemrev, 113, 8981
\bibitem[Sakai et al.(2018)]{Sakai2018} Sakai, N., Yanagida, T., Furuya, K., et al. 2018, \apj, 857, 35
\bibitem[Sch\"oier et al.(2002)]{Schoier2002} Sch\"oier, F. L., J{\o}rgensen, J. K., van Dishoeck, E. F., \& Blake, G. A. 2002, \aap, 390, 1001
\bibitem[Shah \& Wootten(2001)]{Shah2001} Shah, R. Y. \& Wootten, A. 2001, \apj, 554, 933
\bibitem[Sheffer et al.(2007)]{Sheffer2007} Sheffer, Y., Rogers, M., Federman, S. R., et al. 2007, \apj, 667, 1002
\bibitem[Shingledecker et al.(2018)]{Shingledecker2018} Shingledecker, C. N., Tennis, J., Le Gal, R., \& Herbst, E. 2018, \apj, 861, 20
\bibitem[Snyder et al.(1993)]{Snyder1993} Snyder, L. E., Kuan, Y.-J., Ziurys, L. M., \& Hollis, J. M. 1993, \apj, 403, L17
\bibitem[Solomon \& Woolf(1973)]{Solomon1973} Solomon, P. M. \& Woolf, N. J. 1973, \apj, 180, L89
\bibitem[Spezzano et al.(2013)]{Spezzano2013} Spezzano, S., Br\"unken, S., Schilke, P., et al. 2013, \apj, 769, L19
\bibitem[Spezzano et al.(2016)]{Spezzano2016} Spezzano, S., Gupta, H., Br\"unken, S., et al. 2016, \aap, 586, A110
\bibitem[Stahl et al.(2008)]{Stahl2008} Stahl, O., Casassus, S., \& Wilson, T. 2008, \aap, 477, 865
\bibitem[Suzuki et al.(1992)]{Suzuki1992} Suzuki, H., Yamamoto, S., Ohishi, M., et al. 1992, \apj, 392, 551
\bibitem[Tafalla et al.(2000)]{Tafalla2000} Tafalla, M., Myers, P. C., Mardones, D., \& Bachiller, R. 2000, \aap, 359, 967
\bibitem[Takano et al.(1998)]{Takano1998} Takano, S., Masuda, A., Hirahara, Y., et al. 1998, \aap, 329, 1156
\bibitem[Taniguchi et al.(2016)]{Taniguchi2016} Taniguchi, K., Ozeki, H., Saito, M., et al. 2016, \apj, 817, 147
\bibitem[Taniguchi et al.(2017)]{Taniguchi2017} Taniguchi, K., Ozeki, H., \& Saito, M. 2017, \apj, 846, 46
\bibitem[Taniguchi \& Saito(2017)]{Taniguchi-Saito2017} Taniguchi, K. \& Saito, M. 2017, \pasj, 69, L7
\bibitem[Taquet et al.(2017)]{Taquet2017} Taquet, V., Wirstr\"om, E. S., Charnley, S. B., et al. 2017, \aap, 607, A20
\bibitem[Tercero et al.(2010)]{Tercero2010} Tercero, B., Cernicharo, J., Pardo, J. R., \& Goicoechea, J. R. 2010, \aap, 517, A96
\bibitem[Tercero et al.(2015)]{Tercero2015} Tercero, B., Cernicharo, J., L\'opez, A., et al. 2015, \aap, 582, L1
\bibitem[Turner(1989)]{Turner1989} Turner, B. E. 1989, \apj, 347, L39
\bibitem[Turner(1995)]{Turner1995} Turner, B. E. 1995, \apj, 455, 556
\bibitem[Turner(1996)]{Turner1996} Turner, B. E. 1996, \apj, 468, 694
\bibitem[Turner(2001)]{Turner2001} Turner, B. E. 2001, \apjs, 136, 579
\bibitem[Ulich et al.(1977)]{Ulich1977} Ulich, B. L., Hollis, J. M., \& Snyder, L. E. 1977, \apj, 217, L105
\bibitem[van Dishoeck et al.(1995)]{vanDishoeck1995} van Dishoeck, E. F., Blake, G. A., Jansen, D. J., \& Groesbeck, T. D. 1995, \apj, 447, 760
\bibitem[Vastel et al.(2014)]{Vastel2014} Vastel, C., Ceccarelli, C., Lefloch, B., \& Bachiller, R. 2014, \apj, 795, L2
\bibitem[Vastel et al.(2015)]{Vastel2015} Vastel, C., Yamamoto, S., Lefloch, B., \& Bachiller, R. 2015, \aap, 582, L3
\bibitem[Vastel et al.(2016)]{Vastel2016} Vastel, C., Ceccarelli, C., Lefloch, B., \& Bachiller, R. 2016, \aap, 591, L2
\bibitem[Vastel et al.(2018a)]{Vastel2018a} Vastel, C., Kawaguchi, K., Qu\'enard, D., et al. 2018a, \mnras, 474, L76
\bibitem[Vastel et al.(2018b)]{Vastel2018b} Vastel, C., Qu\'enard, D., Le Gal, R., et al. 2018b, \mnras, 478, 5514
\bibitem[Vasyunin et al.(2017)]{Vasyunin2017} Vasyunin, A. I., Caselli, P., Dulieu, F., \& Jim\'enez-Serra, I. 2017, \apj, 842, 33
\bibitem[Vigren et al.(2010)]{Vigren2010} Vigren, E., Hamberg, M., Zhaunerchyk, V., et al. 2010, \apj, 709, 1429
\bibitem[Velusamy et al.(2014)]{Velusamy2014} Velusamy, T., Langer, W. D., \& Thompson, T. 2014, \apj, 783, 6
\bibitem[Widicus Weaver et al.(2017)]{Widicus-Weaver2017} Widicus Weaver, S. L., Laas, J. C., Zou, L., et al. 2017, \apjs, 232, 3
\bibitem[Wilson(1999)]{Wilson1999} Wilson, T. L. 1999, \rpp, 62, 143
\bibitem[Wirstr\"om et al.(2016)]{Wirstrom2016} Wirstr\"om, E. S., Charnley, S. B., Cordiner, M. A., \& Ceccarelli, C. 2016, \apj, 830, 102
\bibitem[Wirstr\"om \& Charnley(2018)]{Wirstrom2018} Wirstr\"om, E. S. \& Charnley, S. B., 2018, \mnras, 474, 3720
\bibitem[Wouterloot et al.(2008)]{Wouterloot2008} Wouterloot, J. G. A., Henkel, C., Brand, J., \& Davis, G. R. 2008, \aap, 487, 237
\bibitem[Yamamoto \& Saito(1990)]{Yamamoto1990} Yamamoto, S. \& Saito, S. 1990, \apj, 363, L13
\bibitem[Yildiz et al.(2013)]{Yildiz2013} Yildiz, U. A., Acharyya, K., Goldsmith, P. F., et al. 2013, \aap, 558, A58
\bibitem[Yoshida et al.(2015)]{Yoshida2015} Yoshida, K., Sakai, N., Tokudome, T., et al. 2015, \apj, 807, 66
\bibitem[Yoshida et al.(2019)]{Yoshida2019} Yoshida, K., Sakai, N., Nishimura, Y., et al. 2019, \pasj, psy136
\bibitem[Ziurys et al.(1994)]{Ziurys1994} Ziurys, L. M., Apponi, A. J., Hollis, J. M., \& Snyder, L. E. 1994, \apj, 436, L181

\end{thebibliography}
\end{document}